\documentclass[onecolumn]{mn2e}
\usepackage{graphicx}
\usepackage{epsfig}
\usepackage{amsmath}
\usepackage{float}                          
\title{A cloudy Vlasov solution.}  
\author[C. Alard \& S. Colombi]{C. Alard$^{1,2}$ \thanks{E-mail: alard@iap.fr} and S. Colombi$^{2}$\thanks{E-mail:colombi@iap.fr}\\
$^{1}$Institut d'Astrophysique de Paris, 98bis boulevard Arago, 75014 Paris France\\
$^{2}$Observatoire de Paris, 77 avenue Denfert Rochereau, 75014 Paris France}
\begin{document}
\date{}
\maketitle
\begin{abstract} 
We propose to integrate the Vlasov-Poisson equations giving the evolution of a dynamical
system in phase-space using a continuous set of local basis functions. 
In practice, the method decomposes the density in phase-space into small 
smooth units having compact support. We call these
small units ``clouds'' and choose them to be Gaussians of elliptical
support. Fortunately, the evolution of these clouds in the local
potential has an analytical solution, that can be used to evolve the whole system
during a significant fraction of dynamical time. In the process, the clouds, initially
round, change shape and get elongated. At some point,
the system needs to be remapped on round clouds once again. 
This remapping can be performed optimally using a small number of Lucy 
iterations. The remapped solution can be evolved again with the cloud method,
and the process can be iterated a large number of times without showing significant
diffusion. Our numerical experiments show that it is possible 
to follow the 2 dimensional phase space distribution during a large number of
dynamical times with excellent accuracy. The main limitation to this accuracy 
is the finite size of the clouds, which results in coarse graining the
structures smaller than the clouds and induces small aliasing effects at these scales. 
However, it is shown in this paper that this method is consistent with an adaptive 
refinement algorithm which allows one to track the evolution of the finer structure 
in phase space. It is also shown that the generalization of the cloud method 
to the 4 dimensional and the 6 dimensional phase space is quite natural.
\end{abstract}
\section{introduction}
The problem we are interested in is solving numerically 
the Vlasov-Poisson system, which writes, in the proper choice of units
\begin{equation}
\frac{\partial f}{\partial t} + {v}.\nabla_{x}f - \nabla_{x} \phi.\nabla_{v}f=0,
\end{equation}
\begin{equation}
\Delta \phi= 2 \, \int f({x},{v},t) \, dv.
\end{equation}
Due to the high dimensionality
of phase-space, $2D$, where $D$ is the dimension of space,
this problem is usually approached with the traditional $N$-body method,
i.e. by approximating the distribution function by a discrete set of particles. However, with modern
supercomputers, it now becomes possible to start envisaging direct phase-space approach with
$D=2$ and $D=3$. In this paper, we are thus interested in solving 
the Vlasov-Poisson equations directly in phase-space.  We consider a new implementation in 
1 dimension, $D=1$, but we shall discuss its extension to higher number of dimensions. 
We now review phase-space methods already studied in the past. After that, we give a sketch of our 
``clouds'' implementation and explain what is new compared to these works. At the end of this introduction,
we shall detail the plan of our paper, which is mostly devoted to the actual technical details involved
in the implementation of our method.

A fundamental property of Vlasov equation
is the Liouville theorem, which states that the phase-space distribution function is conserved
along trajectories of matter elements in phase space:
\begin{equation}
f[{x}(t),{v}(t),t]={\rm constant}.
\label{eq:liouv}
\end{equation}
The first numerical methods used in astrophysics 
to solve Vlasov-Poisson equations in phase space exploited directly this property, using the so-called water-bag
model (De Packh 1962; Hohl \& Feix 1967). The idea of the water-bag model is the following. 
If one assumes that the distribution
function is constant within a patch in phase-space, it is enough to follow 
dynamically the boundary of the patch.
Numerical implementation of the water-bag model is therefore rather straightforward 
(Roberts \& Berk 1967; Hohl \& Feix 1967; Janin 1971; Cuperman, Harten \& Lecar 1971). Even though this isocontour 
method is quite efficient and accurate, it is in fact very costly. Indeed, the distribution
function develops increasing filamentary details during evolution by effects of rolling up in phase-space due
to differential orbital speeds. Therefore, it is in principle necessary to add more and more points
to sample the boundary of the patches as time passes. This is one of the major weaknesses of 
the water bag method, which is fine grained by essence, except for initial conditions. 

Other approaches for solving the Vlasov-Poisson equation are grid based, and a large part of the
technical developments come from plasma physics. One of the most famous numerical implementation, 
since it inspired many subsequent works, is the splitting algorithm of Cheng \& Knorr (1976).  
The splitting scheme consists in exploiting the Liouville theorem in two steps, while evolving the system
during a time step $\Delta t$:
\begin{eqnarray}
f^*({x},{v}) & = & f({x}-{v}\Delta t/2,{v},t), \nonumber \\
f^{**}({x},{v}) &= & f^*({x},{v}+\nabla_{x}\phi \Delta t), \nonumber \\
f({x},{v},t+\Delta t) & = & f^{**} ({x}-{v}\Delta t/2,{v}). \label{eq:splittingmet}
\end{eqnarray}
In the method of Cheng \& Knorr, the distribution function is interpolated on a grid
using either Fourier method or/and splines. It is semi-Lagrangian in the sense that, to
compute the value of the distribution function at a grid site, test particles trajectories
are resolved backwards up to previous time step, where the interpolation is performed.
If the original implementation of Cheng \& Knorr is one dimensional, 
the generalization to higher number of dimensions is straightforward
(e.g. Gagn\'e \& Shoucri 1977; see Sonnendr\"ucker et al. 1999 for a more recent perspective). 
The method of Cheng \& Knorr was applied first in astrophysics by Fujiwara (1981),
Watanabe et al. (1981) and Nishida et al. (1981). 

In principle the algorithm of Cheng \& Knorr can be used as it is, even when the filamentation
effects discussed above occur at resolution scale, 
although it has to be adapted e.g. by using appropriate interpolation
procedure to warrant positivity of the distribution function and mass conservation (e.g. Besse \&
Sonnendr\"ucker 2003 for state of the art latest developments). However, an elegant solution
was proposed by Klimas (1987) to overcome the problem of filamentation. It
consists in writing the exact equation of evolution of
the coarse grained distribution function in velocity space. For that, he assumes a Gaussian smoothing
window. The modified Vlasov equation giving the evolution of the smoothed 
distribution function includes a new source term. 
This method was applied to a splitting algorithm using Fourier decomposition (Klimas \& Farrell 1994).

Other grid based methods include hydrodynamic advection schemes:  the Lax-Wendroff 
integration method (Shlosman, Hoffman \& Shaviv, 1979) or other finite difference methods, 
using, for the interpolation of the fluxes,
standard ``upwind'' and TVD algorithms, convenient to deal with filamentation, such the Van-Leer limited scheme 
and the piecewise parabolic method (PPM), or other schemes
such as the flux corrected transport, the flux balanced method (see Arber \& Vann 2002 for
a review and a comparison between these 4 last methods), or the positive and flux conservative method
(Filbet, Sonnendr\"ucker \& Bertrand, 2001). A finite element method was proposed as well (Zaki, Gardner \&
Boyd 1988), but no further development in that direction was
performed afterwards, probably because this method involves the inversion of coarse but large matrices,
a very costly operation in 6 dimensional phase-space.

It is now worth mentioning two interesting special cases, 
the Lattice method (e.g. Syer \& Tremaine 1995), and
the solver of Rasio et al. (1989). In the lattice method,  the
motion of elements of phase space density is restricted to a set 
of discrete points: the time, positions and
velocities of ``particles'' are restricted to integer values, and forces are rounded to
nearest integer. Such an algorithm has the advantage of solving simplectic equations of motion,
which warrants conservation of geometrical properties of the system and
in the limit of infinite resolution, converges to the ``smooth'' solution and
enforces naturally Liouville theorem. The second solver 
is the spherical code of Rasio et al. (1989), which works in the fully generalistic case. The principle
of this code is to take full advantage of the perfect knowledge of initial conditions:
whenever $f({x},{v},t)$ has to be determined accurately at some point of space, e.g. to compute
accurately the potential, a test particle is followed back in time to find its initial 
position and the value of $f$ associated to it. The needed sampling at each time step is
estimated by a self-adaptive quadrature routine. This code is therefore quite costly, 
since at each time step a full set of backward trajectories has to be recomputed. It has however
the advantage of following with very good accuracy all the details of the distribution function and
it is probably the most accurate code of this kind available. It presents theoretically the same advantages as
the water-bag method, but at variance with the later, it is able to preserve smoothness of the distribution
function. 

Finally, alternate ways of solving Vlasov-Poisson equations consist in 
computing the moments of the phase-space density with respect to velocity space and position space, 
and write partial differential equations for these moments up to some order, with
some recipe to close the hierarchy (e.g. White 1986; Channell 1995). 

A good numerical implementation of Vlasov-Poisson equations 
should stick as close as possible to Eq.~(\ref{eq:liouv}). In particular, it should,
as a direct consequence of Liouville theorem, preserve as much as possible the topology of the distribution function.
More specifically, to render the algorithm TVD (total variation diminishing), and therefore stable,
\begin{itemize}
\item the critical point population i.e. the number of maxima, minima and saddle points 
of various kinds should be preserved, and if it is not possible, should not increase with time. 
Since all Eulerian implementations
use splitting in ${x}$ and ${v}$, this condition reduces in 1D-1D to preservation
of monotonicity in that case.
\item the height of critical points should be preserved, 
or at worse, the height of local maxima should decrease and the height of local minima should increase.
This in particular warrants the positivity of the distribution function.
\end{itemize}
These conditions state that if the solution deviates from the true one, it should only become smoother. This
is the essence of modern advection methods, which try to preserve the features of the distribution without
adding spurious small scale features, such as oscillations around regions with very high gradients.
For instance, Van-Leer and PPM methods are TVD, at variance with most semi-Lagrangian methods. 
The higher is the order of the scheme, the more accurate is the resolution of fine features of the distribution. However,
to preserve the TVD nature of the system, the implementations are lower order in sharp transitions regions [condition
(a)] and in general nearby local extrema [condition (b)]. This as a result can introduce significant diffusive effects.
Clearly, we see that to this respect, grid based methods are thus inferior to the water-bag method, which optimally
fulfills conditions (a) and (b), since it follows isocontours of the phase-space distribution function, 
and thus its topology, by essence. However, the number of sampling points increases with time in the water-bag
model and a fair comparison with grid based methods should allow adaptive mesh refinement. Clearly, the very
particular implementation of Rasio et al. (1989) is very successful in that sense.

The method we propose here is very different from all the works discussed above.
The basic idea is to decompose the density in phase space
into small local units which can be represented
using a continuous function with compact support. The total density
in phase space is the sum of the local functions, to
have a fully analytical and continuous representation of the phase
space density of the system. We will call these small 
units ``clouds'', which will be chosen here to be truncated Gaussians.
The evolution of the distribution function will be followed by solving
the Vlasov equation for each of the individual cloud, plunged in the global
gravitational potential. We allow the clouds to change shape 
during runtime, i.e. to transform from functions with initial
round support (with the appropriate choice of units) 
into functions with elliptic support. These clouds also move: their center
of mass position in phase-space follows standard Lagrangian equations of motion, as
test particles would. If the potential is locally quadratic, our elliptical
shape approximation is exact, and the Vlasov equation can be solved analytically in that case.
It means that as long as the clouds are small compared to the curvature
radius of the force, or in other words, to the scales of variation of
the projected density, this approximation, accurate to second
order in space, is very good. 

At some point, the local quadraticity 
condition is not verified anymore, typically after a fraction of orbital time,
and the analytical solution ceases to be a good approximation of the 
density of the cloud in phase space: the whole system has to be remapped
on a new basis of round clouds that give a smooth description
of the density in phase space. To resample the distribution function with
a new set of clouds (including initial conditions), we use a Lucy or
Van Citter iterative method. This, combined with the fact that remaps are 
not very frequent,  eliminates diffusion almost completely, provided
that the resolution limit of the simulation has not been reached, i.e. as long
as filamentation is not a problem. As a time integrator, we use a predictor
second order corrector with slowly varying time step. 
Our algorithm is thus fully second order in time and in space, and nearly simplectic,
since in the case the time step is constant, it reduces to standard Leapfrog.

As a matter of fact and as we shall see, our method deals very well
with filamentation, as it naturally coarse grains the distribution function
at small scales. It is quite close to finite element methods, except that
the elements are of changing shape and that the sampling grid moves
with time. It is also close in some sense to the water-bag method, but in a coarse
grained way, as by construction equation (\ref{eq:liouv}) is verified
exactly as long as the potential is locally quadratic. The remap procedure
is iterative and costly, that is one of the drawbacks of the method, similarly
as in finite difference methods. However, we do not need to perform it too often
(typically every 10-20 time step). The Lagrangian nature of the method and
the reinterpolation scheme makes our method very weakly diffusive, but it
is not TVD: aliasing effects can appear in regions with high curvature. However,
for the numerical cases we studied, these effects are not critical. Positivity
of the distribution is enforced with Lucy reconstruction, but not with Van-Citter.
Note that our method can be used only in warm
cases, i.e. in cases where the distribution function is smooth and has some width
in velocity space. It is not appropriate for the cold case, e.g. to describe
in phase-space formation of large scale structure in standard cosmological models. 

This paper is organized as follows. In \S~2, we present the theoretical background
intrinsic to our method, i.e. study the dynamics of a phase-space cloud in a quadratic
gravitational potential. We perform a full perturbative stability study, taking into
account small deviations from quadraticity. In \S~3, we discuss the actual numerical
implementation. In \S~4, we test our code by performing simulations of
a stationary solution, an initially Gaussian profile and an apodized water-bag (a top hat in
phase-space). In \S~5, we propose an adaptive refinement procedure, which allows one to
increase resolution where needed, if details of the distribution function need to be followed
at smaller scales. Finally, \S~6 summarizes the results and
discusses some perspectives for the method, in particular its extension to higher number of dimensions
and the treatment of the cold case.

\section{The method: concepts}
%
\subsection{One dimensional equations}
%
In this section we will show that for a quadratic potential, the Vlasov
equation for a cloud has an analytical solution. Let 
$\phi(x,t)$ be the gravitational potential of the system. Locally the quadratic
approximation will read:
\begin{equation}
 \phi({x},t)=\alpha_0(t) \ {x}^2+\alpha_1(t) \ {x}+\alpha_2(t).
\label{eq:quadapp}
\end{equation}
By taking the following cloud equation it is possible to show that the
Vlasov equation forms a closed system:
\begin{equation}
f(x,v,t)=G \left[\lambda_0(x,t)+\lambda_1(x,t) 
\ v+\lambda_2(t) \ v^2 \right],
\label{eq:fapp}
\end{equation}
where $G$ is any smooth function (with continuous derivatives).
The functions $\lambda_i(x,t) {\rm , \ i=0,...,2}$, which 
determine the geometry of the cloud in phase space can be obtained by solving
the Vlasov equation:
\begin{equation}
 \frac{\partial f}{\partial t}+v \frac{\partial f}{\partial x} \ 
 -\frac{\partial \phi}{\partial x} \ \frac{\partial f}{\partial { v}}=0.
\label{eq:labellevlasov}
\end{equation}
It is interesting at this point to separate the general 
motion of the cloud from the evolution of its internal geometry. We
will thus use Lagrangian coordinates, which are defined by 
$(x^* \equiv  x-x_{\rm G},v^* \equiv v-v_{\rm G})$, 
where $({x}_{\rm G}, {v}_{\rm G})$ are the phase
space coordinates of the center of gravity of the cloud. Assuming that
the cloud phase space density $f({x},{v},t)$ become $f(x^*,v^*,t)$
in this referential, we can write:
\begin{equation}
 \frac{\partial f}{\partial t}+\frac{\partial f}{\partial x^*} \ v^*-2 \alpha_0 x^* \ \frac{\partial f}{\partial v^*}=0,
  \label{eq:vlaf}
\end{equation}
\begin{equation}
 f(x^*,v^*,t)=G \left[\Lambda_0(x^*,t)+\Lambda_1(x^*,t) \ v^*+\Lambda_2(t) \ {v^*}^{2} \right], \label{eq:defG1}
\end{equation}
\begin{equation}
 \frac{d^2 {x}_{\rm G}}{d t^2}=-2\alpha_0(t) {x}_{\rm G}-\alpha_1(t).  \label{eq:cg_motion}
\end{equation}

By inserting Eq.~(\ref{eq:defG1}) in Eq.~(\ref{eq:vlaf}), we obtain a quadratic polynomial
in $v^*$, which has to cancel for any value of $v^*$. Hence, we obtain 3 equations:
\begin{eqnarray}
   {\frac {\partial }{\partial t}}\Lambda_0(x^*,t)-2\,\Lambda_1(x^*,t) \ \alpha_0(t) \ x^*&=&0, 
\label{eq:lambda0evol} \\
   {\frac {\partial }{\partial t}}\Lambda_{{1}}(x^*,t)+{\frac {\partial }{
\partial x^*}}\Lambda_{{0}}(x^*,t)-4\,\Lambda_{{2}}(t) \ \alpha_{{0
}}(t) \ x^* &=&0, \label{eq:lambda1evol} \\
{\frac {d}{dt}}\Lambda_{{2}}(t)+{\frac {\partial }{\partial x^*}}\Lambda
_{{1}}(x^*,t)&=&0. \label{eq:lambda2aevol}
\end{eqnarray}
The general solution of Eq.~(\ref{eq:lambda2aevol}) is
\begin{equation}
\Lambda_{{1}}(x^*,t)=-\frac {d \Lambda_2}{dt}x+{\psi_1}(t).
\label{eq:sollambda1}
\end{equation}
By inserting the solution for $\Lambda_1$ in Eq.~(\ref{eq:lambda1evol}) 
and solving for $\Lambda_0(x^*,t)$ we obtain:
\begin{equation}
 \Lambda_{{0}}(x^*,t)=\frac{1}{2}\,\left [{\frac {d^{2}\Lambda_2}{d{t}^{2}}}
+4\,\Lambda_{{2}}(t)\alpha_{{0}}(t)\right] {x^*}^2-{\frac {d\psi_1}{dt}} x^* +{\psi_2}(t).
\label{eq:sollambda0}
\end{equation}
And finally by reporting Eqs.~(\ref{eq:sollambda1}) and (\ref{eq:sollambda0}) into 
Eq.~(\ref{eq:lambda0evol}), we obtain a quadratic polynomial in $x^*$, 
which must be zero for any value of $x^*$. 
The zeroth order of this polynomial gives $\psi_2=$constant. 
The first order of this polynomial reads $d\psi_1/dt=-2\alpha_0 \psi_1$.
However, in the referential of the center of gravity, the first moments of $f$ with
respect to $x^*$ and $v^*$ should be zero. This implies that the form $\Lambda_0(x^*,t) +\Lambda_1(x^*,t)v^*
+\Lambda_2(t) {v^*}^2$, that we know now to be a polynomial of order 2 in 
$x^*$ and $v^*$ with time dependent 
coefficients, does not have any term either in  $x^*$ and $v^*$: 
it is a quadratic form in $x^*$ and $v^*$,
implying $\psi_1(t)=0$. The second order of this polynomial writes
\begin{equation}
\frac{1}{2} \,{\frac {d^{3}\Lambda_{{2}}}{d{t}^{3}}}+4\alpha_{{0}}(t)\,{\frac {d\Lambda_{{2}}}{dt
}}+2{\frac {d\alpha_{{0}}
}{dt}}\,\Lambda_{{2}}(t)=0.
\label{eq:evollambda2}
\end{equation}
Eq.~(\ref{eq:evollambda2}) can be solved with respect to $\alpha_0(t)$:
\begin{equation}
 \alpha_{{0}}(t) \ \Lambda_2(t)^{2}=-\frac{1}{4}\int \,
{\frac {d^{3}\Lambda_{{2}}}{d{t}^{3}}}\Lambda_{{2}}(t){dt}
+ C.
\end{equation}
Since
\begin{equation}
\int \,{\frac {d^{3}\Lambda_{{2}}}{d{t}^{3}}}\Lambda_{{2}}(t){dt}=
{\frac {d^{2}\Lambda_{{2}}}{d{t}^{2}}}\Lambda_{{2}}(t)-\frac{1}{2}\,\left ({\frac {d\Lambda_{{2}}}{dt}}\right )^{2},
\end{equation}
we can write
\begin{equation}
 \frac{1}{4} \,{\frac 
{d^{2}\Lambda_{{2}}}{d{t}^{2}}}\Lambda_{{2}}(t)-\frac{1}{8}\,\left 
({\frac {d\Lambda_{{2}}}{dt}}\right )^{2}+\alpha_{{0}}(t) \ \Lambda_2(t)^{2}=C.
\label{eq:lambda2evol}
\end{equation}
It is possible to simplify this equation a little more by using the following
substitution: $\Lambda_2(t)=w(t)^2$ (here we assume that $\Lambda_2(t)$, which
corresponds to the velocity dispersion, is positive). We obtain an equation which
is easier to use for numerical integration:
\begin{equation}
 \frac{1}{2} \,{\frac {d^{2}w}{d{t}^{2}}}+w(t)\alpha_{{0}}(t)=\frac{C}{w(t)^3}.
  \label{eq:evolw}
\end{equation}
We can now finally rewrite Eq.~(\ref{eq:defG1}) as follows
\begin{equation}
 f(x^*,v^*,t)=G\left[\beta_0(t) \ {x^*}^2+\beta_1(t)\ x^*v^*+\beta_2(t) \ {v^*}^2+\beta_3 \right],
\label{eq:defG2}
\end{equation}
with
\begin{eqnarray}
  \beta_0(t) &=& \frac{1}{2}\,{\frac {d^{2}\Lambda_{{2}}}{d{t}^{2}}}+2\,\Lambda_{{2}}(t) 
  \alpha_{{0}}(t), \label{eq:beta0}\\
  \beta_1(t)&=&-{\frac {d\Lambda_{{2}}}{dt}}, \label{eq:beta1} \\
   \beta_2(t)&=&\Lambda_{{2}}(t), \label{eq:beta2}
\end{eqnarray}
and $\beta_3$ can be set to zero without any loss of generality.
We then see that Eq.~(\ref{eq:lambda2evol}) has a simple geometric interpretation, since it can
now be rewritten
\begin{equation}
 \beta_0(t) \ \beta_2(t) -\frac{1}{4} \beta_1(t)^2=2 \ C,
\end{equation}
which says that the area of the elliptical section in phase-space defined by 
Eq.~(\ref{eq:defG2}), proportional to
\begin{equation}
 {\bar A}\equiv \pi A=\frac{\pi}{\sqrt{{{\beta_0(t) \ \beta_2(t) -\frac{1}{4} \beta_1(t)^2}}}}
\end{equation}
remains constant with time. 
%
%
\subsection{Dynamical properties of the clouds}
\label{sec:perturana}
%
%
In this section, we develop a number of simple analytical calculations
which will help to implement the numerical simulations. These analytical
models will also help us to understand the limits of our approach.
\subsubsection{Small amplitude oscillations}
\label{sec:smallamposc}
The evolution of the cloud is given by equation  Eq.~(\ref{eq:evolw}).  In general, 
the solution of this highly non-linear equation can be approached only by numerical means.
However, if we consider small amplitude oscillations around an equilibrium position,
it is possible to linearize the equations and find a simple analytical solution.
We should consider a stationary equilibrium position with a quadratic potential
given by: $ \alpha_0(t)= q_0 $. In this case, the solution to Eq.~(\ref{eq:evolw}) is:
\begin{equation}
 w(t)=w_0= \left (\frac{C}{q_0} \right)^{1/4}.
\label{eq:statsol}
\end{equation}
Note that due to the Poisson equation,
\begin{equation}
 \frac{d^2 \phi}{d x^2} = 2 \rho,
\label{eq:Poisson}
\end{equation}
it is possible relate the stationary potential $q_0$ 
to the mean density $\rho_0$ around the cloud:
\begin{equation}
 q_0 =  \rho_0.
\label{eq:p_rho}
\end{equation}
Now, we introduce a time dependent perturbation of this stationary regime:
\begin{equation}
 \alpha_0(t)=q_0+\epsilon q_1(t)  \ \ \ {\rm and} \ \ \ w(t)=w_0+\epsilon w_1(t).
\end{equation}
Then  Eq.~(\ref{eq:evolw}) reads:
\begin{equation}
 \frac{1}{2} \frac{d^2 w_1}{d t^2} + 4 \ q_0 \ w_1(t) + w_0 q_1(t) =0.
\label{eq:pert0_w}
\end{equation}
A general solution to Eq.~(\ref{eq:pert_w}) can be obtained by using a Fourier transform.
We define the following Fourier transforms:
\begin{equation}
 \tilde w_1(\omega) = \frac{1}{\sqrt {2 \pi}} \int w_1(t) \ e^{i \omega t} \ dt \ \ \ {\rm and} \ \ \
 \tilde q_1(\omega) = \frac{1}{\sqrt {2 \pi}} \int q_1(t) \ e^{i \omega t} \ dt.
\end{equation}
Using these definitions, the general solution of Eq.~(\ref{eq:pert_w}) writes:
\begin{equation}
\tilde w_1(\omega)= 2 w_0 \ \frac{\tilde q_1(\omega)}{\omega^2-{\omega_0}^2}
+k_0\ \delta(\omega-\omega_0)+k_1\ \delta(\omega+\omega_0),
\label{eq:pert_w}
\end{equation}
where $k_0$ and $k_1$ are two arbitrary constants and
\begin{equation}
\omega_0=2 \sqrt{2 q_0}. 
\label{eq:omega_0}
\end{equation}
Eq.~(\ref{eq:pert_w}) shows clearly that a resonance may occur between the potential and the cloud
oscillations at the frequency $\omega_0$. However for this resonance to occur effectively, it would
 be required that all the nearby clouds which contribute to the local density resonate also
at this frequency. But, the resonant frequency of the nearby clouds can be the same only if
the local density around these clouds is identical, which would require a constant density whatever
the position. A resonance happens only in this case, and we will discuss this very special
 case later. The dynamical properties of the cloud are also dictated by the motion of their center
of gravity which is given by Eq.~(\ref{eq:cg_motion}).
 The solution to Eq.~(\ref{eq:cg_motion}) can be decomposed as
the solution of the homogeneous equation and a given solution to the whole equation. The homogeneous
equation reads:
\begin{equation}
\frac{d^2 x_{\rm G}}{d t^2} + 2 \alpha_0(t) x_{\rm G}(t)=0.
\label{eq:h_eq}
\end{equation}
Setting $x_{\rm G}(t)=x_0+\epsilon x_1(t)$ and noting that  Eq.~(\ref{eq:h_eq}) is similar to 
Eq.~(\ref{eq:evolw}), we find the general solution in the linearized regime:

\begin{equation}
\tilde x_1(\omega)= 2 \bar w_0 \ \frac{\tilde q_1(\omega)}{\omega^2-{\bar \omega_0}^2}+ \bar k_0\ \delta(\omega-\bar \omega_0) + \bar k_1\ \delta(\omega+\bar \omega_0),
\label{eq:pert_x}
\end{equation}
with a resonant frequency, $\bar \omega_0 = \omega_0/2$. Thus all the dynamical frequencies of the
clouds are multiples of the fundamental frequency $\sqrt{2 q_0}$, and thus, with the help of 
 Eq.~(\ref{eq:p_rho}), we infer that the dynamical time of the cloud is:
\begin{equation}
T_{\rm cloud} \simeq \frac{\pi \, \sqrt{2}}{\sqrt{\rho_0}}.
\label{eq:dynatime}
\end{equation}
Obviously this dynamical time has little meaning for a cloud experiencing large potential variations
on a short time scale. In such case the non-linearity is dominant and the linear approximation is not
suitable. However, for most clouds in a given system, this approximation will give an estimate 
of the local time-scale.  
%
%
\subsubsection{Resonant self-oscillations}
As noted in the former section, resonances between the cloud motion and oscillations
would require that the frequency $\omega_0$ is constant. Using Eq.~(\ref{eq:p_rho})
and Eq.~(\ref{eq:omega_0}), we see that this condition $\omega_0=$ constant would require
that the density is constant. The simplest example of a system having
constant projected density is a line in phase space, with a density constant along the line.
the oscillation frequency for any point on the line is the same whatever the position angle of the
line in phase space. Thus such system can have self-oscillation, and out of this case there is
no regime of self-oscillation of the cloud system. Note that this one-dimensional system is 
similar to the case of the sphere of constant density that has been shown to have self-pulsation
in 3 dimensions.
%
%
%
%
%
\subsubsection{Deviation from the quadratic approximation}
\label{sec:pergene}
The cloud approximation assumes that the potential at the scale of the cloud
is quadratic. Even if the quadratic approximation is a good description of the local
potential, when evolving the system,
we expect that the errors to the quadratic approximation will accumulate,
and that at some point the cloud model will deviate from the proper solution.
Our numerical scheme should stop to evolve the system before the deviation is large,
and this is the purpose of this section to estimate the magnitude of this deviation
as a function of the dynamical time describing the system evolution.
We will consider an additional cubic term in the potential, and we will
study the effect of this term in the linear perturbative regime. The new
equation for the potential reads:
\begin{equation}
\phi(x^*,t)=\alpha_0(t) \ {x^*}^2+\epsilon \beta(t) \ {x^*}^3.
\label{eq:newpot}
\end{equation}
We introduce the corresponding perturbation for the density in phase space:
\begin{equation}
 f \left(x^*,v^*,t \right) =f_0 \left( x^*,v^*,t \right) +\epsilon\,f_1
 \left( x^*,v^*,t \right). 
\end{equation}
Using the Vlasov equation [Eq.~(\ref{eq:labellevlasov})], we obtain the following
equations:
\begin{equation}
 \frac {\partial f_0}{\partial t} +
 \frac {\partial f_0}{\partial x^*} \ v^* -
 2\, \alpha_0 \ x^* \ \frac {\partial f_0}{\partial v^*} = 0,
\label{eq:evolf0}
\end{equation}
\begin{equation}
 \frac {\partial f_1}{\partial t} +
 \frac {\partial f_1}{\partial x^*} \ v^* -
  2\, \alpha_0 \  x^* \ \frac {\partial f_1}{\partial v^*} -
  3\, \beta \ {x^*}^2 \ \frac {\partial f_0}{\partial v^*} = 0.
\label{eq:evolf1}
\end{equation}
We already know the solution to Eq.~(\ref{eq:evolf0}) from section 2.1.
Since clouds having gaussian profile are particularly interesting for numerical
applications, we make the following choice in Eq.~(\ref{eq:defG1})
\begin{equation}
 f_0(x^*,v^*,t) = \exp \left [\Lambda_0(x^*,t) + \Lambda_1(x^*,t) \ v^* + \Lambda_2(t) \ {v^*}^2 \right].
\end{equation}
Now, it is possible to find a general solution to Eq.~(\ref{eq:evolf1}) by
using the following functional for $f_1$:
\begin{equation}
f_1 \left( x^*,v^*,t \right) = f_0 \left(x^*,v^*,t \right)  \times \left[ \eta_0
 \left( t \right) +\eta_1 \left( t \right) {x^*}^{3} + \eta_2 \left( t
 \right) {x^*}^{2} v^* + \eta_3 \left( t \right) x^* {v^*}^{2}+ \eta_4 \left( t
 \right) {v^*}^{3} \right].
\label{eq:f1sol}
\end{equation}  
By combining  Eq.~(\ref{eq:f1sol}) with  Eq.~(\ref{eq:evolf1}), we obtain a polynomial
in $v^*$ of order 3. Since the coefficients of this polynomial must
be identically zero, we find the following system of differential equations:
\begin{equation}
\frac {d \eta_4}{dt} + \eta_3  = 0,
\label{eq:toto1} 
\end{equation}
\begin{equation}
\frac {d \eta_3}{dt} + 2 \, \eta_2 - 6 \, \alpha \, \eta_4  = 0,
\label{eq:toto2} 
\end{equation}
\begin{equation}
\frac {d \eta_2}{dt} -6 \,\beta \, \Lambda_2 + 3 \, \eta_1 -4 \, \alpha \, \eta_3  = 0, 
\label{eq:toto3}
\end{equation}
\begin{equation}
\frac {d \eta_1}{dt} -2 \,\alpha \eta_2 -3 \, \beta \Lambda_1  = 0. 
\label{eq:toto4}
\end{equation}
\subsubsection{Effect of time dependent non-quadratic terms: a practical case}
\label{sec:valid}
In practice the potential felt by a cloud moving in
 a given system can be quite different from quadratic.
The Poisson equation,  Eq.~(\ref{eq:Poisson}), shows that the non-quadratic terms are related 
to local gradients in the density distribution. The accurate estimation of the local
potential depends on the details of the density distribution, thus to quantify the
effects we will have to adopt a given density distribution.
However, in general, as the cloud moves, the system evolves and changes its potential,
 but this potential variation is slow and can be neglected for our practical purpose. 
Let us assume that the density distribution is given by the family of profiles:
\begin{equation}
 \rho(x) =  q \left(\frac{x}{h} \right),
\end{equation}
where $h$ is a scale-height. 

Using the properties of this density distribution and the Poisson equation, it is
easy to analyze the scale properties of the cloud equation. The coefficients
of the local potential $\alpha_0$ and $\beta$ in Eq.~(\ref{eq:newpot}) can
be rewritten using the scaled variable $\bar x^*$:
\begin{equation}
 \alpha_0 = \frac{1}{2} \, \frac{d^2 \phi}{d^2 {x^*}} = \rho(x^*) = 
 q(\bar x^*),
\label{eq:alpha0eq}
\end{equation}
\begin{equation}
  \beta = \frac{1}{3} \, \frac{d \rho(x^*)}{dx^*}= 
  \frac{1}{3 h} \, \frac{dq(\bar x^*)}{d \bar x^*}.
\label{eq:betaeq}
\end{equation}
Eq.~(\ref{eq:alpha0eq})  shows that $\alpha_0$  is independent of the scale
of the density distribution. As a consequence,
Eq.~(\ref{eq:lambda2evol}) is also scale independent, and thus the quadratic solution 
$\Lambda_2(t)$ is unaffected by the scale of the density distribution. 
But the behavior of the coefficient $\beta$ is different. According to 
Eq.~(\ref{eq:betaeq}), $\beta$ is scale dependent.
It is possible to evaluate
the effect of this scale dependence by forming an equation for $\eta_4(t)$ similar
to  Eq.~(\ref{eq:lambda2evol}) for $\Lambda_2(t)$. 
By combining Eqs.~(\ref{eq:toto1}), (\ref{eq:toto2}), (\ref{eq:toto3}), and (\ref{eq:toto4}),
it is possible to construct the following equation for $\eta_4(t)$:
\begin{equation}
 -6\, \frac {d \beta}{dt} \Lambda_2
 -15\,\beta \frac {d \Lambda_2}{dt}
 +3\, \frac {d^2 \alpha_0}{dt^2} \eta_4 
 +10\,\frac {d \alpha_0}{dt} \frac {d \eta_4}{dt} + 
 10\,\alpha_0 \frac {d^2 \eta_4}{dt^2}
 + \frac{1}{2}\,\frac {d^4 \eta_4}{dt^4}
 +18\, \alpha_0^2 \, \eta_4 = 0. 
 \label{eq:b4eq}
\end{equation}
It is easy to notice that the scaling property of $\beta$ and the scale
 invariance of $\alpha_0$ and $\Lambda_2$ imply that $\eta_4$ scales like $1/h$.
 Which means that the correction introduced by the cubic term is inversely proportional
 to scale. Thus, we may solve the equation for $\eta_4$ at a given scale and generalize
 the result to any scale using the former scaling rule. The solution for a fixed scale
 can be calculated using Eqs.~(\ref{eq:alpha0eq}), (\ref{eq:betaeq}) and (\ref{eq:b4eq}).

For our numerical implementation, we will adopt a fixed gaussian density
distribution for the system, $q(x)=\exp(-x^2)$.
Given initial conditions in phase space, the evolution of the cloud crossing
this density distribution can be solved numerically by integrating our former system
of equations. The motion of the center of gravity of the cloud itself
can be computed easily by estimating the motion of a point mass in the potential
of the system. The orbit of the cloud
is defined by its initial position $x_0$, and its initial velocity.
We will study the case of zero initial velocity and variable position
$x_0$. Initially the cloud will be a round gaussian of velocity dispersion
equal to 1.  
The numerical integration of these equations shows that the 
deviation from the quadratic approximation is maximal for an initial position
close to $3 h$ (see Fig. 1). The maximum deviation observed in the
system is directly related to the requirement to perform a Lagrangian remap.
Thus we should study the behavior of the error to the quadratic approximation
near the $3 h$ initial position. The relevant plot is shown in Fig. 2.
As discussed before, the curve showing the deviation due to 
the cubic term is generic. The deviation for another scale length $h$
of the density distribution can be obtained by rescaling this curve. 
Since the scaling is inversely proportional to $h$, the error to 
the quadratic approximation will be dominated by the crossing of the
shorter scale density structures.
\begin{figure}
\centerline{\epsfig{file=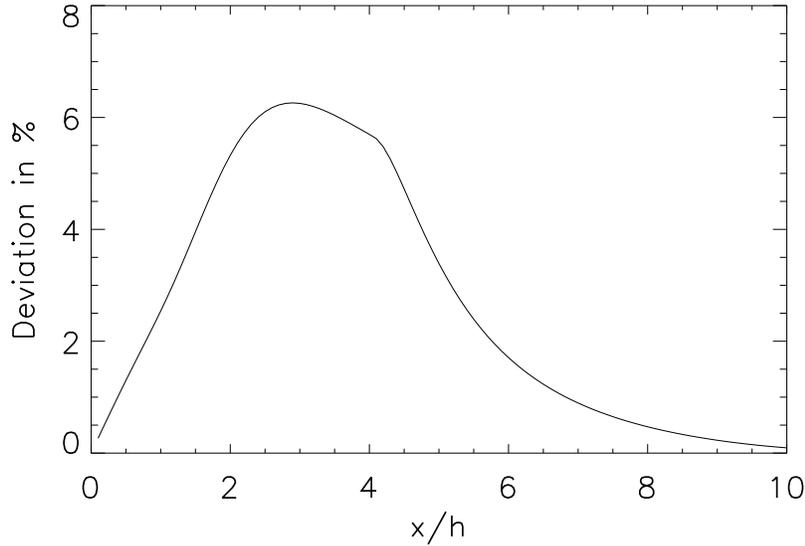,width=10cm}}
\caption{The maximum deviation from the quadratic approximation as
a function of the initial position $x_0$. The distribution has a scale length 
$h$ ten times larger than the cloud. The coordinate $x_0$ has been
normalized by the scale length of the distribution $h$. Note that the
maximum deviation is at about $3 h$ from the center of the distribution.
}
\label{fig:maxdev}
\end{figure}
\begin{figure}
\centerline{\epsfig{file=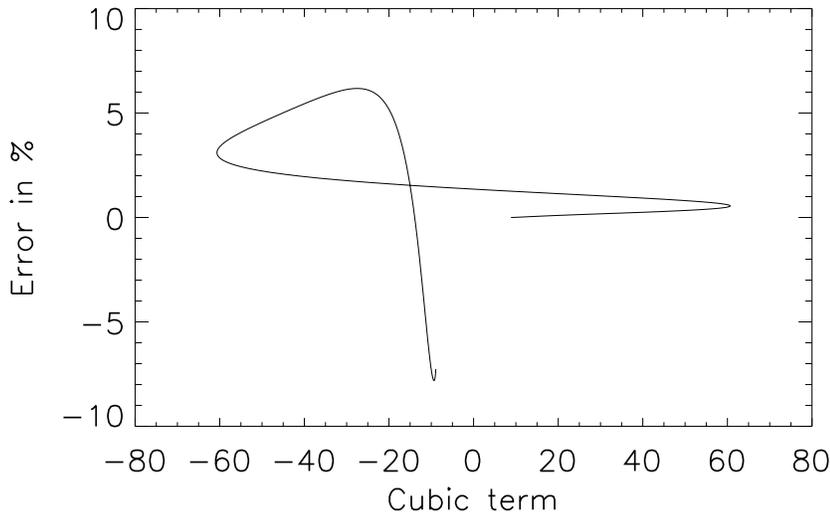,width=10cm}}
\caption{The shape of the deviation due to the cubic term as a function
of the strength of the cubic term. We chose an initial position $x_0$
which maximizes the deviation from the quadratic approximation
 ($x_0 \simeq 3 h$). Note that in case the distribution
has a smaller scale length the error must be rescaled accordingly.
In practice, the minimum size of the distribution is about 2 times
the cloud scale length, for which the deviation can be 5 times larger.}
\label{fig:maxdevevol}
\end{figure}
%
%
\section{The method: algorithm}
\label{sec:algo}
In this section, we study practical implementations of the method. We examined two approaches,
one assuming constant resolution in phase-space, that will correspond to
what we call a ``cloud in mesh'' (CM) code, the other one allowing local refinement
in phase-space using adaptive refinement trees, that we therefore call tree-code.
There are several issues to be addressed while implementing the method. We list them here 
in the same order as they will be treated below:
\begin{itemize}
\item {\em Phase space sampling of the distribution function }
(\S~\ref{sec:phasespacesampling}): the question here
is to decide how to chose our set of clouds such that it reproduces at best
a given distribution function in phase-space. This is necessary not only to
set up initial conditions, but also to resample the distribution function
during run time with a new set of round clouds. Indeed, we know from our
perturbative analysis (\S~\ref{sec:perturana}) that deviations from
local quadratic behavior of the potential increase with time, and that
at some point the clouds will have wrong axis ratio (be too elongated)
and wrong orientation. To (re)sample the distribution function, we propose
to use Gaussian clouds located on a (possibly adaptive) grid, with 
their masses estimated using either Van-Citter or Lucy deconvolution algorithms.
\item {\em Solving the Poisson equation } (\S~\ref{sec:poissonsolver}):
the issue here is to to estimate accurately the forces exerted on each
cloud as well as errors on their determination. These latter will 
indeed be used to quantify deviations from a local quadratic behavior of the potential. 
This is fairly easy in our one dimensional problem, 
since the force exerted on a point of space is simply given by the 
difference between the mass at the right
of the point and the mass at the left of the point. However, we will
try here to experiment methods which can be, in principle, easily generalized
to higher number of dimension without any serious increase
in complexity.
\item {\em Run time implementation and diagnostics} (\S~\ref{sec:timstd}): the
choice of the time step implementation is important. Here we propose a
predictor corrector of second order, which makes our
method full second order both in space and in time. 
We discuss how to compute the slowly varying time step associated
to this integrator. We also examine 
when a remap of the distribution with a new
set of round clouds has to be performed, by using a criterion based on 
the cumulated errors on the forces due to non local quadratic behavior of
the potential.
\end{itemize}
%
%
\subsection{Phase space sampling of the distribution function: practical
implementation}
\label{sec:phasespacesampling}
%
When starting from given initial conditions, $f({x},{v})$, 
it is necessary to set up an ensemble 
of round clouds that all-together reproduce at best $f$.
During runtime, these clouds get more and more elongated and the approximation
for a local quadratic potential breaks: a remapping of the distribution function
is needed with a new set of round clouds. The measured distribution function at
this time becomes new initial conditions that, again, have to be sampled accurately.
Here we explain in detail how we proceed to perform this remapping.
In \S~\ref{sec:funcgchoice} we discuss the cloud shape and
the mean cloud inter-spacing. Our choice is a Gaussian cloud of typical radius 
$R$, truncated at $4 R$ and separated from its nearest
neighbors by a distance $\sqrt{2}R$. In \S~\ref{sec:deconv} we explain how
we compute the cloud masses in order to reproduce at best the distribution
function. The methods proposed are Van-Citter and Lucy deconvolutions algorithms.
In \S~\ref{sec:weightreg}, we propose a weighting scheme for filtering small
scale noise in the estimate of the distribution function obtained from the clouds.
Finally \S~\ref{sec:massconv} examines possibilities of enforcing
conservation of basic quantities such as total momentum and total mass without
adding significant diffusion. In addition to that, Appendix~\ref{sec:appcompf} 
details the algorithms used to compute the distribution
function from the clouds. It shows how to speed up the calculation
in the case the clouds are round, thanks to the separability of the Gaussian
window.
%
\subsubsection{Choice of cloud shape and spacing}
\label{sec:funcgchoice}
\begin{figure}
\centerline{\mbox{\epsfig{file=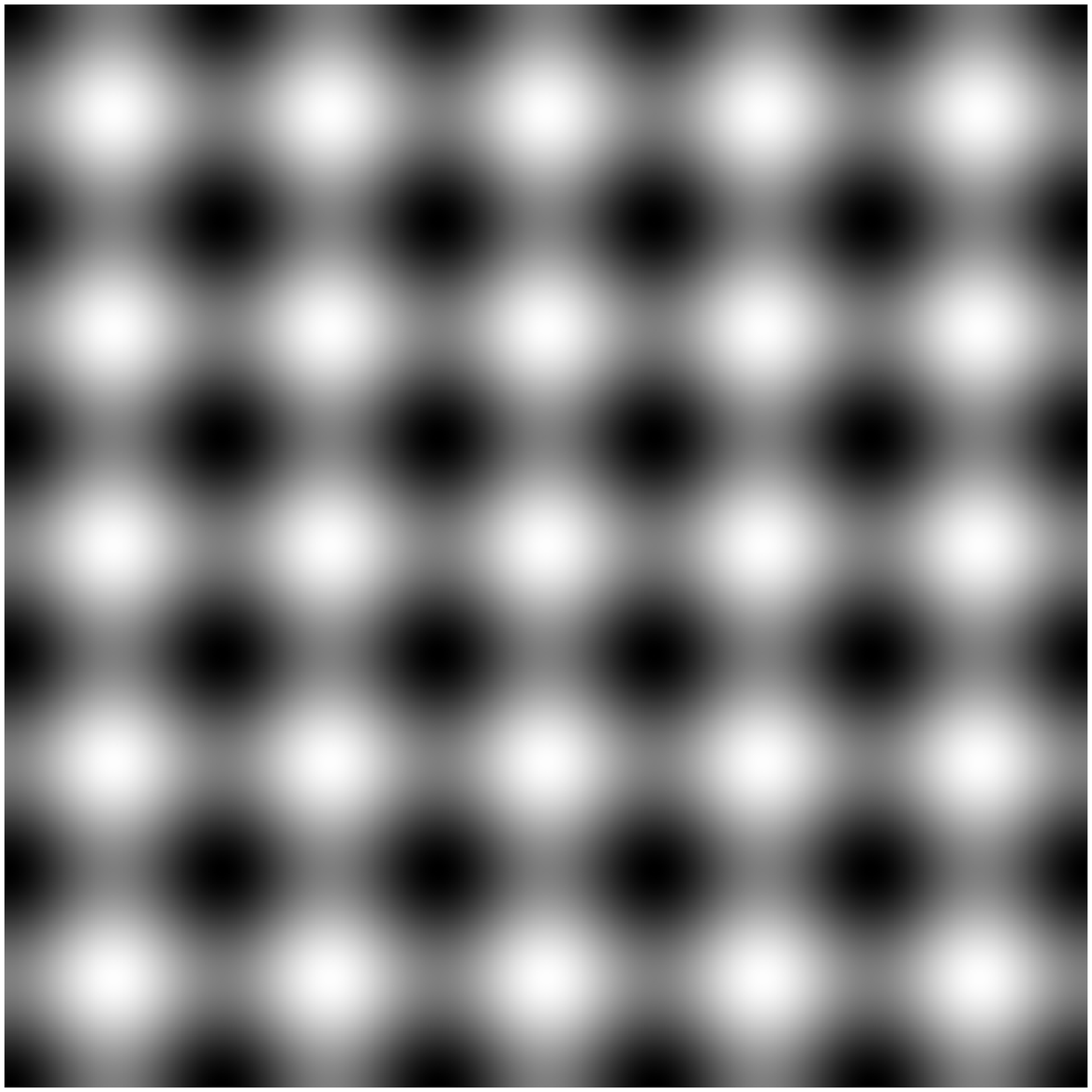,width=8cm}
\epsfig{file=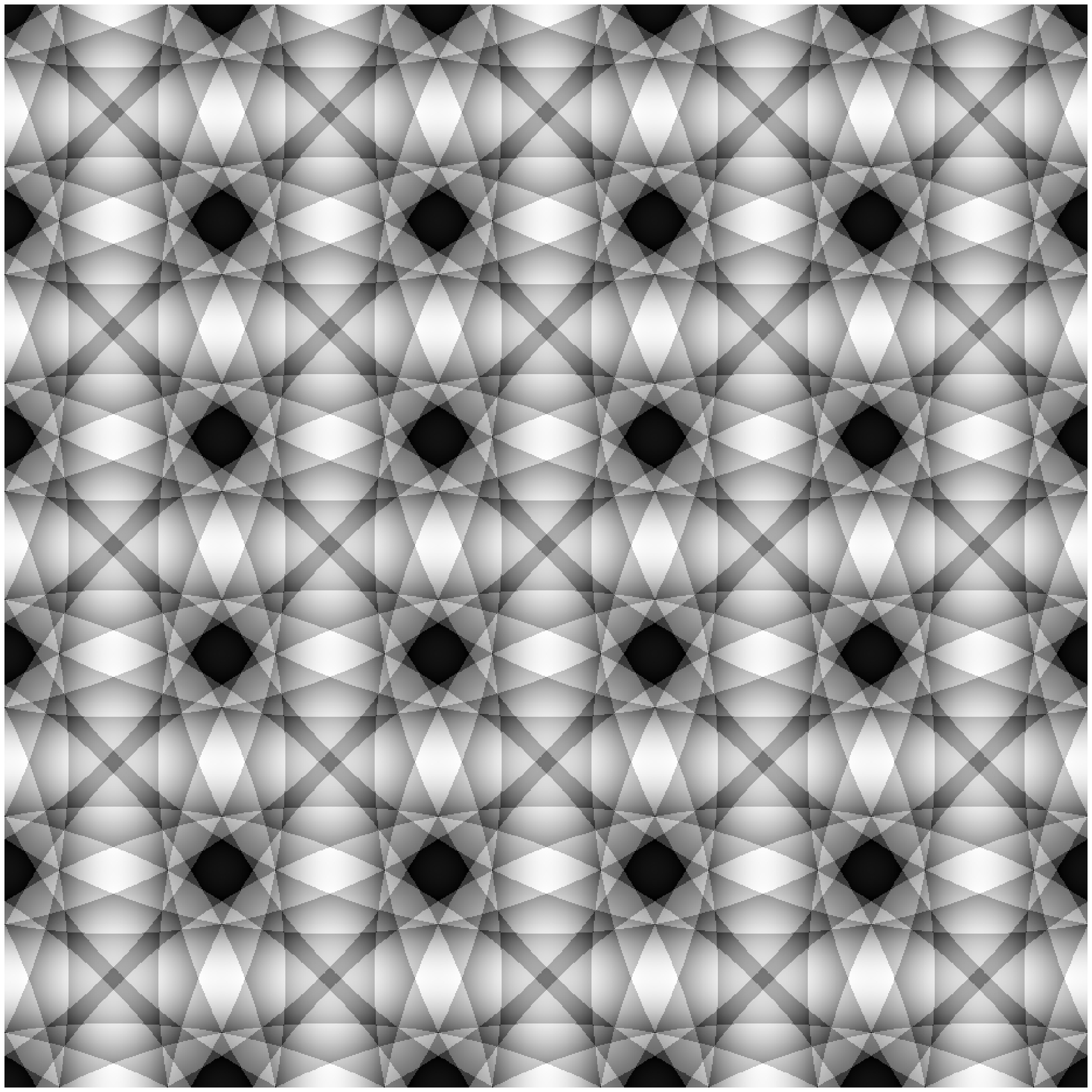,width=8cm}}}
\vskip 0.2cm\centerline{\mbox{\epsfig{file=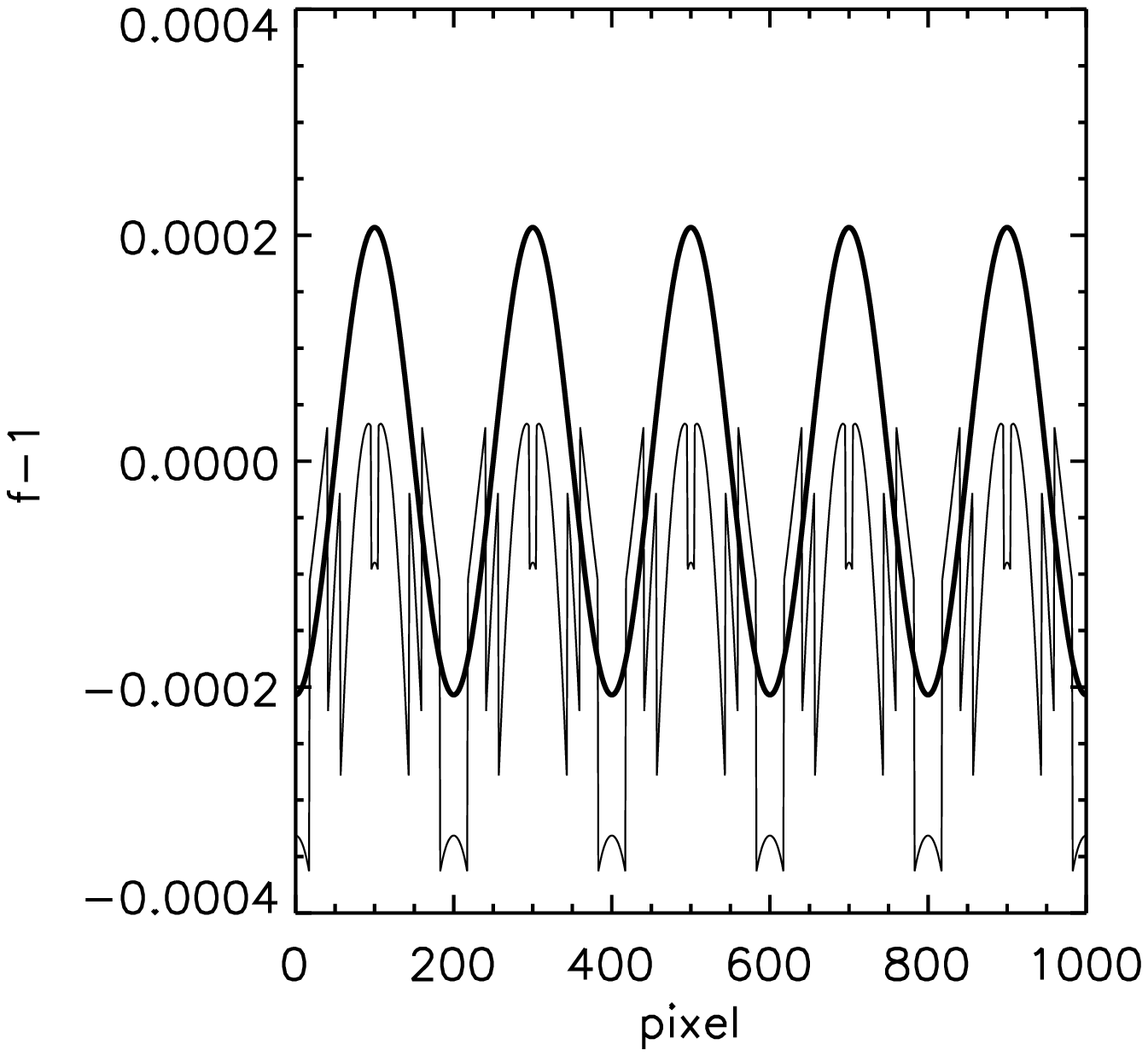,width=8cm}
\epsfig{file=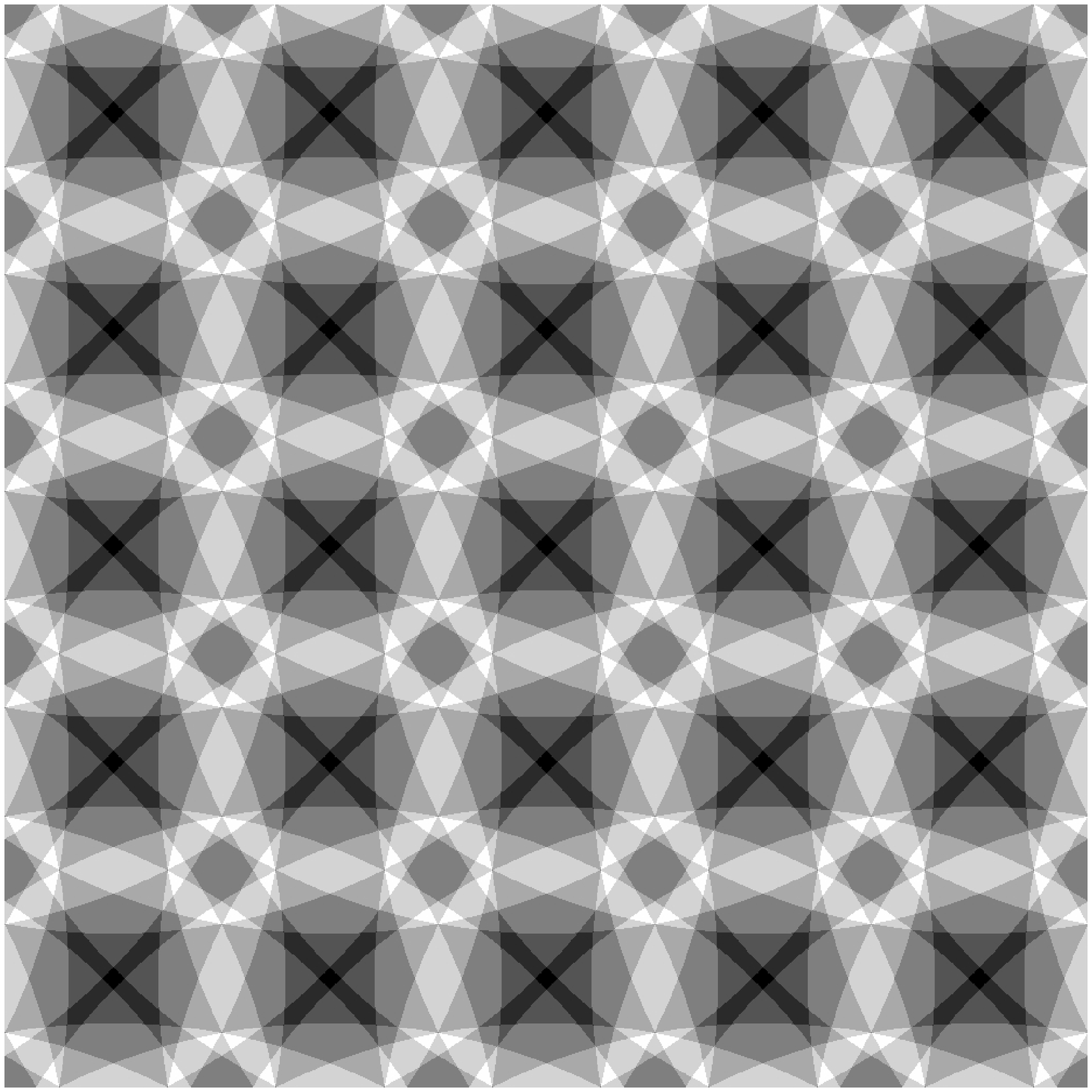,width=8cm}}}
\caption[]{Sampling of $f(x,v)=$ constant $=1$ by truncated Gaussians. We consider
a set of $5\times5$ clouds in a periodic square of size $L=5$. With these units,
each Gaussian is of size $R=\sqrt{2}/2$ and is truncated at $R_{\rm max}=3.95 R$. 
Upper left panel shows the variations
of the sampled distribution function ${\tilde f}$ if the clouds would have infinite extension
(darker and lighter regions correspond to ${\tilde f} > 1$ and ${\tilde f} < 1$ respectively).
They range with a good accuracy from $2 \sigma \simeq 2.1\times 10^{-4}$ to $-2\sigma$ [Eq.~(\ref{eq:intfluc})].
Upper right panel is the same, but the truncation is applied, which explains the discontinuities
observed in ${\tilde f}$. In that case, fluctuations of ${\tilde f}$ are larger and are
not symmetrically distributed, ranging approximately from $-3.8 \times 10^{-4}$ to $2.3 \times 10^{-4}$.
Lower right panel shows the cloud counts, i.e. for each point of space
the number of cloud contributing to it. This count varies significantly, between 21 (light) and 27 (dark). 
Finally, lower left panel displays the values of ${\tilde f}(x,v=x)$ 
(the diagonal of the images) for
the untruncated (thick smooth curve) and truncated case (thin irregular curve). The
truncation adds a significant source of noise with a significant skewness in its distribution,
which can become a source of systematic effects. This effect has to be minimized with the
appropriate weighting scheme.}
\label{fig:figchoicecloud}
\end{figure}

%
%
To sample the distribution function in phase space 
at a given time, we use an ensemble of clouds which are initially 
round (this can be
always true if an appropriate 
normalization for positions and velocities is set).
In this section, we assume that they are disposed on a rectangular
grid of spacing $\Delta_{\rm g}$. The choice of our cloud shape
is entirely determined by specifying function $G$ 
[Eqs.~(\ref{eq:defG1}) and (\ref{eq:defG2})], which should present
the following features:
\begin{description}
\item[(i)] It should have compact support. Indeed, 
we need the cloud to be the least extended as possible, since we use
a local quadratic expansion of the potential within the cloud to compute
the forces exerted on it. 
\item[(ii)] It should be sufficiently smooth in order
to resample a distribution function with continuous first
and second derivatives (e.g. for refinement
procedure as discussed later). This is an essential 
feature of our method and a condition for it to perform well. 
\end{description}
A good choice for function $G$ is a truncated exponential, which
has the great advantage of being separable and makes the cloud Gaussian.
Smoothness condition (ii) forces us to truncate this cloud rather far away
from its center. Our practical choice is a $4$ sigma cut-off, $R_{\rm max}=4 R$.\footnote{More exactly
 $R_{\rm max}=3.95 R$, to avoid the circle of radius $R_{\rm max}$ 
intersecting any of the cloud center.}
This still induces small discontinuities of the order of $3.10^{-4}$, as illustrated by right panel of 
Fig.~\ref{fig:figchoicecloud}. To minimize their effect
we will use a weighted estimator to compute $f$, as detailed in \S~\ref{sec:weightreg}.
The remote nature of the cut-off has another disadvantage, which is one of the main drawbacks
of our method, when it will be extended to higher number of dimensions: 
a large number of clouds contribute to the sampled distribution
function at a given point of phase-space,
as illustrated by lower right panel of Fig.~\ref{fig:figchoicecloud}.

Finally, we have to determine the radius $R$ (corresponding to a one sigma
deviation) of
our cloud as a function of grid spacing for 
best sampling the distribution function. Basically, the choice of $R$
determines by how much the resampled distribution function, ${\tilde f}$, 
will deviate from the true $f$. In order to sample smoothly the variations of $f$ over
the sampling grid, $R/\Delta_{\rm g}$ should be of the order of unity.
To minimize the cost of the sampling, it should be kept as small as possible.
To find $R$, we estimate the quadratic error $\sigma$ due to the representation 
of space by a finite number of Gaussian functions. Since our approximation
requires $f$ to be nearly flat at the scale of the cloud, we
will evaluate the residual in the case where the function to represent
is constant, $f=1$. For an infinite flat distribution $\sigma$ reads
\begin{equation}
 \sigma^2 = \frac{\int_{-\infty}^{\infty} \left[f(x,v)-\tilde f(x,v) \right]^2 dxdv}{\int_{-\infty}^{\infty} dxdv}.
\label{eq:parseval1}
\end{equation}
For gaussian clouds, $f$ is represented by the functional:
\begin{equation}
 {\tilde f}(x,v)=\frac{\Delta_{\rm g}^2}{2 \pi \,R^2} \sum_{i,j}  
\exp \left[- \frac{\left[(x-i\,\Delta_{\rm g})^2+(v-j\,\Delta_g)^2 \right]}{2 R^2} \right].
\end{equation}
Note that the representation $f(x,v)$ can be rewritten as a convolution using Dirac series:
\begin{equation}
 f(x,v)=\frac{\Delta_g^2}{2 \pi \,R^2} \sum_{i,j} \int 
\exp \left[- \frac{\left(u_1^2+u_2^2 \right)}{2 R^2} \right] 
\delta(u_1+x-i\,\Delta_g) \delta(u_2+v-j\,\Delta_{\rm g}) \ du_1du_2.
\end{equation}
This rewriting as a convolution suggests that the equations should be
analyzed using Fourier transform. Since both the numerator and denominator
in Eq.~(\ref{eq:parseval1}) are the norm of a function in real space, the
transformation to Fourier space can be done easily by using Parseval
theorem. Using the symbol $\bar f$ to represent the Fourier transform of $f$, 
Parseval theorem reads:
\begin{equation}
\int_{-\infty}^{\infty} \left[f(x,v)-\tilde f(x,v) \right]^2 dxdv =
\int_{-\infty}^{\infty} \left|\bar f(k_x,k_v)- \bar {\tilde f}(k_x,k_v) \right|^2 dk_xdk_v.
\label{eq:Parseval3}
\end{equation}
And with the help of the convolution theorem:
\begin{equation}
 \bar f(k_x,k_v)=\sum_{ij} \exp \left [-2\,\pi^2\,R^2\,\left(k_x^2+k_v^2 \right) \right] 
\delta(k_x-i/\Delta_{\rm g}) \delta(k_v-j/\Delta_{\rm g}).
\label{eq:Parseval4}
\end{equation}
Thanks to the rapid fall of the exponential, this sum is well
approximated by a small number of terms. In particular a good
approximation at the typical scale of interest is to consider
only the terms $(i,j)$ verifying $|i|+|j| \leq 1$:
\begin{equation}
\int |\bar f(k_x,k_v) - \bar {\tilde f}(k_x,k_v)|^2 \ dk_xdk_v 
\simeq 4 \exp \left[ -\left(2\,\pi\,R/\Delta_{\rm g}\right)^2\right] \ \delta(0)^2. 
\label{eq:Parseval5}
\end{equation}
The divergent factor $\delta(0)^2$ disappears once we apply the normalization
by the denominator in Eq.~(\ref{eq:parseval1}). 
%
%
We thus finally find:
\begin{equation}
\sigma \simeq 4 \, \exp \left[-\left(2 {\pi R}/{\Delta_g}\right)^2 \right].
\label{eq:intfluc}
\end{equation}
Our choice $R=\sqrt{2}/2$ corresponds to $\sigma \simeq 10^{-4}$, consistent
with the discontinuities brought by our $4$ sigma cut-off for the gaussian. 
This is well illustrated by lower-left panel of Fig.~\ref{fig:figchoicecloud}, which
displays the deviations of $f$ from unity for our choice of $R/\Delta_{\rm g}$.
Experimentally, we measure that their mean square is indeed equal to $10^{-4}$,
while the difference between local minima to local maxima is four times larger. 
Notice again that the truncature at $R_{\rm max}$ adds a source of noise at very small scales
which has the bad property of presenting a significant skewness in its distribution.
We shall see in \S~\ref{sec:weightreg} that with appropriate reinterpolation of function
$f$, it is possible to remove these sources of noise to a great extent, in fact
totally in the case $f=$constant considered here.
%
%
%
\subsubsection{Choice of deconvolution scheme}
\label{sec:deconv}
%
%
Once the cloud shape is chosen as well as the size of the sampling grid,
the problem of finding the cloud masses in order to sample correctly
the distribution function remains open. To do that, an iterative procedure
is necessary. We adopted in this paper two simple algorithms, Van-Citter
and Lucy as we discuss now.

The sampled distribution function can be written
\begin{equation}
{\tilde f}(x,v)=\frac{1}{V} 
\sum_i M_i\  G(x- x_{\rm G}^i,v-v_{\rm G}^i),
\label{eq:ffromspl}
\end{equation}
where $M_i$ is the mass of the cloud and function $G(x,v)$ is normalized
in such a way that its integral is $V$. The quantity $V$
is given by $V=\Delta_{x} \Delta_{v}$,
where $\Delta_{x}$ and $\Delta_{v}$ correspond to grid site inter-spacing distance 
along ${x}$ and ${v}$ coordinate respectively. It is the area associated
to each cloud so that $\int f({x},{v}) d{x}d{v}=\sum_i M_i$.
Note that we used implicitly the following notations:
\begin{equation}
G(x,y)\equiv G(\beta_0 x^2 + \beta_1 xy + \beta_2 y^2).
\end{equation}

We have to find $M_i$ such that
\begin{equation}
{\tilde f}({x}_{\rm G}^i,{v}_{\rm G}^i)\equiv {\tilde f}_i=f({x}_{\rm G}^i,{v}_{\rm G}^i)\equiv f_i,
\end{equation}
for each cloud position $({x}_{\rm G}^i,{v}_{\rm G}^i)$ in phase-space. A good first guess is simply
\begin{equation}
M_i^0=f_i V.
\label{eq:firstguess}
\end{equation}
However, with that choice of masses, ${\tilde f}$ will basically be equal to
the convolution of $f$ with function $G$. Some deconvolution algorithm
has to be applied in order to fit better function $f$. 

The Van-Citter algorithm consists simply as follows.
Given ${\tilde f}_i^n$ as computed at iteration $n$, the iteration $n+1$ writes, 
\begin{eqnarray}
\delta_i^n & = & f_i-{\tilde f}_i^n, \nonumber \\
M_i^{n+1} & = &M_i^{n}+\delta_i^n V.
\label{eq:van-citter}
\end{eqnarray}
The algorithm is applied until some convergence criterion is fulfilled,
$\max(|\delta_i^{n}|) \leq \delta_{\rm c}$. To maintain the
domain of calculation as compact as possible 
during runtime, only values of $f_i$ verifying $f_i > \delta_{\rm min}$
are taken into account. In the simulations shown below, 
we take $\delta_{\rm min}=\delta_{\rm c}/2$, but $\delta_{\rm min}$ 
should in principle be kept as small as possible to have best conservation of the 
moments of the phase-space distribution function, e.g. energy. 
Similarly, after convergence, clouds contributing
little to the reconstruction are depleted:
one uses reciprocal of Eq.~(\ref{eq:firstguess}) to estimate approximately 
their contribution, i.e. clouds with $|M_i/V| \leq \delta_{\rm min}$ are eliminated. 
This procedure has the defect of augmenting slightly aliasing effects
in the neighborhood of the regions where $f$ cancels. 

Given the sources of uncertainties associated
to our choice of function $G$, $\delta_{\rm c}$ should be large enough 
compared to $10^{-4}f_{\rm max}$, where $f_{\rm max}$ is the maximum value of $f$,
since at this point we would capture spurious
features and convergence would be made difficult due to the discontinuities
brought by the cut-off. Experience shows that $\delta_{\rm c}/f_{\rm max}$ ranging
from $0.0005$ to $0.002$ is a good compromise. Also, convergence might
become difficult due to other aliasing effects. As well shall see later, 
function $f$ builds finer and finer structures with time, which cannot
be reproduced correctly by our mapping when they appear at scales smaller
than $\sim R$.  Convergence is rendered very difficult in that
case: one might chose therefore to stop iterating when $n$ reaches
some value, typically of the order of 10 according to our practical experiments.
If convergence is not reached at this point, it is clear anyway that 
fine structures of function $f$ wont be reproduced correctly, even
with a large number of iterations. 

The major defect of Van-Citter algorithm is that it does not warrant positivity
of the distribution function. An alternative to Van-Citter is Lucy deconvolution
algorithm:
\begin{eqnarray}
\delta_i^n & = & f_i/{\tilde f}_i^n-1, \nonumber \\
M_i^{n+1} & = & M_i^{n} \sum_j (1+\delta_j^n)\ G(x_{\rm G}^i- x_{\rm G}^j,v_{\rm G}^i-v_{\rm G}^j).
\label{eq:lucy}
\end{eqnarray}
Such an algorithm does not only warrant positivity, it is also in principle 
more accurate
than Van-Citter, since the error $\delta_i^n$ is relative instead of absolute.\footnote{In that
case, $\delta_{\rm c}$ is not anymore expressed in term of $f_{\rm max}$.} 
In practice, Lucy does better than Van-Citter, but is a little more
unstable and is also more subject to aliasing effects. For our fixed
resolution simulations, we adopted Lucy algorithm, but
we used Van-Citter when testing refinement, as discussed later. 

\begin{figure}
\centerline{\mbox{\epsfig{file=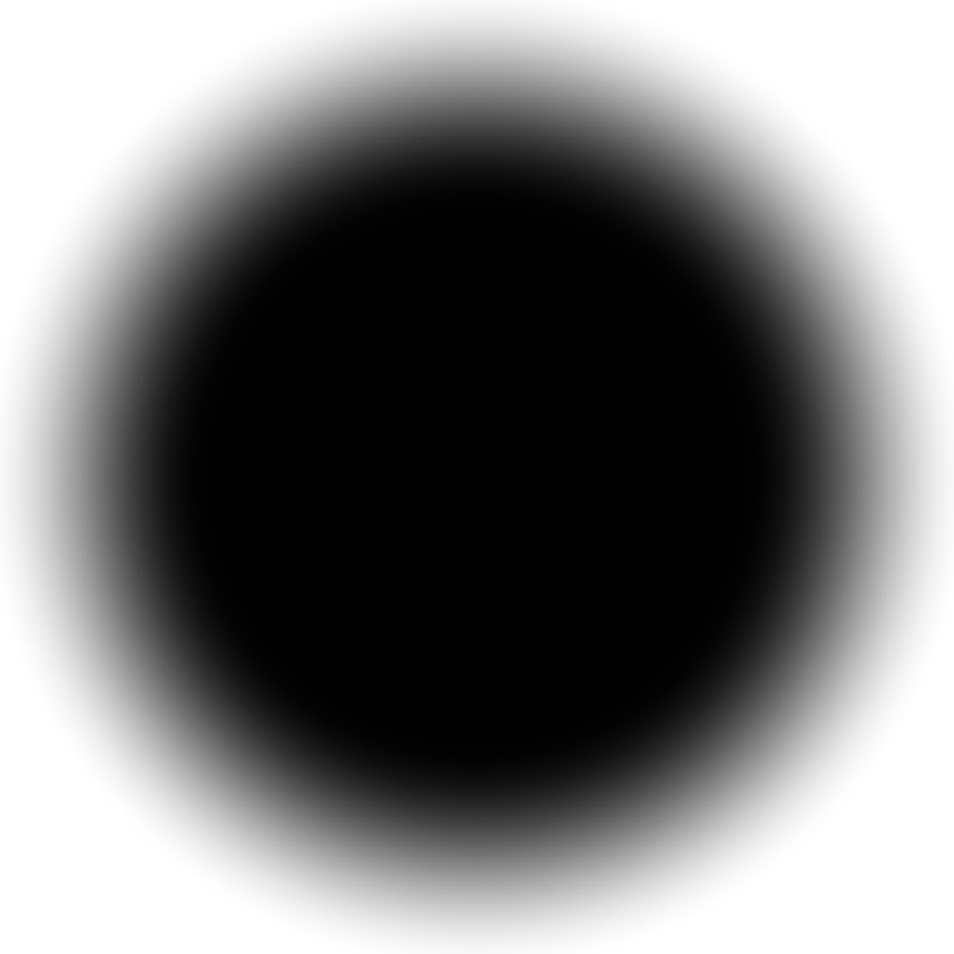,width=7.0cm}
      \hskip 0.5cm\epsfig{file=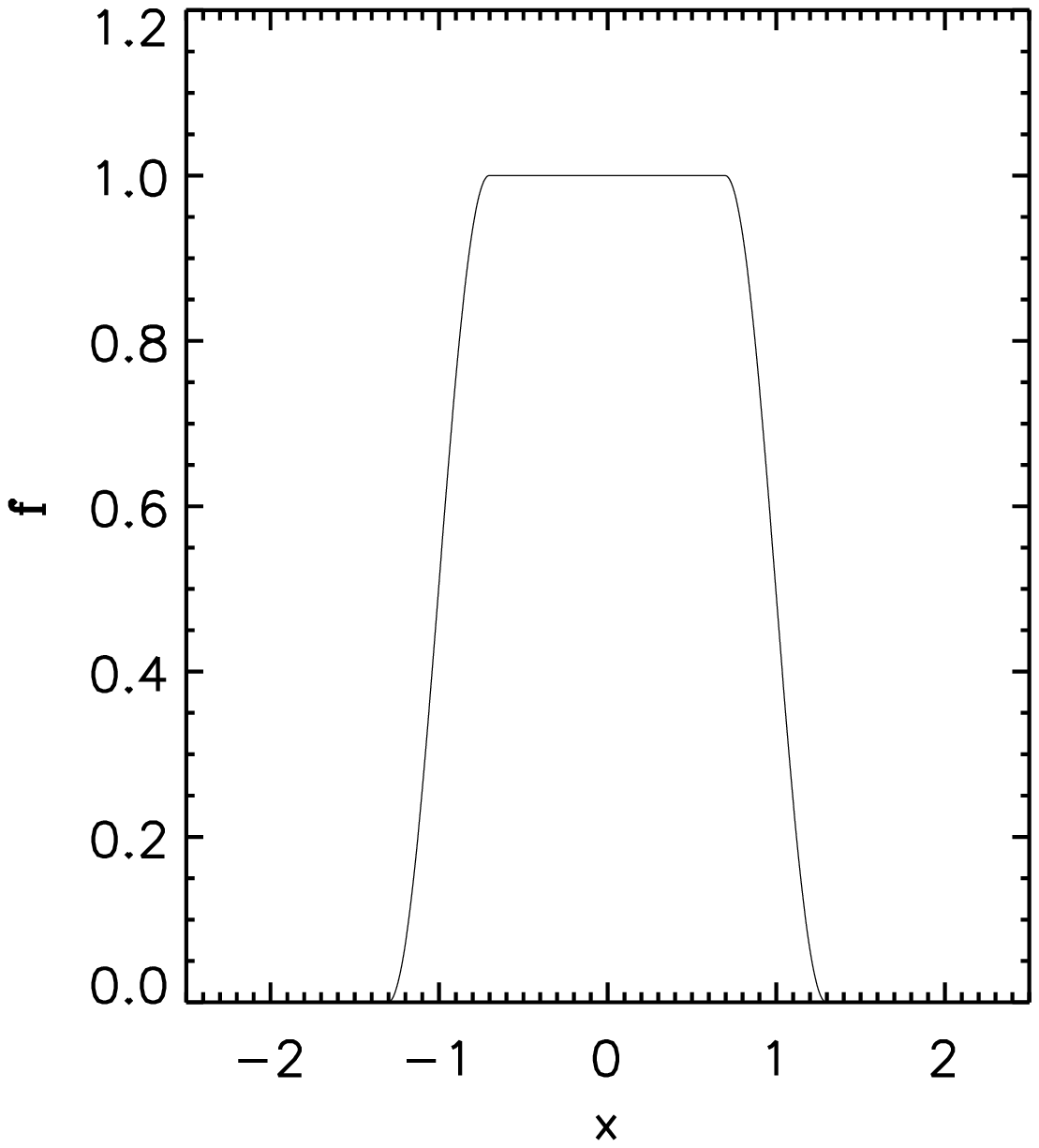,width=7.0cm}}}
\vskip 0.2cm
\centerline{\mbox{\epsfig{file=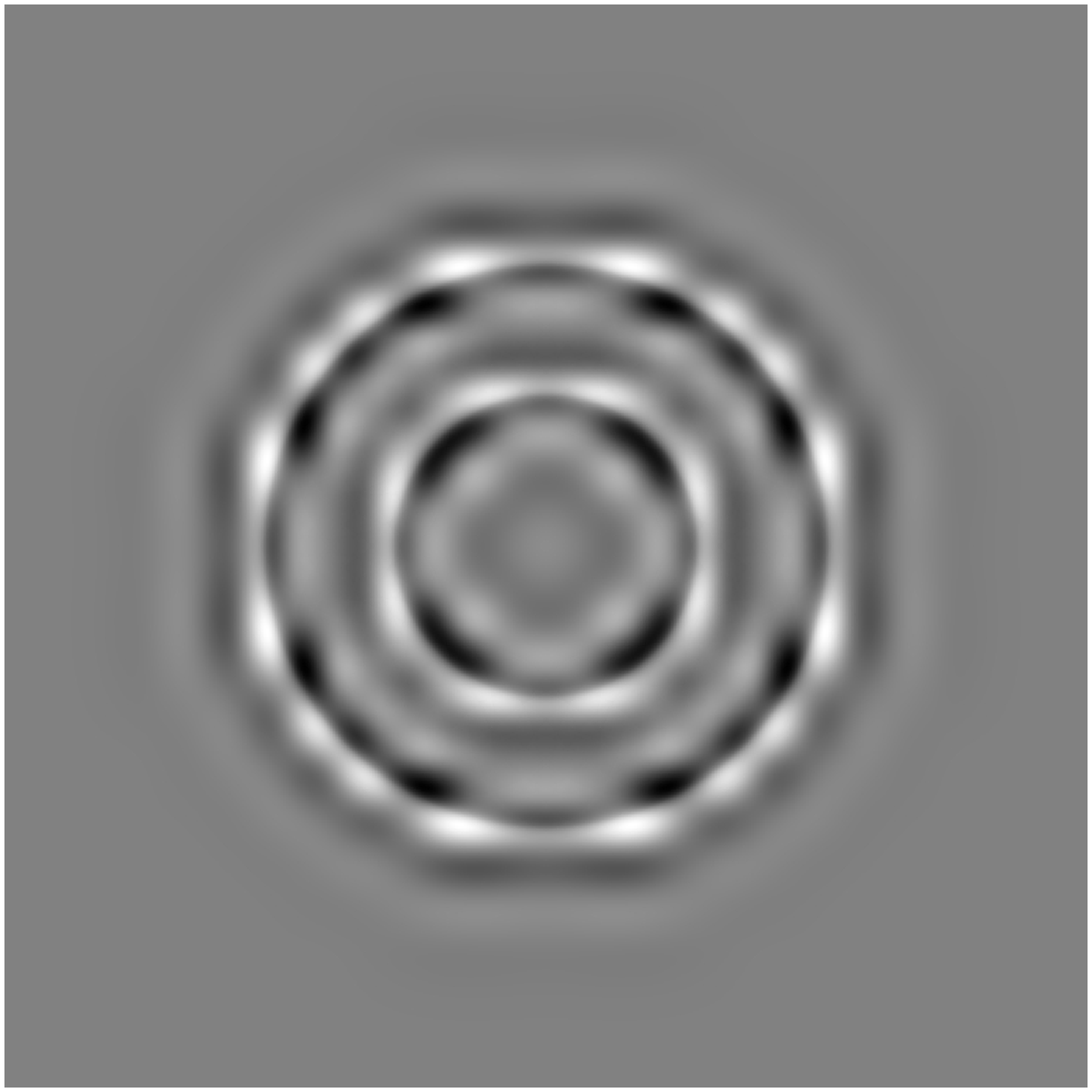,width=7.0cm}
      \hskip 0.5cm\epsfig{file=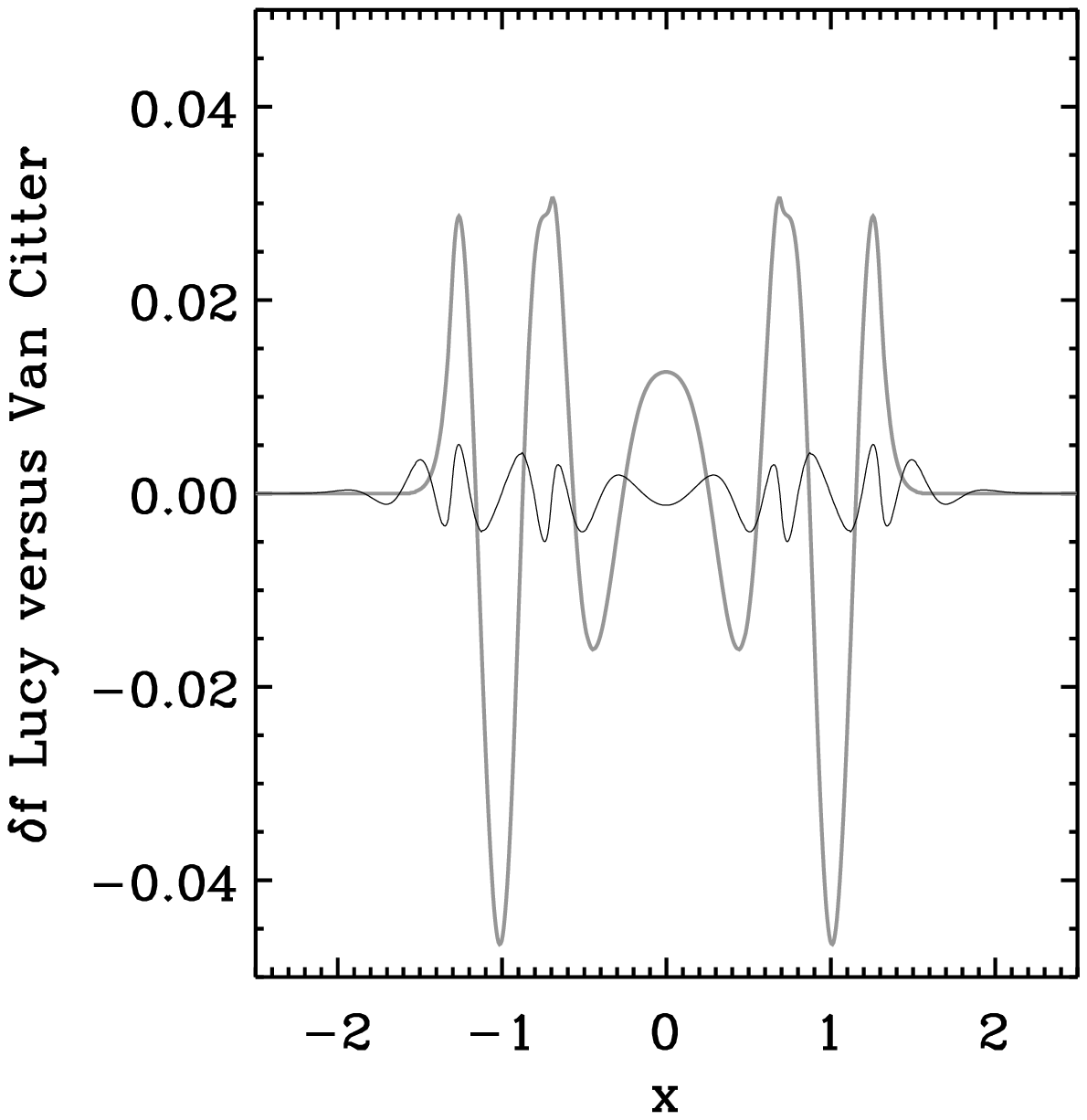,width=7.0cm}}}
\vskip 0.2cm
\centerline{\mbox{\epsfig{file=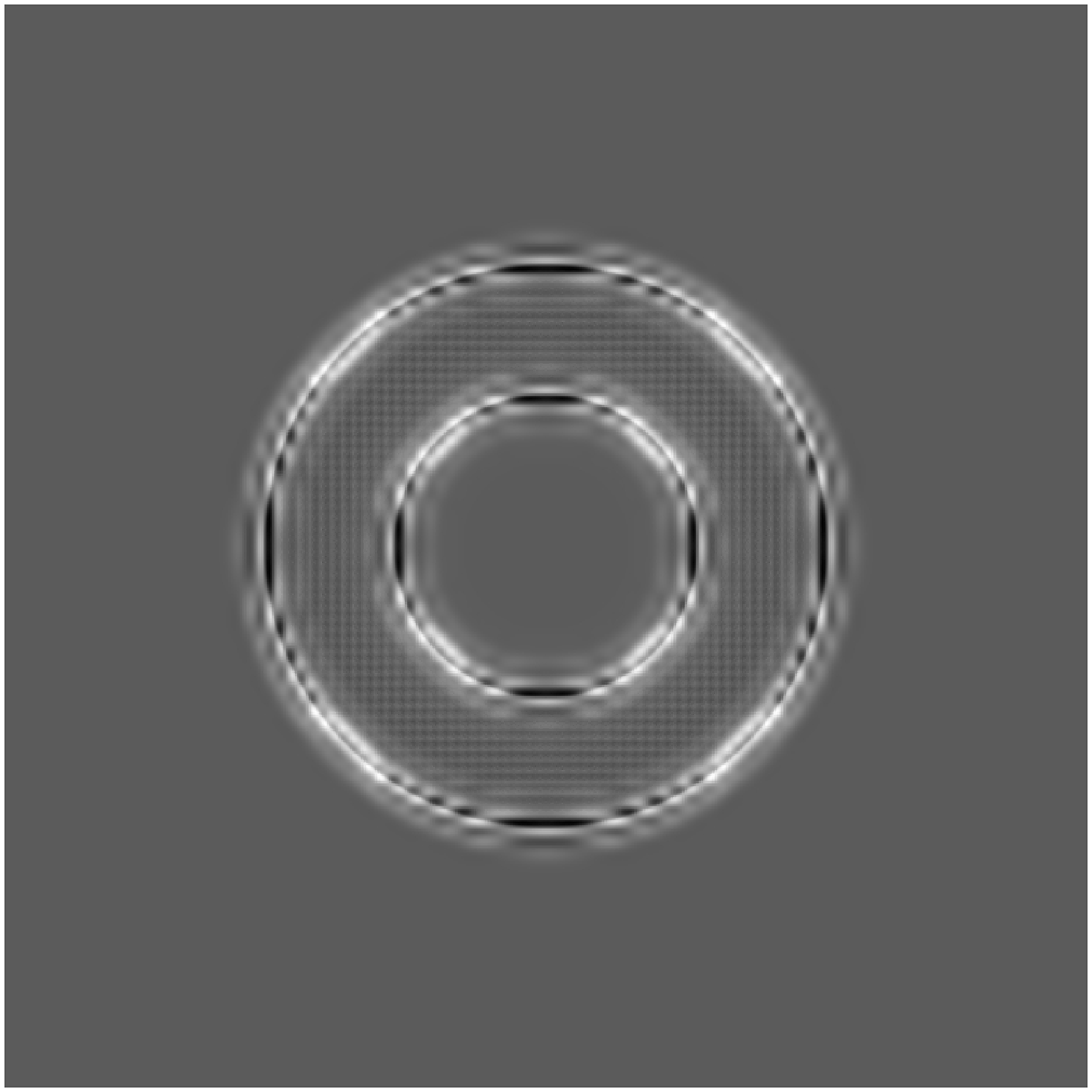,width=7.0cm}
      \hskip 0.5cm\epsfig{file=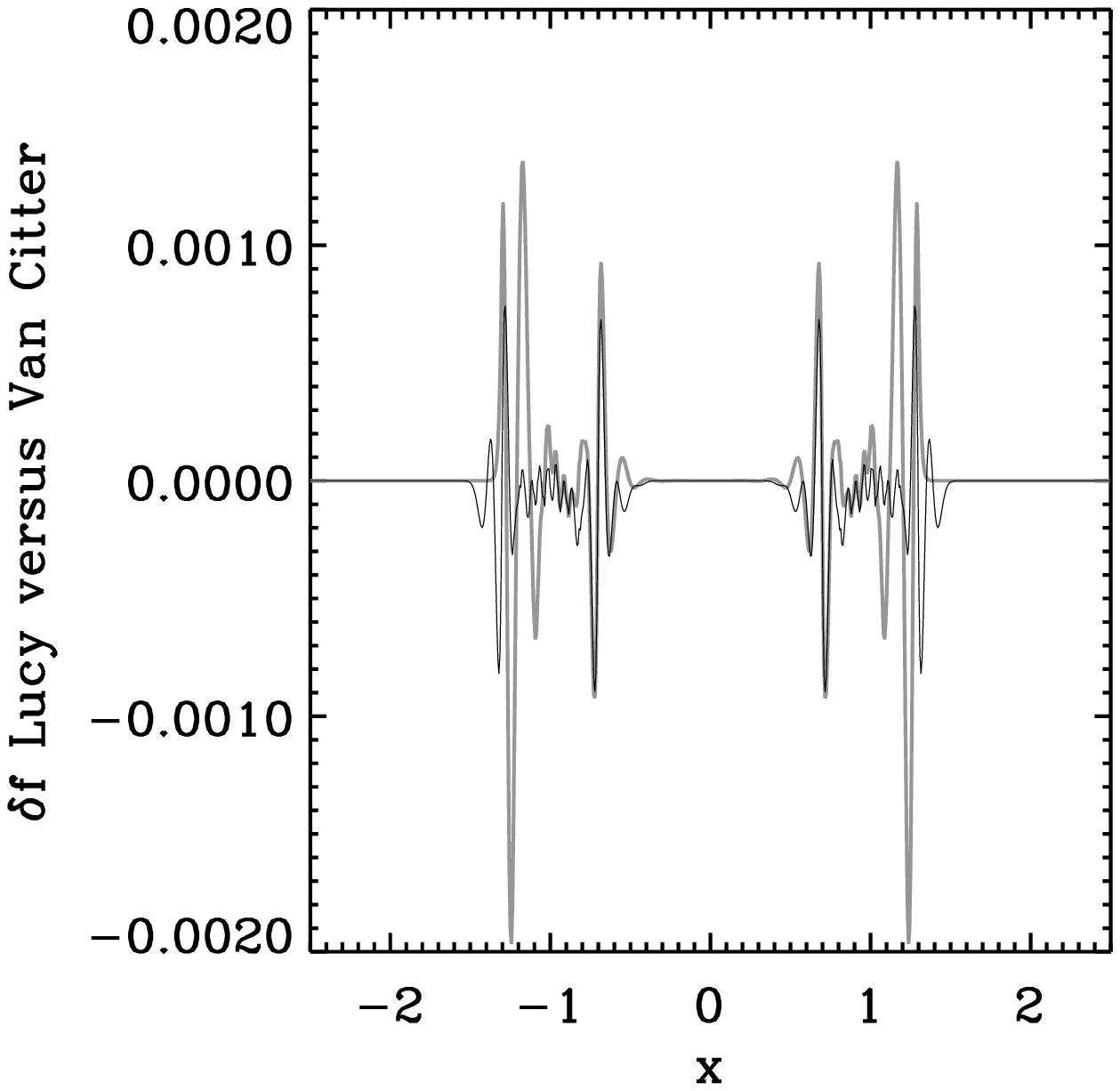,width=7.0cm}}}
\caption[ ]{Comparison of Van-Citter and Lucy algorithms for reconstructing a 
distribution in phase-space. Upper left panel displays the function
$f({x},{v})$ we aim to reproduce, as given by Eq.~(\ref{eq:finitophat}) [Darker
regions correspond to larger values of $f({x},{v})$]. Upper
right panel shows $f({x},0)$ as a function of $x$. Middle left panel displays
$\delta f \equiv {\tilde f}-f$, where ${\tilde f}$ is obtained by using Van-Citter
algorithm. A very similar result would be obtained for Lucy. The middle right
panel gives $\delta f(x,0)$ as a function of $x$, both for Van-Citter (thin
black curve) and Lucy (thick grey curve). To sample the distribution function,
an inter-spacing of $\Delta_{\rm g}=0.2$
and a cut-off $R_{\rm max}=3.95 \Delta_{x}$ was used.
The lower-left and lower-right panels are the same but for higher resolution,
$\Delta_{\rm g}=0.05$.
Note that, for the four lower panels, the weighting scheme discussed in \S~\ref{sec:weightreg}
has been used to estimate ${\tilde f}$, following Eq.~(\ref{eq:fsmoothed}).}
\label{fig:vancitlucy}
\end{figure}

To illustrate the methods, Fig.~\ref{fig:vancitlucy} shows some
results obtained when trying to resample the following distribution
function, used later as initial conditions for some of our simulations:
it is a top-hat apodized with a cosine:
\begin{eqnarray}
  f({x},{v})  &=&  {\bar \rho},   \quad {x}^2+{v}^2 \leq {\cal R}^2 , \nonumber \\
  f({x},{v})  &=&  \frac{1}{2}{\bar \rho} \left\{ \cos\left[ 
   \frac{\pi}{2} \left(\sqrt{{x}^2+{v}^2} - {\cal R}\right)/{\cal R}_{\rm apo} \right]+1 \right\},
    \quad {x}^2+{v}^2 \leq {\cal R}+2 {\cal R}_{\rm apo},  \label{eq:finitophat}
\end{eqnarray}
where ${\bar \rho}=1$, ${\cal R}=0.7$ and ${\cal R}_{\rm apo}=0.3$. Function $f(x,v)$ is
shown and plotted in top panels of the figure. Middle panels of the figure are the
same but the difference $\delta f={\tilde f}-f$ between the reconstructed
distribution function and the true one is considered, when a small number of clouds is used, with
inter-spacing $\Delta_{\rm g}=\Delta_{x}=\Delta_{v}=0.2$. We used 10 iterations in Eqs.~(\ref{eq:van-citter})
and (\ref{eq:lucy})  and a cut-off value $\delta_{\rm min}=0.5\times 10^{-4}$. 
A larger number of iterations would not change significantly the level of
convergence. Since $\Delta_{\rm g}$ is of the same
order of ${\cal R}_{\rm apo}$, aliasing effects are rather significant, particularly
for Lucy where they reach a few percent magnitude, 
because of the positivity constraint intrinsic to this algorithm. Lower panels of the figure
are the same as middle panels, but a larger number of clouds was used, with inter-spacing
$\Delta_{\rm g}=0.05$ now small compared to ${\cal R}_{\rm apo}$. As a result, aliasing effects
are much less significant, of the order of $0.001-0.002$ and the difference between
Lucy and Van-Citter has decreased. This shows that, to sample
correctly variations of $f$ over some length scale $\ell$, we need $\Delta_{\rm g}$ to be sufficiently
small compared to $\ell$, typically $\ell/\Delta_{\rm g} \sim$ a few unities. Reversely, we see that
in runtime, the resampled $f$ can be fully trusted only at coarse graining scales of the order
a few $\Delta_{\rm g}$'s. 

%
%
%
\subsubsection{Filtering small scale noise}
\label{sec:weightreg}
%
%
After using either Van-Citter or Lucy algorithms, one
obtains from interpolation 
(\ref{eq:ffromspl}) a function ${\tilde f}$ which by definition reproduces 
at best the true values of $f$ at the sampling points $({x}_{\rm G}^i,{v}_{\rm G}^i)$, given
some convergence criteria. We just discussed problems of aliasing, which are intrinsic
to the method and cannot be really avoided. They can be 
only reduced by using e.g. refinement procedures
discussed later, or just by decreasing cloud inter-spacing. However, it is possible
to reduce defects discussed in \S~\ref{sec:funcgchoice}, namely small scale variations
due to the finiteness of phase-space sampling (upper left panel of Fig.~\ref{fig:figchoicecloud}) and
discontinuities due to the truncature of the clouds (upper right
panel of Fig.~\ref{fig:figchoicecloud}), to warrant smoothness of the
sampled distribution function. This is indeed necessary to avoid propagating this intrinsic noise
during runtime, when a large number of reinterpolations is performed. 

To minimize these defects, we perform a weighting different from 
Eq.~(\ref{eq:ffromspl}):
\begin{equation}
f({x},{v})=\frac{W}{V} 
\frac{\sum_i G({x}-{x}_{\rm G}^i,{v}-{v}_{\rm G}^i) M_i}
{\sum_i G({x}-{x}_{\rm G}^i,{v}-{v}_{\rm G}^i)},
\label{eq:fsmoothed}
\end{equation}
\begin{figure}
\centerline{\epsfig{file=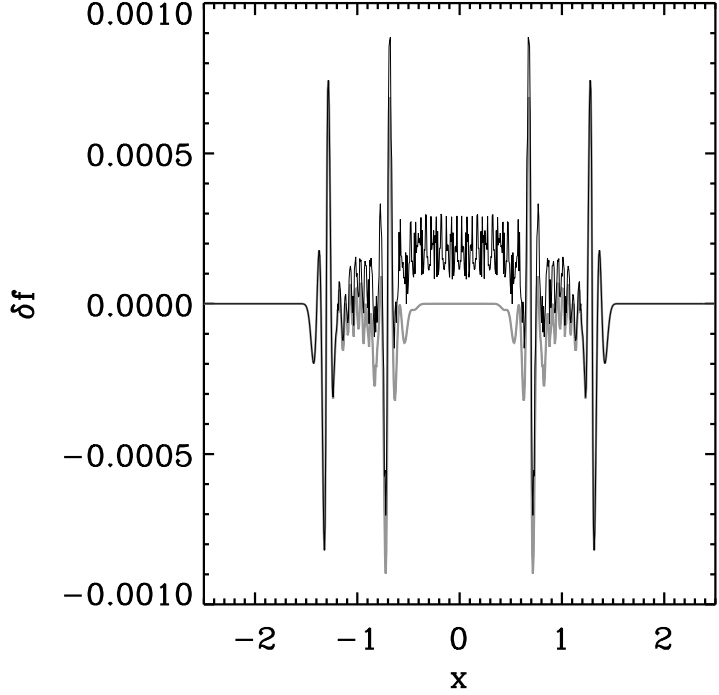,width=10cm}}
\caption[]{The difference $\delta f({x},0)$ as a function of ${x}$,
between reconstructed (with Van-Citter) and true distribution function,
whether Eq.~(\ref{eq:ffromspl}) (thin black curve) or Eq.~(\ref{eq:fsmoothed}) is
used to compute ${\tilde f}$. The function $f$ to sample is the same as in 
Fig.~\ref{fig:vancitlucy} and the resolution is the same as in lower
panels of this figure.}
\label{fig:weightingproc}
\end{figure}
where $W = \sum_{k,l} G(\Delta_{x} k, \Delta_{v} l) \simeq 1$ to ensure proper
normalization. This weighting procedure is nearly equivalent to 
Eq.~(\ref{eq:ffromspl}), since we expect the function $\sum_i G({x}-{x}_{\rm G}^i,{v}-{v}_{\rm G}^i)$
to present small variations with $({x},{v})$
as shown by upper right panel of Fig.~\ref{fig:figchoicecloud}. However,
it has the advantage of smoothing quite efficiently small scale noise
as illustrated by Fig.~\ref{fig:weightingproc}. For instance, in regions where $f=$constant,
it recovers exactly the true value of $f$ within these regions (with
a possible offset due to uncertainties in the convergence of the reconstruction). 
Hence, our weighting scheme would give exactly $f=1$ everywhere in
Fig.~\ref{fig:figchoicecloud}. 

In what follows, when a new mapping with round clouds is necessary
to resample an existing cloud distribution, 
we shall use Eq.~(\ref{eq:fsmoothed}) to compute $f$ from the old 
set of clouds and Eq.~(\ref{eq:ffromspl}) to compute ${\tilde f}$
from the new set of clouds in iterative procedures 
(\ref{eq:van-citter}) and (\ref{eq:lucy}). We decided to do so in the last
case to avoid introducing a bias in the new cloud masses estimate. 

%
%
\subsubsection{Enforcing basic conservations}
\label{sec:massconv}
%
An intrinsic property of deconvolutions methods such as Lucy's
or Van-Citter's is that they conserve very well mass. 
However,  we will have in practice to
perform many remaps of the distribution function: 
any small but systematic deviation
from mass conservation might have, on the long term, dramatic
consequences. As a result, it might be necessary to enforce 
mass conservation, but this is not easy without adding diffusion. 
In order to reduce as much as possible this latter we proceed
as follows. Suppose that the total mass of the system should
be equal to $M$, and let be ${\tilde M}=\sum_i M_i$ the total
mass obtained from the reconstructed clouds. If $M > {\tilde M}$, 
we compute the cumulated positive mass residues:
\begin{equation}
\Delta M_{+}  =  \sum_{i,\delta_i^{\rm last} > 0} \delta M_i,
\end{equation}
where $\delta M_i=\delta_i^{\rm last} V$
for Van Citter algorithm and $\delta M_i=M_i \delta_i^{\rm last}$ for Lucy algorithm:
these quantities correspond to the remaining uncertainty 
on the mass determination for each cloud after last iteration, $n=$last. 
Then, for each cloud having $\delta_i^{\rm last} > 0$, we increase
its mass by a factor $\alpha\ \delta M_i$ with $\alpha=(M-{\tilde M})/\Delta M_{+}$.
We proceed similarly if $M < {\tilde M}$, but by considering 
\begin{equation}
\Delta M_{-}  =  \sum_{i,\delta_i^{\rm last} < 0} \delta M_i.
\end{equation}
Our mass conservation scheme should minimize as much
as possible diffusion effects, since the correction calculated for each cloud is
proportional to the uncertainty on its mass determination. Furthermore, only
clouds for which the mass should be increased (or decreased) in the direction
of  $M - {\tilde M}$ are modified. Note 
that if $\alpha > 1$, it means that our correction will tend
to be above the residues. We can clearly fear for diffusion in that case.\footnote{Also, 
it can be possible but rather unlikely that, e.g.,  $M > {\tilde M}$ and 
$\Delta M_{+}=0$. It is not worth enforcing
mass conservation in that case, since the reconstruction itself is probably already
very biased due e.g. to strong aliasing effects.} Once total mass conservation
is taken care of, we can pay attention to total momentum,
\begin{eqnarray}
  P_v\equiv \sum_i M_i {v}_{\rm G}^i,
\end{eqnarray}
that we always force to be zero by appropriate corrections
of the velocities, if needed, since there is no external force exerted on
the system we consider.

\subsection{Solving the one dimensional Poisson equation.}
\label{sec:poissonsolver}
%
In this section, we explain in details how we solve Poisson equation.
To do that, one needs first to estimate local projected density, which
is rather simple with our Gaussian clouds, as discussed in \S~\ref{sec:projectedensity}. 
Then one has to estimate the force exerted on each cloud, as well as its slope.
To do that, we propose two methods, a tree-code approach based on decomposition
of space on a binary tree (\S~\ref{sec:treecode}) 
and a ``cloud in mesh'' (CM) approach based on sampling of space with a regular grid
(\S~\ref{sec:CMcode}).
%
\subsubsection{Calculation of projected density}
\label{sec:projectedensity}
%
The projected density can be obtained for any value of $x$ 
by summing all the individual contributions
of the clouds. For a given Gaussian cloud, with elliptic
basis and arbitrary orientation, integration over velocity
space of Eq.~(\ref{eq:defG2}) is in fact simple. With
the help of Eq.~(\ref{eq:lambda2evol}),  we obtain
\begin{equation}
\rho(x,t)=M \sqrt{\frac{\pi}{\Lambda_2(t)}} \ 
\exp{\left(-\frac{2 C}{\Lambda_2(t)} \ x^2 \right)}
\label{eq:rhoproj}
\end{equation} 
for a cloud of mass $M$.
In this equation, we thus neglect the truncature of the cloud over velocity space, but this
should not have any significant consequence.
Indeed, in practice, to make sure that the total projected mass of each truncated 
cloud is equal to its
true mass, we renormalize Eq.~(\ref{eq:rhoproj}) by a factor very close to unity. 
%
\subsubsection{Practical implementation: a tree code.}
\label{sec:treecode}
%
Keeping in mind that we want to generalize the code to 2D and 3D, we experimented an implementation
which can be, in principle, easily extended to higher number of dimensions. 
The Poisson equation simply writes, in the appropriate units
\begin{equation}
\frac{\partial^2 \phi}{\partial {x}^2}=2\rho.
\end{equation}
The approximation taken in this paper thus consists in simply 
assuming that $\rho=$ constant within a cloud. The force,\footnote{In this paper, we  indistinctly liken
the quantity $-\nabla \phi$ to a force or an acceleration.}  given by
\begin{equation}
-\frac{\partial \phi}{\partial {x}}=\int_{ y \geq {x}} \rho({y},t) d{y}
                                        -\int_{y \leq {x}} \rho({y},t) d{y}=
                                        M_{\rm right}-M_{\rm left},
\end{equation}
is then locally fitted by a straight line within the cloud.

\begin{figure}
\centerline{\mbox{\epsfig{file=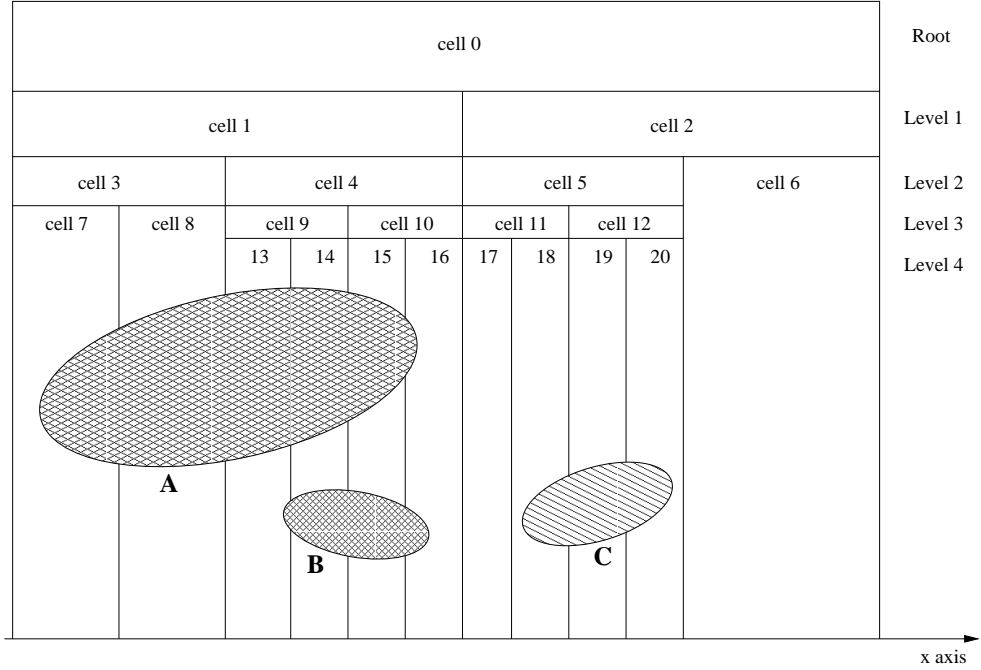,width=13.0cm}}}
\caption[]{Sketch of our tree structure used to compute the forces within each cloud.
Here, we consider the case $N_{\rm s}=3$, i.e. each
cloud must intersect with at least 3 cells of the tree at a given refinement
level. Note that for computing the forces for cloud A, only cells 7,8,9 and 10 of level 3 of refinement
will be used, even if more refinement is available due to cloud B.}
\label{fig:tree}
\end{figure}

\begin{figure}
\centerline{\mbox{\epsfig{file=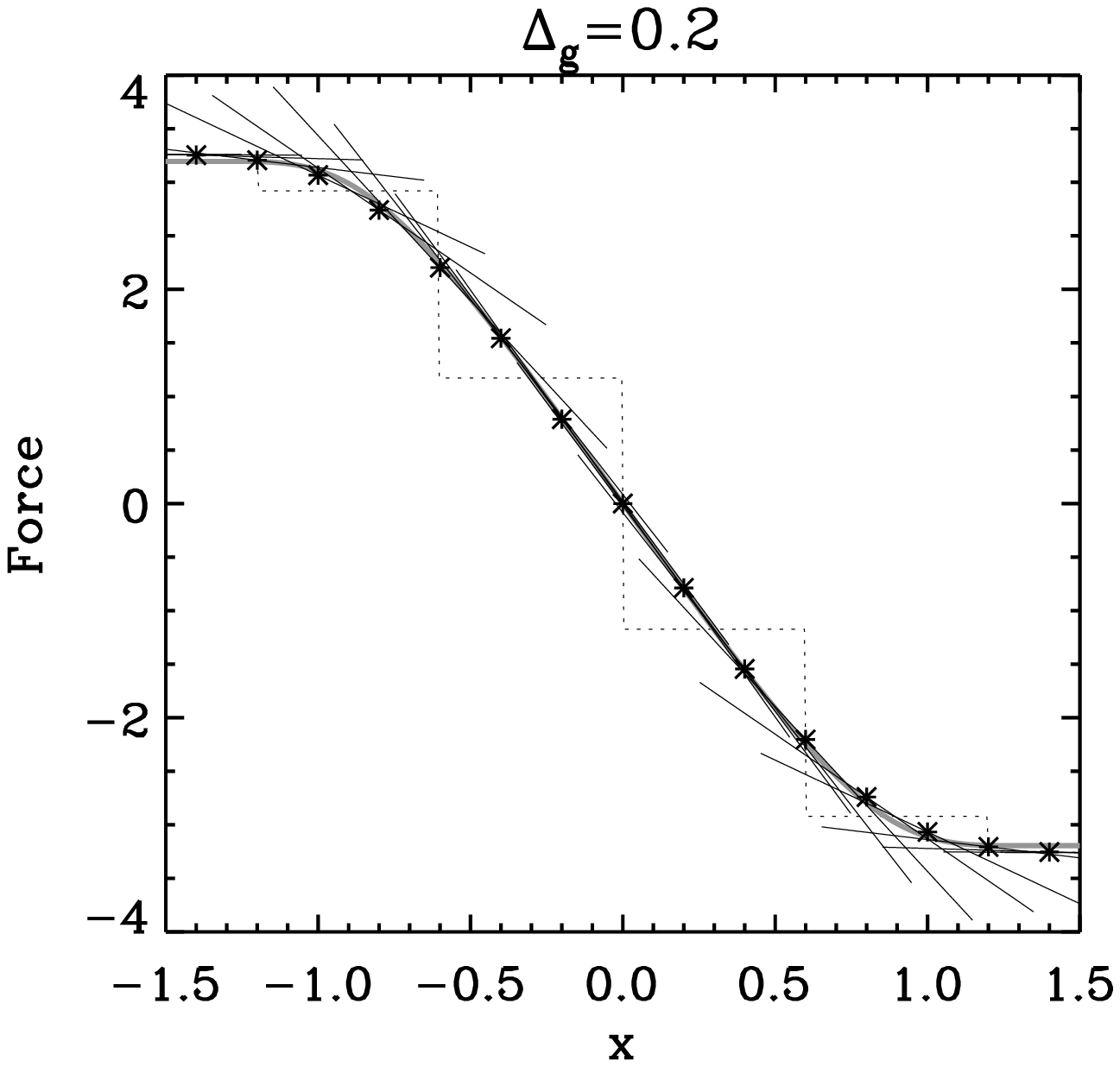,width=8cm}
\epsfig{file=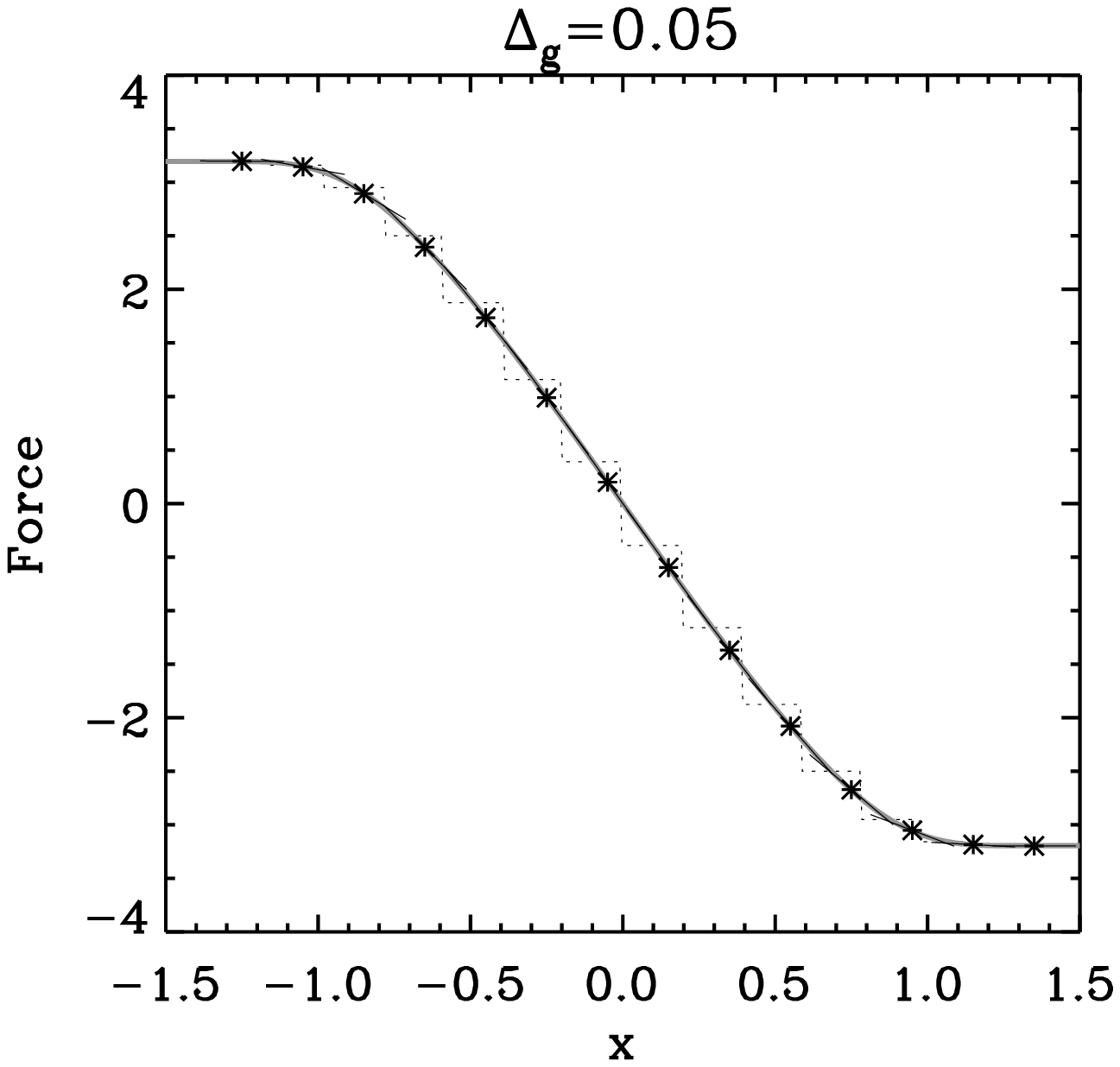,width=8cm}}}
\caption[]{The calculated force for the examples considered in 
Fig.~\ref{fig:vancitlucy}. The left and the right panels 
correspond to the case $\Delta_{\rm g}=0.2$  and $\Delta_{\rm g}=0.05$ respectively.
On each panel, there is a smooth thick grey curve: it corresponds to the exact force derived
directly from integrating Eq.~(\ref{eq:finitophat}). There is a thin step like dotted curve:
it corresponds to the force computed for each cell of the binary tree at the resolution needed
to sample each cloud with a least $N_{\rm s}$ cells. Here we took $N_{\rm s}=2$, and in practice
we obtain $N_{\rm inter}=2$ or 3. There are stars nearly perfectly
fitting the smooth thick grey curve: they give the force calculated at the center
of each selected cloud for doing the figure (a subset of all the clouds). 
A segment passes through each of these stars: it gives
our weighted least square fit $F_{\rm fit}({x})=-2 \alpha_0 {x}-\alpha_1$, which
trivially reduces to the line passing through the two sampling points when $N_{\rm inter}=2$.  
It is interesting to notice that the estimate
we obtain for the force at the cloud centers is very accurate, although one might notice
a slight offset between symbols and the thick smooth grey line on left panel. The projected length
of each segment on ${x}$ axis is equal to the projected length of each cloud (up to cut-off).
As we see, significant deviations from our local quadratic assumption are expected for the potential
on left panel. These deviations are much less visible on the right panel. 
The typical weighted deviation calculated from Eq.~(\ref{eq:errforce}) by using $N_{\rm s}=8$
gives $\Delta F/F \simeq 0.03$ and $0.002$ for left and right panel respectively.  }
\label{fig:forcetest}
\end{figure}

To estimate the force, we decompose the ${x}$ axis hierarchically on a binary-tree
(Fig.~\ref{fig:tree}), by dividing successively segments into two parts of equal length, until
each projected cloud intersects with at least $N_{\rm s}$ tree cells. Of course, there
is an uncertainty on the real number of intersecting cells, $N_{\rm inter} \geq N_{\rm s}$.
It should be at least $N_{\rm s} \geq 2$ and in fact large compared to that if we want to estimate
errors on the forces. To compute the force within a cloud,
given the number $N_{\rm inter}$ of tree cells intersecting with its projection on ${x}$ axis,
we store (i) the force $F_i$ exerted on each cell center, ${x}_C^i$, and (ii) a weight $w_i$ proportional
to $\rho({x}_C^i-{x}_{\rm G})=\rho({x_C^*}^i)$ as given by Eq.~(\ref{eq:rhoproj}). This 
weight is of course more important when the cell center is close
to the projected cloud center. Then, given a list of $(x_i,F_i)$, $i=1,N_{\rm inter}$, we use a simple
weighted least square fit to adjust the force by a local straight line 
$F_{\rm fit}({x})=-2 \alpha_0 {x}-\alpha_1$
[see Eq.~(\ref{eq:quadapp})]. Our estimator for the error on the force then reads
\begin{equation}
\left( \frac{\Delta F}{F_{\rm max}} \right)^2=
\frac{1}{M^2\sum_{i=1}^{N_{\rm inter}} w_i} \sum_{i=1}^{N_{\rm inter}} 
w_i \left[ F_i-F_{\rm fit}({x}_C^i) \right]^2.
\label{eq:errforce}
\end{equation}
This weighted quantity roughly quantifies the deviation of the force from a straight line
within the cloud. It is renormalized by the maximum force, $F_{\rm max}$, which in the units chosen here
is equal to the total mass $M$ of the system. The error on the force is indirectly related to the
projected size of the cloud along ${x}$ axis. If the cloud gets considerably elongated,
the error on the force will become larger in a fixed potential. Therefore, one might use 
$\Delta F/F_{\rm max} \leq$ some small value as a criterion to decide whether a remap of the distribution
function with a new set of round clouds is necessary.
However, as we shall discuss more in detail in \S~\ref{sec:lagrangeremap}, 
what really counts is the cumulated error,
\begin{equation}
E_{\rm cum}\equiv \int \Delta F(t)dt.
\label{eq:ecumul}
\end{equation}
To make sure that the force is estimated consistently
the same way for all the clouds (in particular to preserve approximately local smoothness properties),
we stop refining the tree locally as soon as $N_{\rm inter} \geq N_{\rm s}$, 
even if the tree has many branches due to much smaller
clouds at the same location ${x}$, as shown by Fig.~\ref{fig:tree}.
Since we chose
here $R_{\rm max}=4 R$ for the truncature of the cloud, one might think that 
$N_{\rm s}=8$ should be taken, to have at least two sampling cells per
typical length-scale, $R$, of the clouds.  However, we noticed that 
$N_{\rm s}=2$ is enough to have a very accurate determination of the force with
our choice of $R/\Delta_{x}=\sqrt{2}/2$. This is illustrated by Fig.~\ref{fig:forcetest}. To estimate
the instantaneous error on the force, a larger value of $N_{\rm s}$ is necessary, 
typically $N_{\rm s}\ga 8$, while $N_{\rm s} \ga 4$ is enough to
compute the cumulated error as used in \S~\ref{sec:lagrangeremap}.

%
\subsubsection{Alternate practical implementation: ``CM'' code}
\label{sec:CMcode}
%
The results presented in this paper use a tree-code. The tree-code has the advantage of being
rather flexible and easily generalizable to higher number of dimensions. 
It also represents a natural ground
for adaptive refinement in phase-space, as we shall see later. 
To simplify the approach, it is also possible to assume fixed resolution
in space, which allows one to use FFT or relaxations methods 
to solve Poisson equation, similarly as in particle-in-mesh codes (PM). However,
at variance with PM, a much more accurate  ``cloud-in-mesh'' interpolation (CM) is performed. 
In our 1D case, the resolution of Poisson equation is rather trivial.
The projected density of the clouds is simply calculated on a grid of
inter-spacing $\Delta_{\rm s}$ in such a way that the number of intersecting grid
sites is always larger than $N_{\rm s}$ for any cloud. 
The force is then estimated exactly
the same way as discussed in \S~\ref{sec:treecode}. In the current implementation, 
this fixed resolution code is much faster than the tree-code, especially at the moment
of remap, where a fast convolution method can be used to estimate the phase space distribution
function (Appendix~\ref{sec:fastconvmet}). However, note that the tree-code part can still be 
considerably optimized for the remap part, since the fast convolution method can in principle 
be generalized to a non structured grid. 
%
\subsection{Time stepping implementation and diagnostics}
\label{sec:timstd}
%
In this section, we discuss our run time implementation: a second order
predictor corrector, as
described in \S~\ref{sec:predc}. This makes our approach
second order both in time and in space. The determination of the
slowly varying time step is discussed in \S~\ref{sec:diags},
as well as other diagnostics, such as energy conservation.
Finally, \S~\ref{sec:lagrangeremap} examines the critical issue
of deciding when performing a new sampling of the distribution function
with a set of round clouds, by studying cumulated errors on the determination
of the forces. 
%
\subsubsection{Time stepping implementation: global predictor corrector}
\label{sec:predc}
%
In our approach, we need a number of parameters to describe completely a set
of clouds $i$, $i=1,\ldots,N_{\rm tot}$, interacting through gravity.
These parameters can be chosen as follows:
\begin{description}
\item[-] the mass of the cloud, $M_i$: this one does not change with
time, except when a remap with a new set of round clouds is performed;
\item[-] the position of the center of the cloud, ${x}_{\rm G}^i(t)$;
\item[-] the velocity of the center of the cloud, ${v}_{\rm G}^i(t)$;
\item[-] the acceleration of the center of the cloud, that we write $F_i(t)$;
\item[-] the parameter $\alpha^i_0(t)$ in Eq.~(\ref{eq:quadapp}), which is nothing
but the projected density at position $[{x}_{\rm G}^i(t),{v}_{\rm G}^i(t)]$;
\item[-] the function $w_i(t)=\sqrt{\Lambda^i_2(t)}$;
\item[-] the time derivative of function $w_i(t)$: ${Dw}_i(t)\equiv dw_i/dt$;
\item[-] the area of the elliptical section of the cloud divided by $\pi$: $A_i={\bar A}_i/\pi$:
this one does not change with time, except when a remap with a new set of round clouds
is performed.
\end{description}
Recall that the parameters $\beta_0^i$, $\beta_1^i$ and $\beta_2^i$ defining the shape
of the cloud are entirely determined by function $w_i$, its time derivative $Dw_i$, 
$\alpha_0^i$ and ${\bar A}_i$, namely, from Eqs.~(\ref{eq:beta0}), (\ref{eq:beta1}) 
and (\ref{eq:beta2}), with the help of Eq.~(\ref{eq:evolw}):
\begin{equation}
\beta_0^i  =  ({Dw}_i)^2+\frac{1}{(A_iw_i)^2}, \quad 
\beta_1^i  =  -2 w_i\ {Dw}_i, \quad
\beta_2^i  =  (w^i)^2.
\end{equation}
To evolve the clouds we use standard second-order predictor corrector
algorithm, which is known to preserve quite well simplecticity,
a feature essential in phase-space. To simplify the algorithm, we take
the same global time step for all the clouds.
Our run time implementation can be split up into six main parts:
\begin{enumerate}
\item[(a)]{\em Predictor step:}
\begin{eqnarray}
{x}_{\rm G}^i(t_{n+1/2}) &=& {x}_{\rm G}^i(t_{n})+\frac{1}{2} dt_n {v}_{\rm G}^i(t_{n}),\\
w_i(t_{n+1/2})&=&w_i(t_{n})+\frac{1}{2} dt_n Dw_i(t_n).
\end{eqnarray}
\item[(b)] {\em Force calculation:} the force is calculated as explained
in \S~\ref{sec:treecode} at the ``predicted'' positions
${x}_{G}^i(t_{n+1/2})$  within each cloud of ``predicted'' projected density 
given by Eq.~(\ref{eq:rhoproj}) using $\Lambda_2(t_{n+1/2})=[w_i(t_{n+1/2})]^2$.
This gives, for each cloud, the parameters $F_i(t_{n+1/2})$ and $\alpha^i_0(t_{n+1/2})$. 
\item[(c)]{\em Corrector step:}
\begin{eqnarray}
{x}_{\rm G}^i(t_{n+1}) &=& {x}_{\rm G}^i(t_{n+1/2}) + \frac{1}{2} dt_n {v}_{\rm G}^i(t_{n})
  +\frac{1}{2} dt_n^2 F_i(t_{n+1/2}), \\
{v}_{\rm G}^i(t_{n+1}) &=& {v}_{\rm G}^i(t_n) + dt_n F_i(t_{n+1/2}), \\
w_i(t_{n+1}) &= &w_i(t_{n+1/2})+\frac{1}{2} dt_n {Dw}_i(t_n) + \frac{1}{2} dt_n^2 
\left[ \frac{1}{A_i^2 w_i^3(t_{n+1/2})} - 2 \alpha_0(t_{n+1/2}) w_i(t_{n+1/2}) \right], \\
Dw_i(t_{n+1}) &= &Dw_i(t_n)+dt_n \left[ \frac{1}{A_i^2 w_i^3(t_{n+1/2}) }- 2 
\alpha_0(t_{n+1/2}) w_i(t_{n+1/2}) \right].
\end{eqnarray}
\item[(d)]{\em Outputs:} various quantities such as all the informations on the clouds,
the density in phase space, etc, can be output at this point.
\item[(e)]{\em Diagnostics: calculation of next time step, test for Lagrange remap,
energy conservation check:} as we are going to
describe more in detail below, we use 
a time-step constrained by the slope of the force, $-2 \alpha^i_0(t_{n+1/2})$.
The fact that the next time step, $dt_{n+1}$, depends on a quantity
calculated half a time step before does not affect significantly the quasi-simplecticity
of our integrator. The important thing here is that $dt_n$ should vary slowly with time.
At this point, we also test if it is necessary to remap the distribution function
with a new set of round clouds, due to the accumulation of deviations
from local harmonicity of the potential, as studied in \S~\ref{sec:lagrangeremap}.
\item[(f)]{\em Lagrange remap, if needed}, as explained in \S~\ref{sec:phasespacesampling}.
\end{enumerate}
Note importantly, thus, that our algorithm is second order both in space and in time.
It is quasi-simplectic in the sense that it would be time reversible, equivalent
to generalized leap frog, if the time step was constant.
%
\subsubsection{Diagnostics: ``Courant'' condition and energy conservation}
\label{sec:diags}
%
There is a list of diagnostics to perform during runtime. The most important
are those related to time stepping. 
What matters in our Lagrangian approach is that orbits
in phase-space should be sampled with sufficient number of points as well
as cloud shape variations during the trajectory. From the analysis of \S~\ref{sec:smallamposc},
that means, using Eq.~(\ref{eq:dynatime}),
\begin{equation}
dt \leq \frac{\pi{\cal C}}{\max_i\sqrt{2 \alpha_0^i}},
\end{equation}
where ${\cal C}$ is a ``Courant'' parameter small compared to unity.
In practice we find that we should have ${\cal C}\la 0.01$.
For all the simulations made in this paper, we took
${\cal C}=0.01$.

Another diagnostic is of course total energy conservation, which
writes, in our units,
\begin{equation}
{\cal E}_{\rm tot} \equiv \frac{1}{2} \int {v}^2 f({x},{v},t)\  d{x}\ d{v}
- \frac{1}{4}\int \left[ \left( \frac{d\phi}{d{x}} \right)^2 - M^2 \right]d{x}={\rm constant}.
\label{eq:conservener}
\end{equation}
This equation assumes that the system is finite of total mass $M$.
We added a term proportional to $M^2$ to compensate for the divergence in
the integral $\int (d\phi/d{x})^2 d{x}$ so that the constant in
Eq.~(\ref{eq:conservener}) is finite.

One might use instantaneous deviations from energy conservation to constrain
the time step, but we chose here to test energy conservation only as a consistency 
check. We shall indeed see in  \S~\ref{sec:simuset} 
that energy is very well conserved in our code. 

Finally notice that mass is conserved by definition, except when a resampling
of the distribution is performed with a new set of round clouds. During this
step, mass conservation is enforced as explained in \S~\ref{sec:massconv},
as well as total momentum.
%
\subsubsection{Lagrange remap}
\label{sec:lagrangeremap}
%
The assumption for a quadratic local potential is in general always violated to
some extent during runtime since it corresponds to variations in the projected
density $\rho({x})$ (\S~\ref{sec:valid}). 
What counts, as already mentioned in \S~\ref{sec:treecode},
is the cumulated error on the force, which translates in clouds with
increasingly wrong ellipticity parameters.  This effect, even if always small during 
a time-step (see \S~\ref{sec:valid}), cumulates with time. In particular, the
clouds get inevitably spuriously elongated and some ``hairy'' structure
appears during evolution (Fig.~\ref{fig:showerr}). This is why, we have
at some point to remap the distribution function with a new set of round clouds, an operation
that we call ``Lagrange remap''.

To illustrate this point, we consider here the same example as in \S~\ref{sec:valid}:
the evolution of a distribution function initially Gaussian, i.e. given by
\begin{eqnarray}
  f({x},{v})  &=&  {\bar \rho}\ \exp\left( -\frac{1}{2}\frac{{x}^2+{v}^2}{h^2} \right),   
   \quad {x}^2+{v}^2 \leq {\cal R}^2 , \nonumber \\
  f({x},{v})  &=&  \frac{1}{2}  {\bar \rho}\ \exp\left( -\frac{1}{2}\frac{{x}^2+{v}^2}{h^2} \right)
   \left\{ \cos\left[ 
   \frac{\pi}{2} \left(\sqrt{{x}^2+{v}^2} - {\cal R}\right)/{\cal R}_{\rm apo} \right]+1 \right\},
    \quad {x}^2+{v}^2 \leq {\cal R}+2 {\cal R}_{\rm apo}.  \label{eq:finigaussian}
\end{eqnarray}
This initial profile will be used for the set of simulations studied in \S~\ref{sec:simuset}. 
The same apodization than in Eq.~(\ref{eq:finitophat}) is performed to regularize the function at its edges:
Eq.~(\ref{eq:finigaussian}) approaches Gaussian initial conditions only if $h \ll {\cal R}$.
For the simulation considered here, we take $h=0.2$,  ${\bar \rho}=1/(2\pi h^2)$,
${\cal R}=1$, ${\cal R}_{\rm apo}=0.1$.
The cloud size is $R=h/10=0.02$, similarly as in \S~\ref{sec:valid} and
the inter-cloud spacing is thus given by $\Delta_{\rm g}=\sqrt{2}R=0.028$. To compute the force
and the error on it [Eq.~(\ref{eq:errforce})],
we take $N_{\rm s}=8$ as advocated in \S~\ref{sec:treecode}.

\begin{figure}
\centerline{\mbox{\epsfig{file=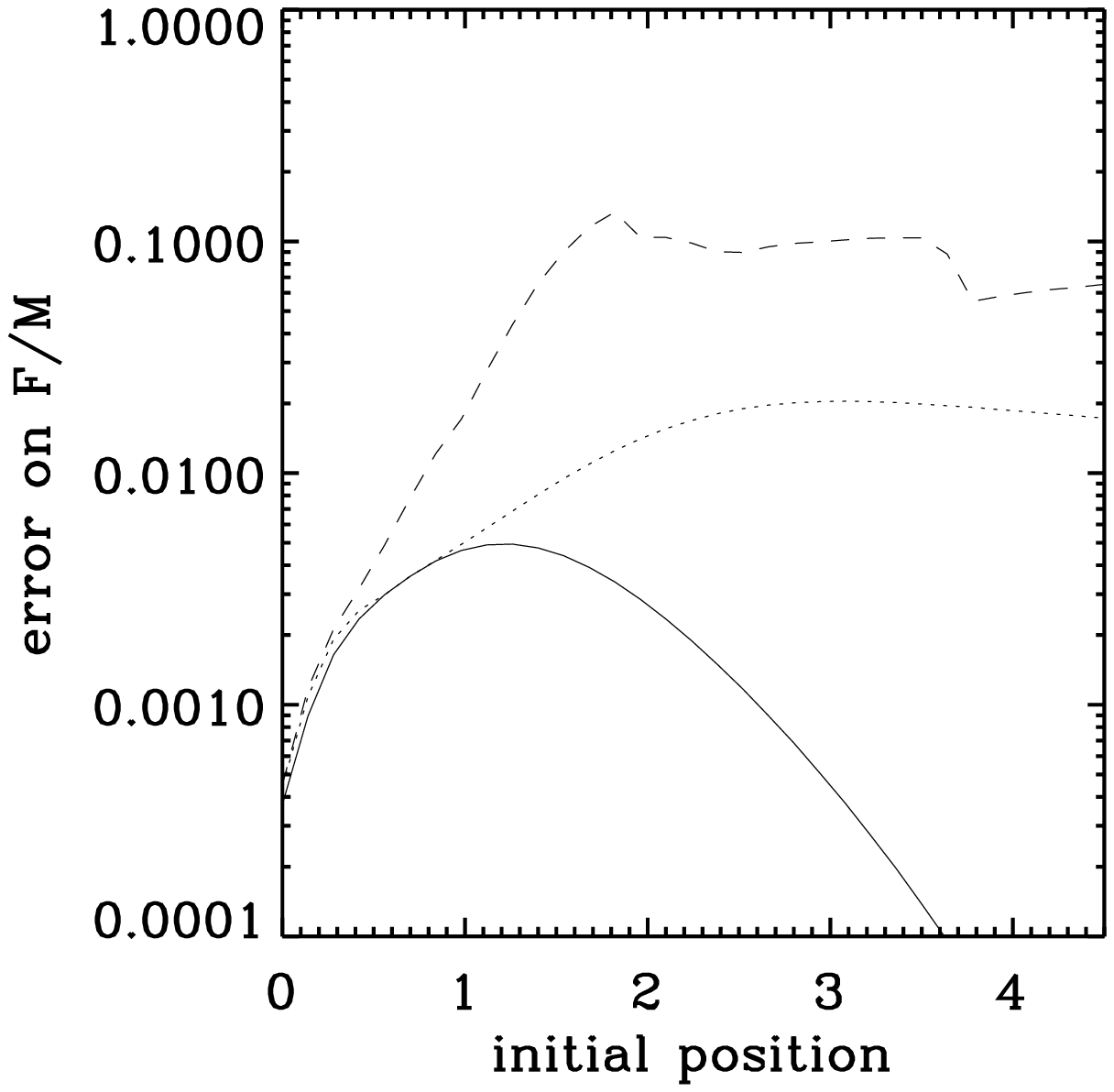,width=6cm}
                  \epsfig{file=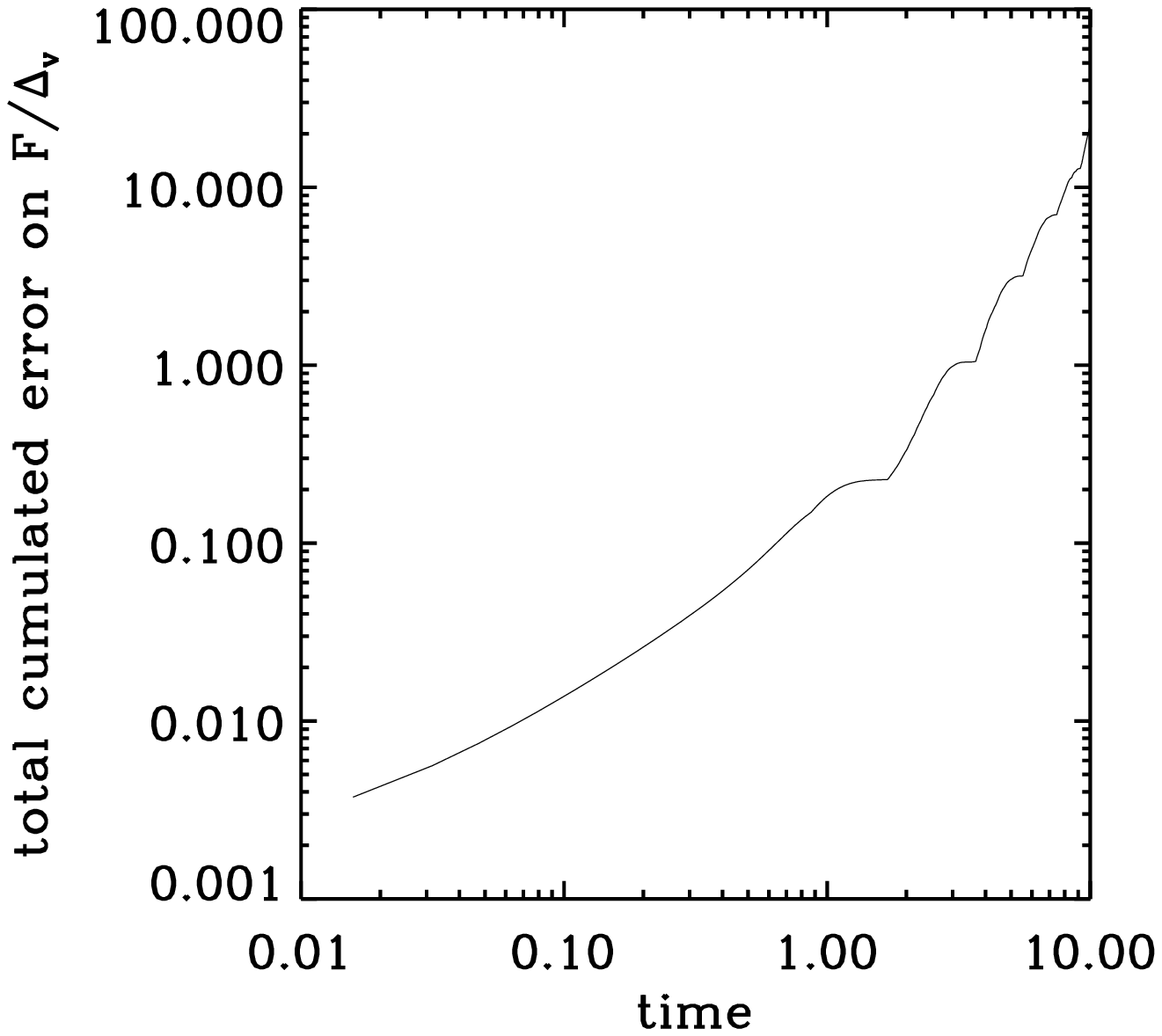,width=6cm}
                  \epsfig{file=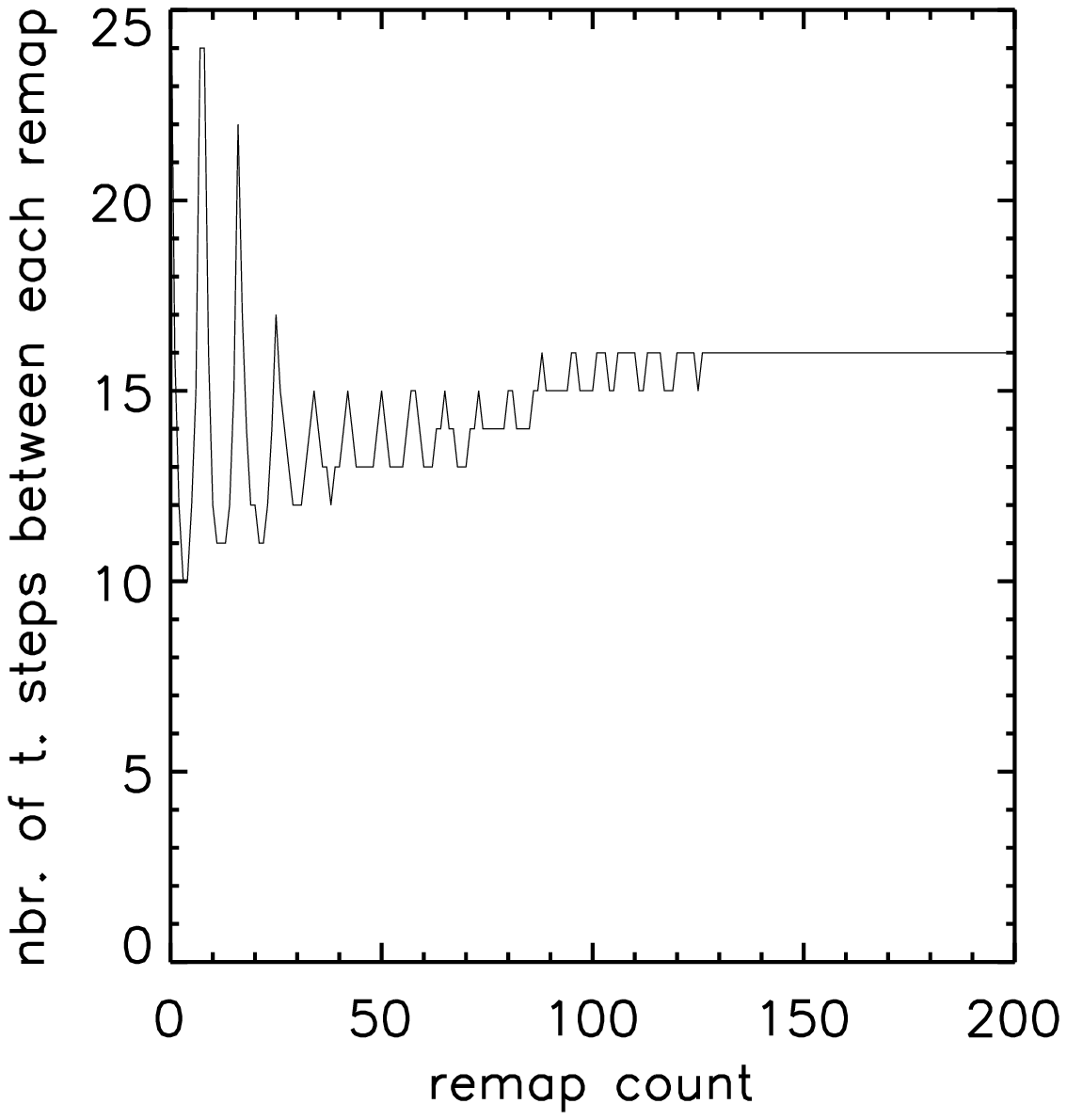,width=6cm}}}
\caption[]{{\em Left panel:} the maximum 
instantaneous error, $\Delta F/F_{\rm max}$, on force calculation 
[Eq.~(\ref{eq:errforce})] as a function of initial cloud position, measured
in a simulation without remapping. 
The initial conditions considered here are a truncated Gaussian of size ${\cal R}$
as given by Eq.~(\ref{eq:finigaussian}). The initial position, ${x}_{\rm ini}$, of the
cloud is expressed in units of $h$. To reproduce a case similar to the one studied
in Fig.~\ref{fig:maxdev}, the cloud size $R$ is given by $R/h=0.1$. 
Three dynamical times are considered, $t=0.5$ (solid), $1.5$ (dots) and $6.0$ (dashes). 
It is difficult to make an exact comparison to Fig.~\ref{fig:maxdev}, given that fact
that our estimate of $\Delta F/F_{\rm max}$ is a quadratic dispersion. However, the
general order of magnitude found on the error agrees with the results of \S~\ref{sec:valid}. The dotted
curve which corresponds approximately, dynamically, to the parameters chosen to make
Fig.~\ref{fig:maxdev}, agrees well, at least qualitatively with Fig.~\ref{fig:maxdev}.
{\em Middle panel:} the maximum cumulated error, $E_{\rm cum}^{\rm max}/\Delta_{v}$,
as a function of time. A remapping should be performed each time this error exceeds some threshold.
{\em Right panel:} the number of time-steps, $n_{\rm remap}$, between each remap as a function of
remap count, when the condition (\ref{eq:critlagrem})
is fulfilled, with $E_{\rm max}=0.05$. The simulation used to make this plot is exactly
the same as the one used for left and middle panels, except that Lagrange remap was
enforced. The simulation was stopped after 3000 time steps, at $t \simeq 39$, corresponding
to 35 dynamical times according to Eq.~(\ref{eq:dynatime}) with $\rho_0={\bar \rho}$. 
Note that the variations of $n_{\rm remap}$ decrease with time:
$n_{\rm remap}$ seems to converge progressively to some fixed value,
as expected, while the system relaxes to a stationary state.}
\label{fig:anaerr}
\end{figure}

\begin{figure}
\centerline{\mbox{\epsfig{file=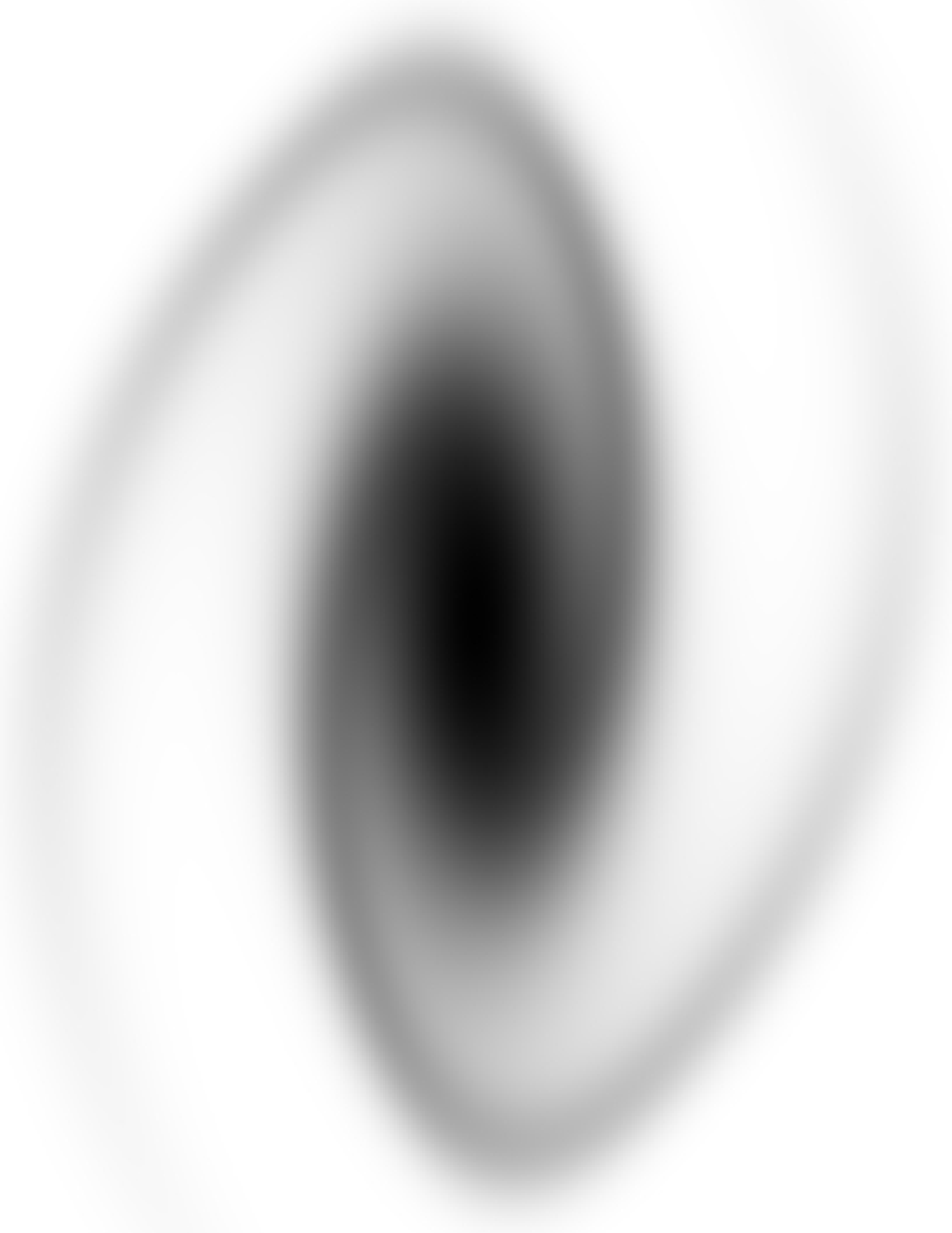,width=9cm}
\epsfig{file=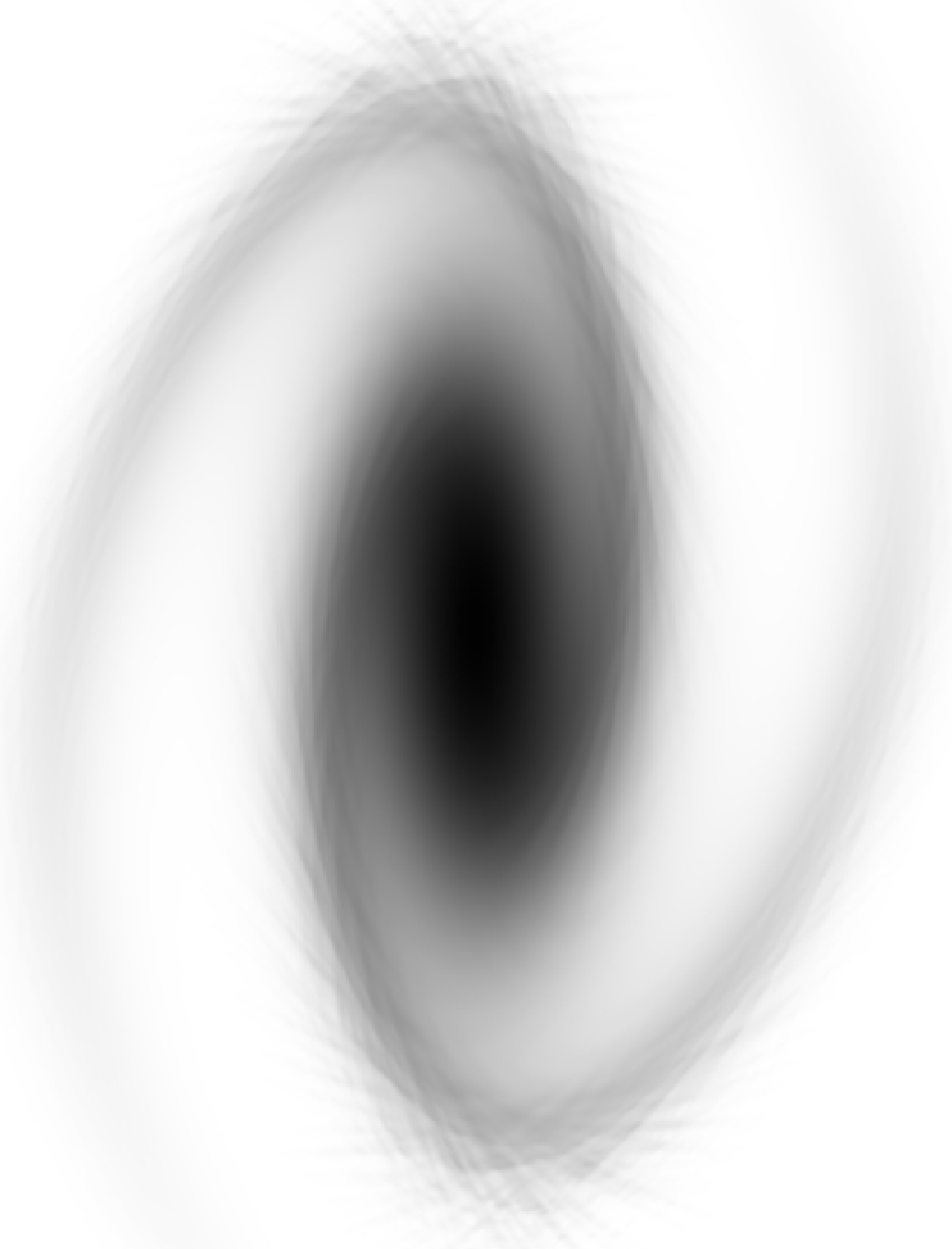,width=9cm}}}
\centerline{\mbox{\epsfig{file=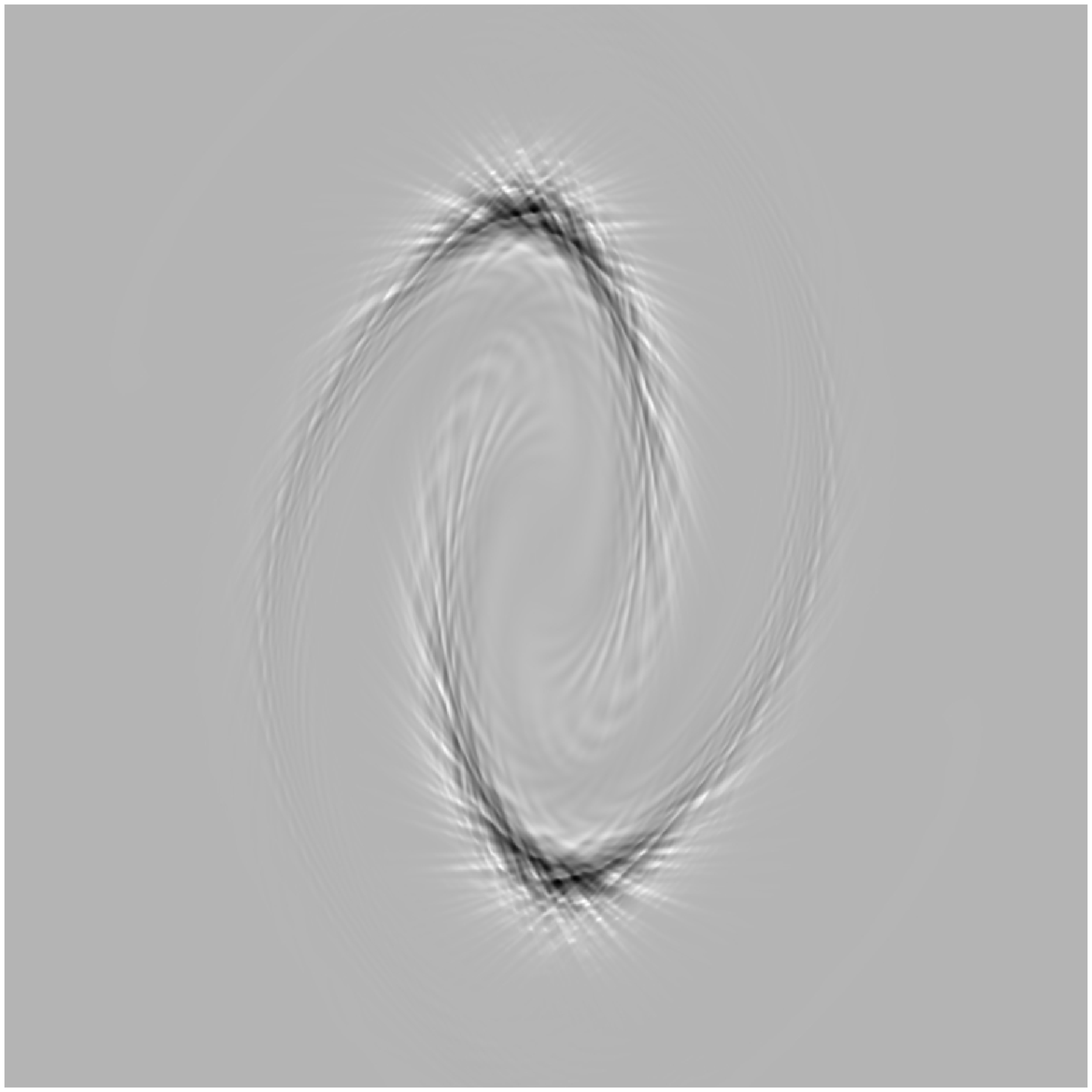,width=9cm}}}
\caption[]{Apparition of a hairy structure after a few dynamical times. 
The simulation considered is the same as the one as in the two left panels of Fig.~\ref{fig:anaerr},
at $t=6$. The top left panel gives the ``correct'' answer, i.e. $f({x},{v})$
obtained by remapping the density field with new round clouds using
condition (\ref{eq:critlagrem}) with $E_{\rm max}=0.05$. The top right
panel shows the ``wrong'' answer obtained without remapping.
The bottom panel displays the difference between left and right panels, which shows the
hairy structure appearing due to the divergence of the axis ratios of the clouds.}
\label{fig:showerr}
\end{figure}

Left panel of Fig.~\ref{fig:anaerr} gives the maximum instantaneous error on the force
as a function of initial position, ${x}_{\rm ini}$.
It shows that the deviations from local quadratic approximation globally increase with time, as expected.
What is important here, though, is the accumulation, $E_{\rm cum}$, 
of little kicks at each time-step due these deviations, as given by Eq.~(\ref{eq:ecumul}).
This quantity, shown as a function
of time in middle panel of Fig.~\ref{fig:anaerr},
is homogeneous to a velocity. It should remain small compared to velocity resolution:
\begin{equation}
\frac{E_{\rm cum}^{\rm max}}{\Delta_{v}} \leq E_{\rm max} \ll 1.
\label{eq:critlagrem}
\end{equation}
If condition (\ref{eq:critlagrem}) is violated, a new sampling of the distribution function
with round clouds should be performed.
We notice that a typical value of the threshold, $E_{\rm max}=0.05$, i.e. a 5 percent cumulated
error, corresponds, as expected, to a fraction of dynamical time, $t \simeq 0.4$. 
Interestingly, but not surprisingly, if condition (\ref{eq:critlagrem}) is enforced, 
the number $n_{\rm remap}$ of time-steps between each remap is pretty stable,
as shown on right panel of Fig.~\ref{fig:anaerr}.
For our Courant parameter choice ${\cal C}=0.01$ and for $E_{\rm max}=0.05$ we find $n_{\rm remap}\simeq 15$.
Its variations decrease with time, as expected, while the system converges to a stationary state. As a result,
it is a very good approximation to keep $n_{\rm remap}$ fixed, which is quite useful, since it is not necessary
in that case to estimate the cumulated error, and as a consequence, $N_{\rm s}=2$ can be taken to speed up considerably 
the calculation of the force, without any significant loss in accuracy. 

One would thus be tempted to fix $n_{\rm remap}$ as a function of our Courant condition using the following formula,
\begin{equation}
n_{\rm remap}({\cal C})=\frac{n_{0.01}}{({\cal C}/0.01)},
\end{equation}
with, typically, $n_{0.01} \simeq 15$ to produce a cumulated error on the force of the order of 5 percent.
This estimate is only valid for the initial conditions considered here, but according to arguments
of \S~\ref{sec:valid}, the same result should be roughly found for other configurations: with ${\cal C}\simeq 0.01$,
a Lagrange remap should be performed every 10-15 time-steps.

One might be worried that a cumulated error of a few percent per Lagrange remap is too much: accumulation of errors
after a few thousand remaps might still be too large. However, on has to take into account the fact 
that the cumulated error, as given by Eq.~(\ref{eq:ecumul}), 
does not take into account possible global cancellations during the dynamics. In particular, 
if a given symmetric potential is fixed, we see that a cloud plunged in the right side of the
potential will be the object of next to quadratic distortions opposite to those in the other side: they will
cancel each other, which makes the solution rather stable on the long term, even if our absolute
cumulated error becomes very large. This is illustrated by Fig.~\ref{fig:showerr}, which compares
the results obtained for the simulation of Fig.~\ref{fig:anaerr} without remapping 
at $t=6$ to the exact solution. The ``hairy'' effect is quite visible, but the solution 
is rather close to the correct answer at the coarse level.

%
\section{Some applications}
\label{sec:simuset}
%
In this section, we show how our tree-code performs for three different kinds
of initial conditions. Similar results should be obtained for the CM code,
so we do not feel necessary to show them here. 

The first model, considered in \S~\ref{sec:stationary},
is a stationary profile, verifying 
$\partial f/\partial t=0$. This is a crucial test: the code should be able
to maintain a stationary profile during many dynamical times to a very good accuracy. 
We shall see that our code passes this test with great success.

The second model, considered in \S~\ref{sec:gaussian}, 
is a Gaussian distribution function: this kind
of very smooth initial conditions is well adapted to our method, that should
be able to follow their evolution quite accurately until details appearing during
runtime become too small to be resolved by our sampling clouds.
We shall see that at this point our numerical solution still gives a very
good coarse level version of the full resolution one.

The third model, considered in \S~\ref{sec:tophat},
is a top hat apodized by a cosine. The evolution
of such a kind of distribution function 
allows us to test directly 
to which extent our code can maintain a region
with $f=$constant. In particular, effects
of aliasing can be quantified easily. 

For each of the models considered above, we perform
one ``low resolution'' simulation with 
$\Delta_{x}=\Delta_{v}=0.02$, and one ``high resolution'' simulation
with $\Delta_{x}=\Delta_{v}=0.005$. We run these simulations  
using a Courant condition ${\cal C}=0.01$ and 
a Lagrange remap every 15 time steps, following the conclusions of the analyses above. 
To perform the resampling, we use Lucy algorithm with 10 iterations exactly, whatever
the residue obtained between the resampled distribution and the previous one, following
discussions of \S~\ref{sec:deconv}. To maintain the domain of contributing
values of $f$ finite, we proceed as explained in the paragraph just after 
Eq.~(\ref{eq:van-citter}) with $\delta_{\rm min}=5.10^{-5}$.
In all the cases, we aim to evolve the system during approximately 30 dynamical times
(corresponding to 60 orbital times), a goal that we achieve in the following 
way: we set up initial conditions 
such that $\rho_0$ is of order two in  Eq.~(\ref{eq:dynatime}) 
for the top-hat and the Gaussian case, and evolve them up to $t=100$.
For the stationary solution, the choice of $\rho_0$ is of order unity and we
evolve the simulation up to $t=150$.
With our Courant condition, this amounts finally to
approximately 7000 time steps for the stationary simulation and 8000 time steps
for the simulations with Gaussian and top hat initial conditions, corresponding to
a large number of remaps, about 470 and 530, respectively.

This way we are able to quantify the
limitations of our approach, namely coarse graining and
aliasing effects due to the finite size of the sampling clouds at the moment
of remap. We shall see that, as expected,
the low-resolution simulations give a very accurate
coarse grained version of the high resolution ones.

\subsection{First application: stationary profile}
\label{sec:stationary}
A family of stationary solutions is given by (Spitzer, 1942)
\begin{equation}
f_{\rm s}({x},{v})={\bar \rho} 
                     \left[ \cosh\left( \frac{x}{\sigma_x} \right) \right]^{-2}
                     \exp\left(-\frac{1}{2}\frac{{v}^2}{\sigma_v^2} \right),
\label{eq:stationarys}
\end{equation}
with
\begin{equation}
\sigma_x = \sqrt{\frac{\sigma_v}{\sqrt{2 \pi}{\bar \rho}}}.
\end{equation}
For further reference, the total mass writes
\begin{equation}
M=2 (2\pi)^{1/4} {\bar \rho}^{1/2} \sigma_v^{3/2},
\label{eq:masssta}
\end{equation}
and the total energy [Eq.~(\ref{eq:conservener})] reads
\begin{equation}
{\cal E}_{\rm tot}=3 (2\pi)^{1/4} {\bar \rho}^{1/2} \sigma_v^{7/2}.
\label{eq:etotsta}
\end{equation}
To perform the simulations, we set up initial conditions with same 
apodization as in Eqs.~(\ref{eq:finitophat}) and (\ref{eq:finigaussian}), to make the
support of the function compact. To achieve an 
initial set up sufficiently close to the stationary solution,
we must have $\sigma_v$ and $\sigma_x$ sufficiently small compared
to apodization radius, ${\cal R}$. Our choice here is ${\cal R}=1$, ${\cal R}_{\rm apo}=0.2$, 
${\bar \rho}=2$, $\sigma_v=0.2$ [hence, $\sigma_x \simeq 0.2 \simeq \sigma_v$]. 
With this set up, the number of clouds contributing during runtime is 
about 80000--120000 and  4200--7500 in the high and the low resolution simulation respectively.
This number decreases slowly with time as a result of the truncation procedure used to maintain
the support of $f$ compact (see \S~\ref{sec:deconv}). In principle we could fine tune this procedure
in order to maintain the number of sampling clouds approximately constant with time, but we decided
not to do so at this point, because it affects only the tails of the distribution function.

\begin{figure}
\centerline{\mbox{\epsfig{file=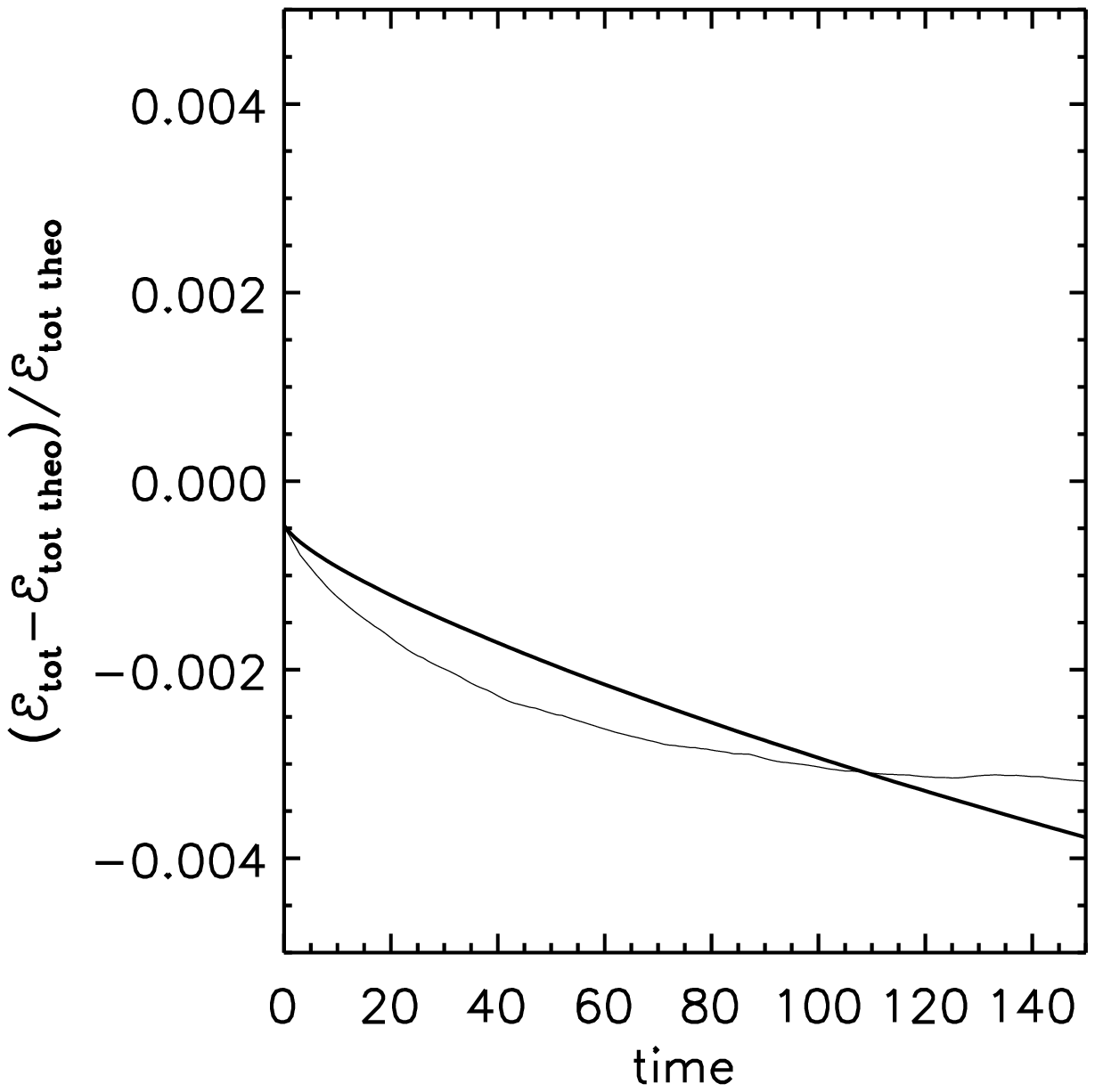,height=6cm}
\epsfig{file=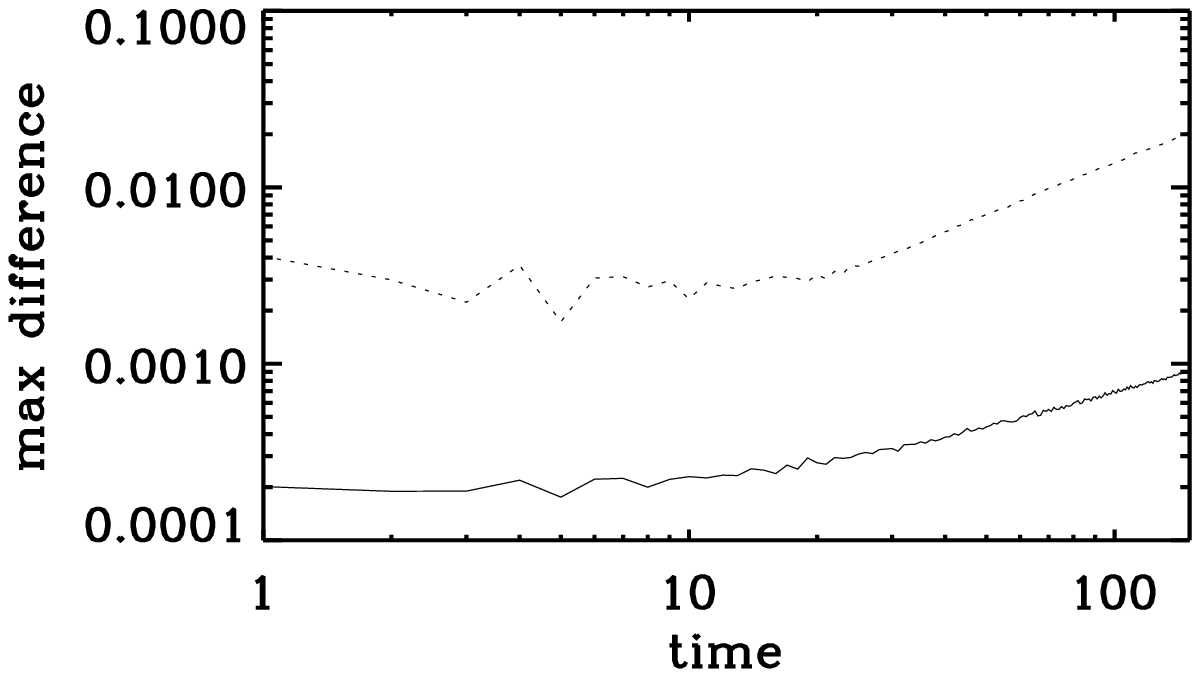,height=6cm}}}
\caption[]{{\em Left panel:} test of energy conservation for a stationary solution:
The quantity $({\cal E}_{\rm tot}-{\cal E}_{\rm tot\ theo})/{\cal E}_{\rm tot\ theo}$
is displayed as a function of time, where ${\cal E}_{\rm tot}$ is measured in our simulation
according to Eq.~(\ref{eq:conservener}) and ${\cal E}_{\rm tot\ theo}$ is given by Eq.~(\ref{eq:etotsta}).
The parameters used to do the simulations are ${\bar \rho}=2$ and $\sigma_v=0.2$. 
The thick and the thin curves correspond to the high and the low resolution simulation, respectively.
{\em Right panel:} deviations from stationarity: the quantity $\max|{\tilde f}(x,v,t)-f_{\rm s}(x,v)|$ is
represented as a function of time, where ${\tilde f}$ is the simulated $f$ and $f_{\rm s}(x,v)$ is given
by Eq.~(\ref{eq:stationarys}) with additional apodization as explained in the text. The continuous and
the dotted curve correspond to the high and the low resolution simulations, respectively.}
\label{fig:studystation1}
\end{figure}

Fig.~\ref{fig:studystation1}
examines energy conservation and deviations from the stationary solution as functions
of time. As expected, energy conservation is
very good both for the high and for the low resolution simulation, better than $0.5$ percent.
It however worsens with time. This is mainly a consequence of our truncation 
of the phase-space distribution function at $\delta_{\rm min}=5.10^{-5}$. 
Note that the maximum difference between the true distribution
function and the simulated one tends to augment linearly with time at a certain point,
even for the high resolution simulation, as illustrated by the right panel of
Fig.~\ref{fig:studystation1}. For the high resolution simulation, this is again mainly due
to the truncature of the tails of $f$. Indeed, the difference $\delta f$ between
the numerical and the analytical solutions measured at the center of the fluctuation, 
$(x,v)=(0,0)$,  is only of the order of $-2.10^{-4}$ for $t=150$. 
For the low resolution simulation, an other effect adds up to the truncation of the tails,
namely a weak diffusion at the maximum of $f$ at the moment of Lagrange remap.
When fluctuations become of the order of the cloud size, one
can indeed expect two competing effects, according to the details of the shape of the sampled
distribution function: (i) a coarse graining effect, where small scale 
fluctuations tend to be progressively smeared out, (ii) 
an aliasing effect already discussed in \S~\ref{sec:deconv}, where, 
on the contrary, artificial contrasts are
created with the appearance of spurious oscillations. Since the distribution function
we sample here is smooth, effect (ii) is not present.
However, although a deconvolution method such as Lucy's (or Van-Citter's) aims to minimize effect (i), 
it cannot reduce it completely. This effect is very small, since it induces a rather weak
diffusion, even for the low resolution simulation, of the order of 2 percent only at the center
of the fluctuation at $t=150$. 

\begin{figure}
\centerline{\mbox{\epsfig{file=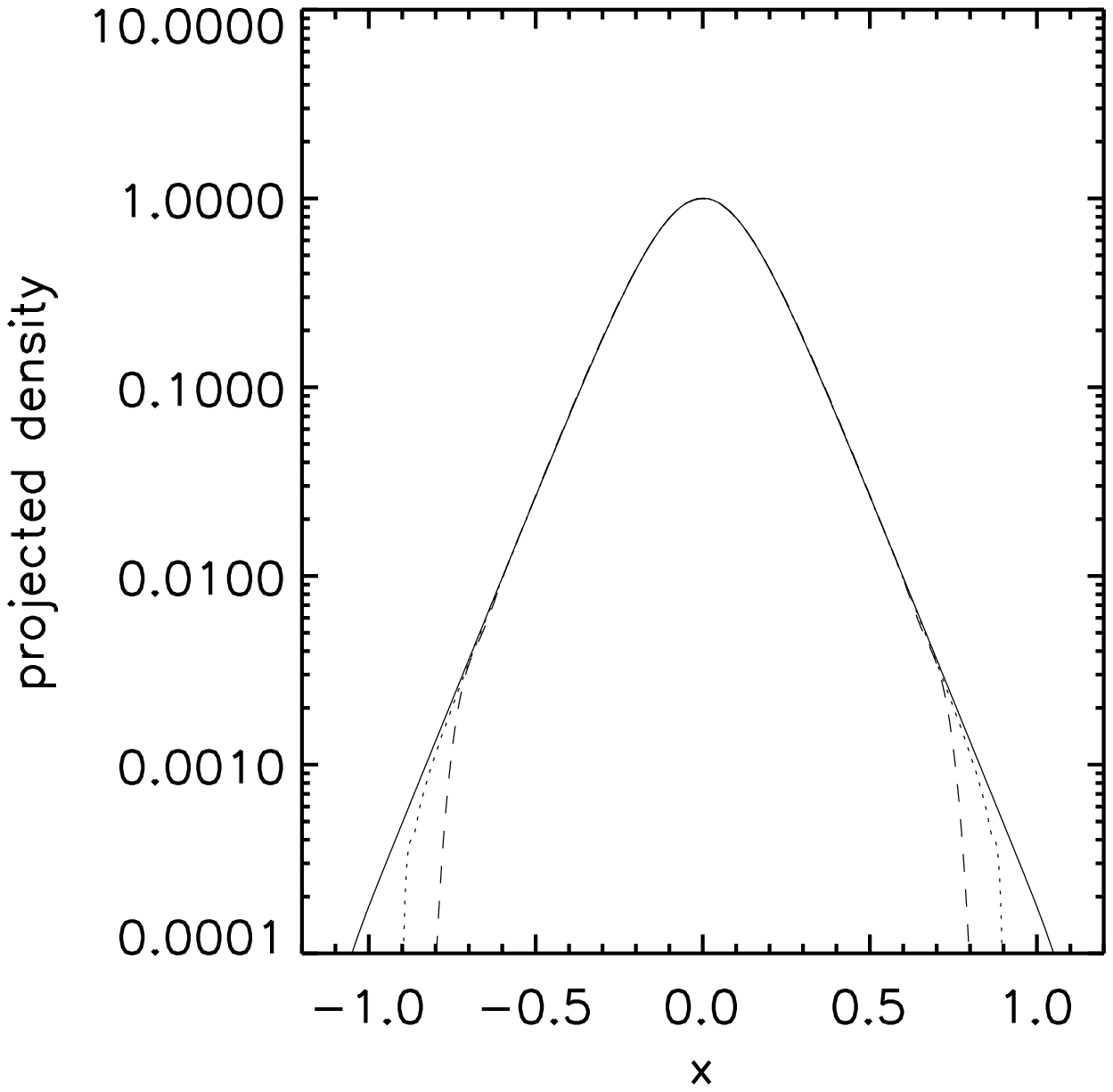,width=8cm}
\epsfig{file=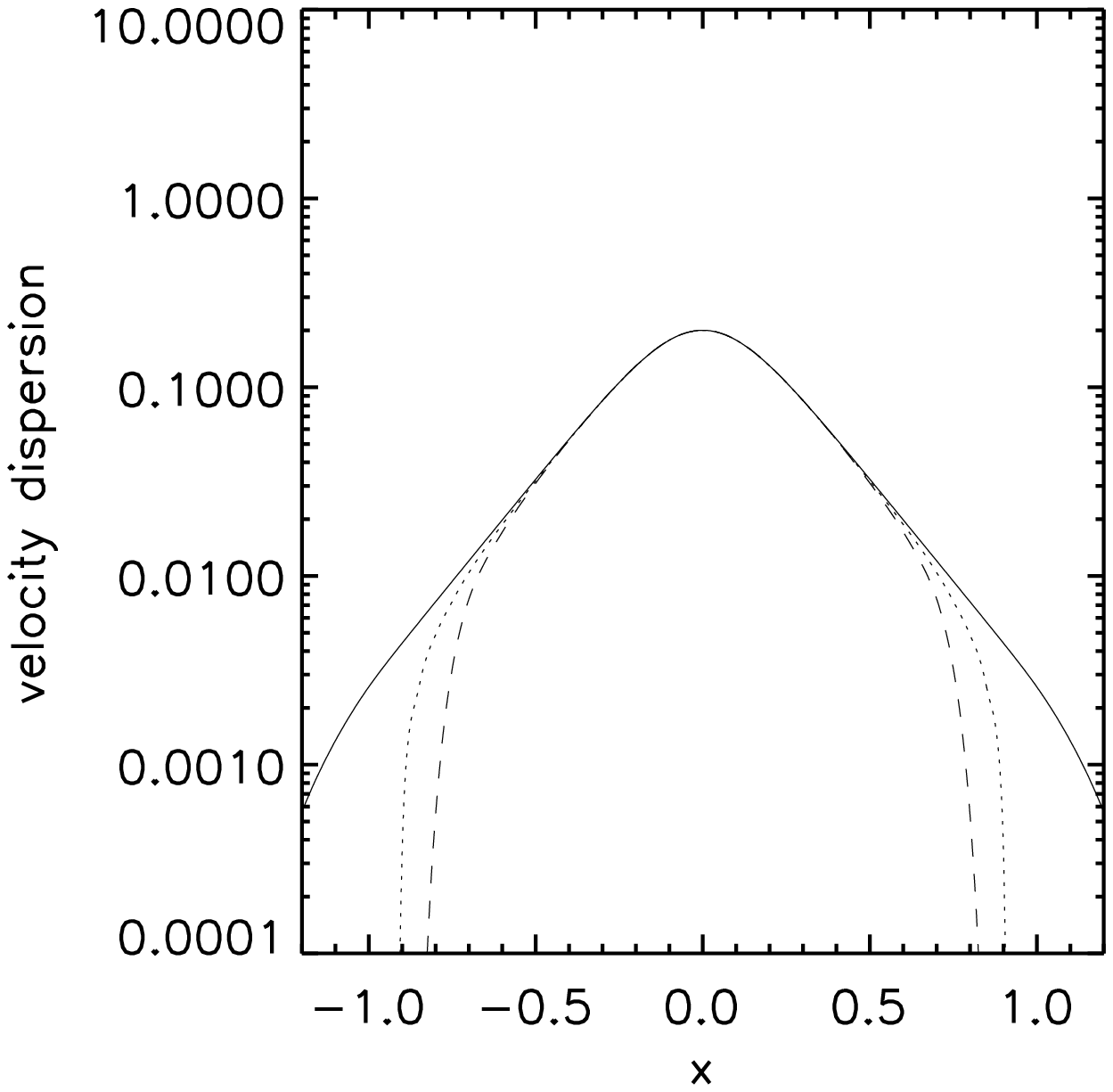,width=8cm}}}
\caption[]{Mean projected density, $\rho({x})=\int f({x},{v})d{v}$ 
(left panel) and velocity dispersion, $v_{\rm rms}(x)$, 
$v_{\rm rms}^2(x) \equiv \langle \delta {v}^2 \rangle=\int f({x},{v}) {v}^2 d{v}
- [\int f({x},{v}) {v} d{v}]^2$ (right panel),  as functions
of position, for the stationary solution examined in Fig.~\ref{fig:studystation1}. 
The solid curve corresponds to Eq.~(\ref{eq:stationarys}) with apodization
as described in the text, which explains the bending of the tails visible in
the right panel. The dotted and the dashed curves give the results
obtained from the high and the low resolution simulations, respectively.}
\label{fig:studystation2}
\end{figure}

Fig.~\ref{fig:studystation2} shows the zeroth and second moments of
$f(x,v)$ with respect to velocity, namely the projected density and the velocity dispersion, 
as functions of $x$, at latest output, $t=150$. Even though the low resolution simulation
has already significantly diffused, with a two percent depletion on the maximum of $f$,
the overall solution is still pretty close to the correct one.  The
deviations observed in the tails both in the high and the low resolution simulation are
a consequence of the truncation procedure used to maintain the computing domain finite.


\subsection{Second application: a gaussian as initial conditions}
\label{sec:gaussian}
The gaussian initial conditions we consider are specified by
Eq.~(\ref{eq:finigaussian}), with ${\cal R}=0.8$, ${\cal R}_{\rm apo}=0.2$, ${\bar \rho}=4$
and $h=0.2$. With this set up, the number of clouds contributing is roughly 95000--120000
and 6600--7700 for the high and the low resolution simulation respectively.

\begin{figure}
\centerline{\mbox{\epsfig{file=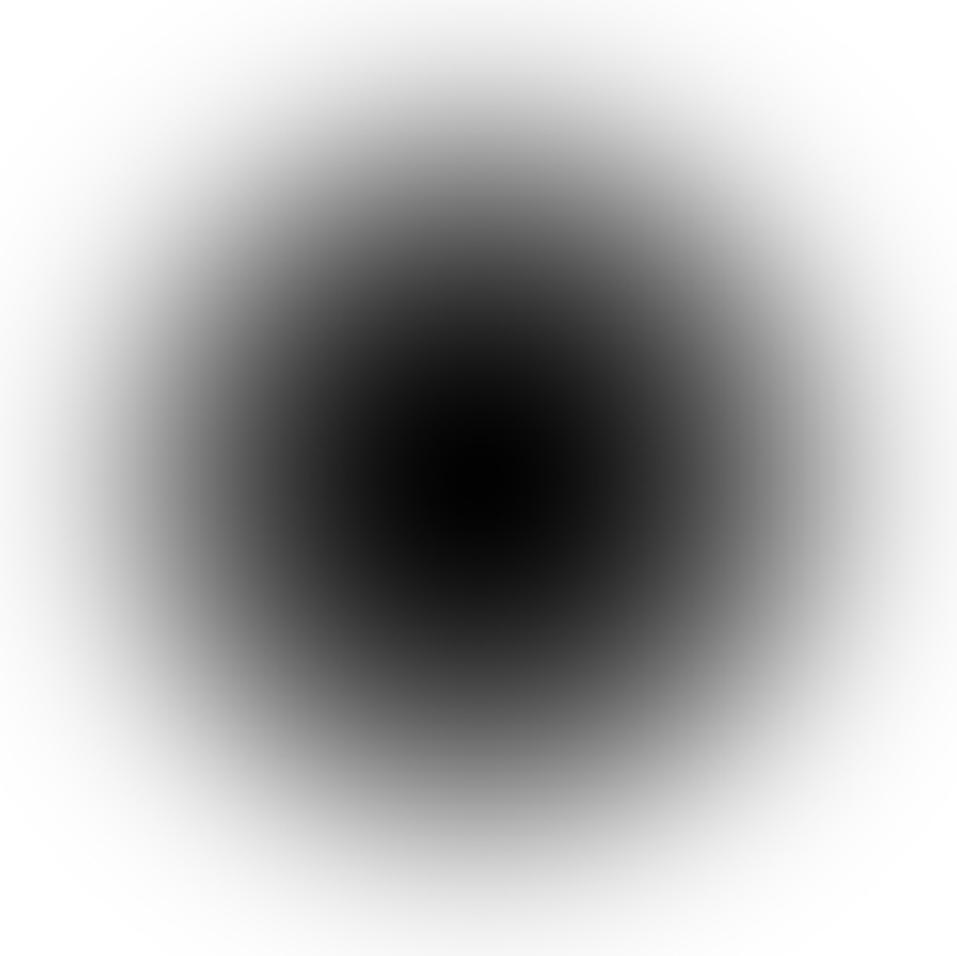,width=11cm}
}}
\centerline{t=0}
\centerline{\mbox{
\epsfig{file=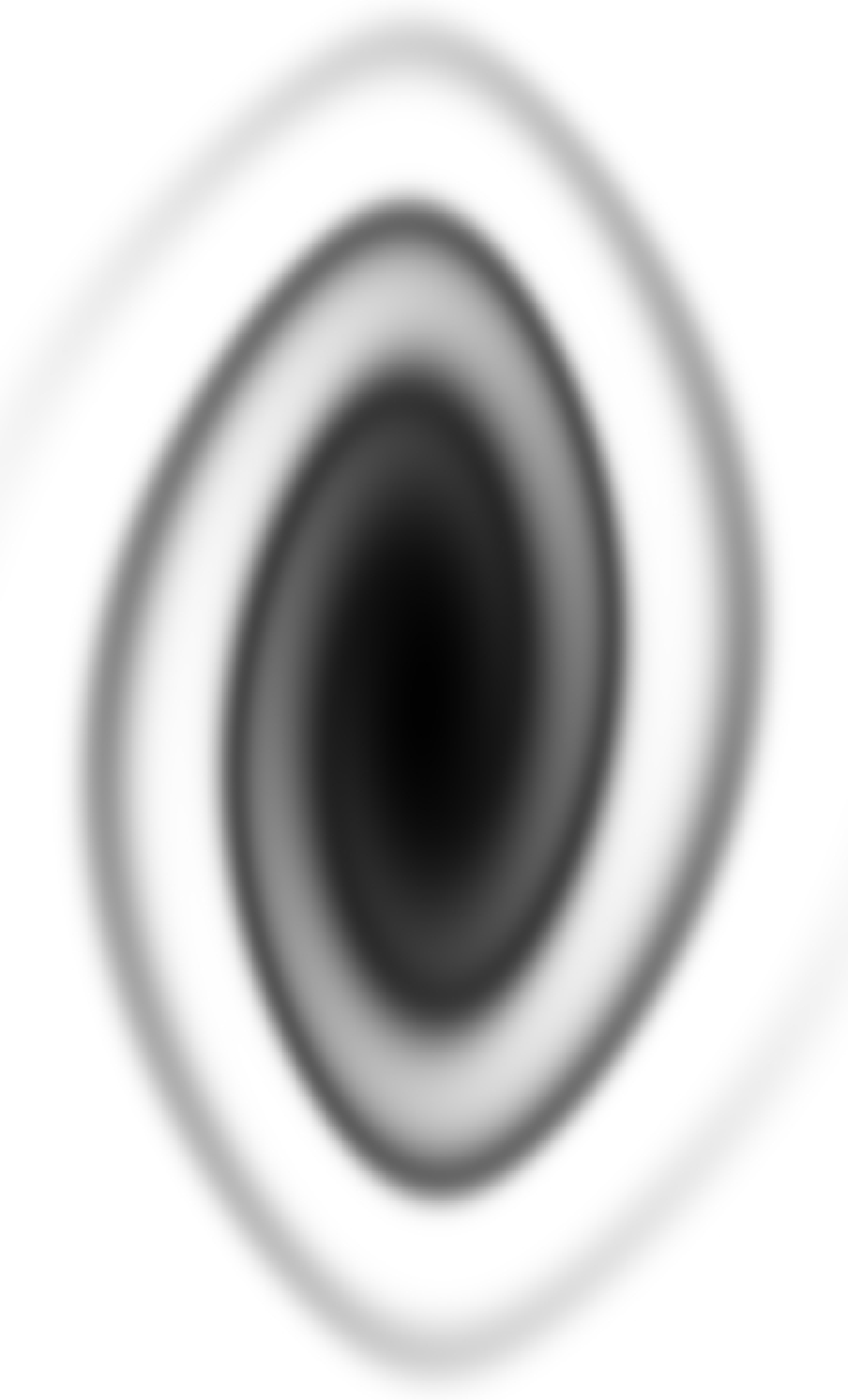,width=11cm}
\epsfig{file=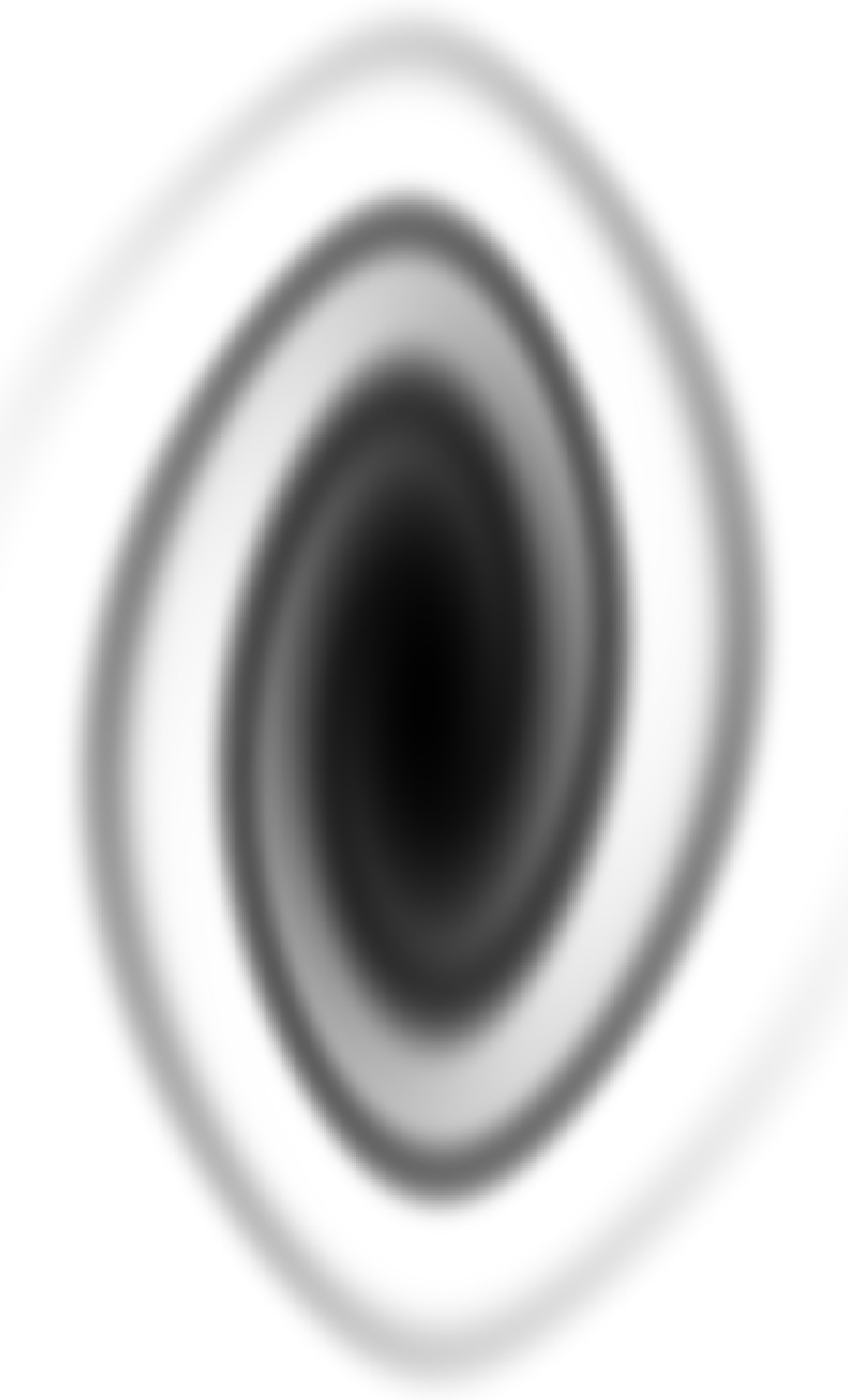,width=11cm}
}}
\centerline{high resolution\mbox{ \hskip 4cm} t=10 \mbox{\hskip 4cm} low resolution}
\caption[]{Phase space distribution function at various times, for Gaussian initial
conditions. Left and right panels correspond
to the high and the low resolution simulation, respectively. Darker regions correspond to higher
values of $f$.}
\label{fig:studygaussian1}
\end{figure}

Fig.~\ref{fig:studygaussian1} shows the distribution function in phase-space at various
times, $t=0$, 10, 40 and 100. As expected, a spiral structure appears,
which roles up with time. At some point, it becomes so thin that it disappears due
to finite resolution. Of course, this event happens earlier for the low resolution simulation.
Visually, this latter seems to represent a very good coarse grained version of the high resolution
simulation. This can be examined more in details on Fig.~\ref{fig:studygaussian2}, which
displays projected density and velocity dispersion at the same instants as in Fig.~\ref{fig:studygaussian1}.
At $t=10$, it is not possible to distinguish yet between low and high resolution. The difference between
the two simulations is the most significant for $t=40$: in that case, details
are lost in the low resolution simulation, but it gives a rather accurate coarse grained
version of the high resolution one. At late time, $t=100$,
details have nearly disappeared even in the high resolution simulation and the agreement between
low and high resolution remains excellent. 

Fig.~\ref{fig:studygaussian3} displays the relative deviation from energy conservation as a function
of time. For the high resolution simulation, results are very similar to what was obtained for
the stationary solution: total energy decreases slowly with time, mainly because of
our truncature of the tails of the phase-space distribution function, but is conserved with
a precision better than $0.5$ percent at $t=100$.
The behavior observed for the low-resolution simulation is more complex. Indeed,
energy first increases up to $t\simeq 50$, due to coarse graining effects which introduce
a slight bias in the tails of the distribution function. At variance with the high-resolution
simulation, these effects dominate over those due to the truncature until details
in the distribution have completely disappeared. Note that 
the maximum of $f$, that should stay constant, presents small variations between initial and
final time. The net result is an increase 
of $0.1$ and $2$ percent for the high and the low resolution simulation, respectively.
This is probably due to aliasing effects, which are however not expected to
affect significantly energy conservation.

Examination of the second part of Fig.~\ref{fig:studygaussian2} suggests
that the system is relaxing towards a stationary regime. 
Fig.~\ref{fig:studygaussian4} compares the phase space distribution function
at the latest stage of the high resolution simulation to a stationary
solution given by Eq.~(\ref{eq:stationarys}). To compute the parameters of the stationary solution,
we use Eqs.~(\ref{eq:masssta}) and (\ref{eq:etotsta}). For the total energy, we take the initial value
obtained from the (apodized) Gaussian profile and obtain ${\bar \rho}\simeq 3.79$ and $\sigma_v=0.298$,
which corresponds to the dotted curve. It globally agrees well with the simulated solution (thick
grey curve), if ones keeps initial conditions as a reference for comparison (solid thin curve). 
Note that the agreement  is improved in the high $f$ 
part if ${\bar \rho}$ is changed to $4$ (dashed curve), 
but remains unperfect: the dashed curve
is not able to reproduce the clear transition between a plateau at 
low values of $f$ and the central belt shape, which suggests a more complex stationary
regime, with two components.
\begin{figure}
\centerline{\mbox{
\epsfig{file=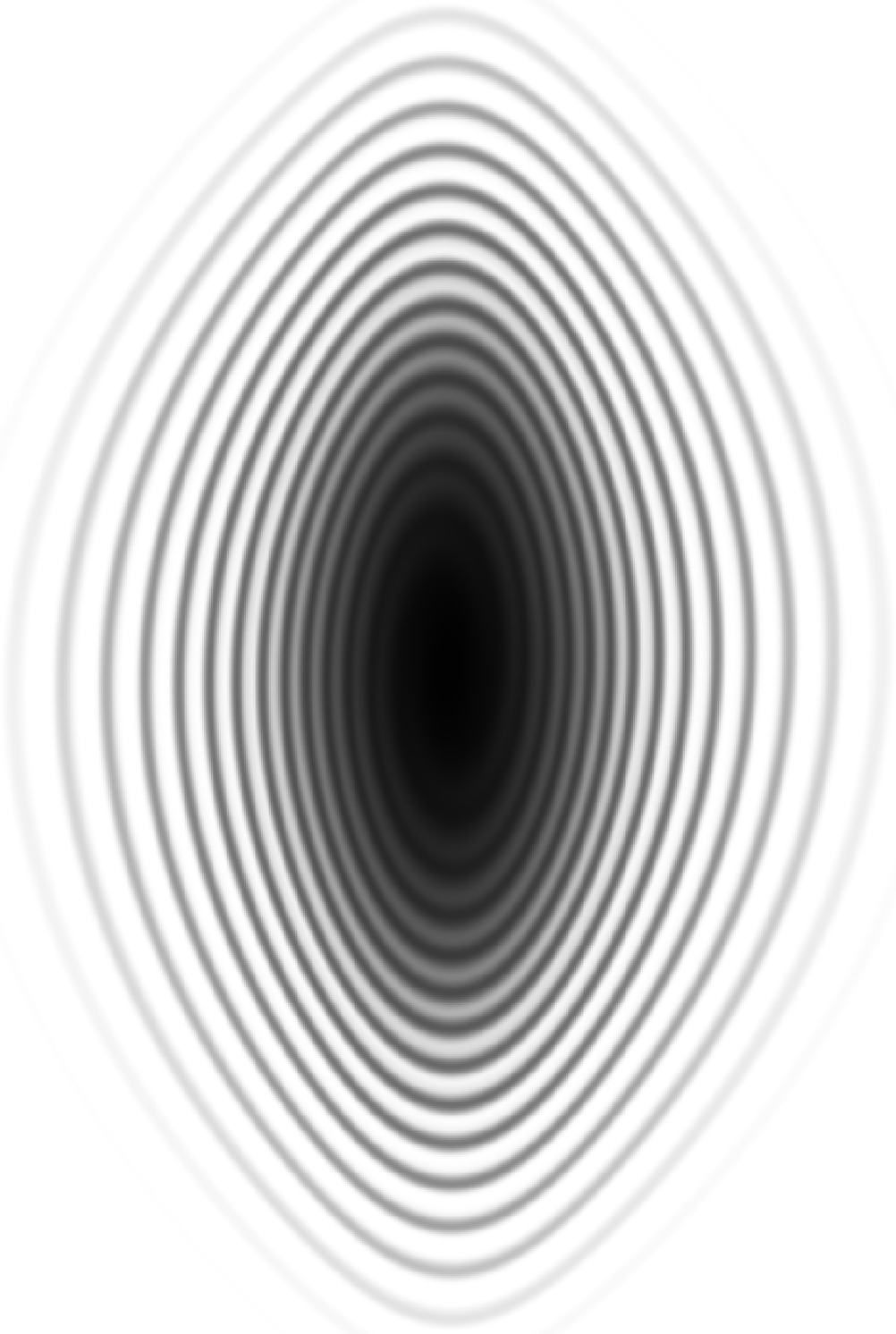,width=11cm}
\epsfig{file=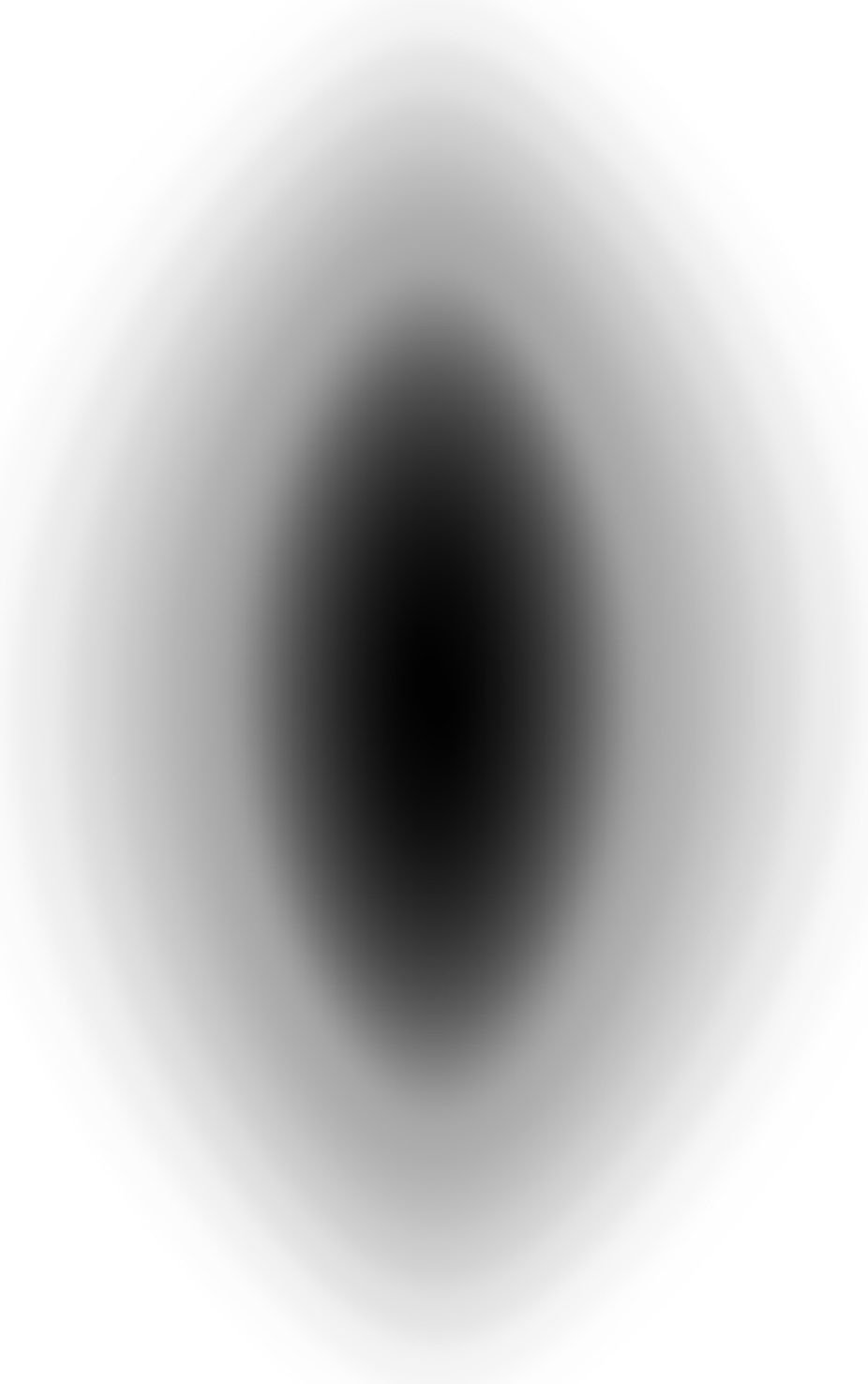,width=11cm}
}}
\centerline{t=40}
\centerline{\mbox{
\epsfig{file=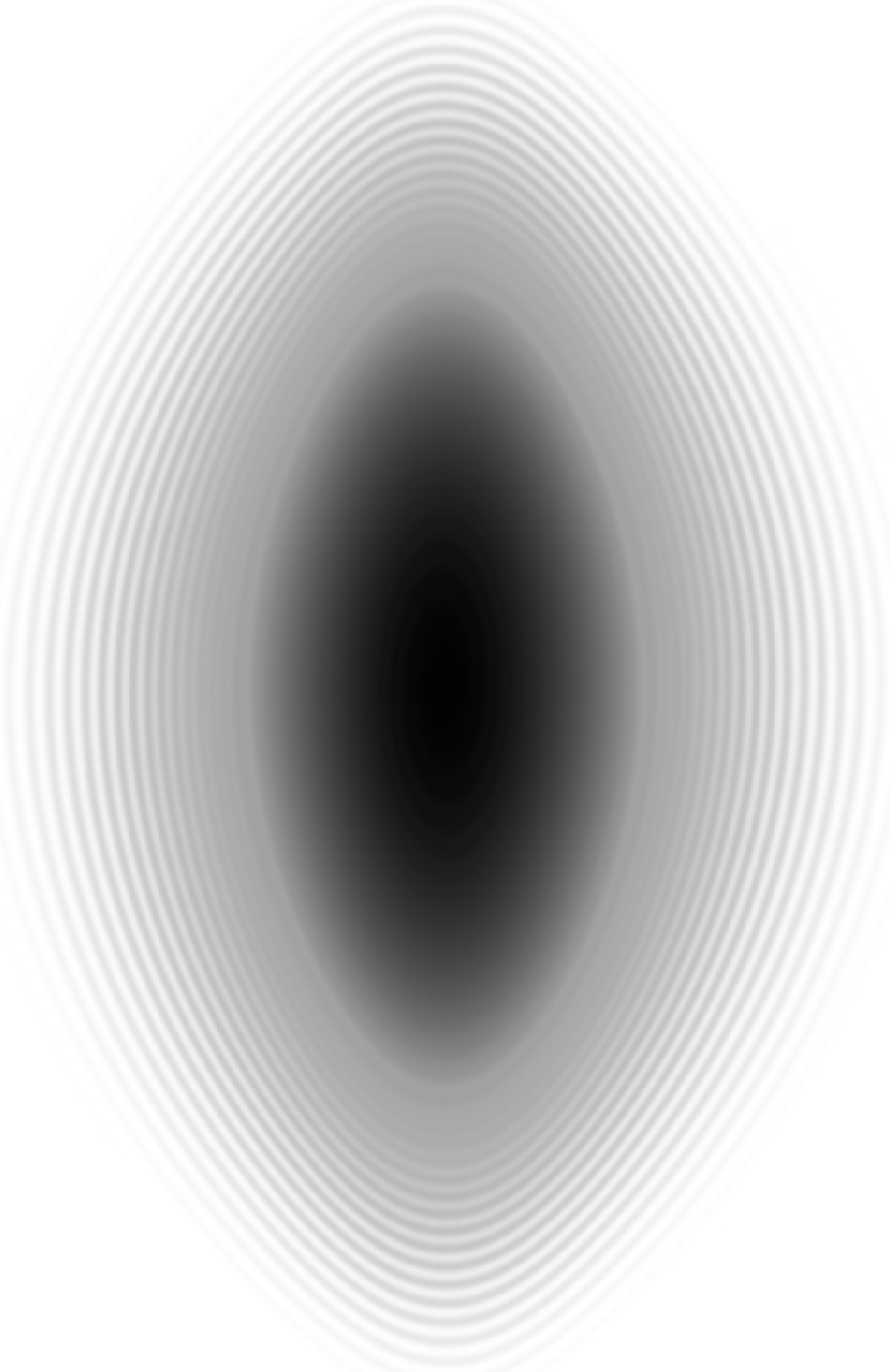,width=11cm}
\epsfig{file=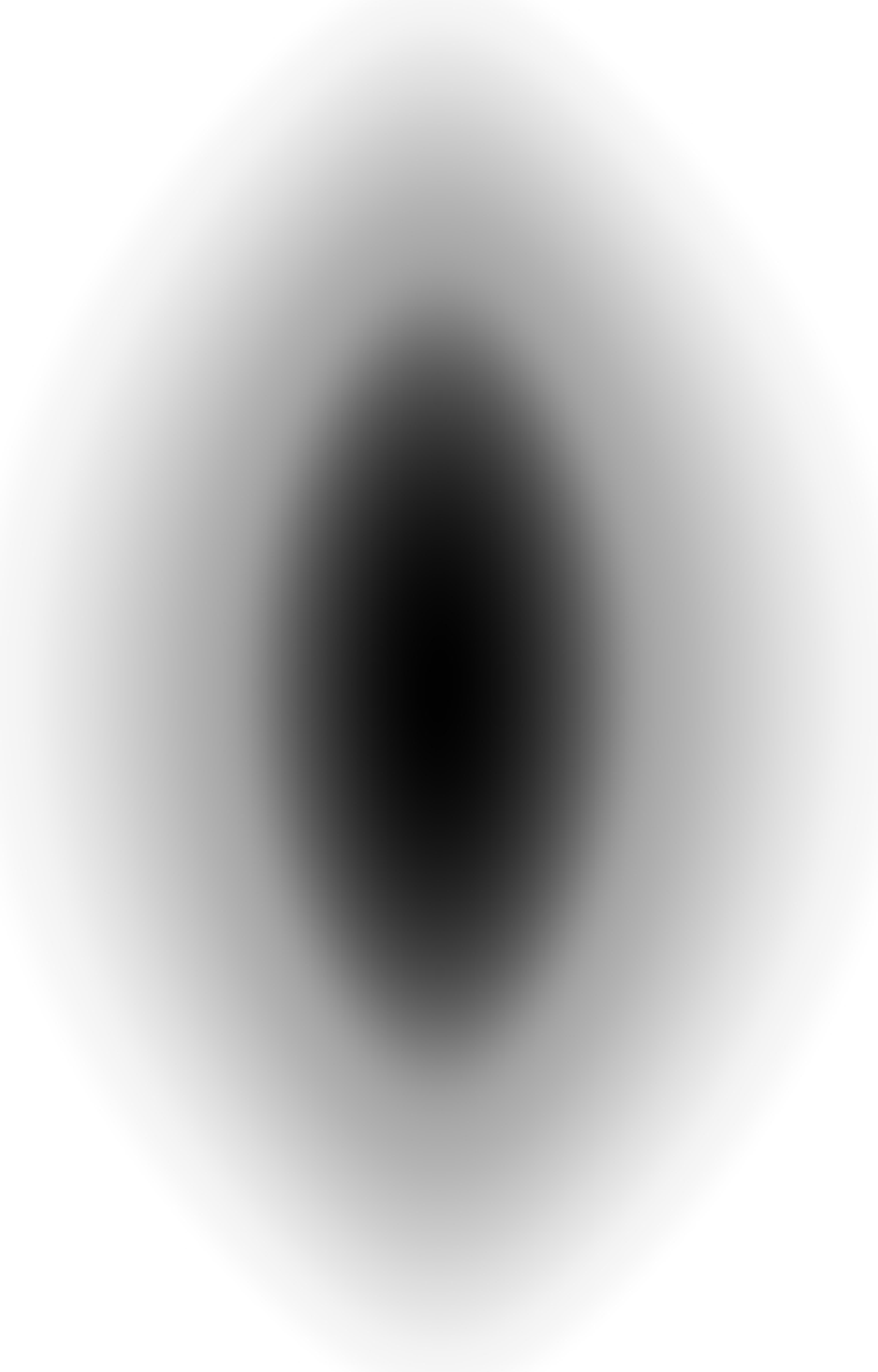,width=11cm}
}}
\centerline{high resolution\mbox{ \hskip 4cm} t=100 \mbox{\hskip 4cm} low resolution}
\centerline{\small {\bf Fig.~\ref{fig:studygaussian1}} (continued).}
\end{figure}
\begin{figure}
\centerline{\fbox{\rule[-0.1cm]{0cm}{0.5cm}\hskip 0.3cm \ t=0\   \hskip 0.3cm}}
\centerline{\mbox{
\epsfig{file=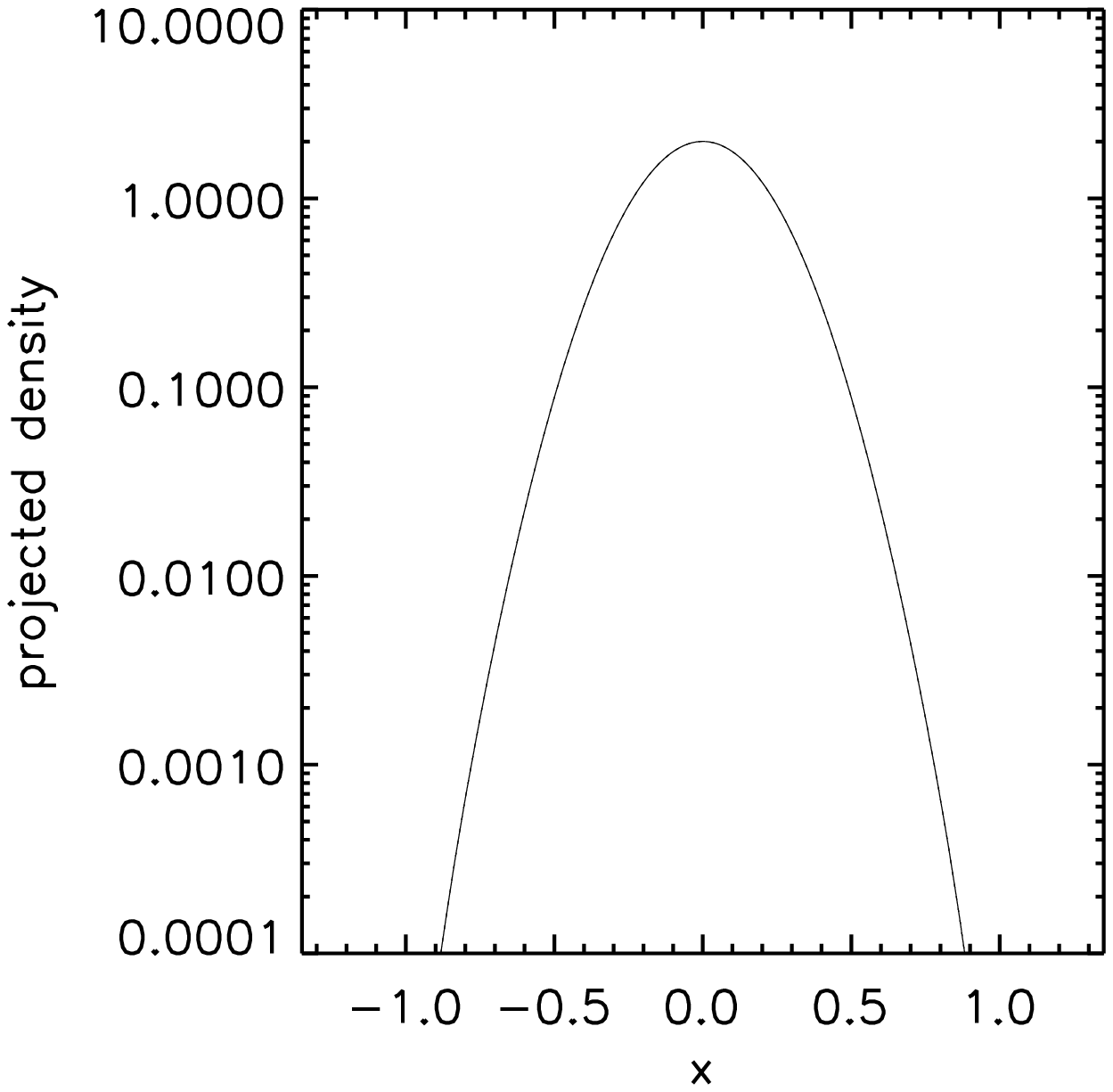,width=8cm}
\epsfig{file=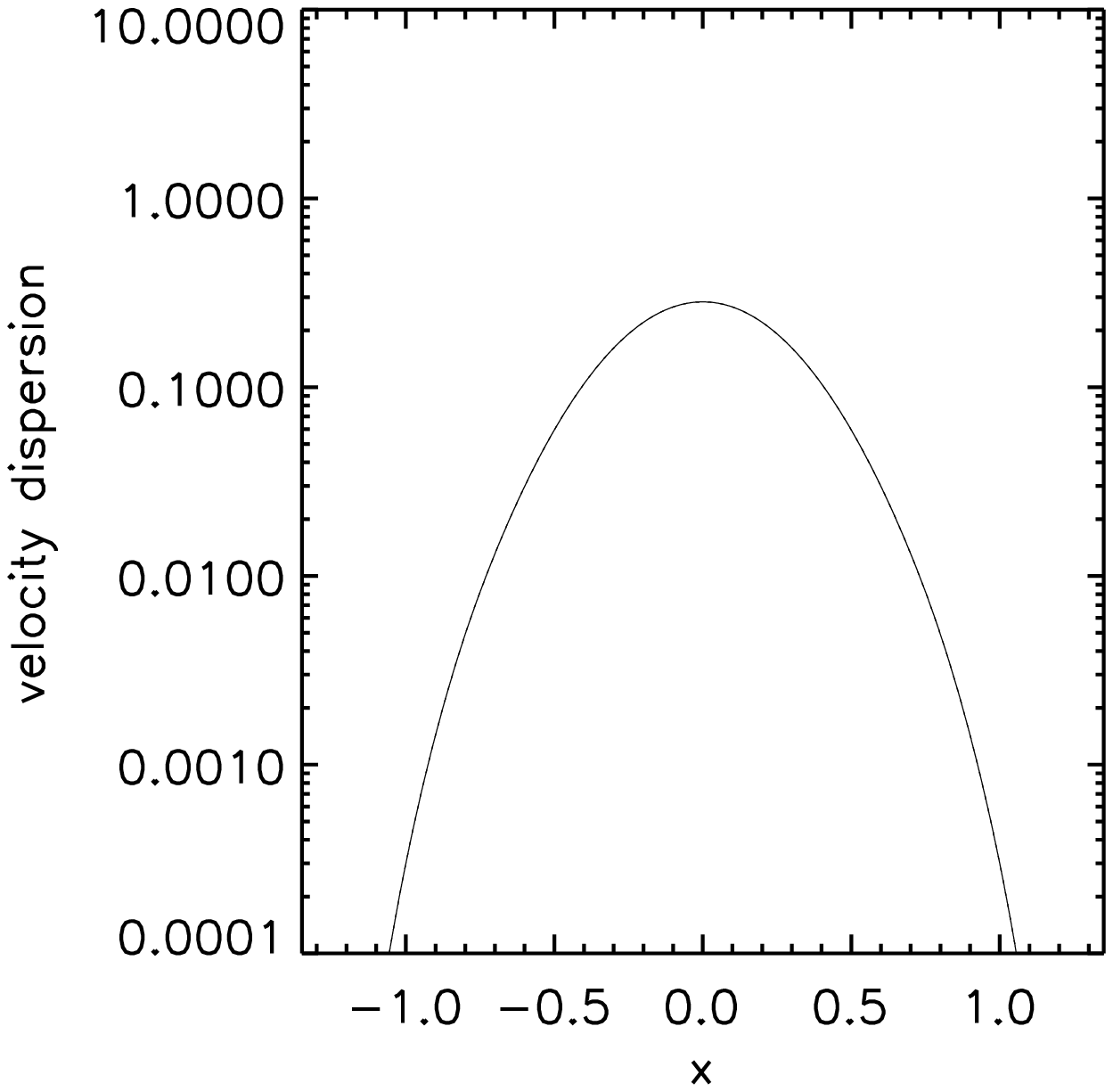,width=8cm}
}}
\vskip 0.3cm
\centerline{\fbox{\rule[-0.1cm]{0cm}{0.5cm}\hskip 0.3cm t=10\  \hskip 0.3cm}}
\centerline{\mbox{
\epsfig{file=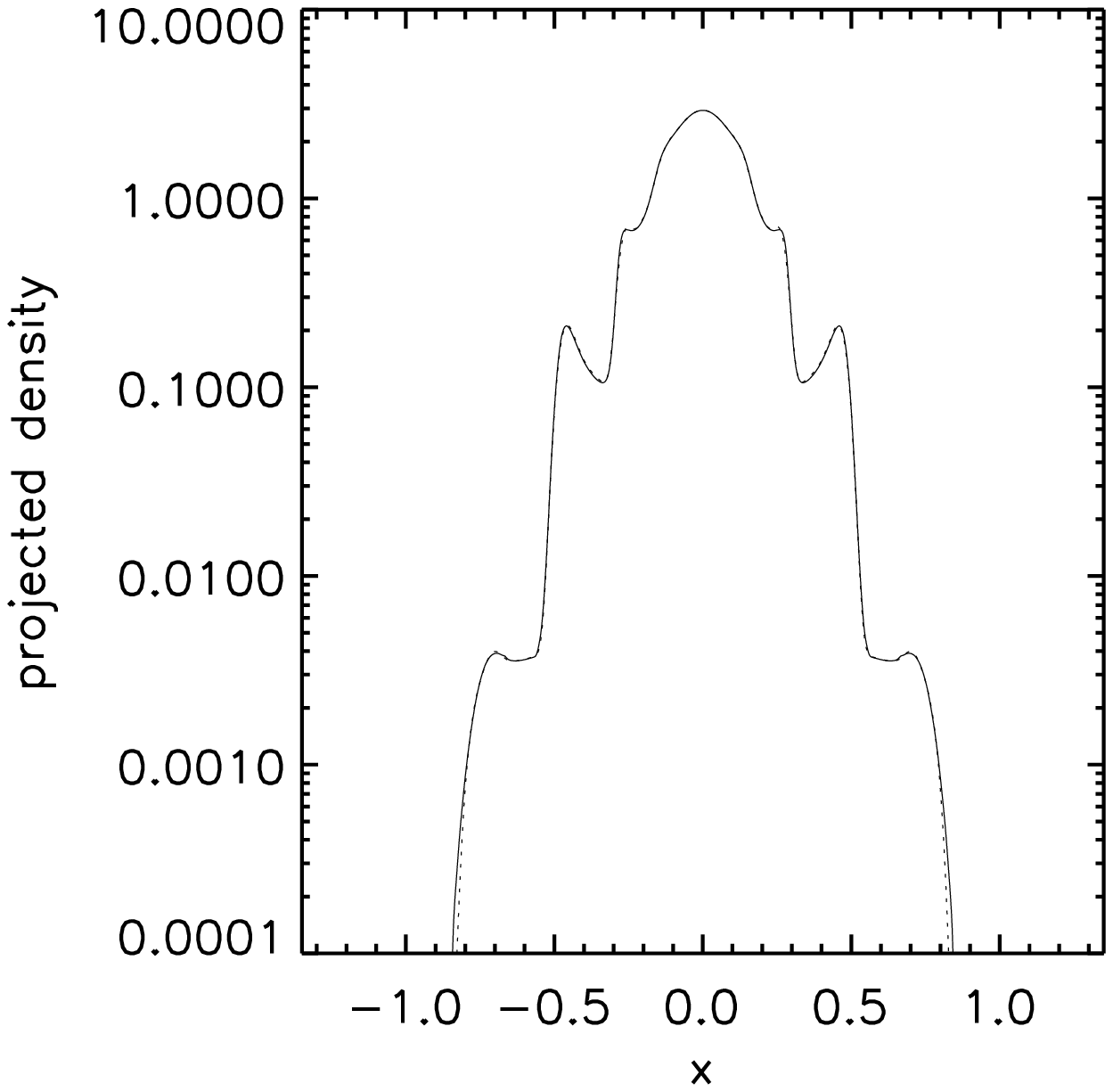,width=8cm}
\epsfig{file=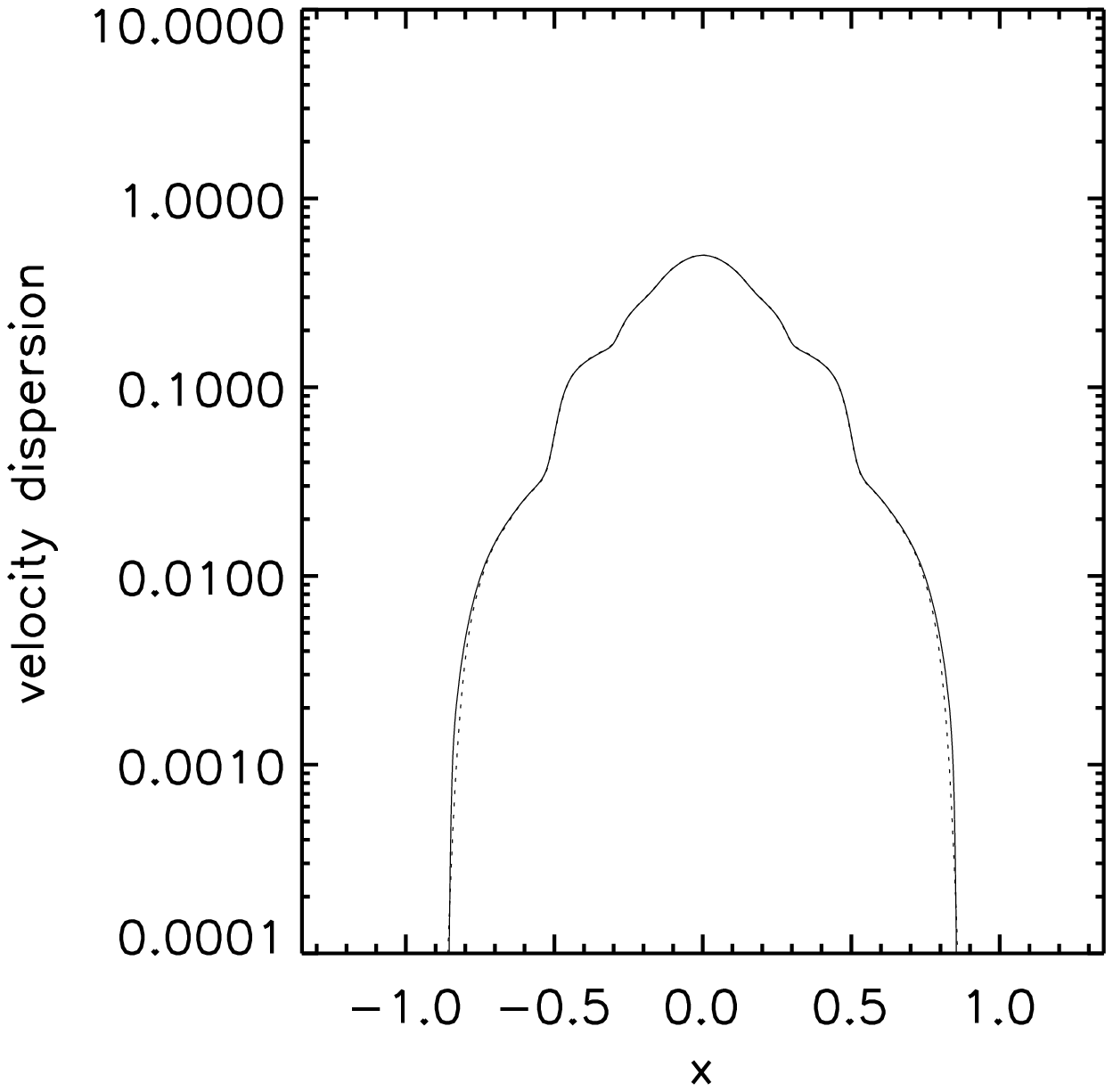,width=8cm}
}}
\caption[]{Mean projected density
(left panels) and velocity dispersion (right panels),  as functions
of position, for the simulations
with Gaussian initial conditions. Each row of panels
corresponds to a given time. The solid curve and the dotted curve are for
the high and the low resolution simulation, respectively, except
for $t=0$: in that case, the curves are calculated according to 
Eq.~(\ref{eq:finigaussian}).
}
\label{fig:studygaussian2}
\end{figure}

\begin{figure}
\centerline{\fbox{\rule[-0.1cm]{0cm}{0.5cm}\hskip 0.3cm t=40\   \hskip 0.3cm}}
\centerline{\mbox{
\epsfig{file=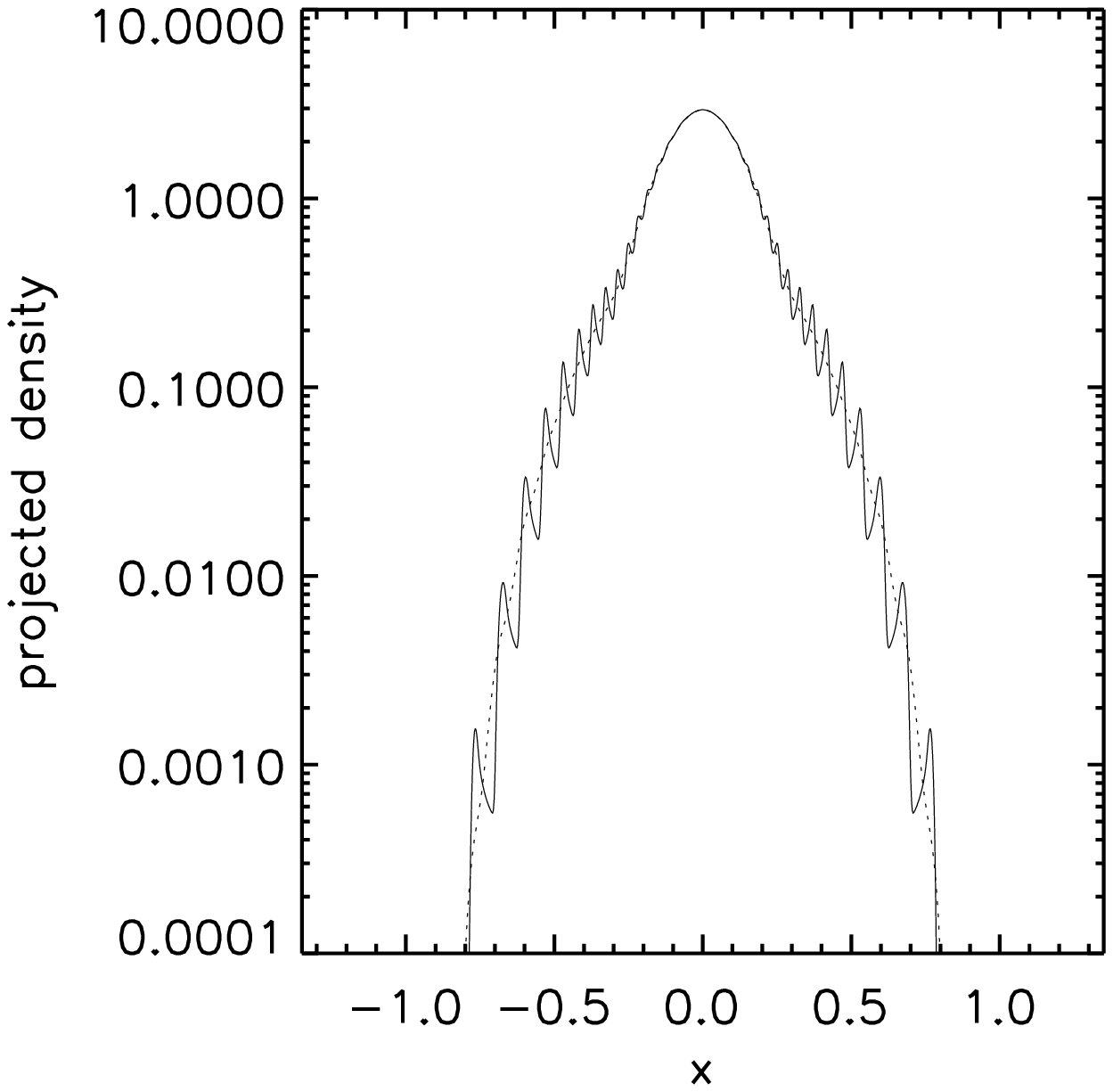,width=8cm}
\epsfig{file=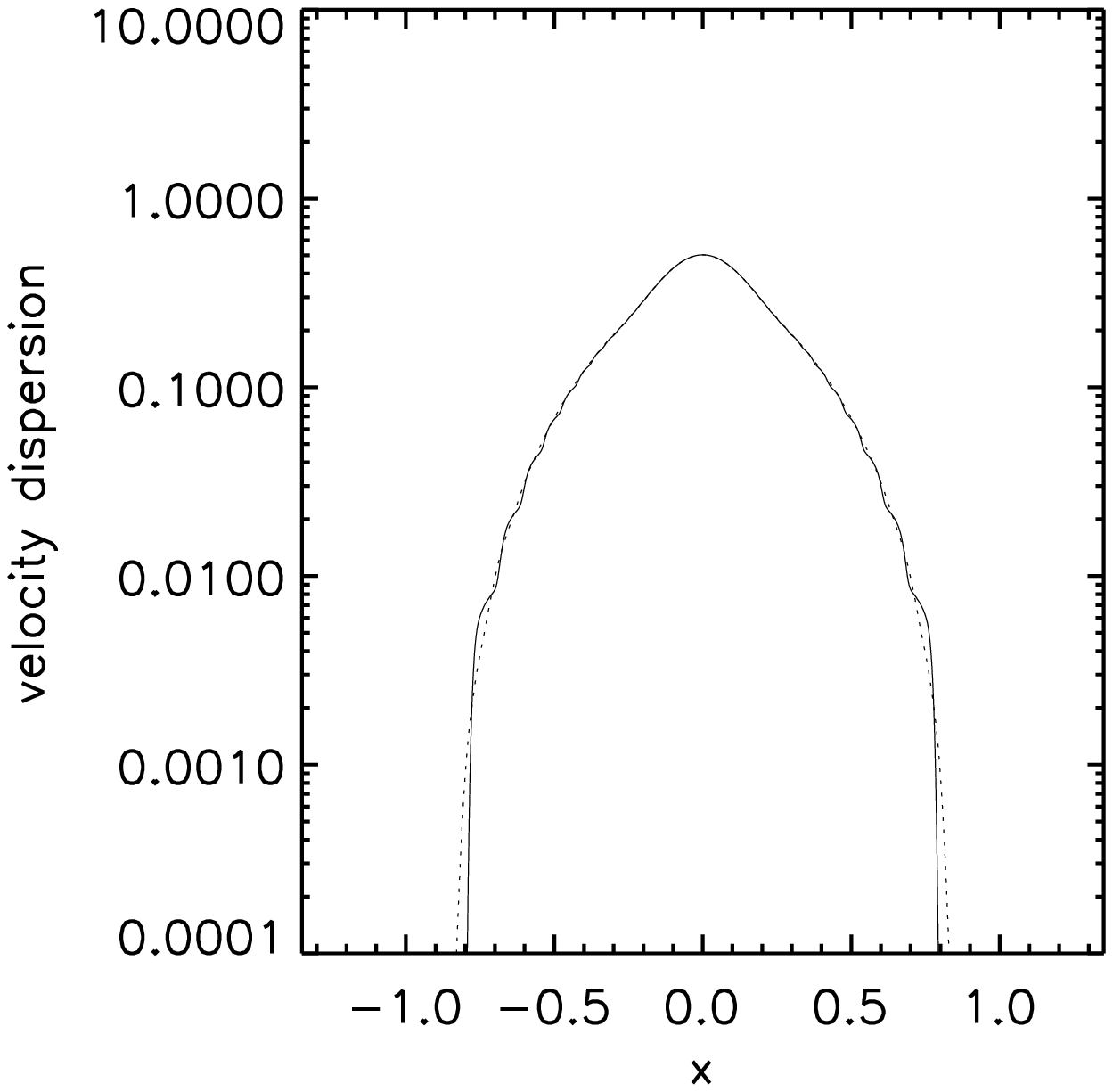,width=8cm}
}}
\centerline{\fbox{\rule[-0.1cm]{0cm}{0.5cm}\hskip 0.3cm t=100   \hskip 0.3cm}}
\centerline{\mbox{
\epsfig{file=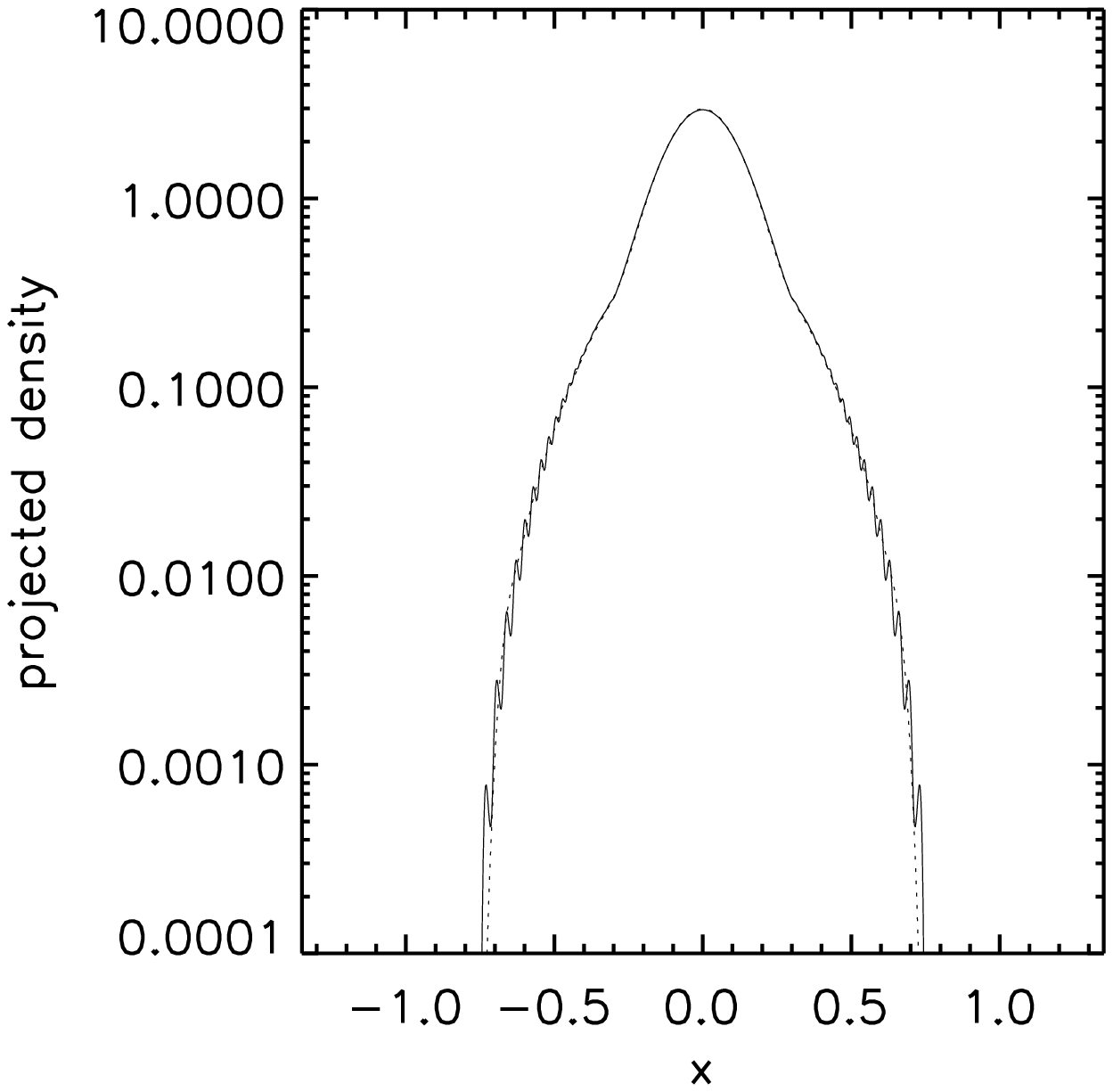,width=8cm}
\epsfig{file=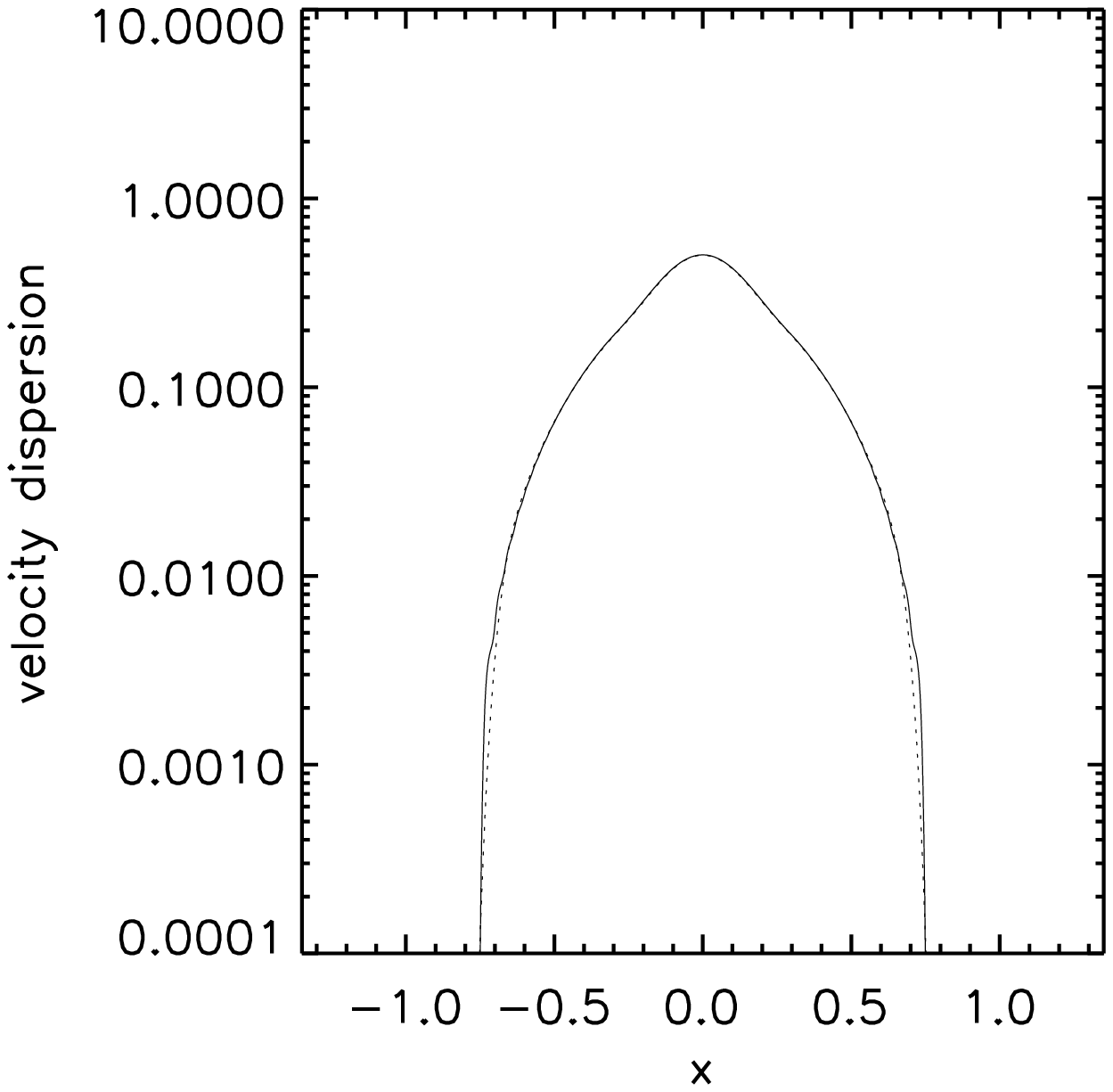,width=8cm}
}}
\centerline{\small {\bf Fig.~\ref{fig:studygaussian2}} (continued).}
\end{figure}
\begin{figure}
\centerline{\mbox{
\epsfig{file=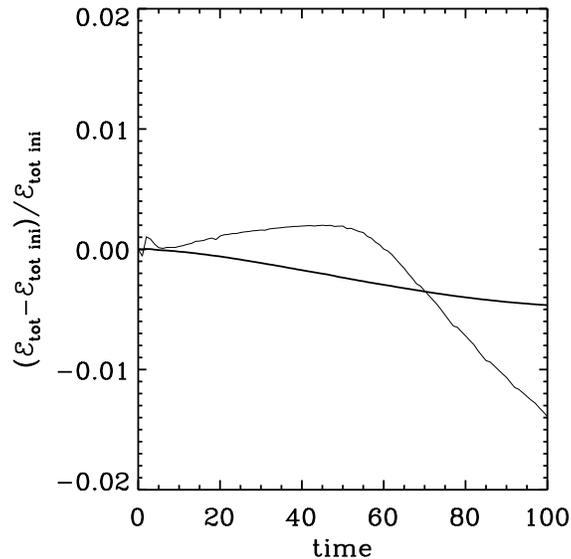,width=8cm}}}
\caption[]{Energy conservation for the simulations with Gaussian initial conditions.
The relative deviation from energy conservation is displayed as a function of time
for the high resolution simulation (thick curve) and the low 
resolution simulation (thin curve): ${\cal E}_{\rm tot}$ and
${\cal E}_{\rm tot\ ini}$ correspond to total
energy and initial total energy, respectively, as measured in the simulations according
to Eq.~(\ref{eq:conservener}). }
\label{fig:studygaussian3}
\end{figure}
\begin{figure}
\centerline{\mbox{
\epsfig{file=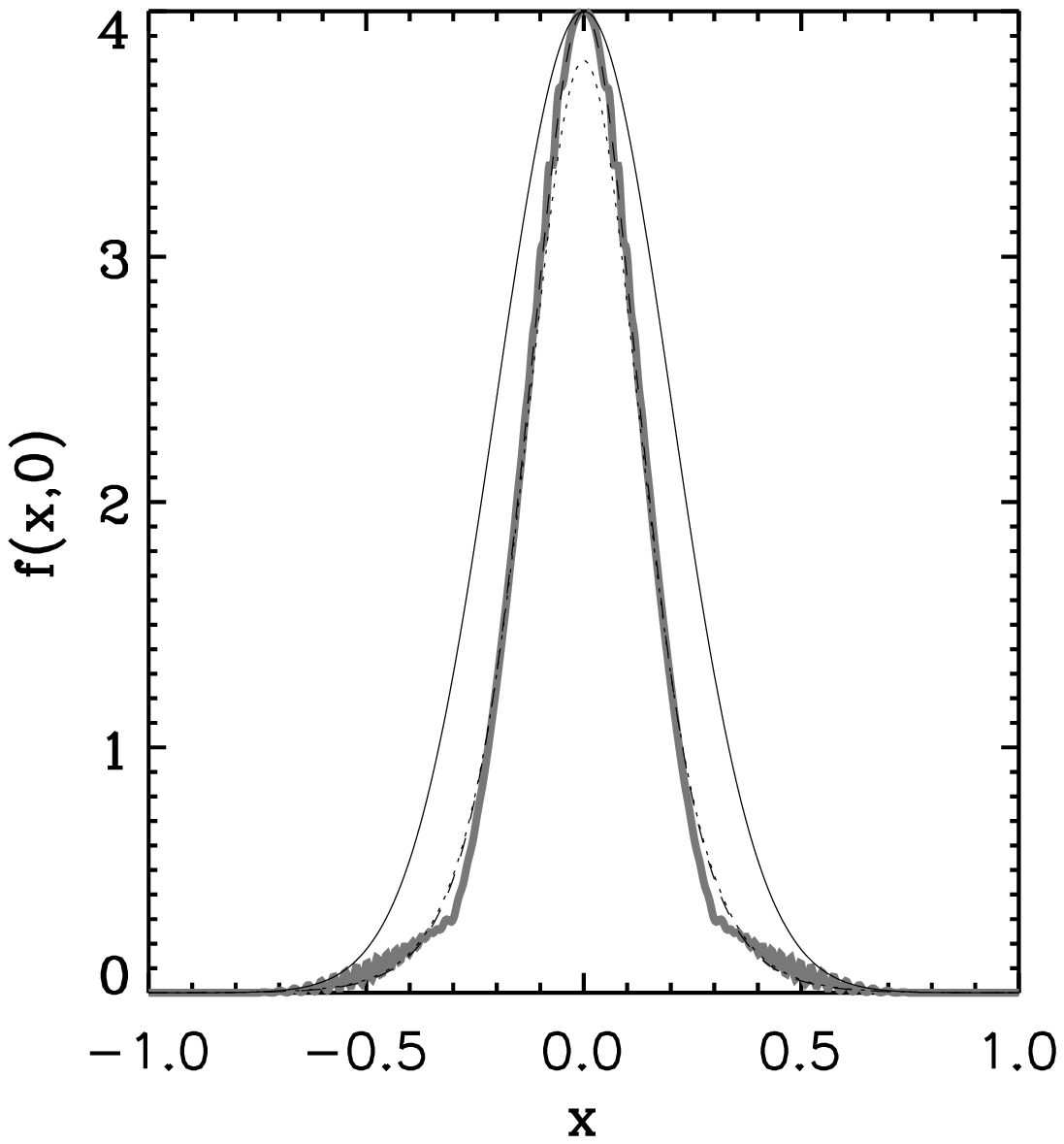,width=8cm}
\epsfig{file=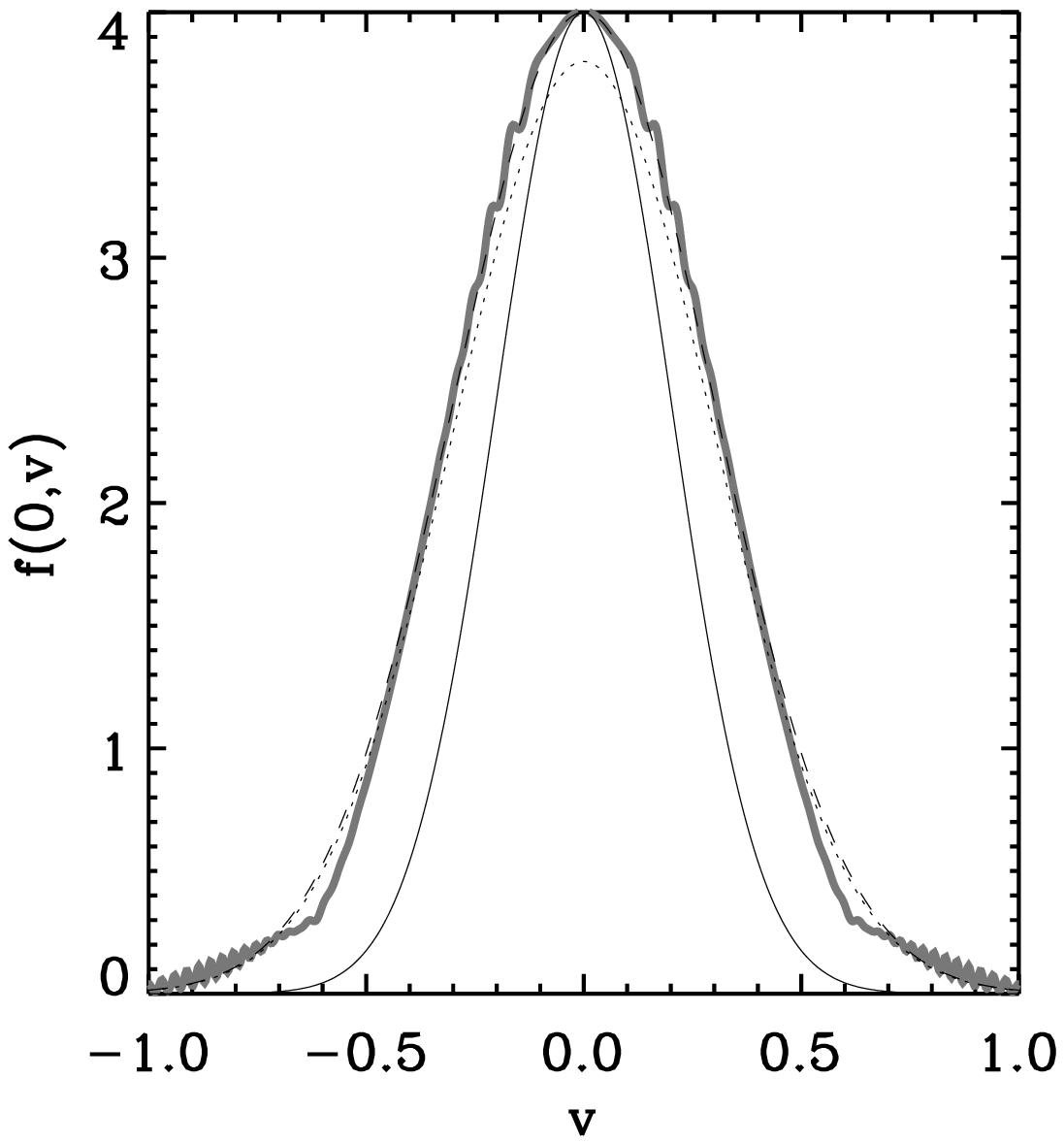,width=8cm}}}
\caption[]{Comparison of the simulation with Gaussian initial conditions with a stationary
solution.  Function $f(x,0)$ is displayed in the left panel, while function
$f(0,v)$ is displayed in the right panel. On each panel, 
the simulation (thick grey curve) is compared
to the stationary solution given by Eq.~(\ref{eq:stationarys}) 
with same energy and same mass (dots).
The dashes are the same as the dots, except that the stationary solution
has been normalized to match the maximum of $f$. For reference, the thin solid curve
corresponds to initial conditions.}
\label{fig:studygaussian4}
\end{figure}
\subsection{Third application: a top hat as initial conditions}
\label{sec:tophat}
%
To set up initial conditions, we use Eq.~(\ref{eq:finitophat}) with
${\cal R}=0.8$, ${\cal R}_{\rm apo}=0.2$ and ${\bar \rho}=1$.
The number of clouds contributing to $f$ 
varies between 190000 and  350000 and between 11000 and 29000
for the high and the low resolution simulation, respectively. 

\begin{figure}
\centerline{\mbox{\epsfig{file=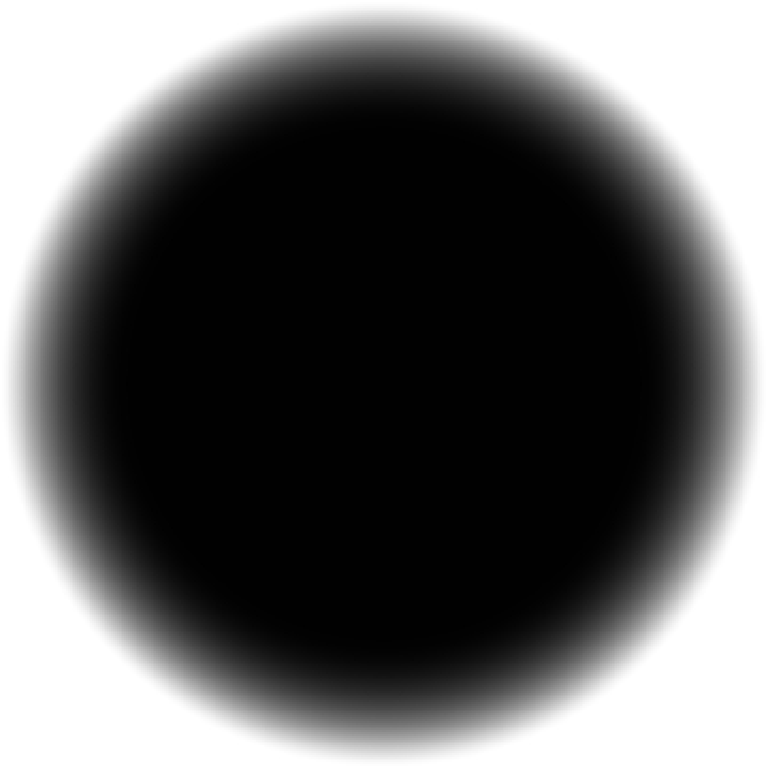,width=11cm}
}}
\centerline{t=0}
\centerline{\mbox{
\epsfig{file=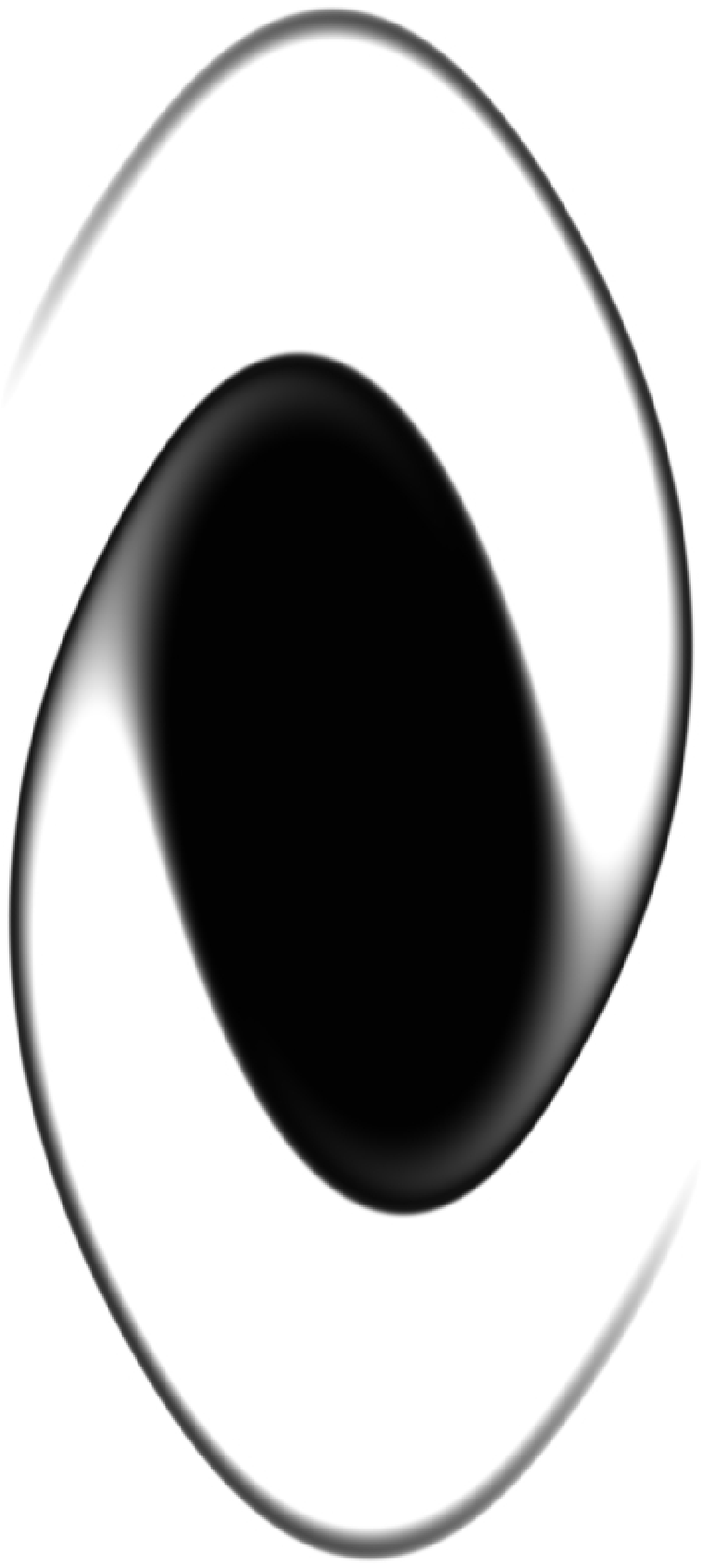,width=11cm}
\epsfig{file=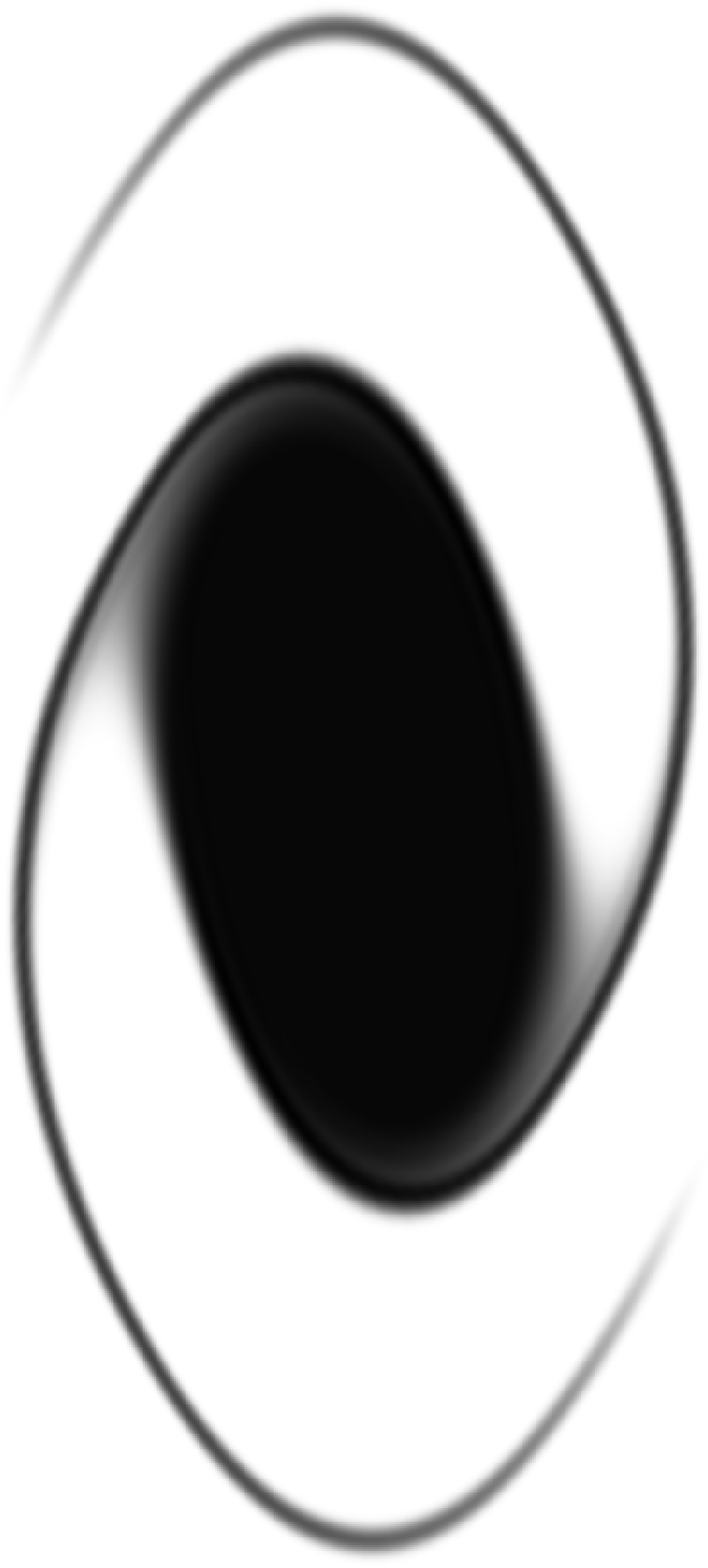,width=11cm}
}}
\centerline{high resolution\mbox{ \hskip 4cm} t=10 \mbox{\hskip 4cm} low resolution}
\caption[]{The phase-space distribution
function for top hat initial conditions, similarly as
in Fig.~\ref{fig:studygaussian1}.}
\label{fig:studytophat1}
\end{figure}

Figs.~\ref{fig:studytophat1} and \ref{fig:studytophat2} show 
the resulting distribution function
in phase space and its zeroth and second moments, 
at various times, $t=0$, 10, 40 and 100, both
for the low and the high resolution simulation. 
The systems builds
with time a two component structure: most of the region $f={\bar \rho}$ 
remains in a compact structure with a roughly elliptic shape
rotating and pulsating around the center, while a ring appears around
it by an effect of rolling up. The appearance of this ring, formed of a very fine
spiral structure transporting a small fraction of the total mass
(see Fig.~\ref{fig:studytophat2}),  explains the significant 
increase of the number of clouds contributing to the distribution function.
At $t=100$, our pulsating core+ring structure has not converged to
any stationary solution, as expected, but remains topologically 
stable on the coarse level, if one takes into account the fact 
that mass migrates slowly from the edges of the central
patch to the external ring.

A careful examination of Figs.~\ref{fig:studytophat1} 
and \ref{fig:studytophat2} shows that the agreement between 
low and high resolution is pretty good,
although not as impressive as for the Gaussian case: some details
differ even at the coarse level, particularly in the region of transition
between the external ring and the central patch. These  differences
do not affect significantly energy conservation, as illustrated
by Fig.~\ref{fig:studytophat3}: in both simulations,
energy decreases slowly with time, mainly as a result of our truncation of the tails
of the phase-space distribution function, and is conserved with an accuracy better
than $0.2$ percent.

Note finally that aliasing effects 
seem to be more significant here 
than for the Gaussian simulation, although still quite
reasonable: at $t=100$, we find for both simulations that 
the maximum of $f$ is about 1--2 percent larger than it should be. 
This happens at the edges of the central patch. 
At the center, we measure $f(0,0)\simeq 1.005$
and $1.0007$ for the low and the high resolution simulation, respectively, which represents
a very small deviation from unity. 
\begin{figure}
\centerline{\mbox{
\epsfig{file=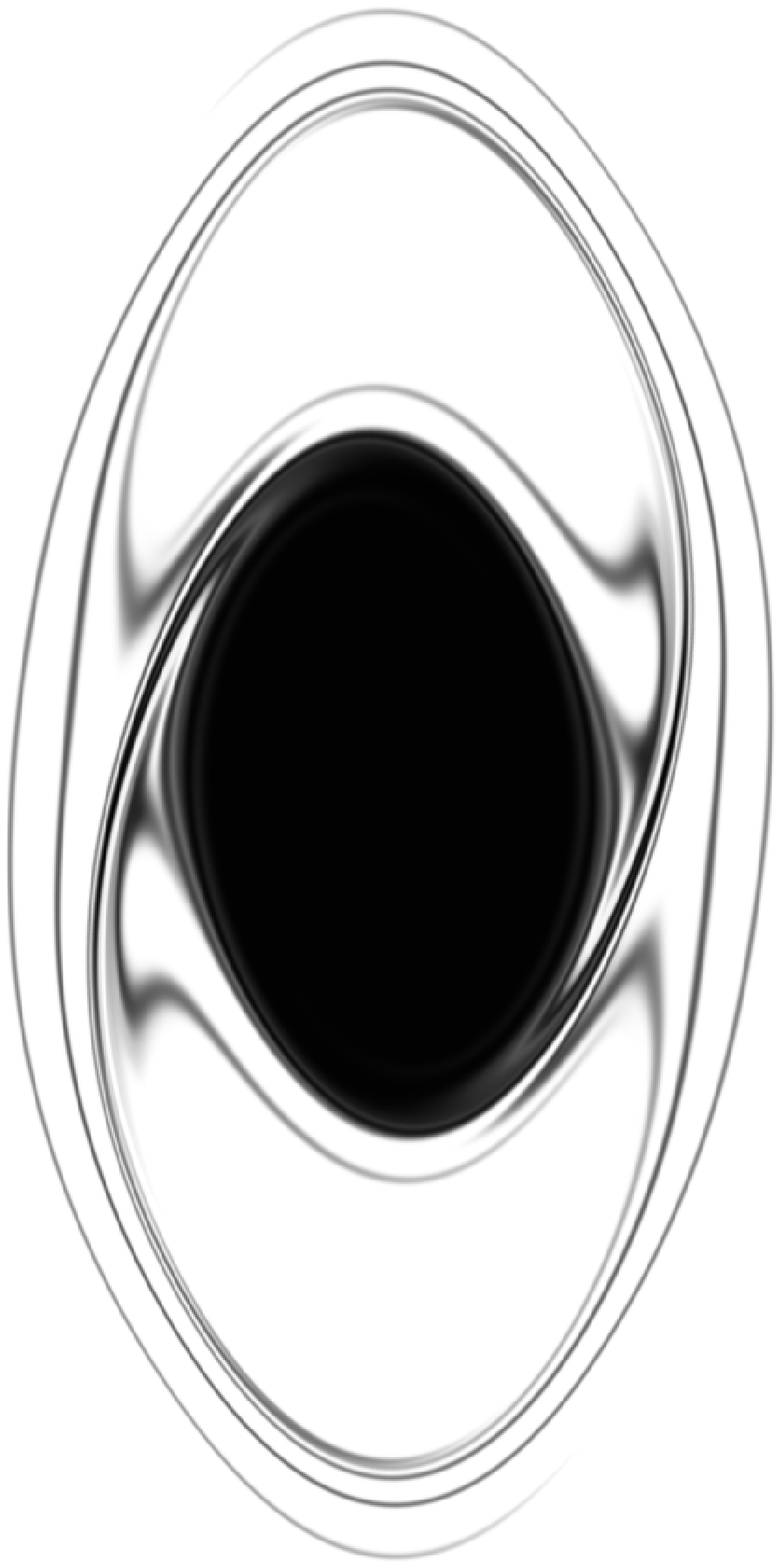,width=11cm}
\epsfig{file=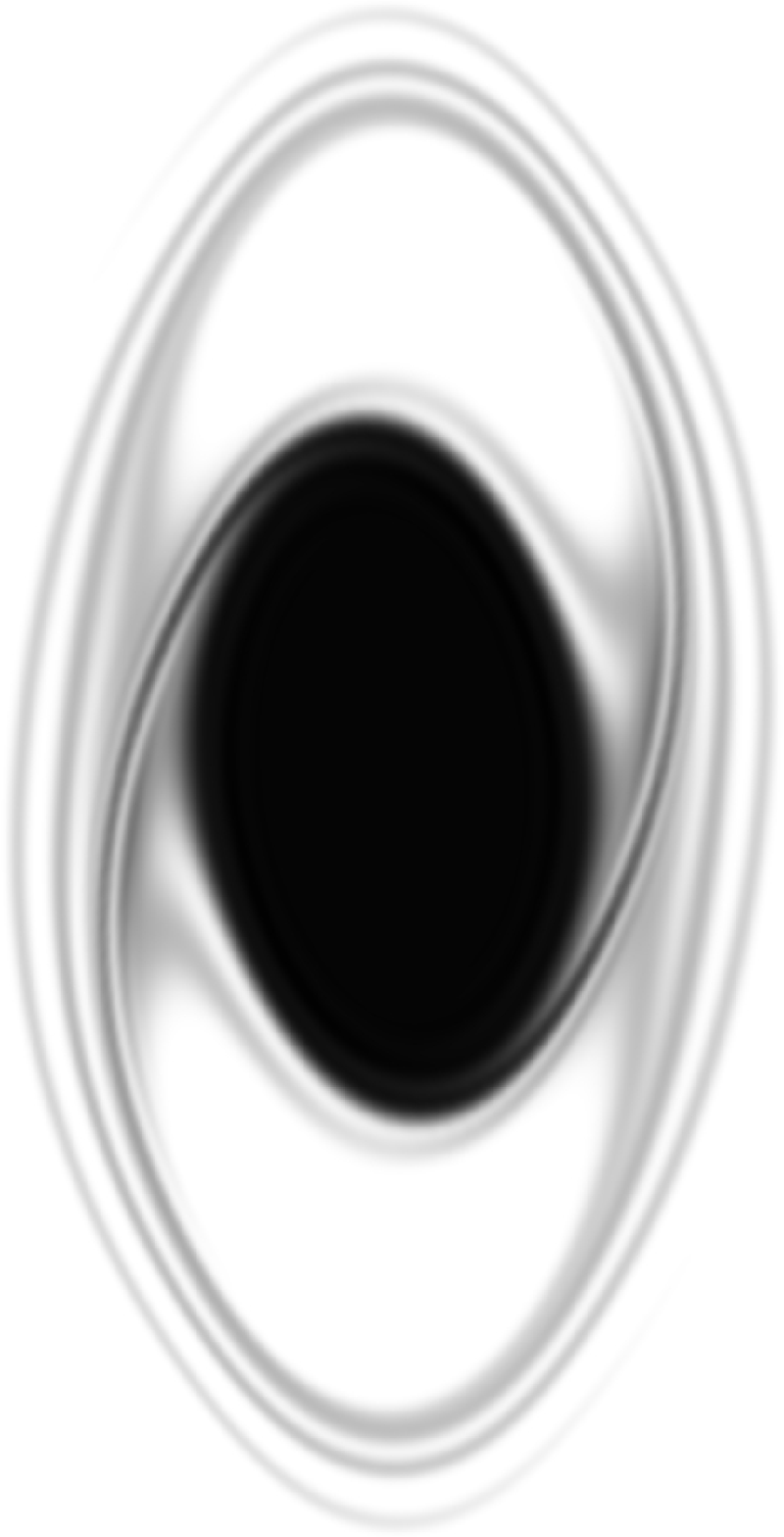,width=11cm}
}}
\centerline{t=40}
\centerline{\mbox{
\epsfig{file=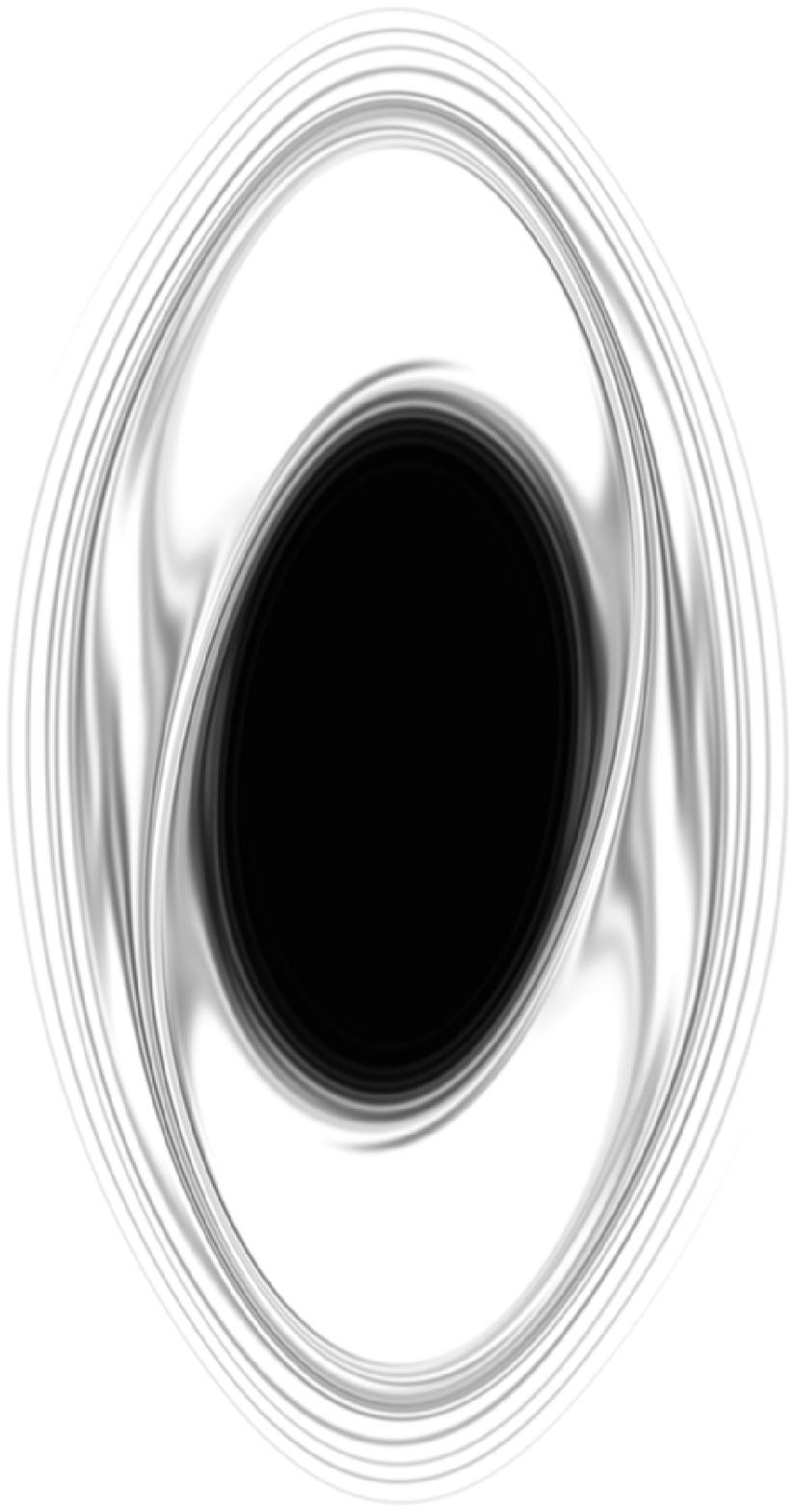,width=11cm}
\epsfig{file=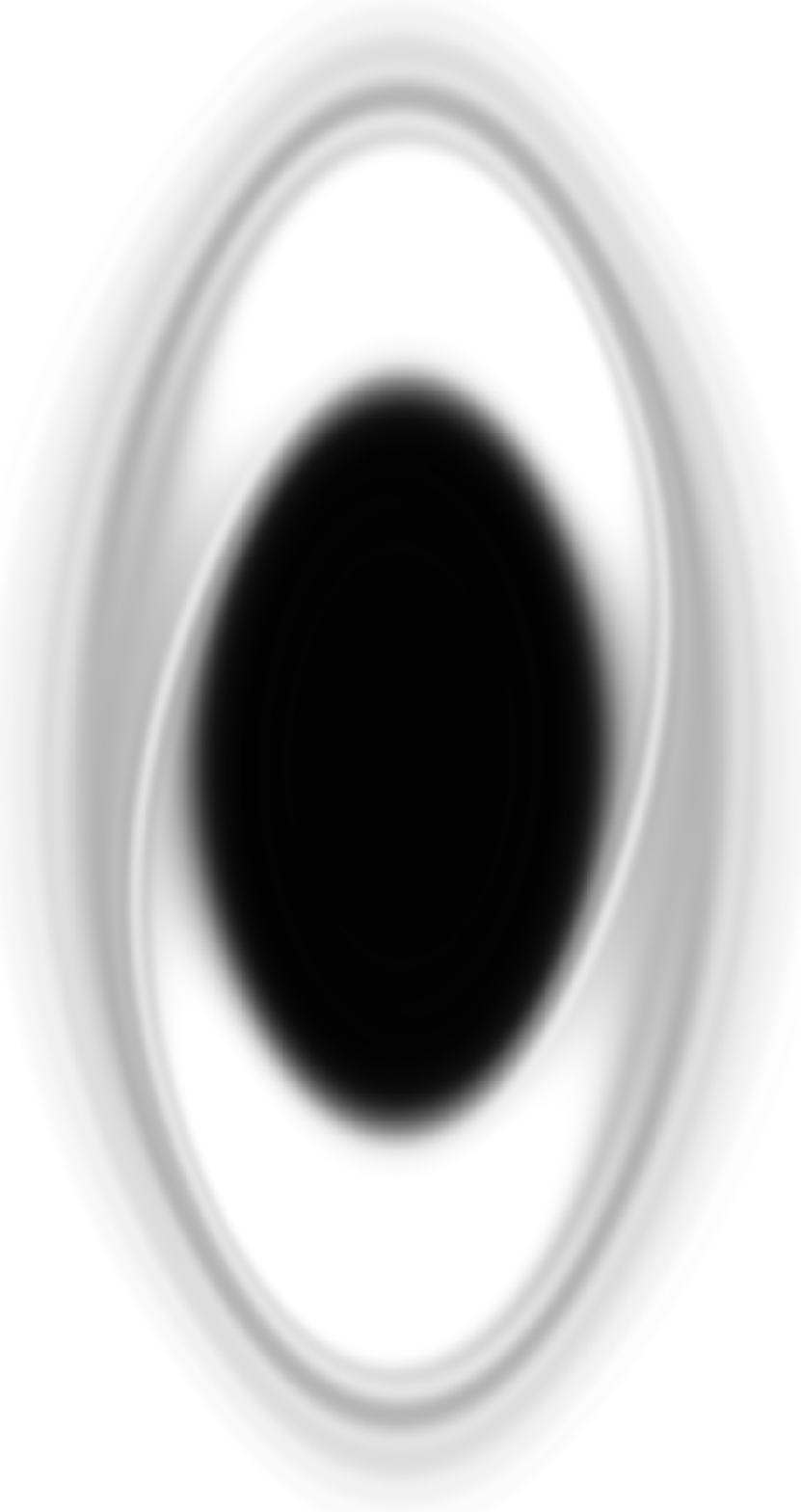,width=11cm}
}}
\centerline{high resolution\mbox{ \hskip 4cm} t=100 \mbox{\hskip 4cm} low resolution}
\centerline{\small {\bf Fig.~\ref{fig:studytophat1}} (continued).}
\end{figure}
\begin{figure}
\centerline{\fbox{\rule[-0.1cm]{0cm}{0.5cm}\hskip 0.3cm \ t=0\   \hskip 0.3cm}}
\centerline{\mbox{
\epsfig{file=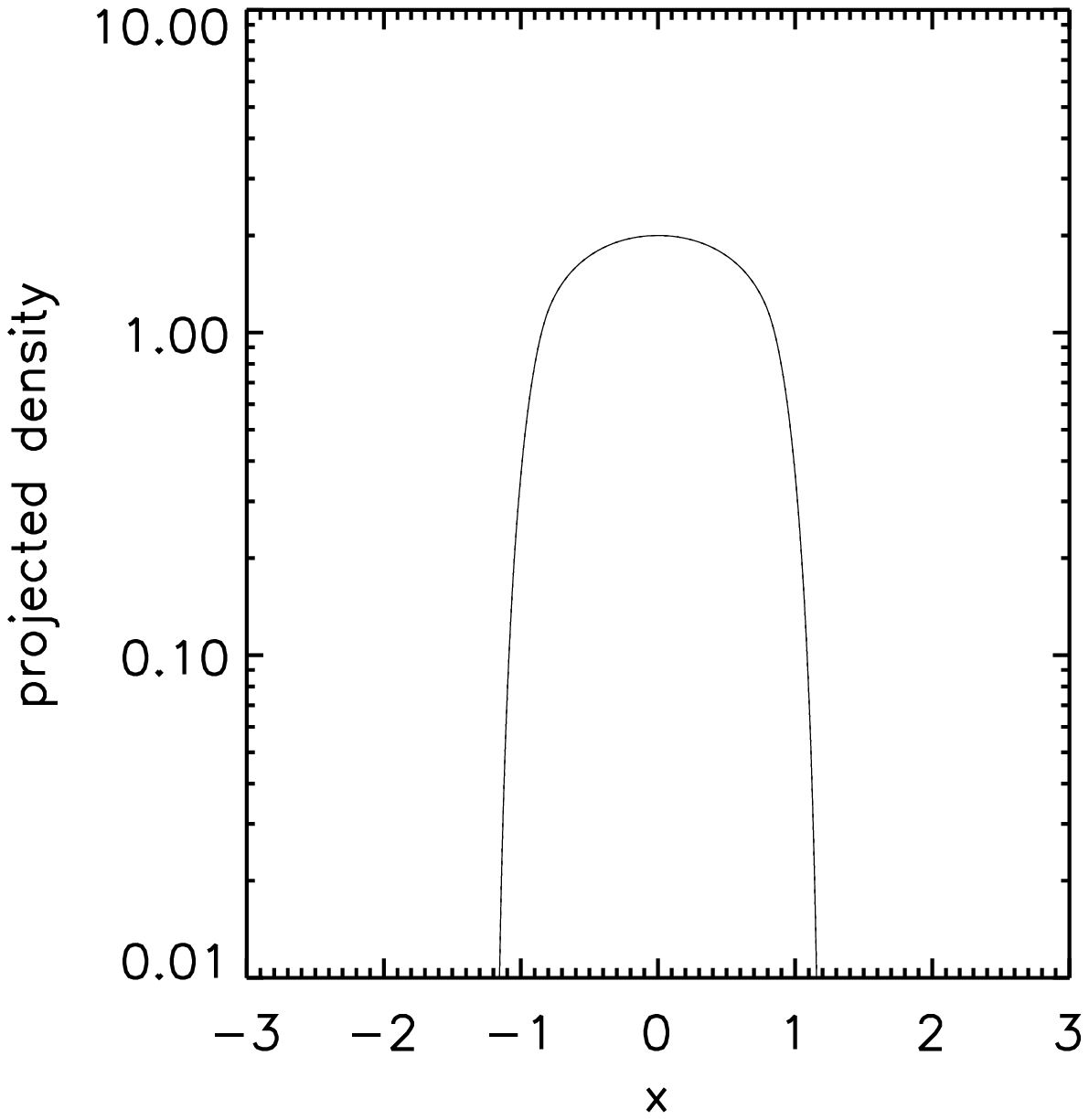,width=8cm}
\epsfig{file=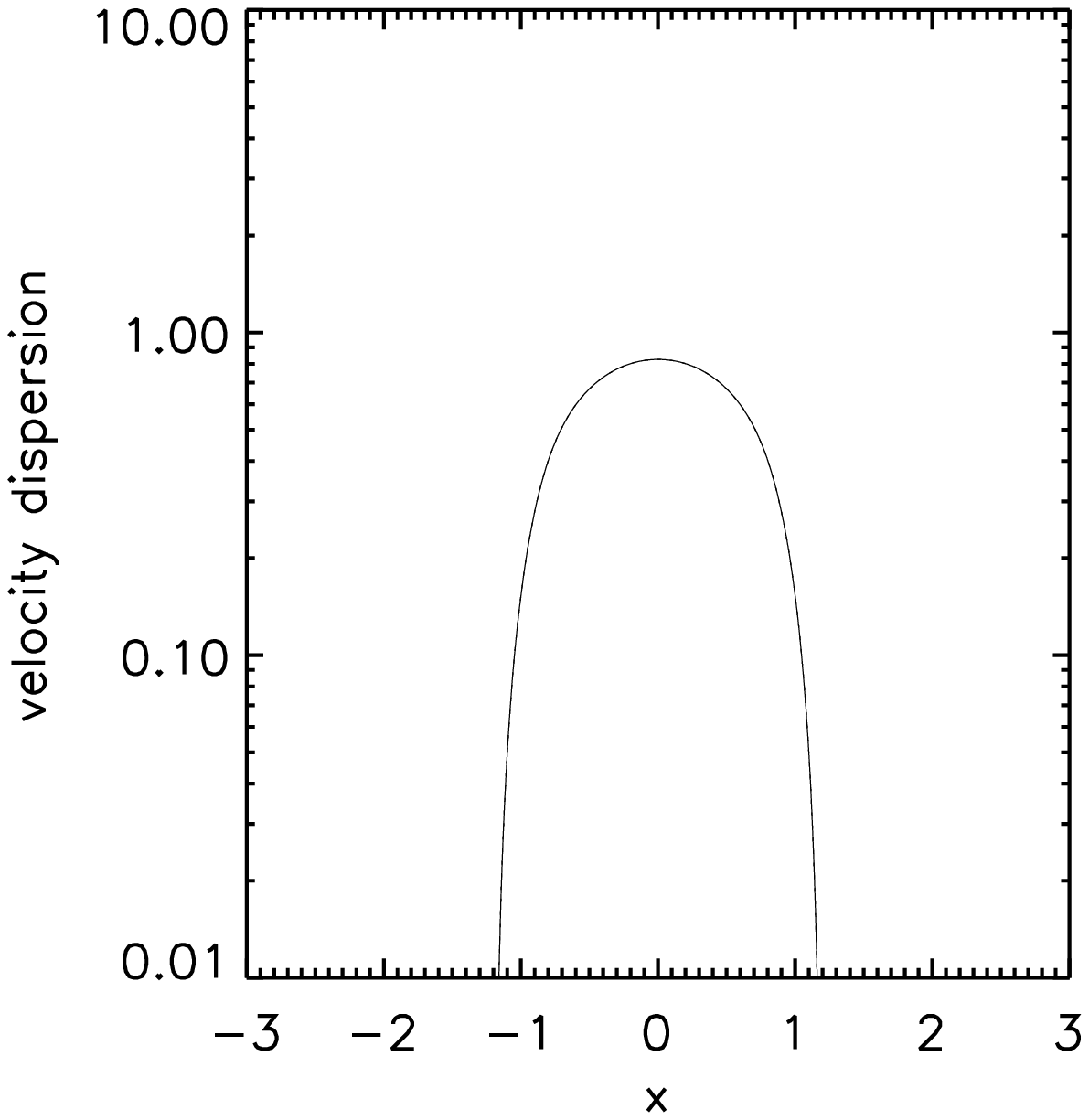,width=8cm}
}}
\vskip 0.3cm
\centerline{\fbox{\rule[-0.1cm]{0cm}{0.5cm}\hskip 0.3cm t=10\  \hskip 0.3cm}}
\centerline{\mbox{
\epsfig{file=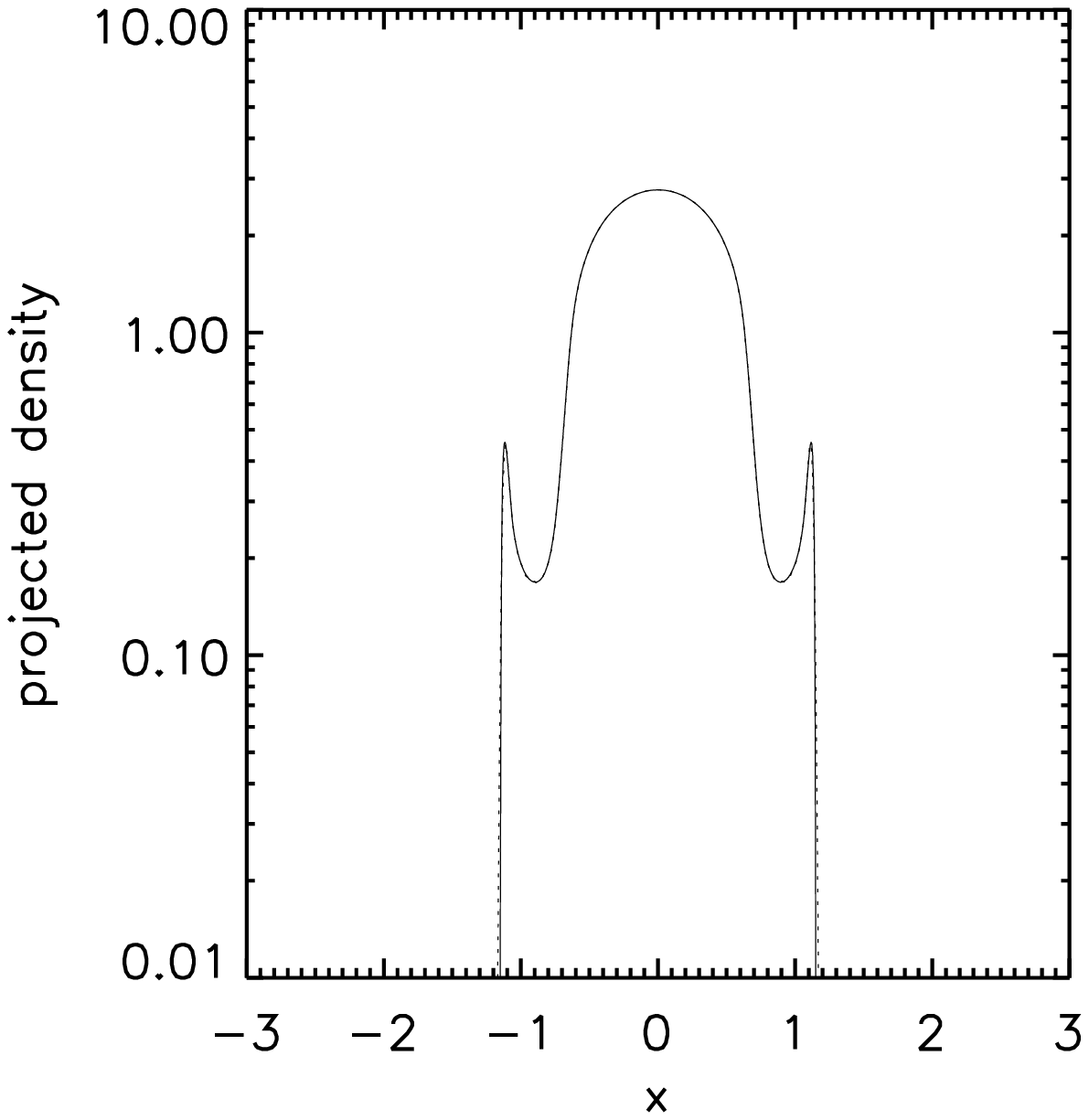,width=8cm}
\epsfig{file=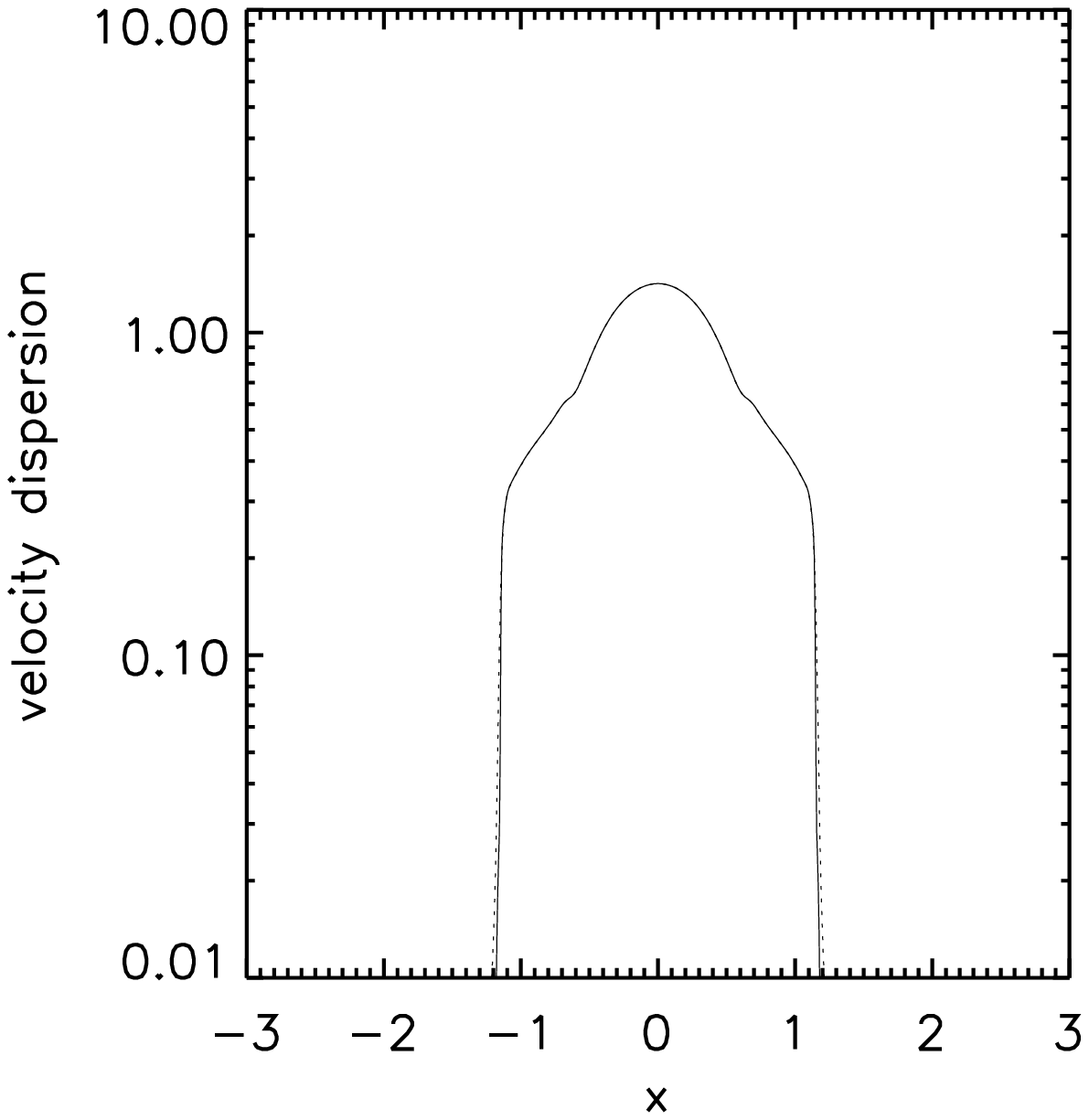,width=8cm}
}}
\caption[]{Mean projected density
 and velocity dispersion, for the simulations with
top hat initial conditions, similarly
as in Fig.~\ref{fig:studygaussian2}.}
\label{fig:studytophat2}
\end{figure}
\begin{figure}
\centerline{\fbox{\rule[-0.1cm]{0cm}{0.5cm}\hskip 0.3cm t=40\   \hskip 0.3cm}}
\centerline{\mbox{
\epsfig{file=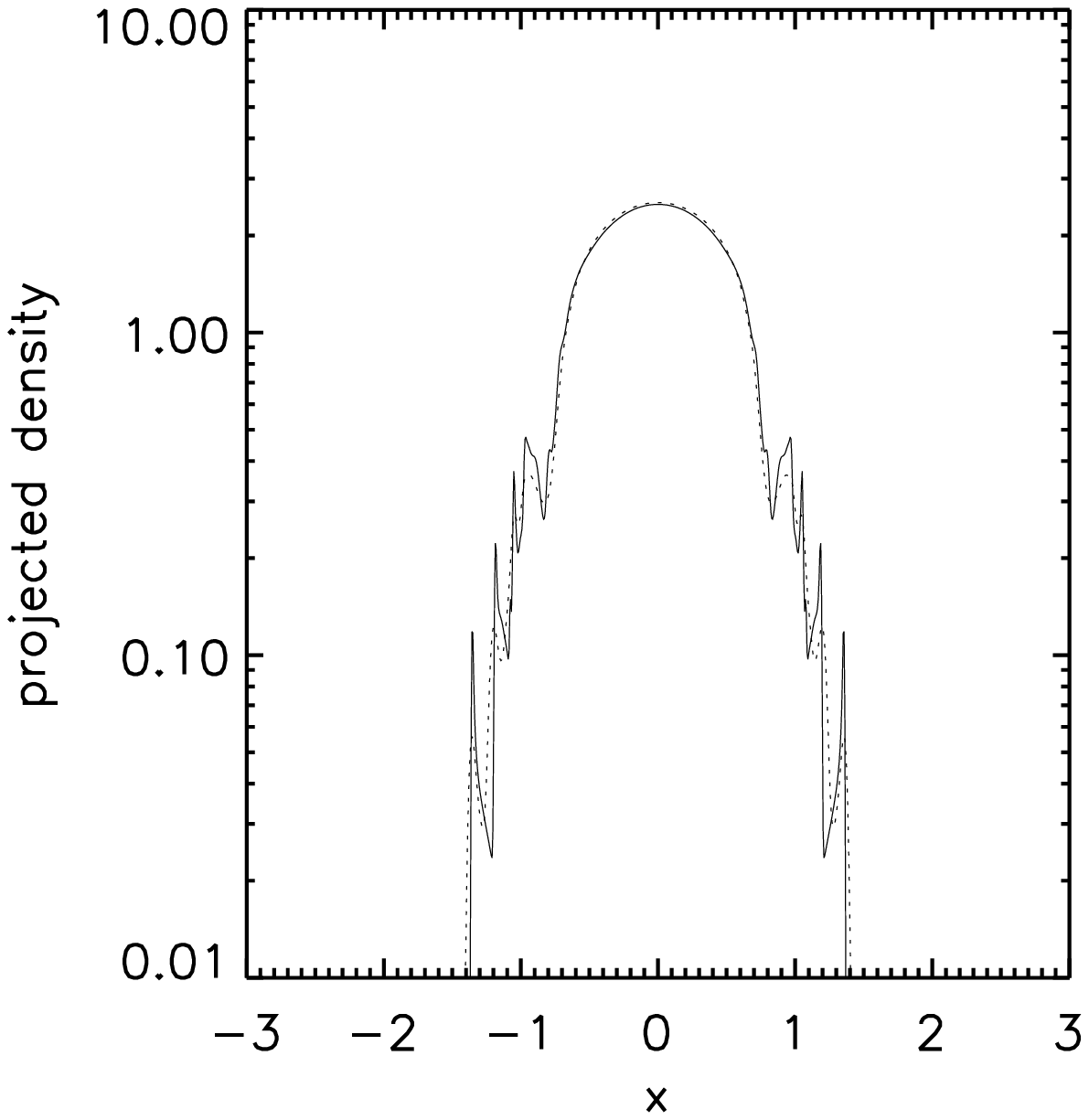,width=8cm}
\epsfig{file=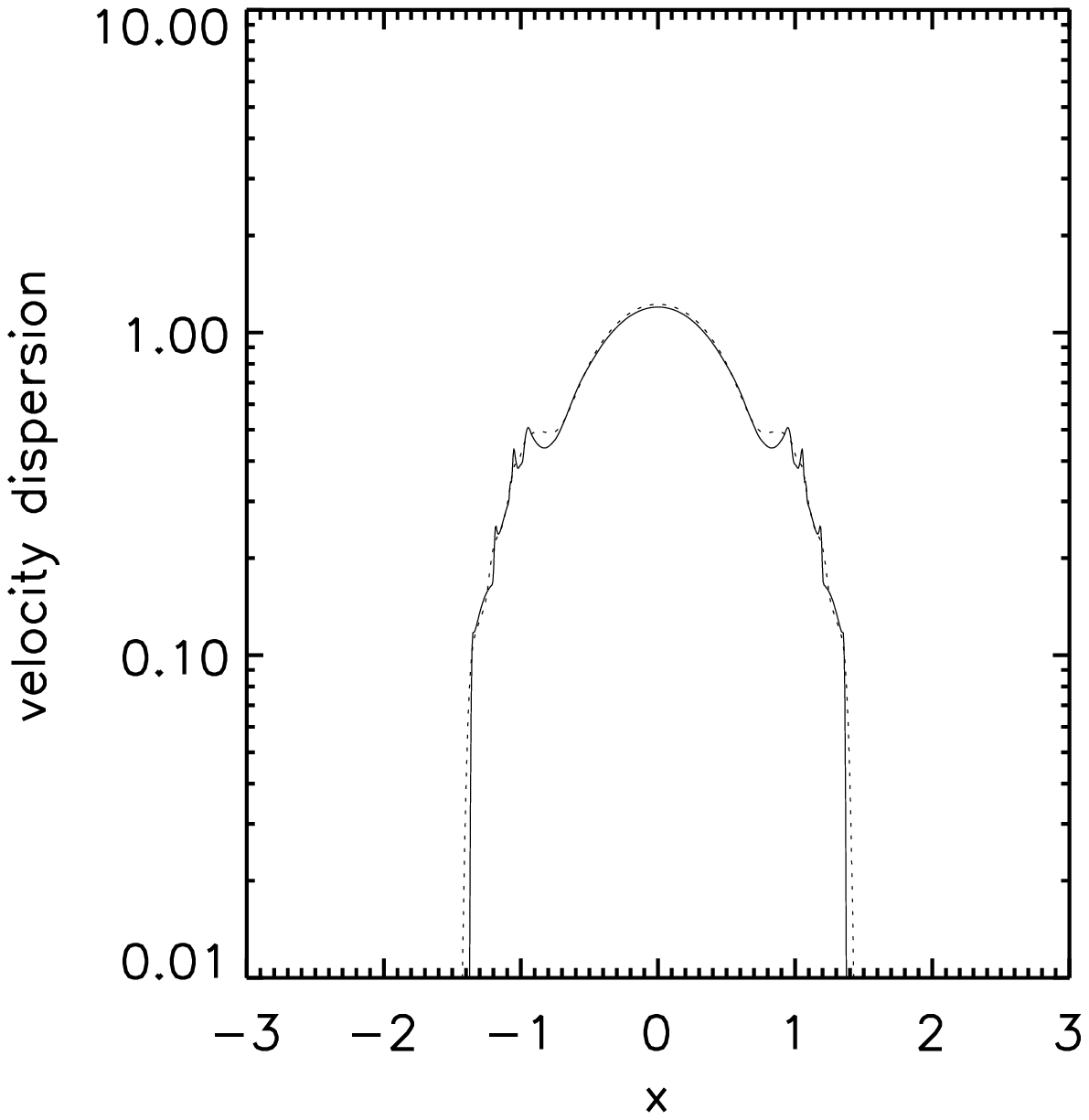,width=8cm}
}}
\centerline{\fbox{\rule[-0.1cm]{0cm}{0.5cm}\hskip 0.3cm t=100   \hskip 0.3cm}}
\centerline{\mbox{
\epsfig{file=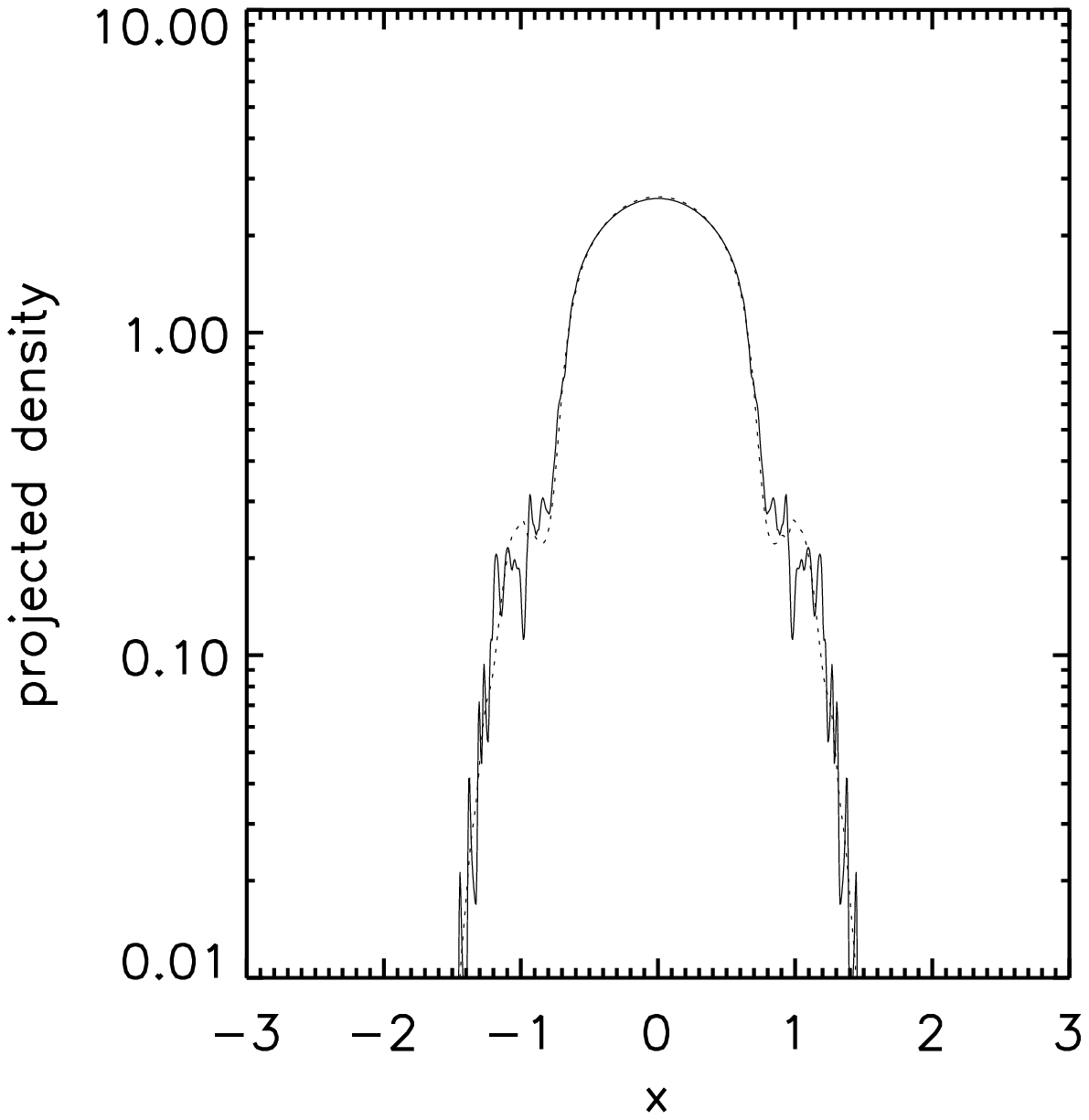,width=8cm}
\epsfig{file=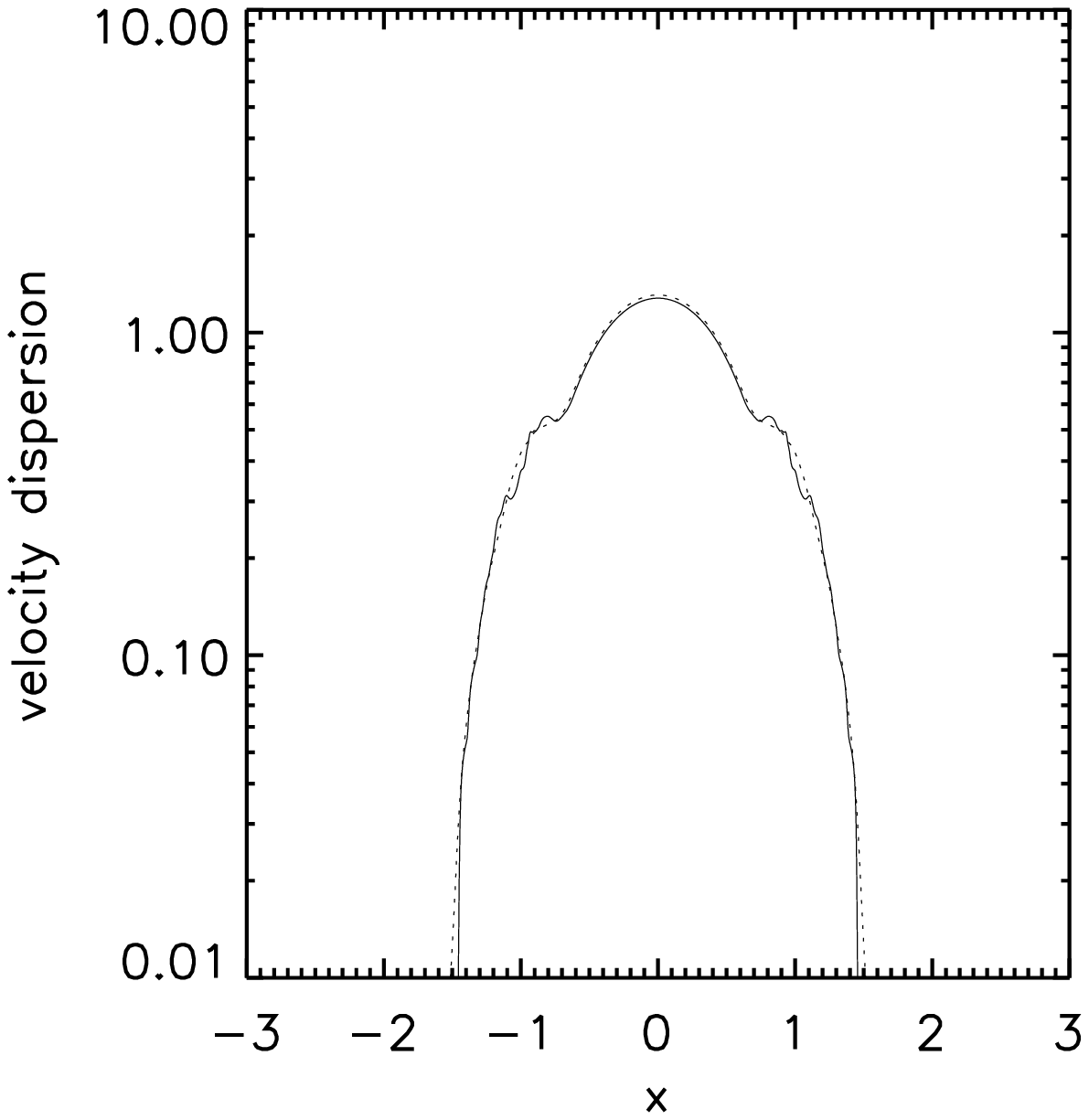,width=8cm}
}}
\centerline{\small {\bf Fig.~\ref{fig:studytophat2}} (continued).}
\end{figure}
\begin{figure}
\centerline{\mbox{
\epsfig{file=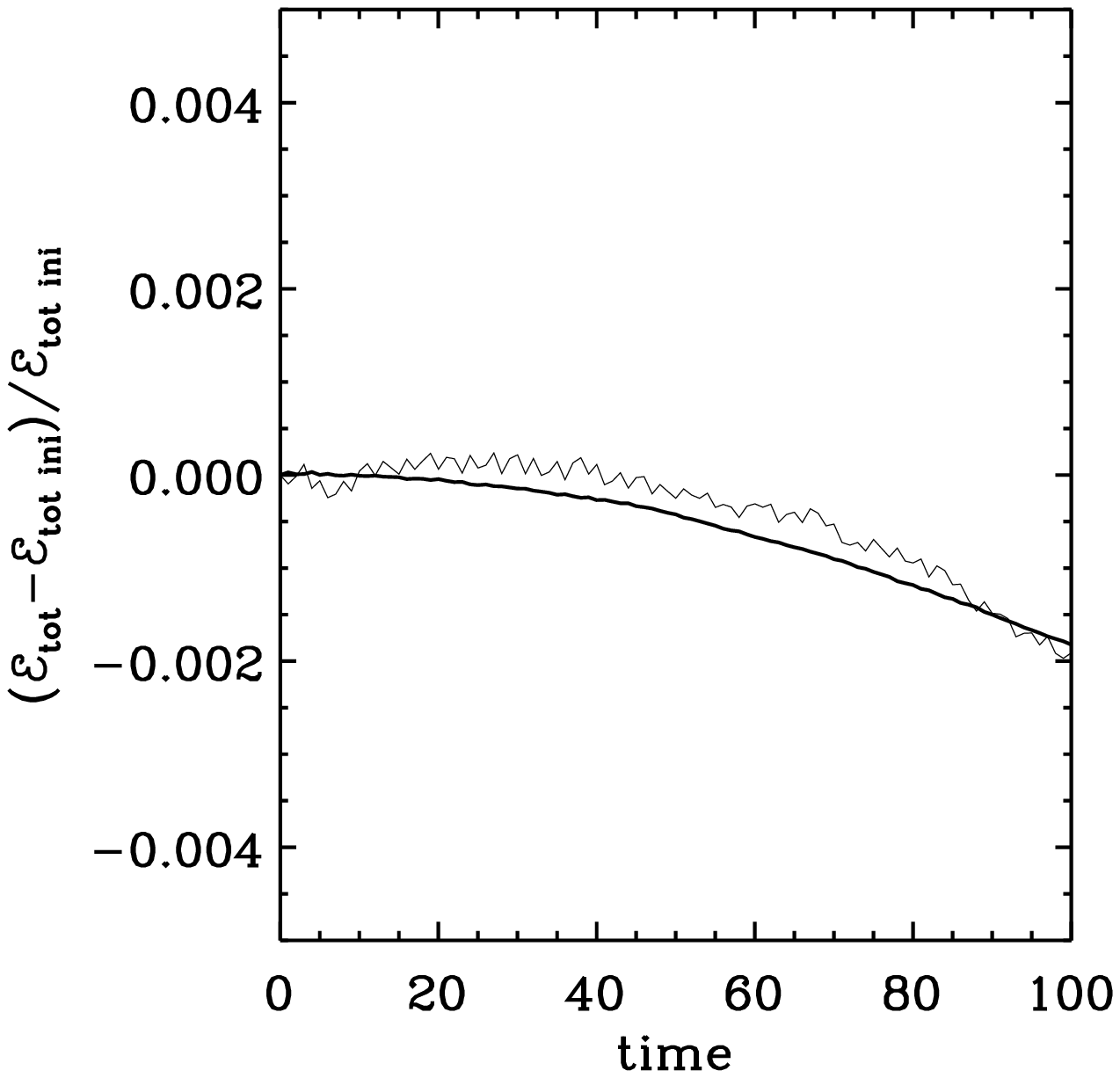,width=8cm}
}}
\caption[]{Energy conservation for the simulations with top hat initial conditions,
similarly as in Fig.~\ref{fig:studygaussian3}.}
\label{fig:studytophat3}
\end{figure}
\section{Adaptive refinement}
\label{sec:adaptiveref}
%
In the simulations with Gaussian and top hat initial conditions studied in previous section,
the complexity of the phase-space distribution function augments with time, due to a well known
effect of rolling up. In the top hat case, details are 
significant only in the external ring. This calls for local refinement, i.e. for increasing 
resolution only where needed: in principle this should reduce significantly the cost of the
simulation. The gain obtained from local refinement in 2D phase-space might be
of course quite questionable. For instance, such a gain is expected to be rather small 
in the Gaussian case, where complexity appears everywhere in the computing domain 
with approximately the same level of detail.
However, refinement is expected to be much more relevant in higher number
of dimensions, especially in the cosmological case, where only a very small fraction of
phase-space is occupied.  

Here, we examine a simple approach based on standard Adaptive Refinement Tree 
(ART) methods. Our goal is to only demonstrate that refinement is possible with the
cloud method. This section is organized as follows. In \S~\ref{sec:sketchref}, we give
a sketch of the method. Technical details are discussed in Appendix~\ref{sec:techdet}.
In \S~\ref{sec:simuraf}, we show how the method performs for the top hat 
initial conditions used  in \S~\ref{sec:tophat}: we start from low
resolution initial set up with $\Delta_{x}=\Delta_{v}=0.02$, allow for two levels
of refinement, and compare the results obtained 
to the high resolution simulations of \S~\ref{sec:tophat}, with $\Delta_{x}=\Delta_{v}=0.005$.
%
\subsection{Method of refinement}
\label{sec:sketchref}
In the practical implementation that we test on the tree-code, the
phase-space is decomposed hierarchically on a quad-tree at the moment of remap. 
Each cell of the quad-tree is associated
to a cloud. If needed, the cell is splitted equally into four sub-cells corresponding to four
sub-clouds. The process is performed as long as necessary, i.e. until the size of the
corresponding (sub-)clouds obey some criteria based on local properties of the phase-space
distribution function. 

At each successive level of refinement, we use Van-Citter algorithm to
reconstruct the distribution function. We first start from the coarse level, where we
reconstruct $f$ the same way as for the fixed resolution code, as described in \S~\ref{sec:deconv}. 
We then consider the set of clouds corresponding to first level of refinement
as residues to reconstruct $f$ more accurately by applying Van-Citter algorithm again
on these clouds. And so on, until last level of refinement.

We consider two criteria of refinement. 
The first one is based on the measurement of local curvature of the phase-space distribution function:
local curvature is indeed a key quantity for preserving
details in the distribution function: the cloud size should be always small compared
to the local curvature radius. The second criterion is based on convergence of
the reconstruction at successive refinement levels: clouds where $f$ is poorly
reconstructed are refined. We shall see that in practice, both curvature and convergence criteria 
give similar refinement structure.

In our implementation of refinement, even when splitted into smaller sub-clouds, clouds at
a given refinement level are thus kept since their sub-clouds counterparts 
are considered as residues. In principle, it would be advisable from 
the dynamical point of view to remove these clouds from the hierarchy, 
i.e. to set their mass to zero. Indeed, because of the structures in the gravitational
potential brought by the small clouds, the quadratic approximation is expected at
some point to become invalid for the larger clouds. We notice as well that one
of our refinement criteria should rely as well on variations of the projected density,
which account for deviations from quadraticity of the potential. Since our
goal is just to demonstrate that refinement is possible with our method,
we decided in the present work to put aside removal of coarser level clouds and refinement based on
projected density. As it is, thus, our refinement procedure is quite improvable and 
will work only if the gravitational potential remains sufficiently quadratic 
at the coarse level, which should be hopefully the case if the number of 
refinement levels is not too large.
%
%
\subsection{Example: top hat initial conditions}
\label{sec:simuraf}
%
To check how our refinement procedure performs, we realized again simulations with the
same top hat initial conditions as in \S~\ref{sec:tophat}, but starting from low
resolution initial conditions with $\Delta_{\rm g}=0.02$ and allowing refinement
until high resolution is reached, $\Delta_{\rm g}=0.005$. This represents a significant
increase in mass resolution, a factor 16, corresponding in total to a coarse level plus
two levels of refinement. We performed two simulations, based on local curvature and 
local convergence criteria, respectively.
All the parameters defining the simulations are the same
as in \S~\ref{sec:simuset}, except for those depending on refinement, 
which are given in Appendices~\ref{sec:refinementprinciple}
and \ref{sec:critraf}. There is also a difference in the choice of $\delta_{\rm min}$
(see \S~\ref{sec:deconv}) that we set to $\delta_{\rm min}=0.0005$ for simulations based
on adaptive refinement with local convergence criterion.

\begin{figure}
\centerline{\mbox{
\epsfig{file=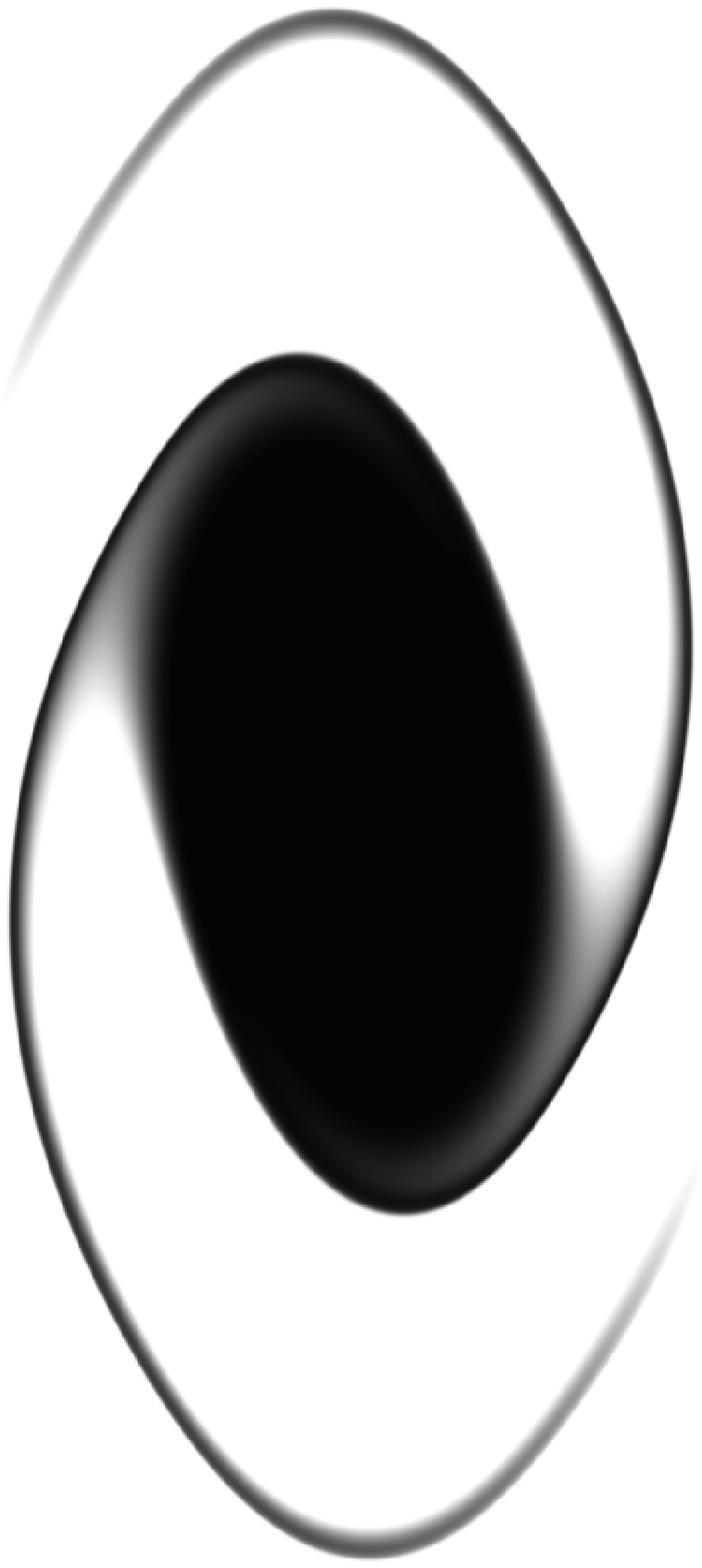,width=6.8cm}
\epsfig{file=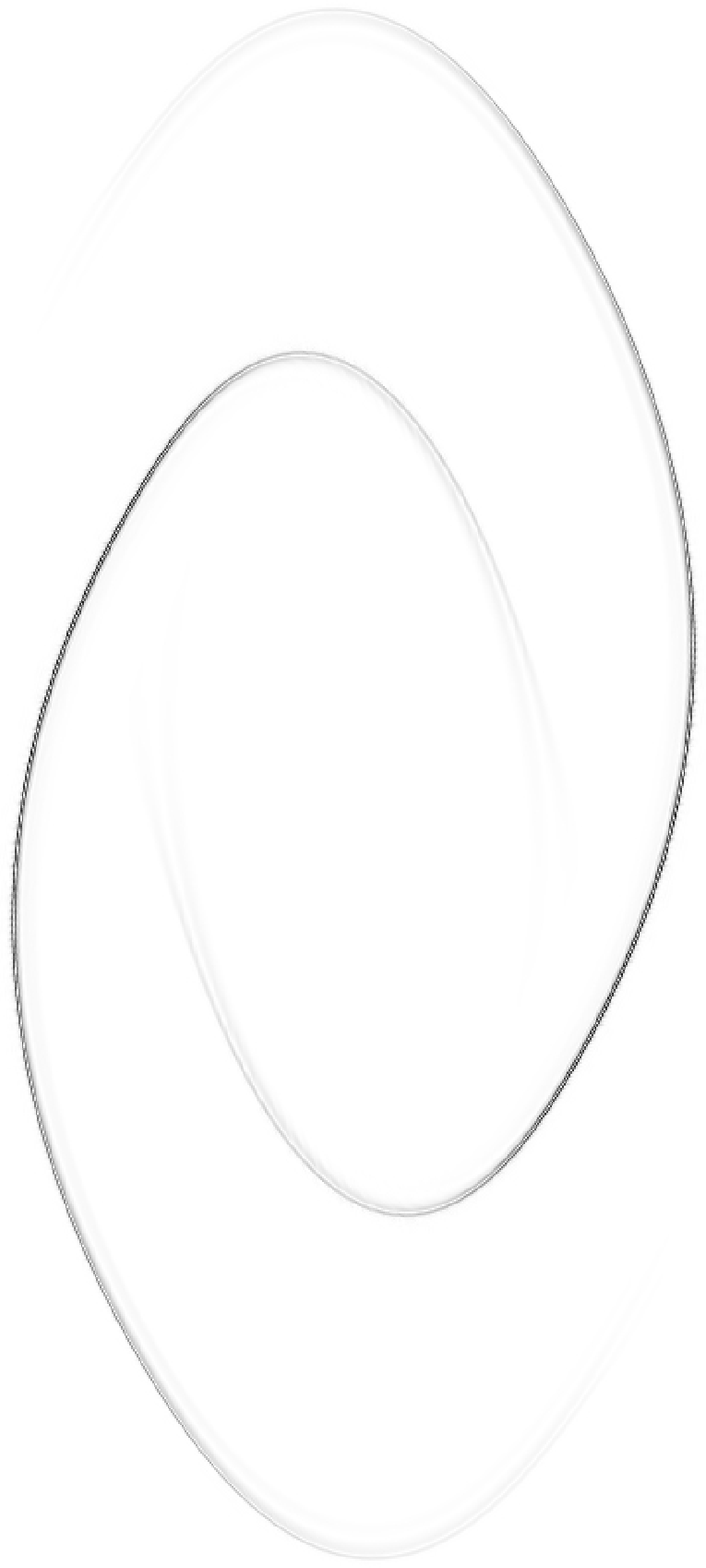,width=6.8cm}
}}
\centerline{full resolution\mbox{ \hskip 5cm} local curvature}
\vskip 0.3cm
\centerline{\mbox{
\epsfig{file=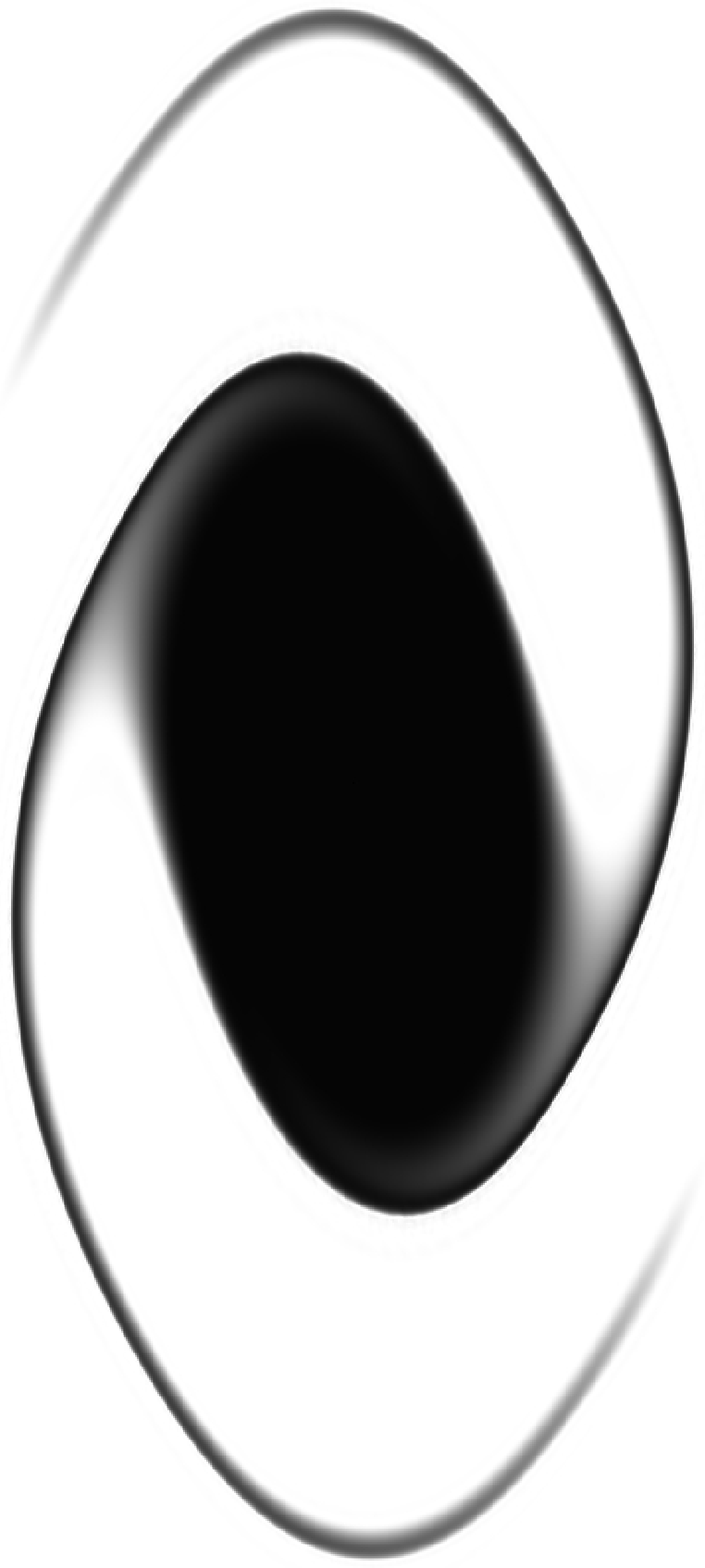,width=6.8cm}
\epsfig{file=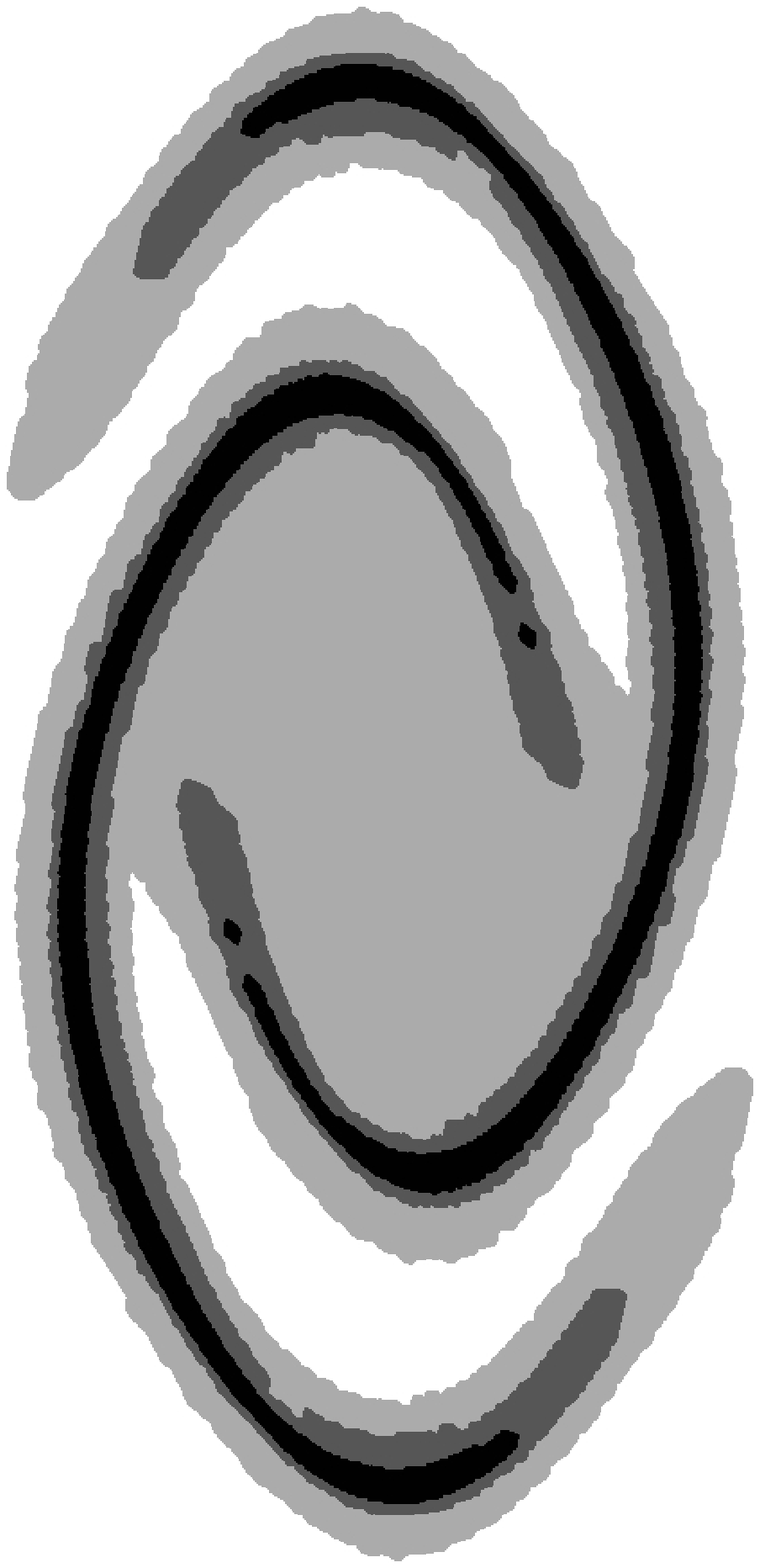,width=6.8cm}
}}
\centerline{curvature criterion\mbox{ \hskip 4.8cm} curvature levels}
\vskip 0.3cm
\centerline{\mbox{
\epsfig{file=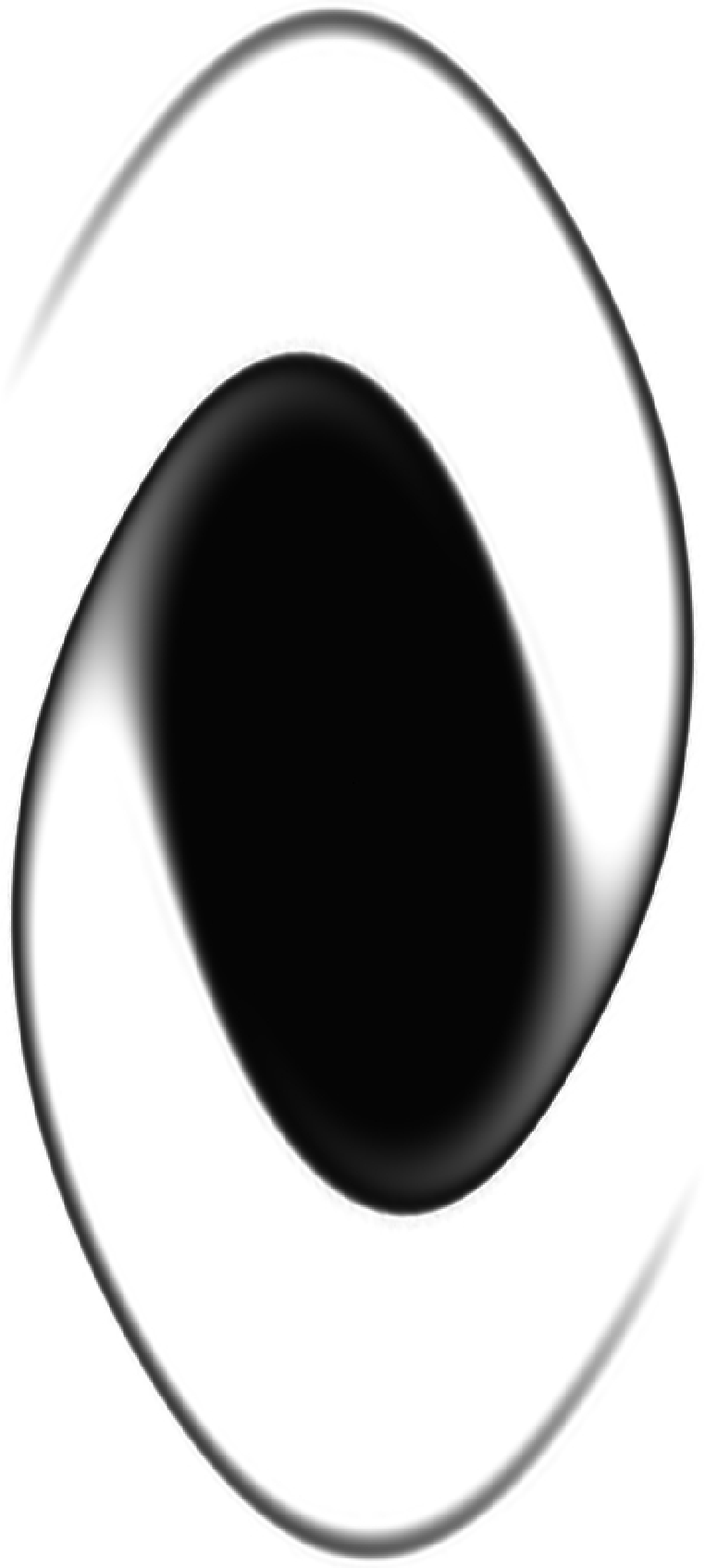,width=6.8cm}
\epsfig{file=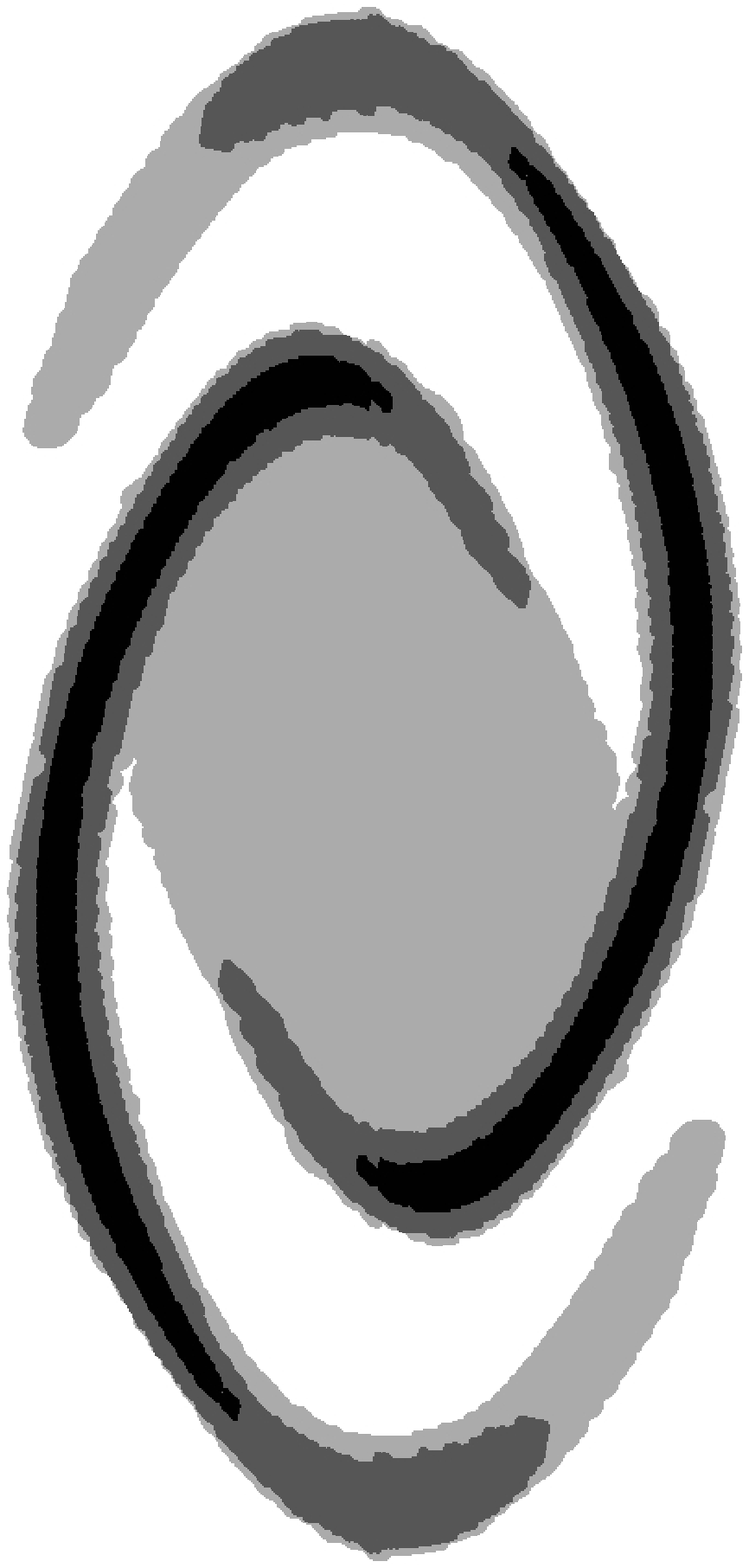,width=6.8cm}
}}
\centerline{convergence criterion\mbox{ \hskip 4.5cm} convergence levels}
\caption[]{Top hat simulations at $t=10$.
{\em Left panels:} phase-space distribution function for 
the full resolution simulation (top), the simulation with 
refinement based on local curvature (middle) and the simulation with refinement based on local
convergence (bottom). {\em Top right panel:} the local curvature measured in the simulation
with refinement based on local curvature, as explained in the text of \S~\ref{sec:simuraf}.
{\em Two bottom right panels:} levels of refinement in the simulations with
local curvature criterion (middle) and local convergence criterion (bottom). Darker regions
corresponds to higher level. }
\label{fig:studytophatref1}
\end{figure}

\begin{figure}
\centerline{\mbox{
\epsfig{file=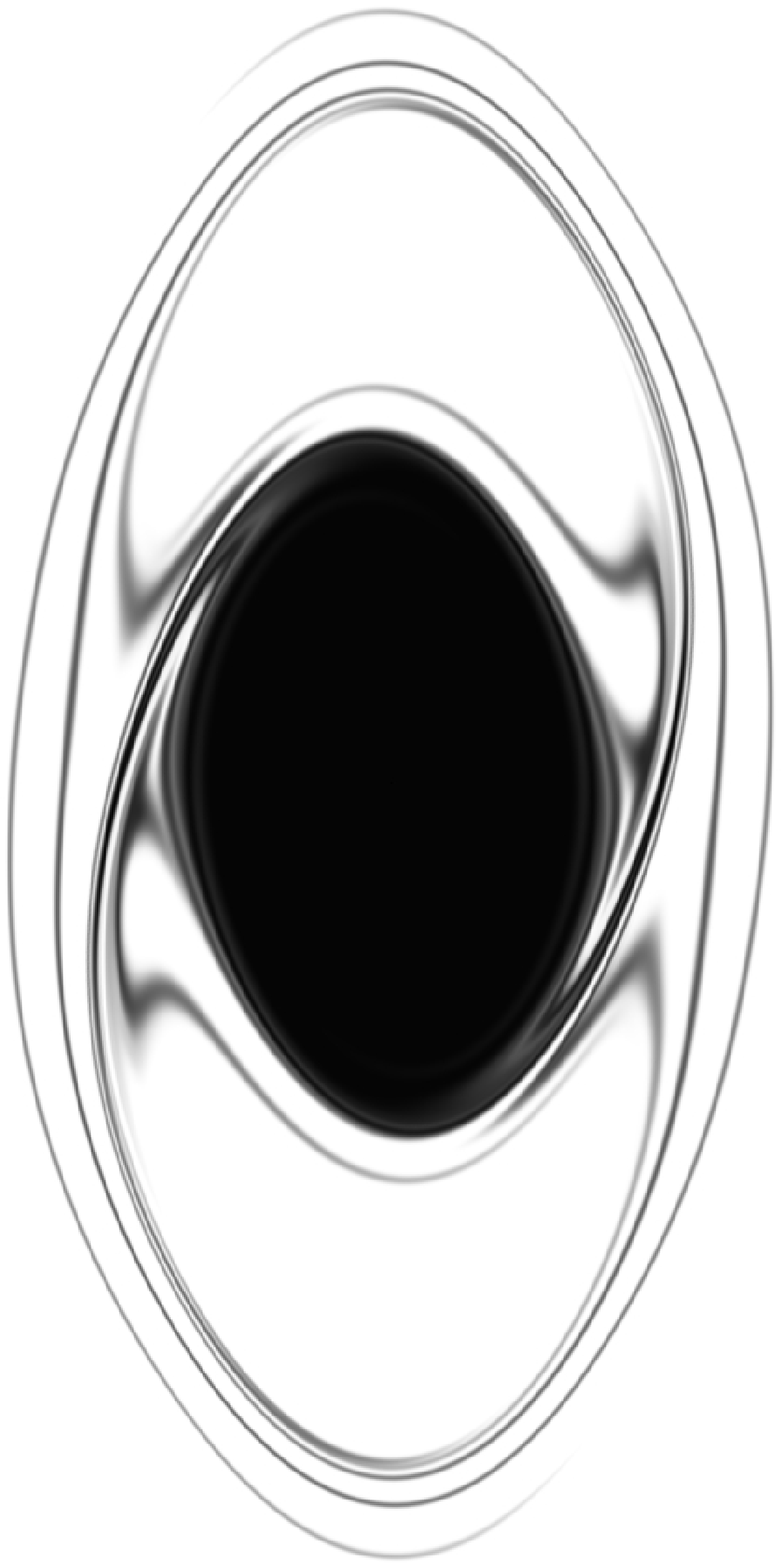,width=6.8cm}
\epsfig{file=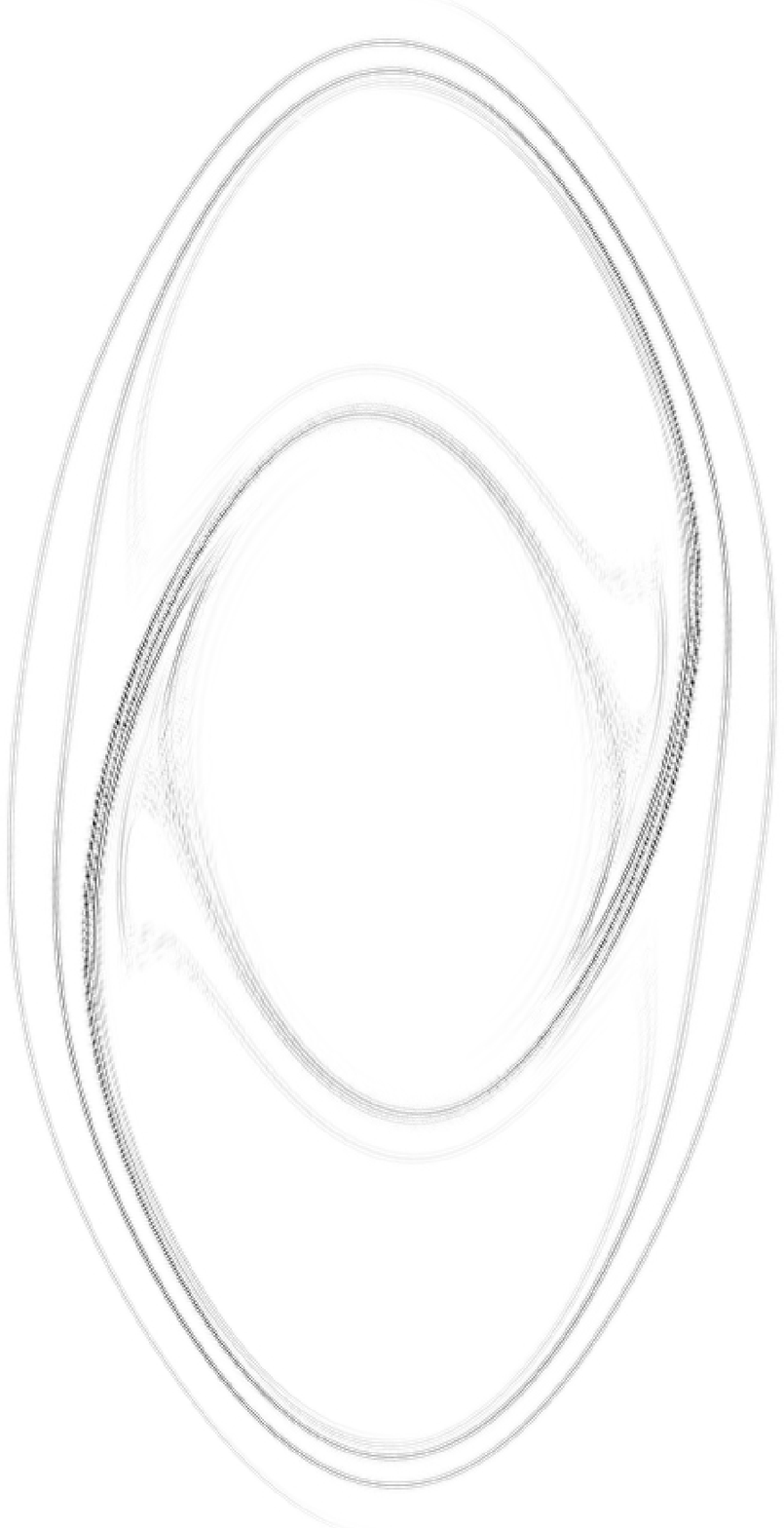,width=6.8cm}
}}
\centerline{full resolution\mbox{ \hskip 5cm} local curvature}
\vskip 0.3cm
\centerline{\mbox{
\epsfig{file=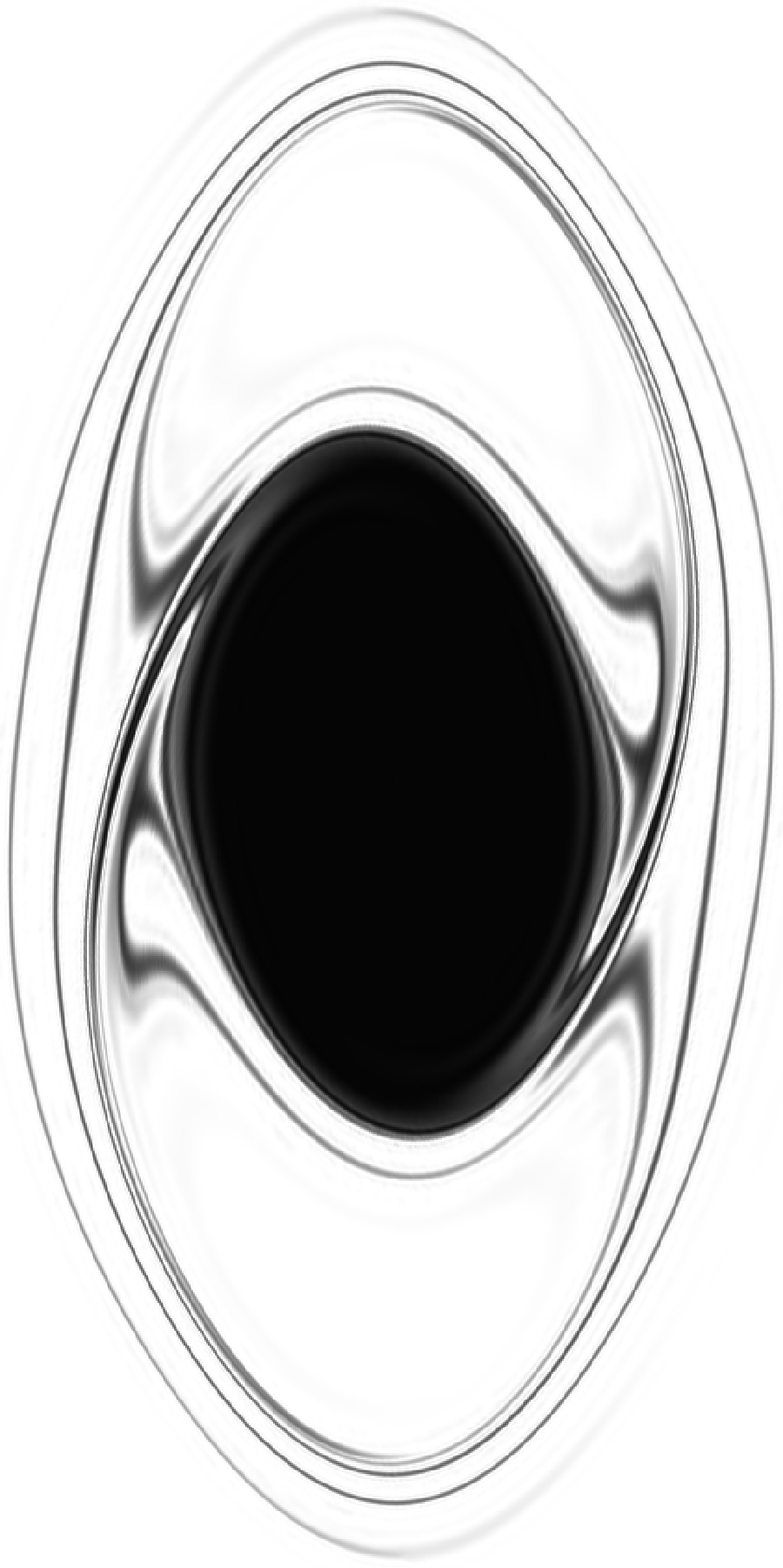,width=6.8cm}
\epsfig{file=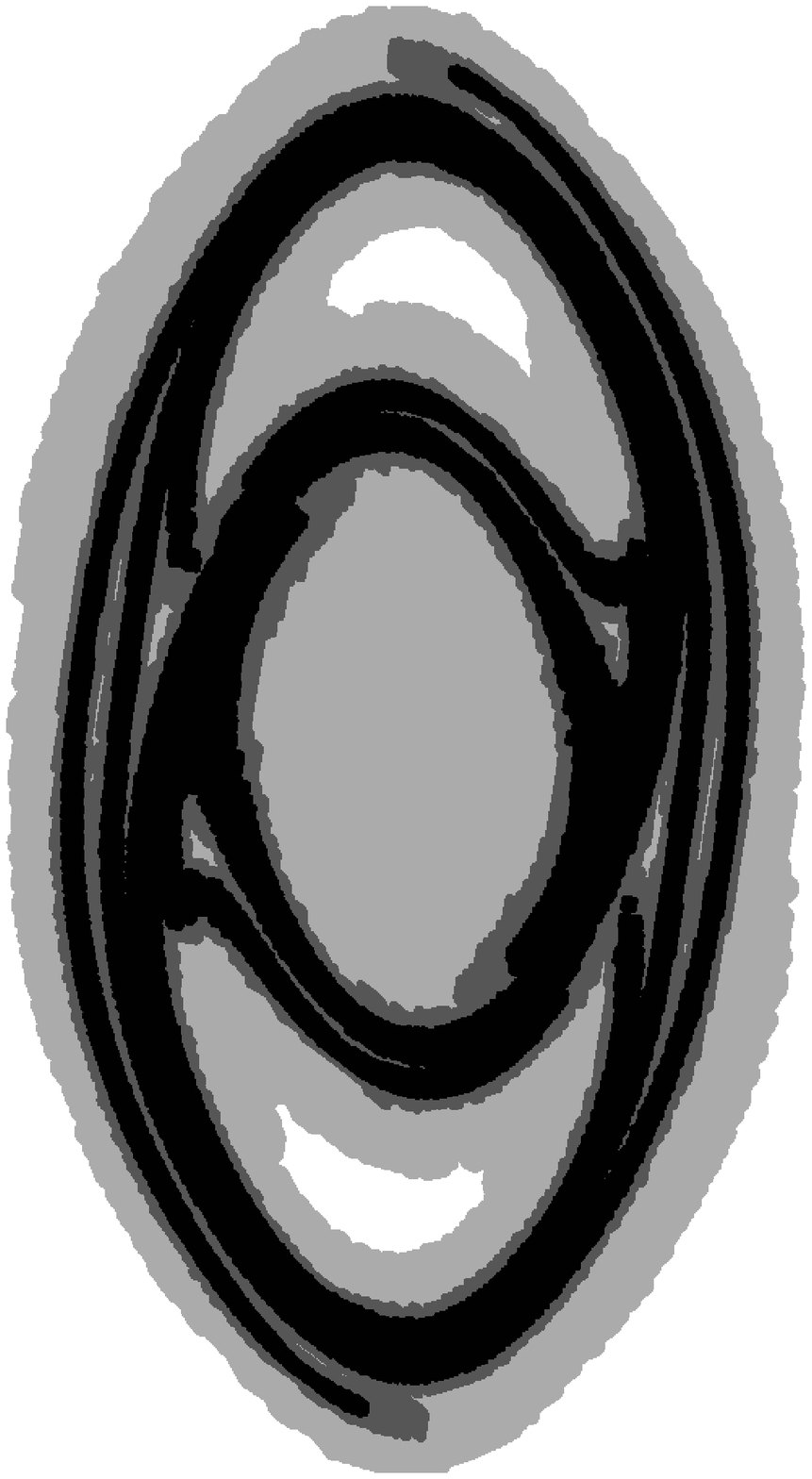,width=6.8cm}
}}
\centerline{curvature criterion\mbox{ \hskip 4.8cm} curvature levels}
\vskip 0.3cm
\centerline{\mbox{
\epsfig{file=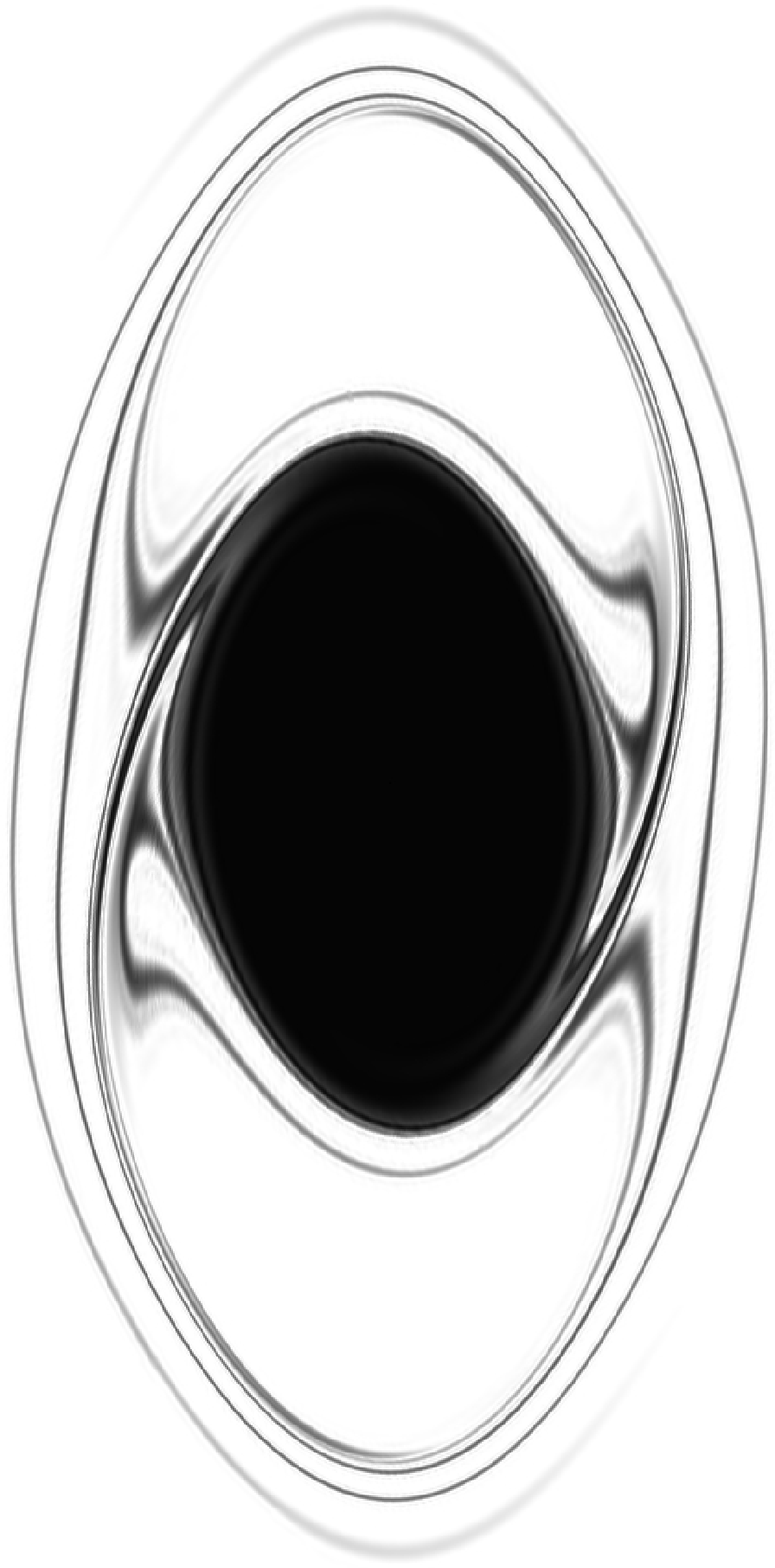,width=6.8cm}
\epsfig{file=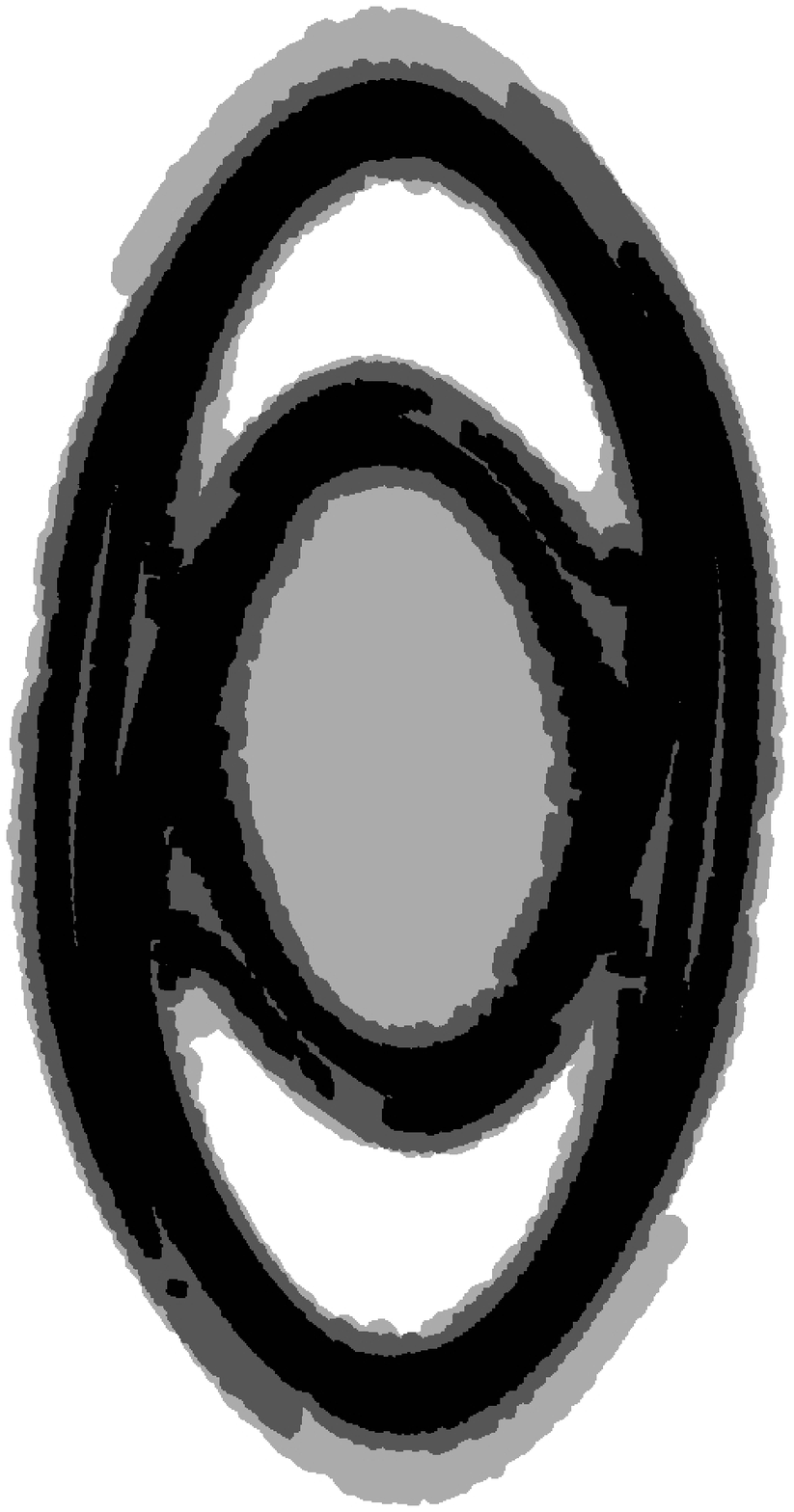,width=6.8cm}
}}
\centerline{convergence criterion\mbox{ \hskip 4.5cm} convergence levels}
\caption[]{Same as in Fig.~\ref{fig:studytophatref1}, but for $t=40$.}
\label{fig:studytophatref2}
\end{figure}

\begin{figure}
\centerline{\mbox{
\epsfig{file=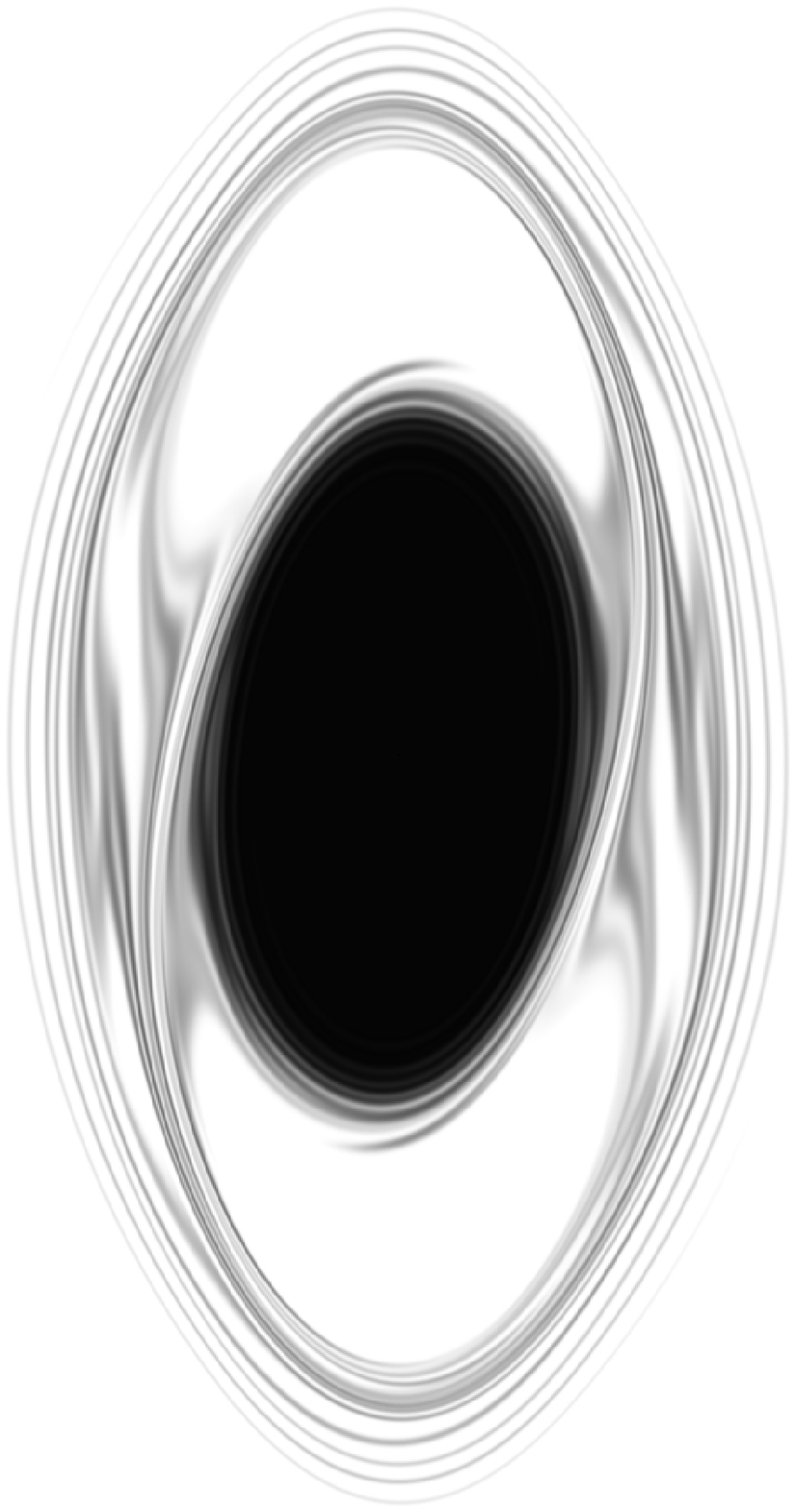,width=6.8cm}
\epsfig{file=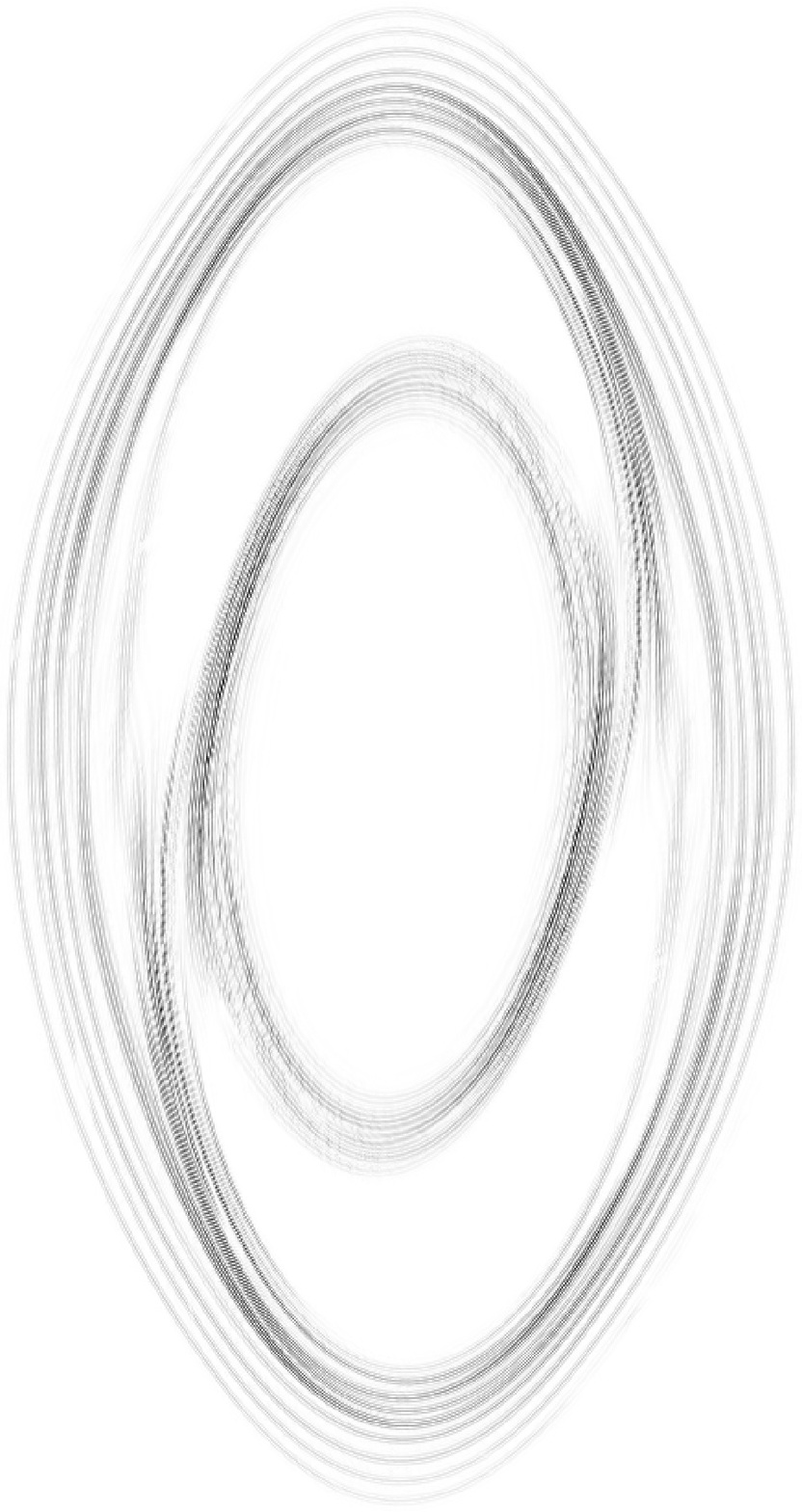,width=6.8cm}
}}
\centerline{full resolution\mbox{ \hskip 5cm} local curvature}
\vskip 0.3cm
\centerline{\mbox{
\epsfig{file=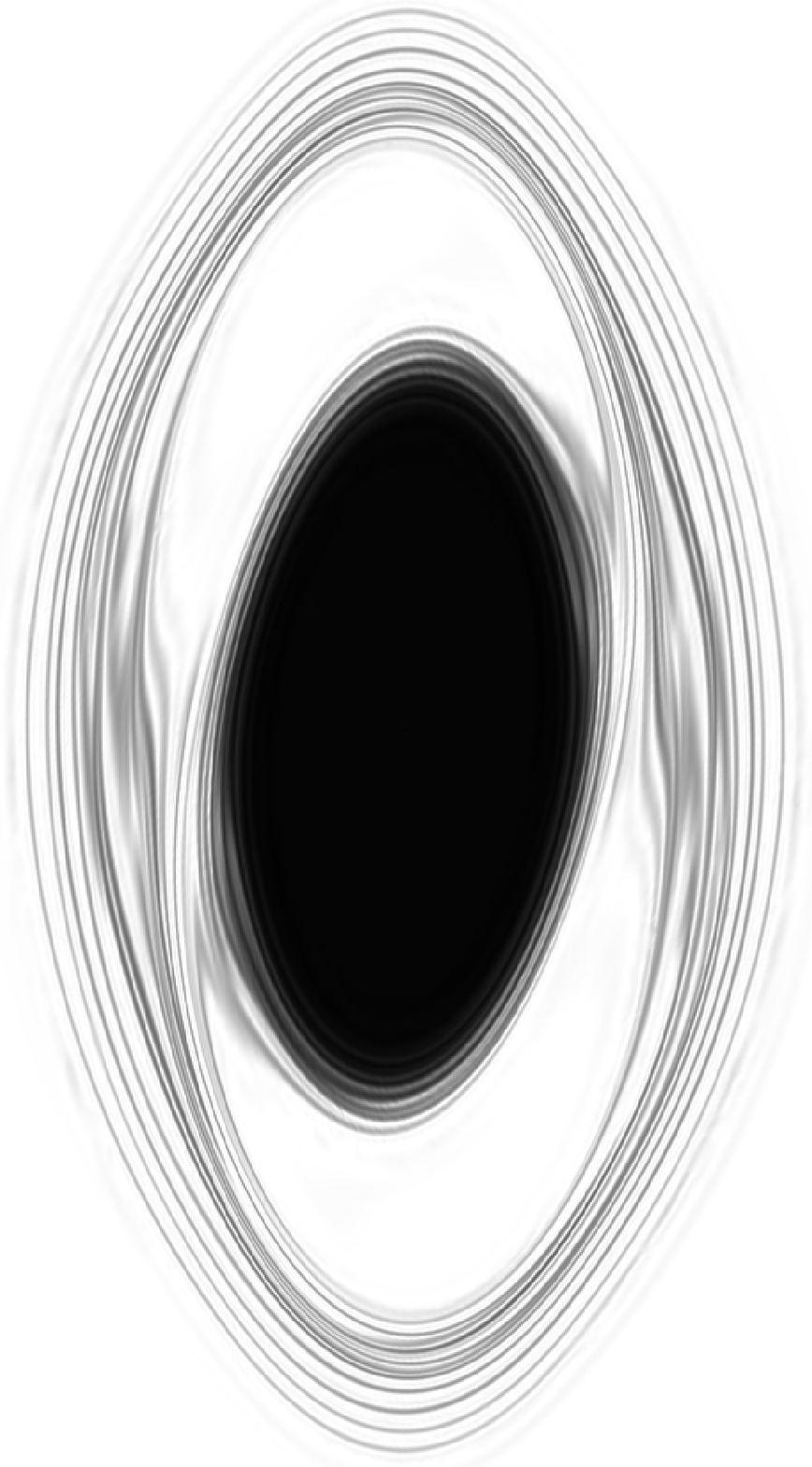,width=6.8cm}
\epsfig{file=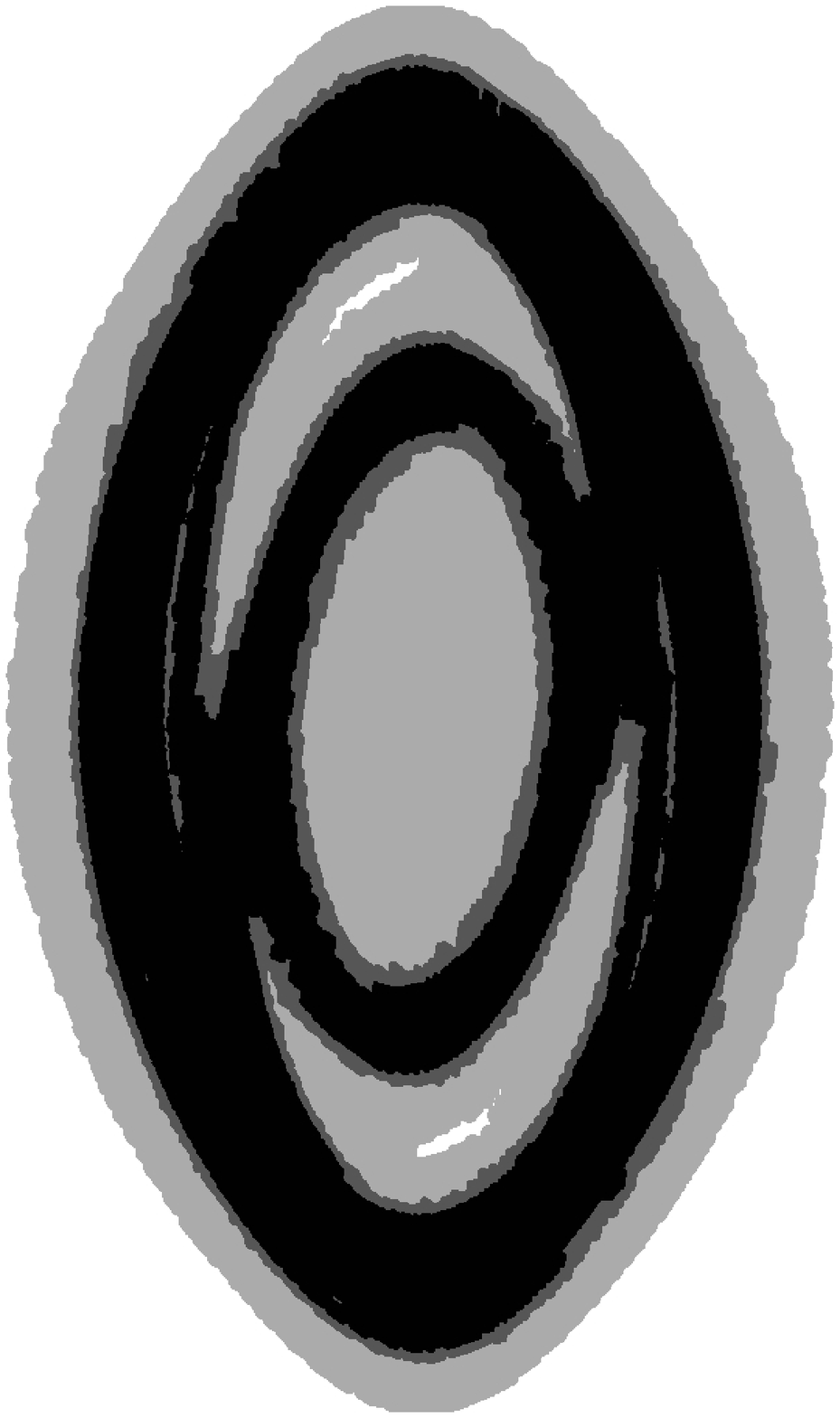,width=6.8cm}
}}
\centerline{curvature criterion\mbox{ \hskip 4.8cm} curvature levels}
\vskip 0.3cm
\centerline{\mbox{
\epsfig{file=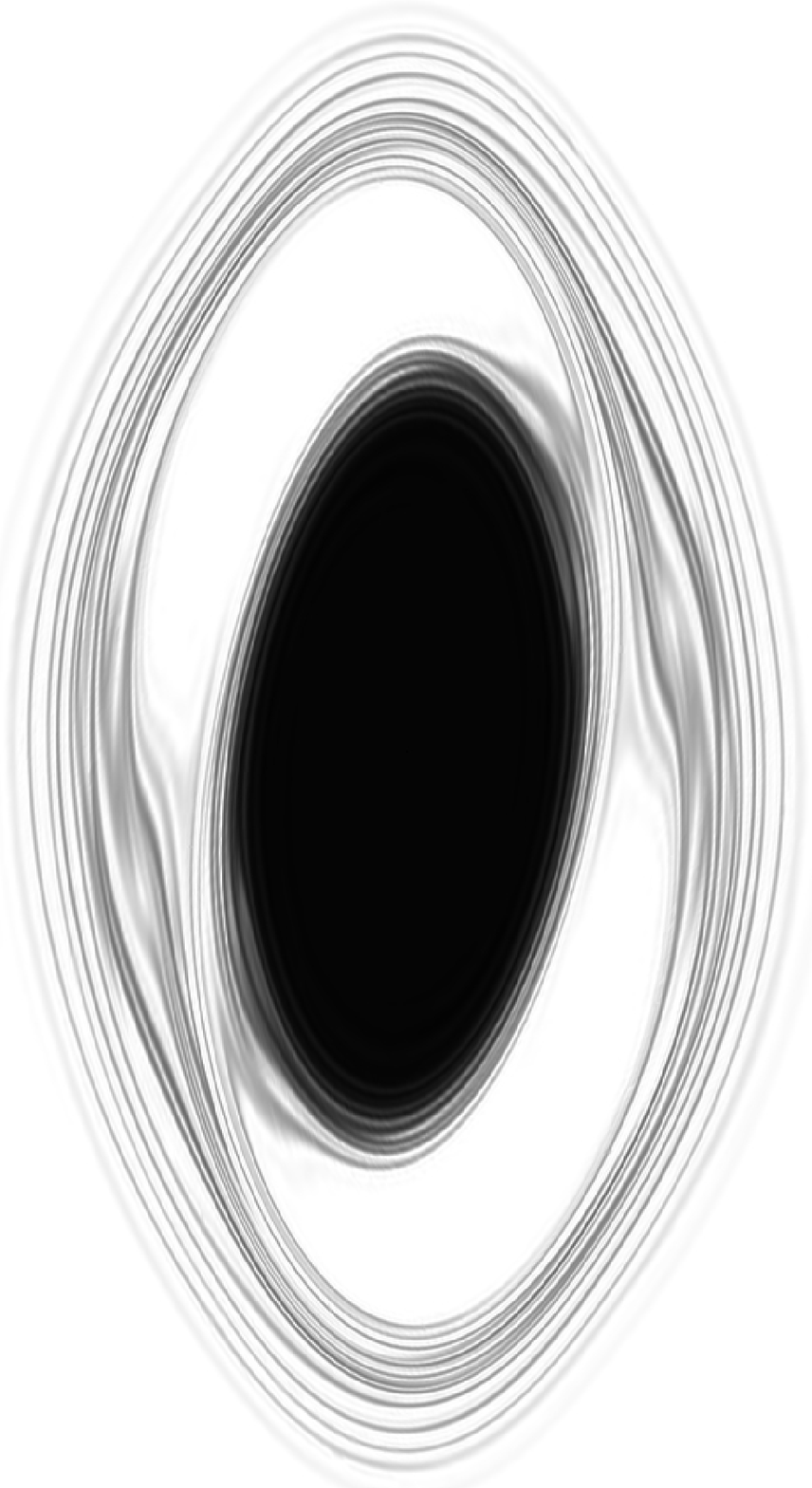,width=6.8cm}
\epsfig{file=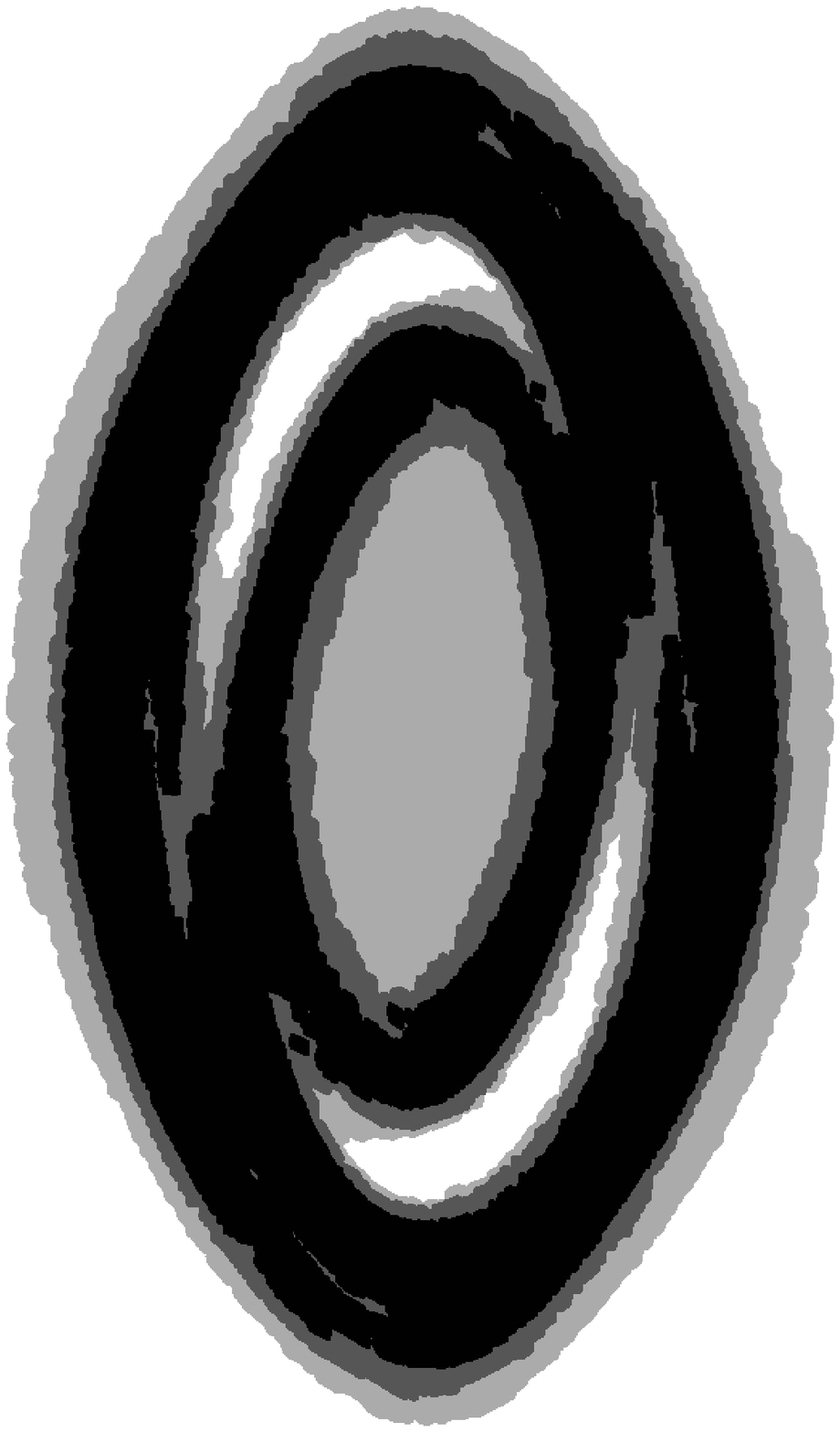,width=6.8cm}
}}
\centerline{convergence criterion\mbox{ \hskip 4.5cm} convergence levels}
\caption[]{Same as in Fig.~\ref{fig:studytophatref1}, but for $t=100$.}
\label{fig:studytophatref3}
\end{figure}

Figures~\ref{fig:studytophatref1}, \ref{fig:studytophatref2} and \ref{fig:studytophatref3}
show the phase-space distribution function at various
times, $t=10$, 40 and 100. The refined simulations seem to reproduce
rather well the results of the full resolution simulation, up to $t=40$. For $t=100$,
the refined simulations clearly differ from the full resolution one, 
even though they present the same level of detail.
The two lower right panels on each of the figures
show the refinement levels. As expected, the result obtained from local curvature
criterion is very similar to that obtained from 
local convergence criterion. For reference, the upper right panel
shows an estimate of the local curvature, more exactly the quantity
${\cal C}=\max(|\lambda_1|,|\lambda_2|)$ intervening in Eq.~(\ref{eq:refcurva})
where $\lambda_1$ and $\lambda_2$ are the eigenvalues of 
the Hessian of the phase-space distribution function. Note that
${\cal C}$ is quite noisy despite our sophisticated weighting scheme 
to compute $f$ and its derivatives, as it still captures
some small scales defects, but it seems to be determined accurately enough to set up
correctly refinement based on local curvature. 

\begin{figure}
\centerline{Refinement based on local curvature}
\centerline{\mbox{
\epsfig{file=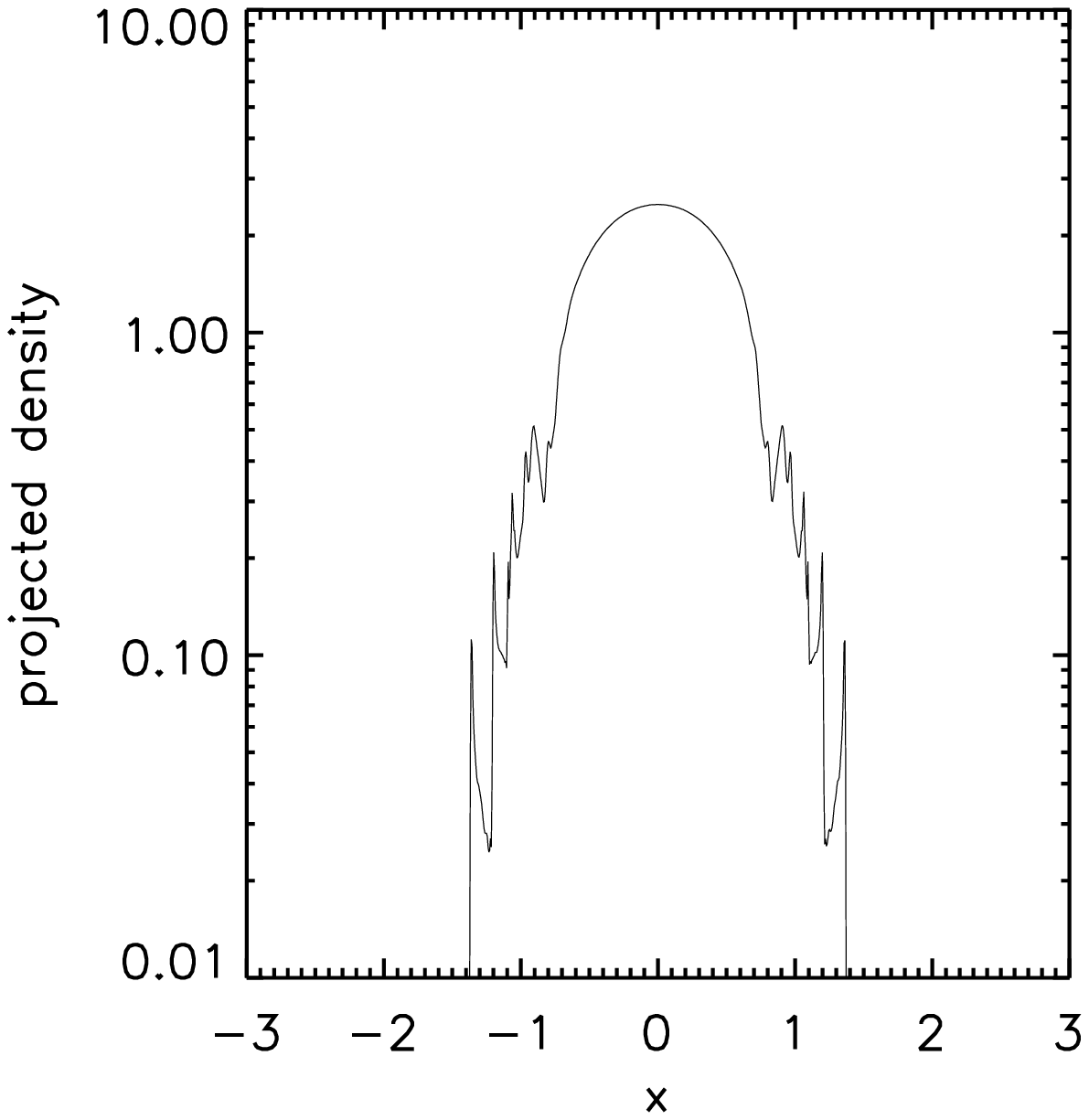,width=7cm}
\epsfig{file=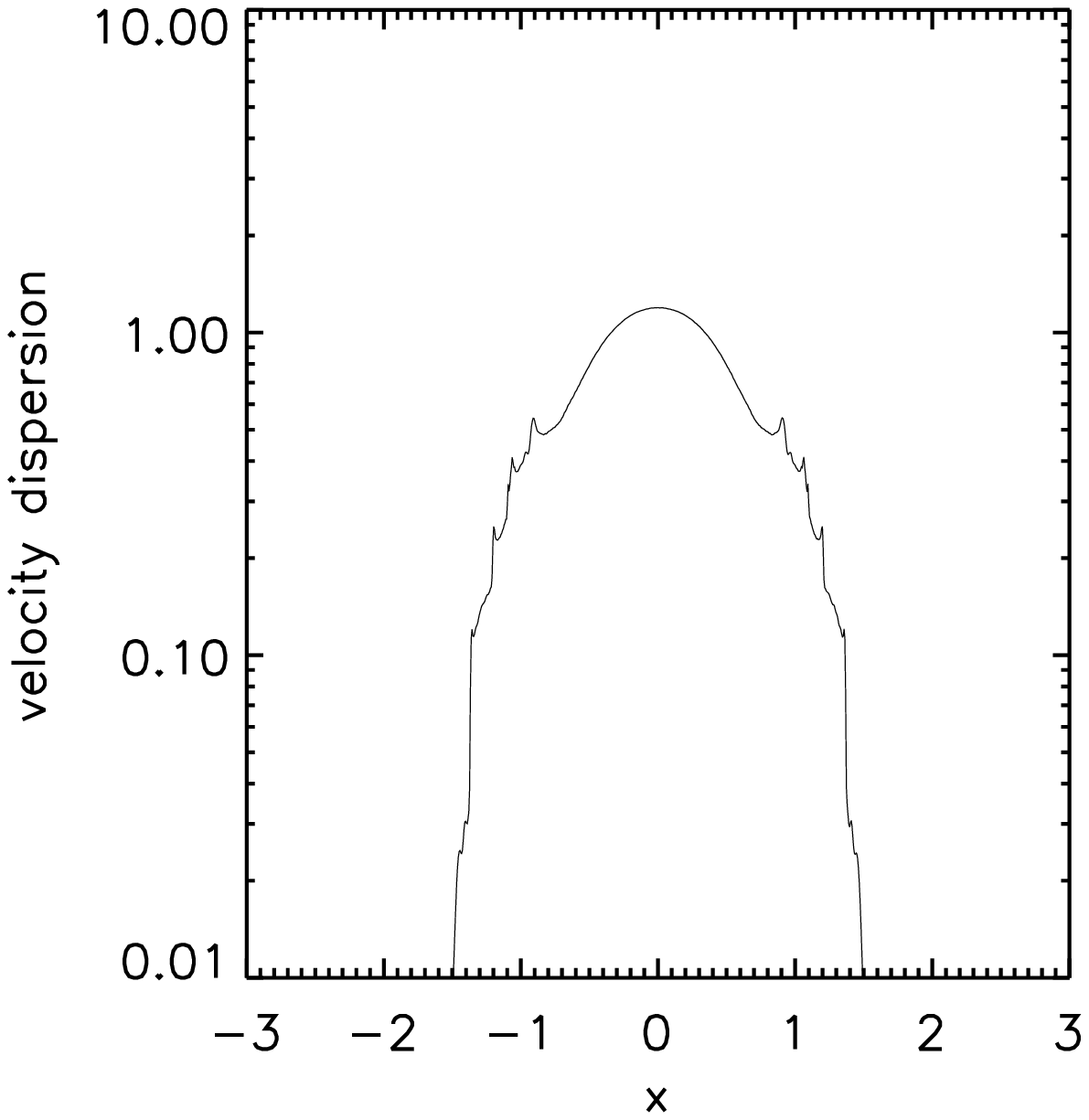,width=7cm}
}}
\vskip 0.3cm
\centerline{Full resolution}
\centerline{\mbox{
\epsfig{file=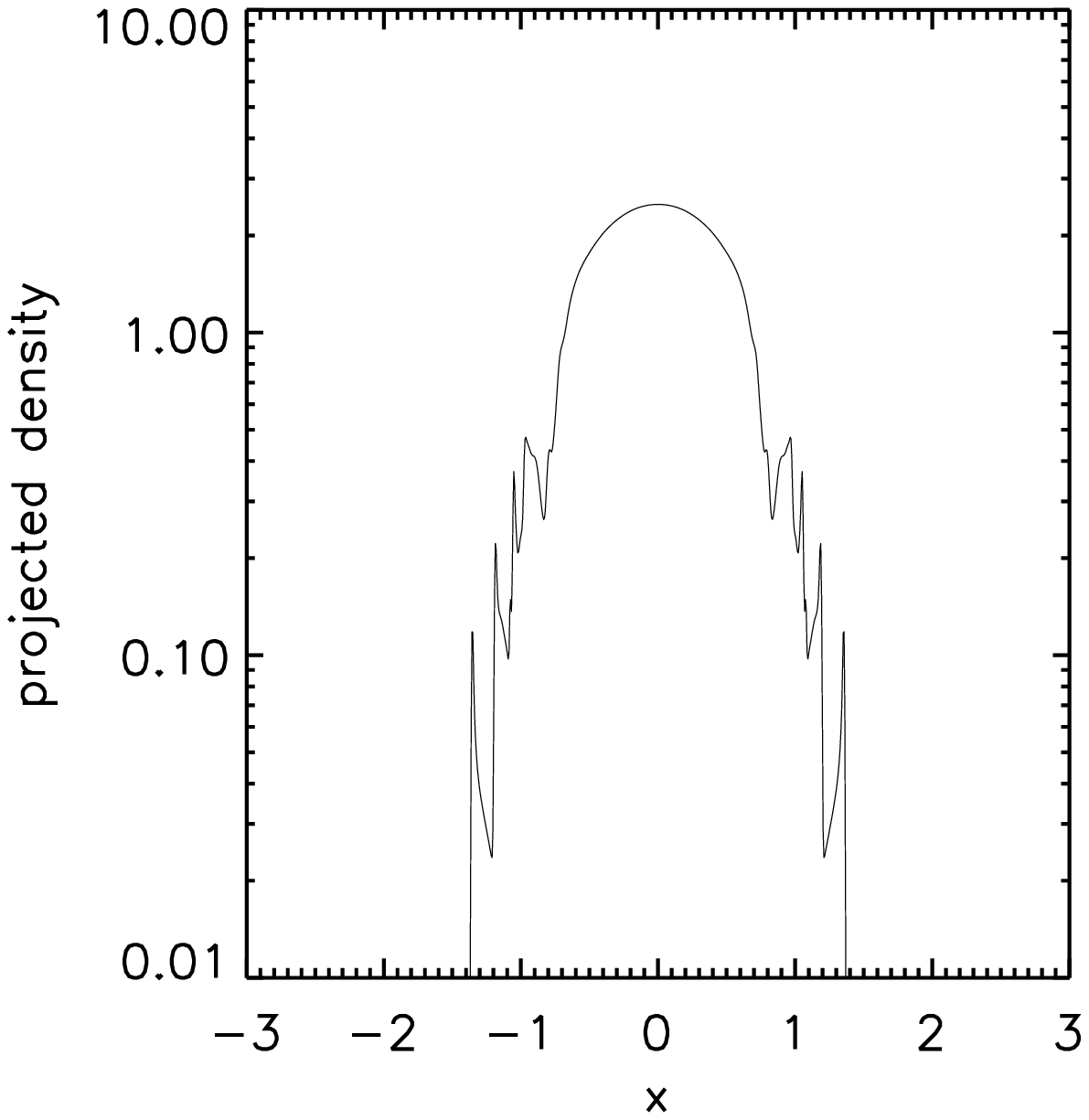,width=7cm}
\epsfig{file=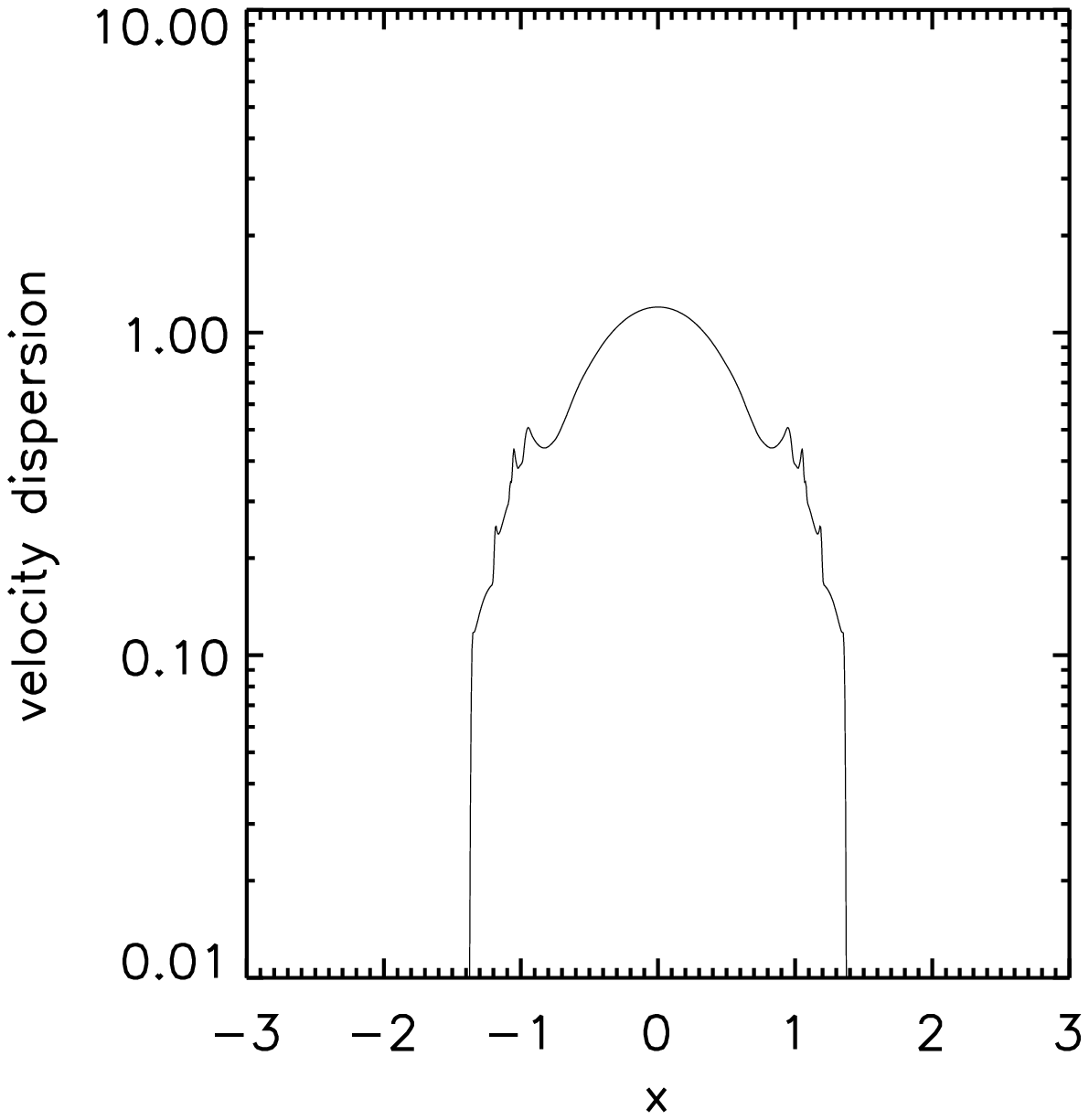,width=7cm}
}}
\vskip 0.3cm
\centerline{Refinement based on local convergence}
\centerline{\mbox{
\epsfig{file=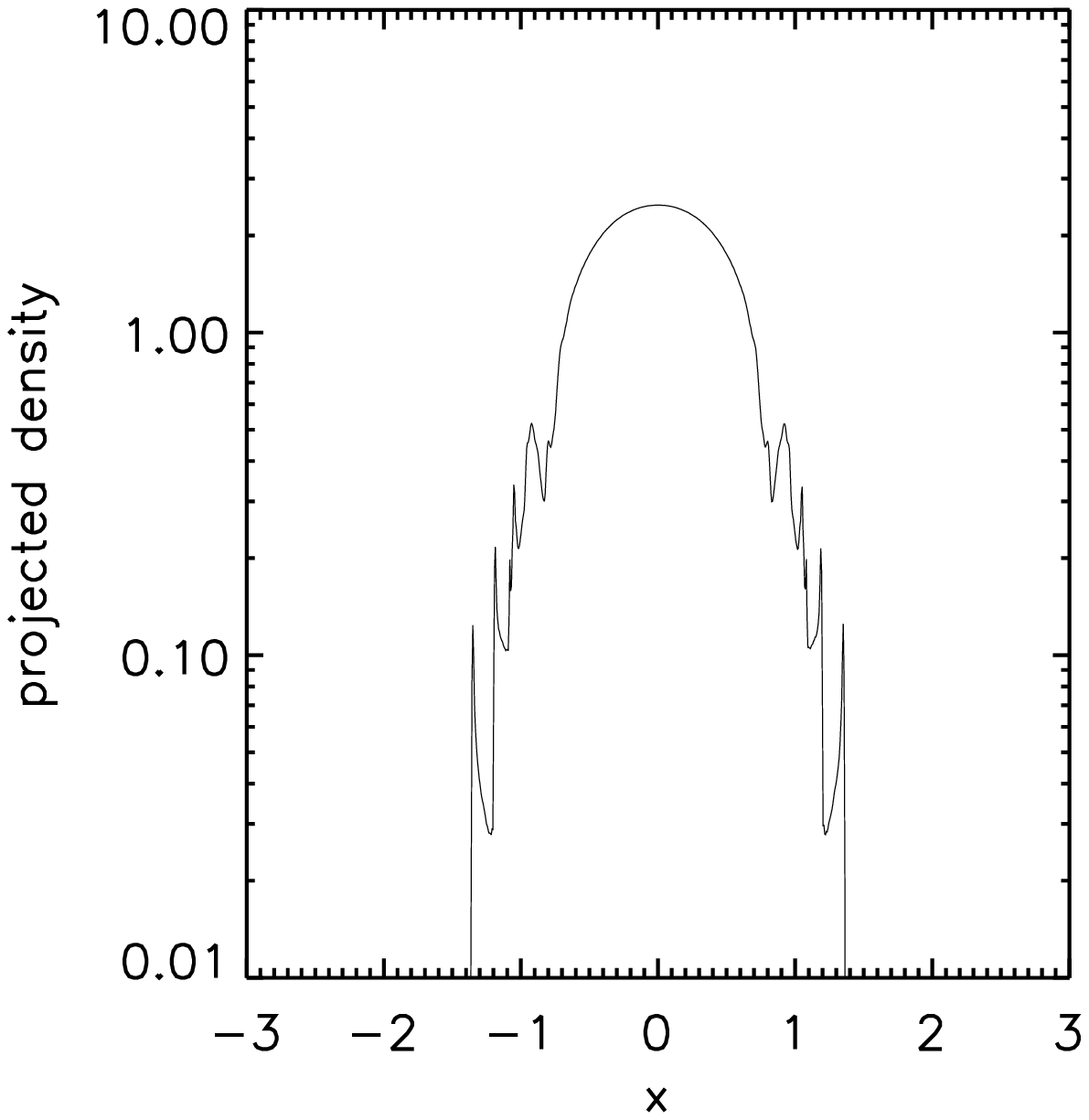,width=7cm}
\epsfig{file=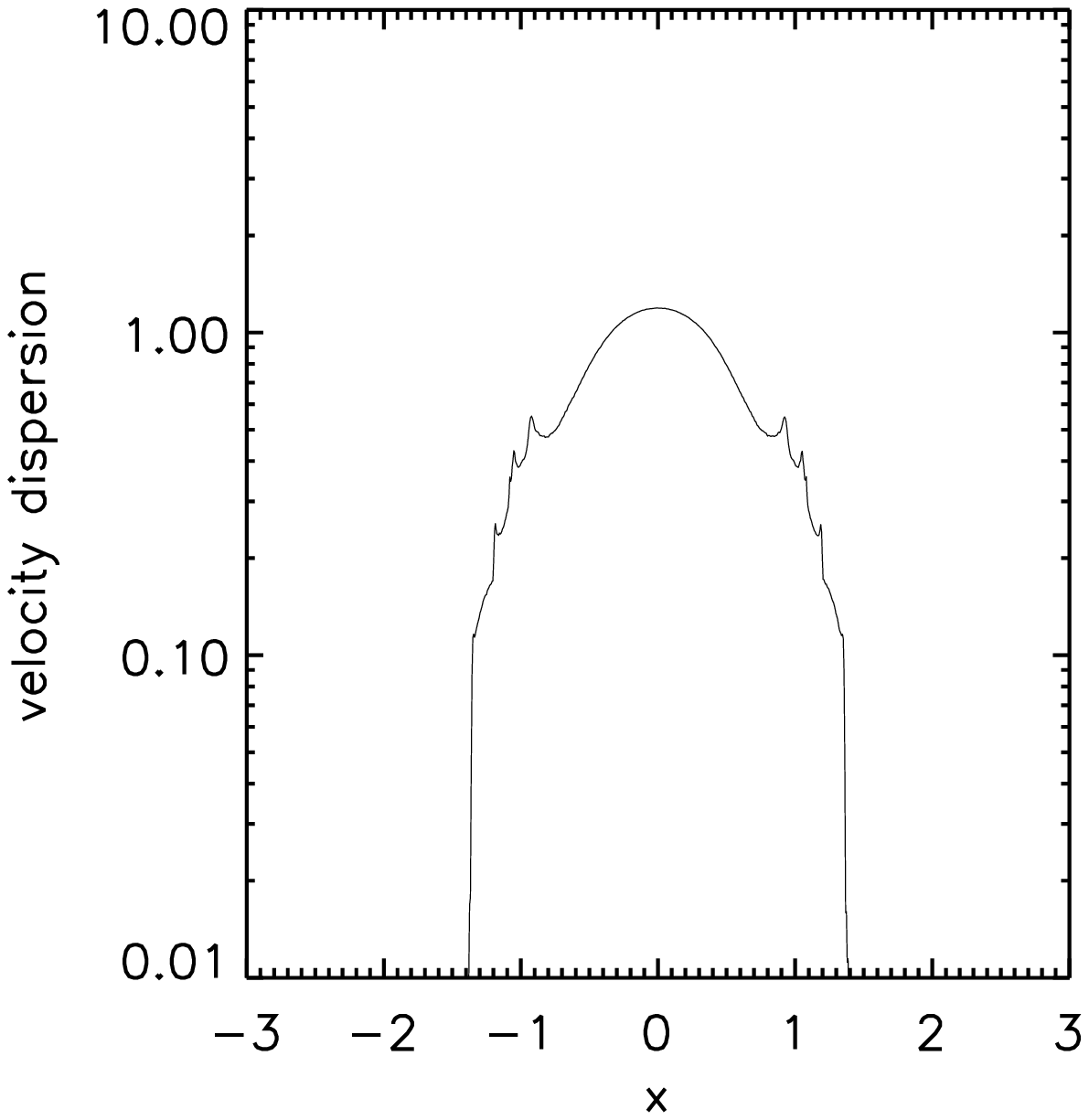,width=7cm}
}}
\vskip 0.3cm
\caption[]{Mean projected density
(left panels) and velocity dispersion (right panels),  as functions
of position, for top hat simulations with refinement based on 
local curvature (top row) and on local convergence (bottom row),
compared to full resolution (middle row), at $t=40$.}
\label{fig:studytophatref4}
\end{figure}

\begin{figure}
\centerline{Refinement based on local curvature}
\centerline{\mbox{
\epsfig{file=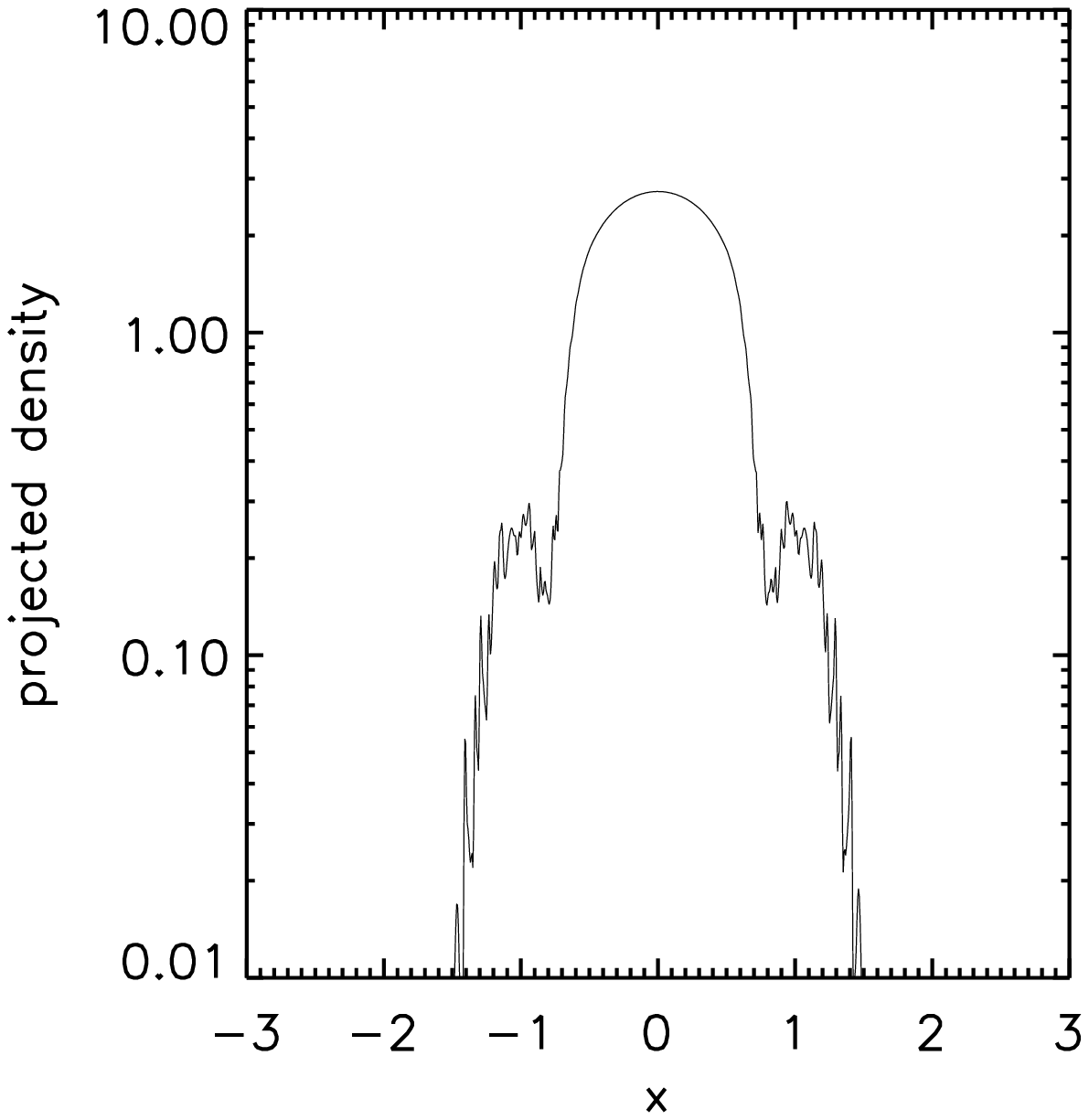,width=7cm}
\epsfig{file=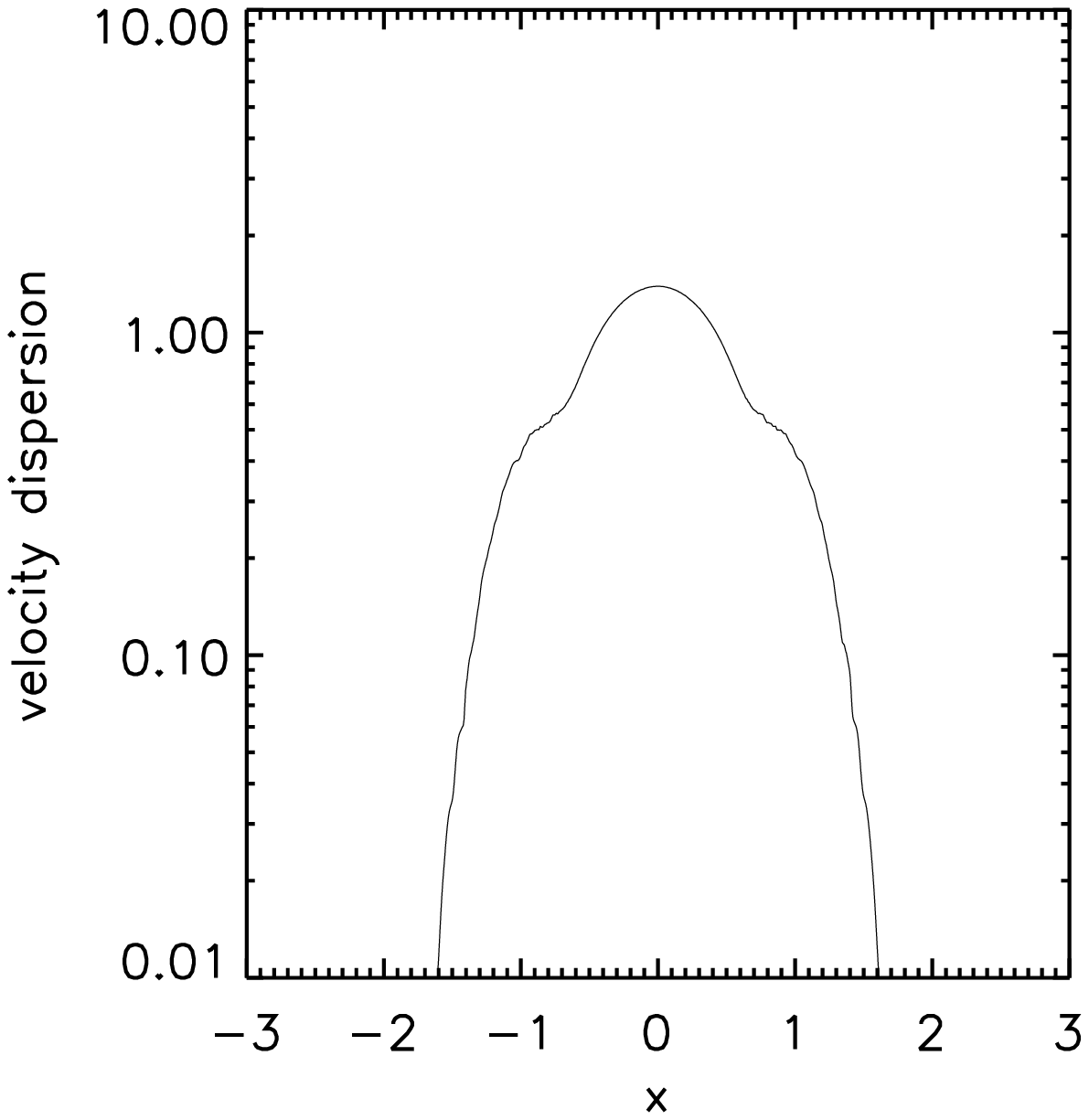,width=7cm}
}}
\vskip 0.3cm
\centerline{Full resolution}
\centerline{\mbox{
\epsfig{file=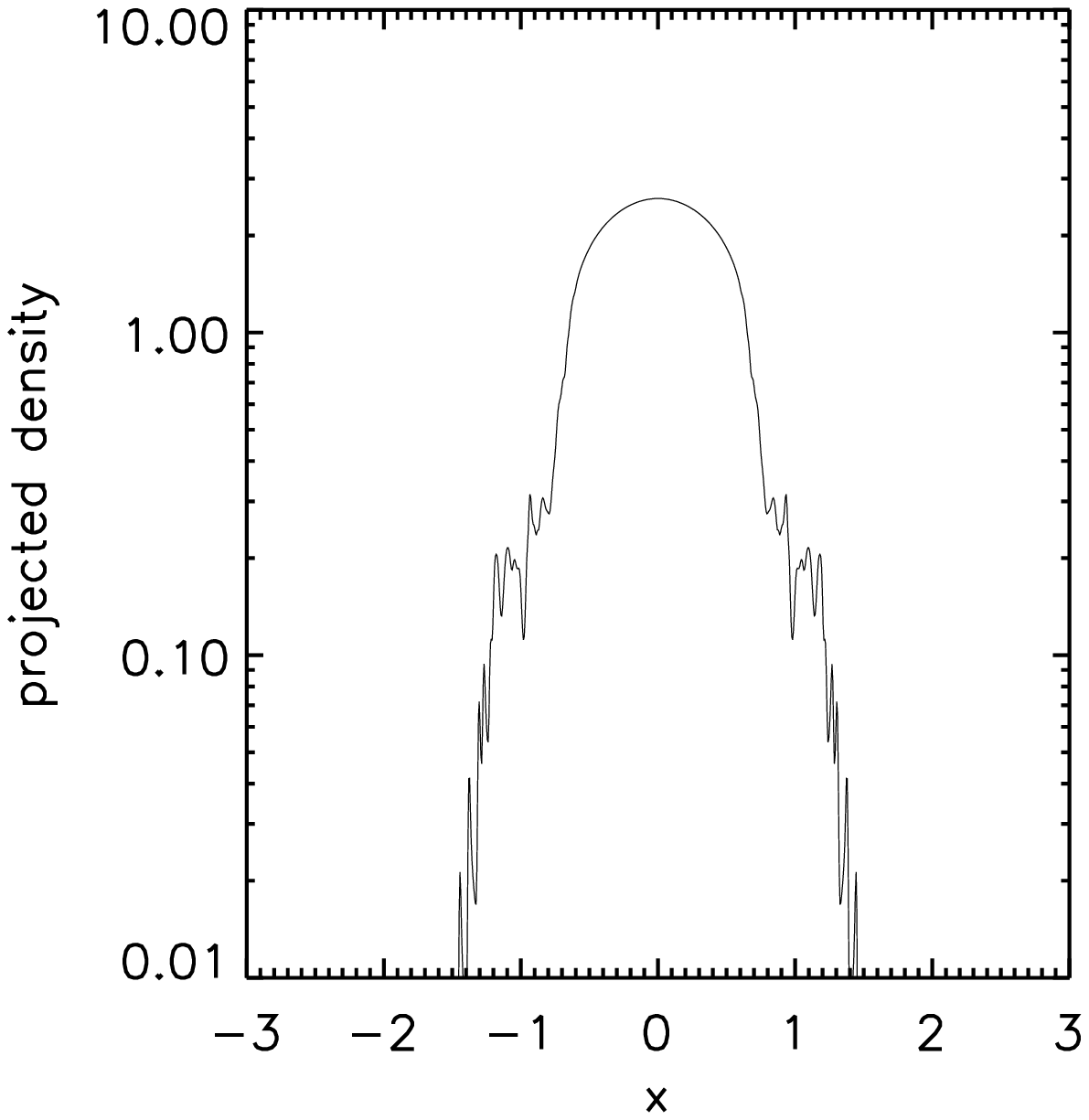,width=7cm}
\epsfig{file=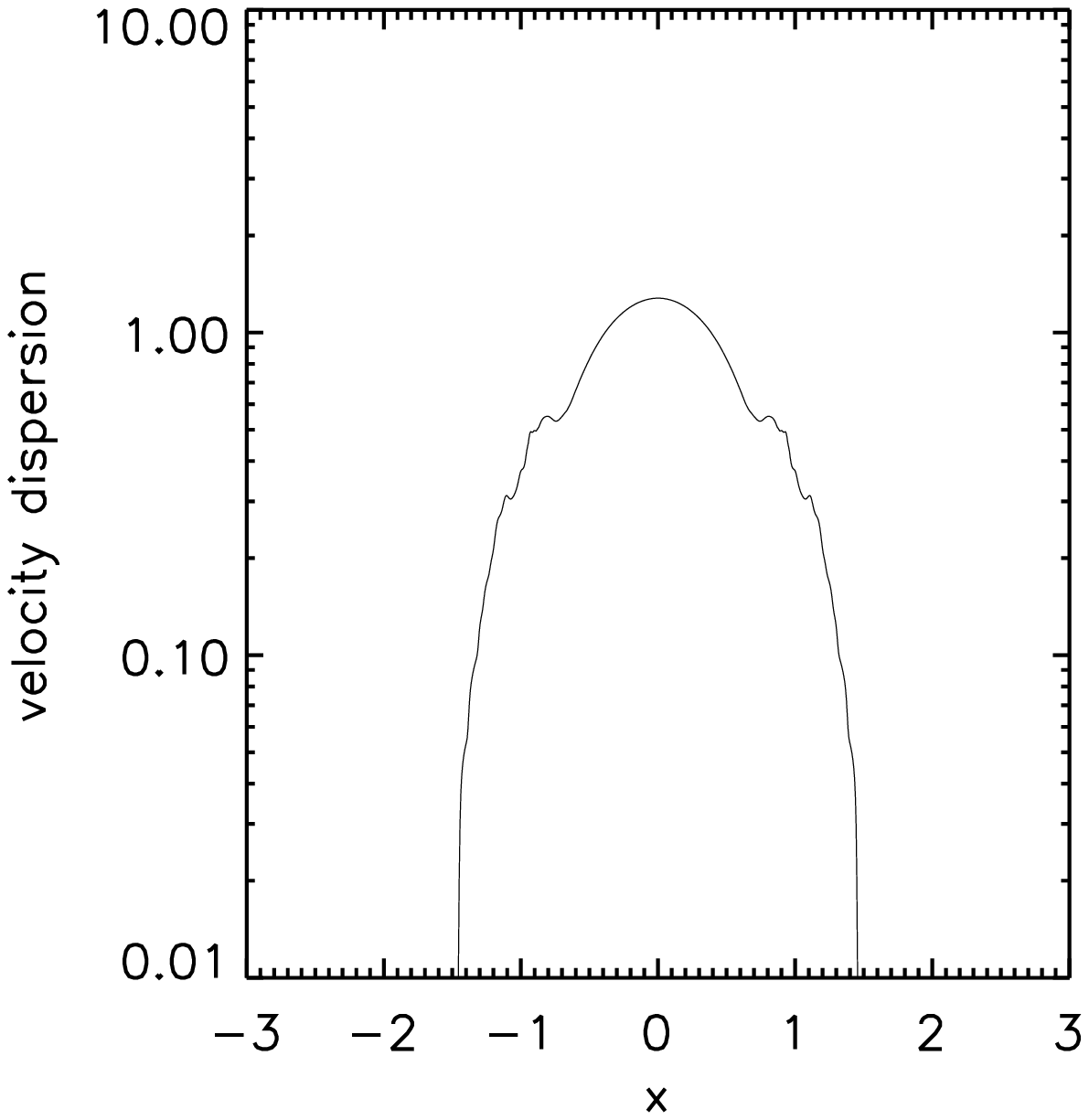,width=7cm}
}}
\vskip 0.3cm
\centerline{Refinement based on local convergence}
\centerline{\mbox{
\epsfig{file=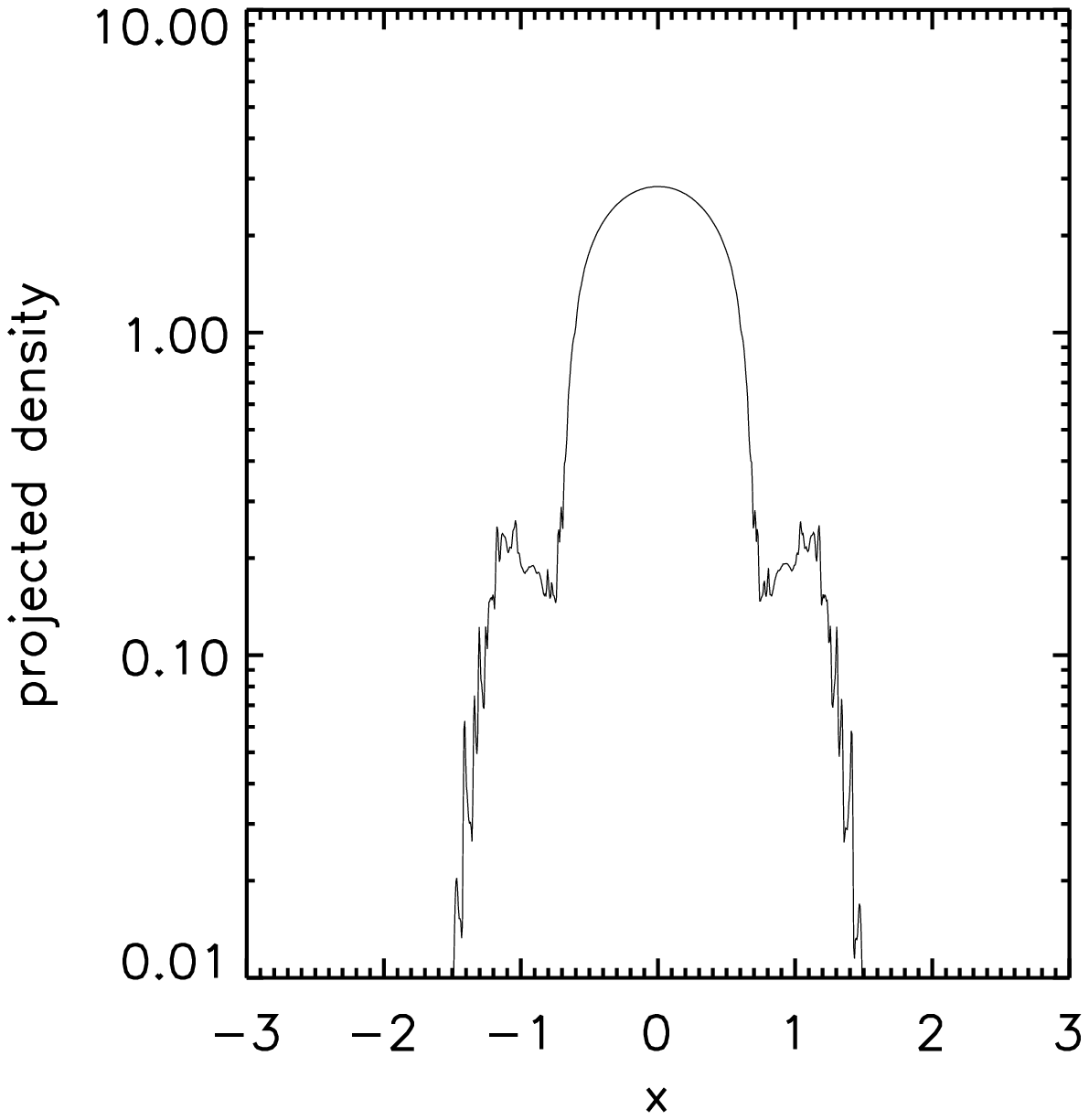,width=7cm}
\epsfig{file=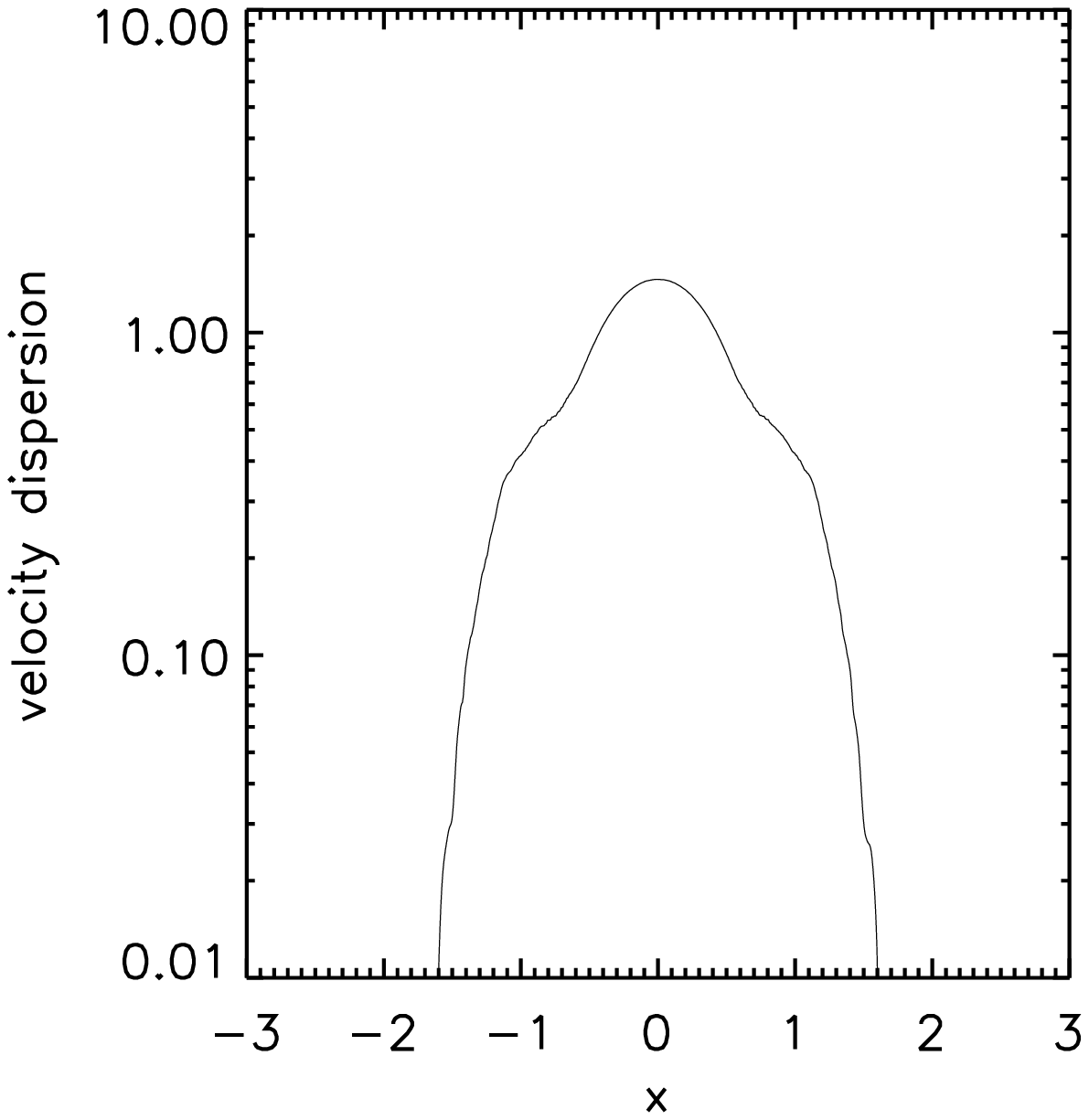,width=7cm}
}}
\caption[]{Same as Fig.~\ref{fig:studytophatref4}, but for $t=100$.}
\label{fig:studytophatref4bis}
\end{figure}

Figures~\ref{fig:studytophatref4} and \ref{fig:studytophatref4bis}
show the projected density and the velocity dispersion for the
snapshots $t=40$ and $t=100$. Their careful examination confirms the visual impression
from Figs.~\ref{fig:studytophatref2} and \ref{fig:studytophatref3}: the refined simulations
reproduce well the fine features of the full resolution one for $t=40$, but not for $t=100$.

\begin{figure}
\centerline{\mbox{\epsfig{file=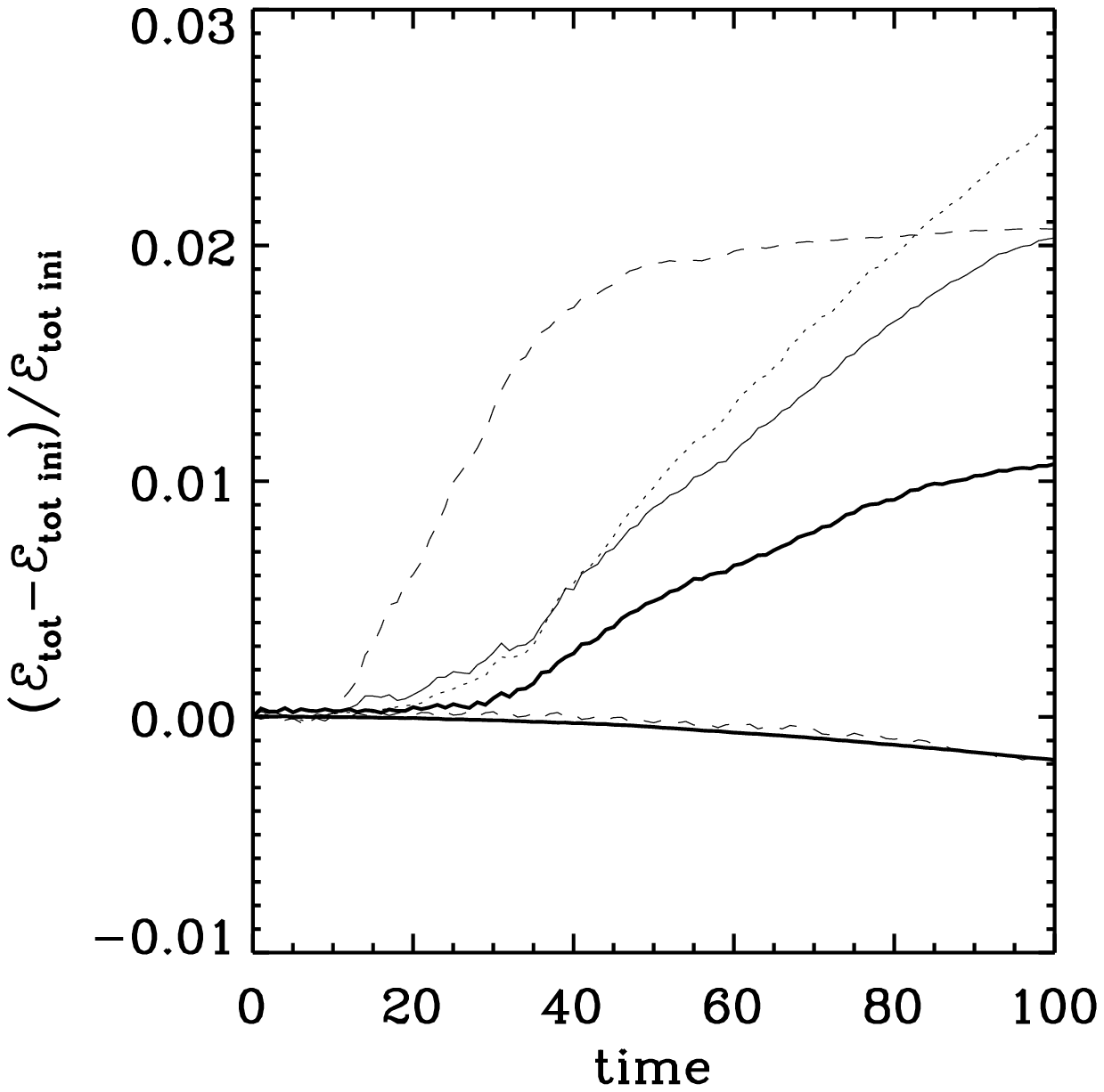,width=12cm}
}}
\caption[]{Energy conservation in top hat simulations with refinement, compared
to fixed resolution simulations.
The solid and dotted thin curves correspond to simulations with refinement based
on local curvature and local convergence, respectively. The lower thick curve and nearly
superposed dashed one correspond respectively to the high and low resolution simulations
of \S~\ref{sec:tophat}. These simulations use Lucy deconvolution. 
The upper thick and dashed curves
correspond to similar simulations, but that were performed using Van-Citter reconstruction.}
\label{fig:studyenerconsref}
\end{figure}

Finally, Fig.~\ref{fig:studyenerconsref} examines energy conservation for various
simulations. Refined simulations present the same behavior, whether
local curvature or local convergence criterion is used:
excellent energy conservation up to $t\simeq 40$,
then a nearly linear increase of energy with time with a significant but still reasonable
final violation of energy conservation of the order of 2 percent. 
The curves corresponding to the high and low resolution simulations of \S~\ref{sec:tophat}, 
performed with Lucy deconvolution, are displayed for reference, but 
it would be more fair to make the comparison with fixed resolutions 
simulations performed with Van-Citter. 
Indeed, in our current implementation,
simulations using Van-Citter deconvolution conserve energy less well
than those using Lucy: 
on Fig.~\ref{fig:studyenerconsref}, the high resolution simulation using Van-Citter
presents a similar behavior to our simulations with refinement, except that the
final violation  of energy conservation is twice smaller. 
The low resolution simulation does significantly worse. First it conserves well
energy, up to $t\simeq10$, then energy increases rather fast up to $t\simeq 40$
where it reaches a plateau corresponding to a global violation of energy conservation
of the order of 2 percent.

This behavior is related to the fact that the
positivity of the distribution function is not warranted in Van-Citter algorithm and
that we perform a truncature of the tails of $f$: if $f$ is smaller than $\delta_{\rm min}$,
it is set to zero. If this truncature, which is meant to keep the computing
domain finite and $f$ as positive as possible, was not present, energy conservation would 
be by construction much better. The cut-off of the negative contributions of $f$ implies that energy
increases with time. The effect is all the stronger since resolution is low. Refinement
is expected to improve energy conservation compared to low resolution, 
at least during some time, but since it keeps
coarse and intermediary levels of refinement, it is not expected to do as well as
the full resolution simulation, hence the result observed in Fig.~\ref{fig:studyenerconsref}.

However, a two 2 percent violation of energy conservation is probably not enough
to explain the significant disagreement at $t=100$ between simulations
with refinement and the full resolution one. There is indeed at least one 
other effect intervening, as discussed in last paragraph of
\S~\ref{sec:sketchref}: namely that at the coarser levels, local quadraticity of
the potential is likely to be violated during runtime. 
This can produce cumulative effects that show up only after
a large number of orbits. 

Despite the limitations of the current implementation, we think having
demonstrated here that the cloud method is compatible with
adaptive refinement.
%
%
%
%
\section{Perspectives and extension to a higher number of dimensions}
%
We demonstrated the ability of the cloud method to solve the 
Vlasov-Poisson system in 1 dimension. An interesting feature of this
method, is that its structure allows a simple generalization to the
$n$-dimensional case. We will show in the next section that the $n$-dimensional 
cloud equations can be solved quite easily, as a natural generalization
of the one dimensional equations. It is also quite obvious that the tree-code
method that we used to solve the Poisson equation can be extended to a higher
number of dimensions.
%
\subsection{$n$-dimensional equations.}
%
Provided that we keep in mind the results obtained in 1 dimension,
the generalization of the cloud equation to $n$ dimensions is straightforward.
In the 1-dimensional case, it was shown that a general solution
cloud be written as a function of second order polynomials in
$(x_1...x_n,v_1...v_n)$. We will adopt this result and
show that a closed system of equations can be obtained
for a $n$-dimensional cloud. To simplify the writing of
 the equations, we define the vector $\bf u$ such that:
\begin{equation}
  u_i =
   \left\{ \begin{array}{l}
       x_i \ \ \ 0 < i \leq n, \\  v_i \ \ \ n < i \leq 2n. \\
       \end{array} \right. \
\label{eq:Xdef}
\end{equation}
Using this notation, we are now able to write the generalized
Vlasov equation in Lagrangian coordinates for a cloud of density 
$f({\bf u},t)$ evolving in the potential $\phi({\bf u},t)$:
\begin{equation}
 {\frac {\partial }{\partial t}} f({\bf u},t) +
 {\frac {\partial }{\partial u_i}} f({\bf u},t) \ u_{i+n} \  \eta(i) -
 {\frac {\partial }{\partial u_i}}\phi({\bf u},t) \
 {\frac {\partial }{\partial u_{i+n}}} f({\bf u},t) \ \eta(i)  =0,
\label{eq:Boltz}
\end{equation}
with the following definition for the function $\eta$:
\begin{equation}
 \eta(i)=
   \left\{ \begin{array}{l}
       1 \ \ \ 0<i \leq n, \\  0 \ \ \ i<1 \ {\rm or} \ i>n.  \\
       \end{array} \right. \
\end{equation}
We will look for a general solution in the case of a quadratic
potential. An extrapolation of the 1 dimensional case
suggests that the general cloud equation can be written as a
function of a quadratic polynomial.
In this case:
\begin{eqnarray}
 \phi({\bf u},t) &=& \alpha_{ij} \ \eta(i) \eta(j) \  u_i u_j,\\
 f({\bf u},t) &=& F \left(\lambda_{ij} \ u_i u_j \right).
\label{eq:fdef}
\end{eqnarray}
Note that the coefficients $\alpha_{ij}$ and $\lambda_{ij}$ depend
on time, in general, $\alpha_{ij} \equiv \alpha_{ij}(t)$ and
$\lambda_{ij} \equiv \lambda_{ij}(t)$. \\\\

By inserting this expression for $f({\bf u},t)$ in the Vlasov 
equation, Eq.~(\ref{eq:Boltz}), we
will obtain a closed system of differential equations for the 
coefficients $\lambda_{ij}(t)$. We will now illustrate the relevant 
calculations step by step.
By taking each term in the Vlasov equation from left to right we 
have: \\\\
First term in the Vlasov equation:
\begin{equation}
 {\frac {\partial }{\partial t}} f({\bf u},t) = {\frac {\partial  
\lambda_{ij}}{\partial t}} \  u_i u_j.
\label{eq:firstterm}
\end{equation}
Second term in the Vlasov equation:
\begin{equation}
{\frac {\partial }{\partial u_i}} f({\bf u},t) \eta(i) = 2  \lambda_{ij} 
\eta(i) \ u_j.
\end{equation}
Thus,
\begin{equation}
{\frac {\partial }{\partial u_i}} f({\bf u},t)  \ u_{i+n} \eta(i) = 2  
\lambda_{ij} \  \eta(i)  \ u_j u_{i+n}.
\end{equation}
Which can be rewritten:
\begin{equation}
{\frac {\partial }{\partial u_i}} f({\bf u},t)  \ u_{i+n} \eta(i) = 2  
\lambda_{(k-n)j}\ \eta(k-n) \ u_j u_k.
\label{eq:secondterm}
\end{equation}
The last term of the Vlasov equation can be evaluated using the same 
method.
We arrive at the following result:
\begin{equation}
 {\frac {\partial }{\partial u_i}}\phi({\bf x},t)
 {\frac {\partial }{\partial u_{i+n}}} f({\bf u},t) \ \eta(i) =
 4 \phi_{ik} \ \lambda_{(k+n)j} \ \eta(i) \eta(k) \ u_j u_k.
\label{eq:thirdterm}
\end{equation}
We are now ready to tackle the Vlasov equation itself. We have to 
write that
each coefficient of the second order polynomial is zero. For each second 
order
term $u_i u_j$ we have to consider the contribution from the pair of
indices $(i,j)$ and $(j,i)$. In the case $i=j$, there is only one contributor,
but we will still add the symmetric term, this will result in a factor of 2
on the left side, and since the right side is equal to 0 it does not change
the equation. Thus by adding the contribution of each term in Eqs.~(\ref{eq:firstterm}),
(\ref{eq:secondterm}) and (\ref{eq:thirdterm}),
and taking into account the symmetric term, we find that:
\begin{equation}
  {\frac {\partial  \lambda_{ij}}{\partial t}} +
   \lambda_{(i-n)j}\ \eta(i-n) +
   \lambda_{(j-n)i}\ \eta(j-n) -
 2 \phi_{ki} \ \lambda_{(k+n)j} \ \eta(k) \eta(i) -
 2 \phi_{kj} \ \lambda_{(k+n)i} \ \eta(k) \eta(j) = 0.
\label{eq:Ndimensions}
\end{equation}
\subsection{Applications of the method.}
Considering the current capabilities of the largest computers it seems
clear that the 2 dimensional case could be undertaken rapidly, and that
in astrophysics it would offer a new look on a few interesting problems,
like for instance the dynamics of galactic disks. Due to the particular
interest of the 2D case we present the detailed cloud equations in
 Appendix~\ref{sec:2deqs}. Due to the high dimensionality of phase space
in the 3 dimensional case, it seems that trying to integrate directly
the Vlasov-Poisson system using the cloud method may be too costly. However,
it is clear that a most interesting case, namely the cold dark matter model in
cosmology, may not require the same amount of resources and could be solved
using an appropriate version of the cloud method. It is important to notice
that in this case the density in the 6 dimensional phase space is adequately
represented by the extended folding of a 3 dimensional sheet 
having a nearly constant density. In particular, this case
may not require the sophisticated Lucy deconvolution scheme and the 
general remapping technique.
\section*{acknowledgements}
We thank B. Paczy\'nski, R. Teyssier, J. Touma and S. Tremaine for useful discussions.
This work was performed within the Numerical
Investigations in Cosmology group (NIC) as a task of the HORIZON project.

\appendix
%
\section{Sampling of the distribution function from a set of clouds}
\label{sec:appcompf}
%
%
%
\subsection{Arbitrary set of clouds to a regular grid}
\label{sec:simpleaffec}
Assume we want to compute  $f_{i,j}=f({x}_i,{v}_j)$
where $({x}_i,{v}_j)$ are sampling points on a square grid of spacing
$\Delta$, for an ensemble of clouds with arbitrary axis ratio and orientation, using
either Eq.~(\ref{eq:ffromspl}) or Eq.~(\ref{eq:fsmoothed}).
Since the clouds are rather extended and
overlap significantly, the calculation of $f_{i,j}$,  is expected to
be rather CPU time consuming, in general, but simple. 
The natural way of performing it
is simply to project each cloud independently on the grid. 
The only difficulty is to find the subsample of grid points 
where a given cloud contributes.  To do that, one can estimate 
the projected size of the cloud on $x$ and $v$ axis:
\begin{equation}
R_{{\rm max}, x}=\frac{R_{\rm max}}{R}\sqrt{\frac{\beta_2}{2C}}, \quad 
R_{{\rm max}, v}=\frac{R_{\rm max}}{R}\sqrt{\frac{\beta_0}{2C}},  
\label{eq:rmaxxv}
\end{equation}
where $\beta_0$ and $\beta_2$ are given by Eqs.~(\ref{eq:beta0}) and (\ref{eq:beta2}), 
respectively. This defines the rectangle where the ellipse 
corresponding to the cloud is inscribed. To speed up the calculation, 
one can find, for fixed values of  $x\in [-R_{{\rm max},x},R_{{\rm max},x}]$ 
in the cloud coordinate frame, the segment $[v_1,v_2]$ given by the intersection between the 
ellipse covered by the cloud and the vertical line of abscissa $x$, that we
do not feel necessary to write here for simplicity.
\subsection{Round set of clouds to a grid: fast convolution algorithm}
\label{sec:fastconvmet}
When one considers a round set of clouds set up on a regular grid, it
is possible to use a simple trick to speedup the calculation of $f_{i,j}$ compared
to the simple algorithm described in \S~\ref{sec:simpleaffec},
by noticing that the convolution of a function by a Gaussian can
be factorized in that case:
\begin{equation}
\int dx^*\ dv^*\ f({x}+x^*,{v}+v^*)\ 
 \exp\left[-\frac{1}{2} \left(\frac{{x^*}^2}{R_x^2}+\frac{{v^*}^2}{R_v^2} \right) \right]
=\int dx^*\ \exp\left( -\frac{1}{2} \frac{{x^*}^2}{R_x^2} \right) \int dv\ f({x}+x^*,{v}+v^*)
  \ \exp\left( -\frac{1}{2} \frac{{v^*}^2}{R_v^2} \right).
\end{equation}
In practice, that means the following: assume for simplicity that the center,
$({x}_{\rm G}^n,{v}_{\rm G}^n)$, of 
each cloud, $n$, coincides exactly with a grid sampling point $({x}_{i_n},{v}_{j_n})$,
and furthermore that cloud inter-spacing, $\Delta_{\rm g}$, 
is a multiple of the target mesh inter-spacing $\Delta$.
With these assumptions, first create an array $f^{**}_{i,j}$, which is zero everywhere,
except at cloud positions:
\begin{equation}
f^{**}_{i_n,j_n}=M_n/V,
\end{equation}
where $M_n$ is the mass of the cloud $n$, and $V=\Delta_{\rm g}^2$.
Then perform an operation equivalent to the convolution in velocity space. For doing that,
it is just necessary to propagate vertically the initial values $f^{**}_{i_n,j_n}$ using
the weights given by function $G(x,y)$, on the columns of the array $f^{**}_{i,j}$
which contains non zero elements:
\begin{equation}
f^*_{i_n,j}=\sum_{m, |{v}_j-{v}_{\rm G}^m|\leq R_{\rm max}} 
f^{**}_{i_n,j_m} G(0,{v}_j-{v}_{\rm G}^m).
\label{eq:convol1}
\end{equation}
The last operation consists of convolving along ${x}$ axis:
\begin{equation}
f_{i,j}=\frac{1}{G(0,0)}\sum_{m, |{x}_j-{x}_{\rm G}^m|\leq R_{\rm max}} 
f^*_{i_m,j} G({x}_j-{x}_{\rm G}^n,0).
\label{eq:convol2}
\end{equation}
Here we used the separability of function $G(x,y)$: $G(x,y)=G(x,0) G(0,y)/G(0,0)$.
We made the approximation, to speed up the calculation, that the truncation
is not $x^2+v^2 \leq R_{\rm max}^2$, but rather $|x| \leq R_{\rm max}$ and $|v|\leq R_{\rm max}$,
but this should not have any significant consequence if the value of $R_{\rm max}$ is large
enough, e.g. $R_{\rm max}=4 R$ as used in this paper.
For simplifying the argumentation, we
introduced the arrays $f^{**}_{i,j}$ and $f^{*}_{i,j}$ with the same size as
the target mesh size, but one can see that only $f^{*}$ is necessary, and it can be
reduced to an array of dimension $(N_{{\rm c},x},N_{{\rm g},v})$, where $N_{{\rm c},x}$
and $N_{{\rm g},v}$ are the size of the cloud mesh along ${x}$ axis and of the target
mesh along ${v}$ axis, respectively. 
Typically, the first step of the calculation, Eq.~(\ref{eq:convol1}),
takes $2 N_{\rm tot} R_{\rm max}/\Delta$ operations, where $N_{\rm tot}$ is
the total number of clouds, while the second step, 
Eq.~(\ref{eq:convol2}), takes $2 N_{\rm sites} R_{\rm max}/\Delta$, where $N_{\rm sites}$
is the total number of target grid sites. The total number of operations,
$(N_{\rm tot}+N_{\rm sites}) 2R_{\rm max}/\Delta$, is to be compared to 
$\pi N_{\rm sites} (R_{\rm max}/\Delta)^2$, as expected from the method
presented in \S~\ref{sec:simpleaffec}, a tremendous gain in time. 
Indeed, if we use Eqs.~(\ref{eq:convol1}) and (\ref{eq:convol2}) to compute ${\tilde f}_i$
in Lucy or Van-Citter algorithms, we see that we gain a factor of order of
$(4/\pi)(R_{\rm max}/\Delta)^2 \simeq 10$ for the deconvolution and our parameters choice
($R_{\rm max}=4 R$, $R=\Delta/\sqrt{2}$). As a result, we used this fast convolution
method in the CM code to speed up reconstruction and the method described in \S~\ref{sec:simpleaffec}
in other cases, when the clouds have arbitrary shape and orientation.

In principle, we could have used this efficient implementation as well for our tree-code.
However, this program is designed for adaptive mesh refinement (\S~\ref{sec:adaptiveref}).
It would be rather involved, algorithmically, although possible, to adapt this method
to an unstructured grid such as the one used in our refinement procedure. 
Instead, for all the simulations presented in this paper, which are done with the tree-code, 
we used the very general but much slower method described below, 
at a significant cost of CPU time.

Note finally that Eqs.~(\ref{eq:convol1}) and (\ref{eq:convol2}) apply
to Eq.~(\ref{eq:ffromspl}), but the method can easily generalized to Eq.~(\ref{eq:fsmoothed}). 

\subsection{Arbitrary set of clouds to arbitrary point: the quad-tree algorithm}
In general, if one aims to estimate $f({x},{v})$ at an arbitrary point
of phase-space, it is necessary to find rapidly the clouds contributing to this
point. There is a very general and standard method to do that, based on hierarchical
decomposition of phase-space on a quad-tree structure, until there is zero or
one cloud center per cell of this tree. The list of clouds contributing
to point $({x},{v})$ is constructed by walking into this tree, from
root to leaves. To build the tree properly, one has take into account of the extension
of the clouds. While the tree is constructed, for each cell it contains, defined by some
coordinate range, $[({x}_1,{v}_1),({x}_2,{v}_2)]$, one computes 
the effective potential range $C_{\rm eff}\equiv[({x}_{\rm eff,1},{v}_{\rm eff,1}),
({x}_{\rm eff,2},{v}_{\rm eff,2})]$, where
\begin{eqnarray}
{x}_{\rm eff,1} &=& \min_{\mbox{\small clouds\ $i$\ in\ the\ cell}} 
\left( {x}_{\rm G}^i-R_{{\rm max}, x}^i \right),\\
{x}_{\rm eff,2} &=& \max_{\mbox{\small  clouds\ $i$\ in\ the\ cell}} 
\left( {x}_{\rm G}^i+R_{{\rm max}, x}^i \right),\\
{v}_{\rm eff,1} &=& \min_{\mbox{\small  clouds\ $i$\ in\ the\ cell}} 
\left( {v}_{\rm G}^i-R_{{\rm max}, v}^i\right),\\
{v}_{\rm eff,2} &=& \max_{\mbox{\small  clouds\ $i$\ in\ the\ cell}} 
\left( {v}_{\rm G}^i+R_{{\rm max}, v}^i\right).
\end{eqnarray}
In these equations, $R_{{\rm max}, x}^i$ and $R_{{\rm max}, v}^i$ are given by Eqs.~(\ref{eq:rmaxxv}) 
for each cloud belonging to the cell, 
$({x}_{\rm G}^i,{v}_{\rm G}^i) \in [({x}_1,{v}_1),({x}_2,{v}_2)]$.
To construct the list of clouds contributing to point $({x},{v})$, one starts from the root
and walks down into the tree. During the walk, only cells verifying  $({x},{v}) \in C_{\rm eff}$ 
are opened, i.e. decomposed in four sub-cells, until convergence is achieved, i.e.
when the cell contains zero or
one cloud. This algorithm is quite fast: the loss of speed compared to the trivial method
explained in \S~\ref{sec:simpleaffec} is of the order of $\ln N_{\rm tot}$, where $N_{\rm tot}$ is
the total number of clouds. It has the advantage of being very general since it allows to
compute $f({x},{v})$ at any point. This explains why we adopted it from the very beginning
when we started to develop our tree-code, in order to keep this latter as flexible as possible.
\section{Refinement procedure: technical details}
\label{sec:techdet}
This appendix deals with technical issues of our refinement procedure. 
In \S~\ref{sec:refinementprinciple}, we explain in detail
how the refinement structure is implemented. In particular, as in standard ART methods, 
to keep the method stable, we make sure  that transitions between
refinement levels are not too abrupt. We also take into account the large extension of the clouds.
In \S~\ref{sec:critraf}, we describe the way we set our criteria of refinement, 
namely the one based on local curvature and the other one based on local convergence. 
\subsection{Principle of refinement}
\label{sec:refinementprinciple}
\begin{figure}
\centerline{\mbox{\epsfig{file=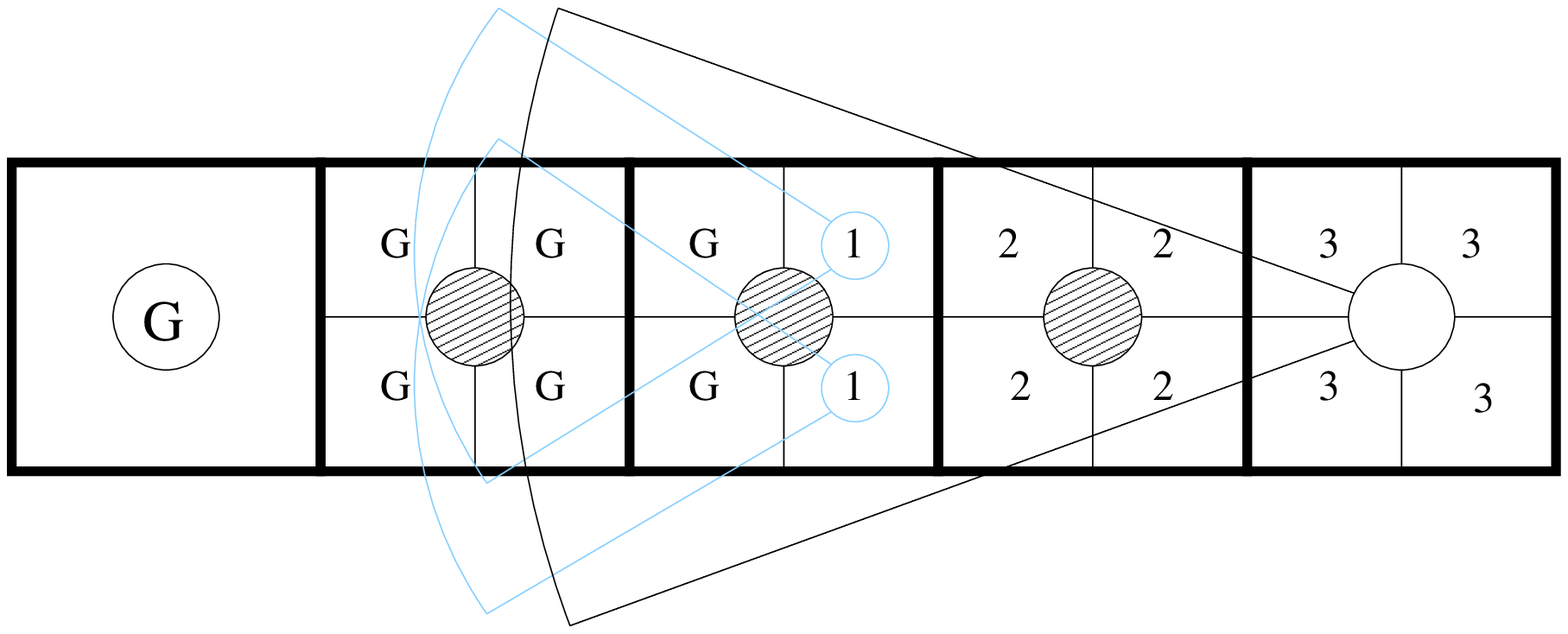,width=12cm}}}
\vskip 0.5cm
\centerline{\mbox{\epsfig{file=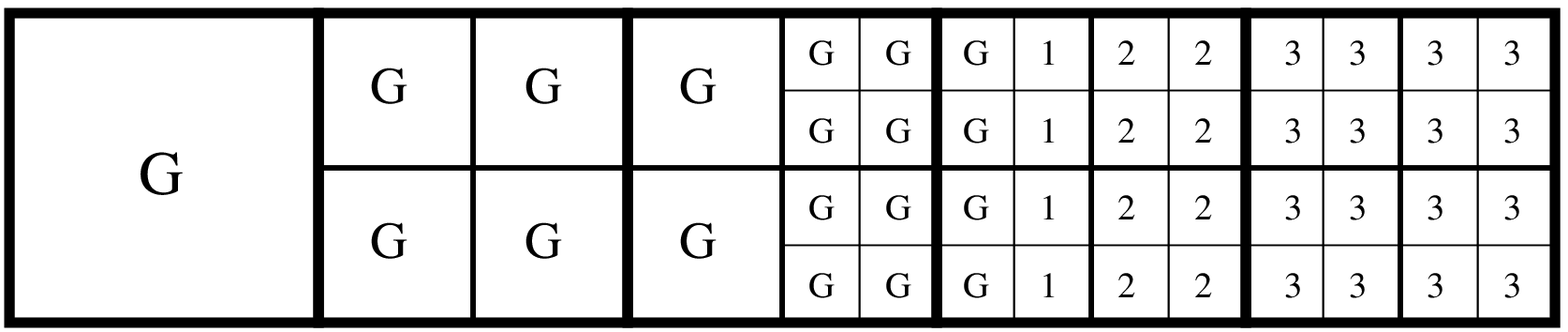,width=12cm}}}
\caption[ ]{Sketch of our refinement procedure, for our choice of
cloud size, $R_{L}$, compared to local cloud inter-spacing, 
$\Delta_{\rm g}/2^{L}$: $2^{L} R_{L}/\Delta_{\rm g}=\sqrt{2}/2$,
which implies a $4R_{L}$ cut-off at $2^{L} R_{\rm max}/\Delta_{\rm g}=2\sqrt{2}$. 
Suppose we decide to refine the right most cloud of upper panel, 
of level $L-1$. This cloud is supplemented with four twice smaller
sub-clouds of level $L$, labeled by 3. It however influences a region
of radius $R_{\rm max}/2^{L-1}$, represented by the black arc of circle. To enforce
our refinement criterion on all the region influenced by this cloud, we must
add supplementary layers of sub-clouds around it, 
labeled by 2 and by 1. We can stop adding corrections at sub-clouds labeled by one, 
since their extension covers the zone of influence (light arcs of circle). 
For easy implementation of regularization (\ref{eq:fsmoothedadap}),
we create as well ghosts clouds of zero mass designed by the ``G'' letter in order 
to compute correct weights for the refined level. With this refinement procedure, 
we see that the cloud to be refined has to be sufficiently far away from ghosts. The closest
possible ghost for the rightmost cloud at level $L-1$ is the left most one, which was
labeled by a G surrounded by a circle. 

Our refinement can be performed recursively to higher level, as illustrated by lower panel, by obeying
the following simple rules:\\
\null\hskip 0.5cm(1) it is forbidden to refine ghosts; \\
\null\hskip 0.5cm(2) clouds labeled by 1 can only be refined in ghosts; \\
\null\hskip 0.5cm(3) clouds labeled by 2 can only be either refined in ghosts, 
in sub-clouds labeled by 1 or in sub-clouds labeled by 2 according to their 
distance to closest ghosts, as illustrated by the upper and lower panel; \\
\null\hskip 0.5cm(4) clouds labeled by 3 can be fully refined. \\
Note that, by construction, the region labeled with 3's can be refined to arbitrary level $L_{\infty}$, while
keeping smooth transition between coarse level $0$ and $L_{\infty}$.}
\label{fig:complexityofref}
\end{figure}

Let $L$ be the level of refinement, equal to $L=0$ for the coarse
level. For simplicity, we assume that the coarse level is a grid with fixed
inter-spacing, $\Delta_{x}=\Delta_{v}=\Delta_{\rm g}$. 
In order to avoid instabilities, function $L({x},{v})$
should vary smoothly in phase-space: abrupt transitions between two levels $|L_1-L_2| > 1$ are
forbidden. Also, special care has to be taken of the large extension
of the clouds. For example, assume that we split a cloud of size $R=\sqrt{2}/2\Delta_{\rm g}$
with cut-off radius $R_{\rm max}=4 R$ into 4 twice smaller clouds. Since the 
cloud influences remote regions beyond the cut-off scale of the 4 sub-clouds, 
refinement should be performed as well in these regions. 
Fig.~\ref{fig:complexityofref} shows how to handle this problem and 
to naturally preserve smoothness of function $L({x},{v})$.
While following the rules dictated by the procedure 
detailed in Fig.~\ref{fig:complexityofref},  we use a criterion based
on local convergence or/and on measurement of local curvature to decide 
if a cloud has to be refined or not, as explained later in \S~\ref{sec:critraf}.

Once a full hierarchy of clouds and sub-clouds has been created, we proceed as follows to reconstruct the
phase-space distribution function:
\begin{enumerate}
\item[(o)] Setup coarse level, $L=0$: compute $f_i^0$ from the
           old set of clouds (prior to remap) at sampling points 
           $({x}_{\rm G}^{i,0},{v}_{\rm G}^{i,0})$.
           Then use Van-Citter algorithm to compute
           the masses, $M_i^0$, of the coarse level clouds, as explained in 
           \S~\ref{sec:deconv};
\item[(i)] If maximum level of refinement is not reached, increase $L$ by one, otherwise stop.
\item[(ii)] Compute $f_i^L$ at the positions of clouds of level $L$,  
           $({x}_{\rm G}^{i,L},{v}_{\rm G}^{i,L})$, from the old set of clouds. 
           Now, consider clouds of level $L$ like residues
           and use Van-Citter algorithm to compute the masses of these clouds, $M_i^L$.
           To estimate ${\tilde f}_i^L$ in Eq.~(\ref{eq:van-citter}), clouds 
           of lower levels $\ell \leq L-1$ contribute, but with fixed masses 
           computed previously. Start again with step (i).
\end{enumerate}
In step (ii), once the cloud masses have been determined up to some level, they
remain unchanged while determining the masses of clouds of upper levels.
It is indeed expected that the corrections brought by residues
improve the reconstruction of the distribution function
within the corresponding refinement region, without contaminating 
the remaining part of phase-space.  

In our implementation of refinement, we use Van Citter deconvolution algorithm. 
Alternatively, one might use Lucy method. In that case, clouds with $L \geq 1$ 
cannot be considered anymore as residues, and, due to the multiplicative nature 
of the Lucy algorithm, we would have to
reiterate at all levels to compute the overall distribution of masses, $M_i^{\ell}$, $\ell\leq L$. 
This would complicate considerably the algorithm, with no warranty of easy
convergence, so we decided for this work not
to test such refinement scheme.

To compute ${\tilde f}_i$ in Eqs.~(\ref{eq:van-citter}) and (\ref{eq:lucy})
from the new set of round clouds, one can still use Eq.~(\ref{eq:ffromspl}), 
as discussed in end of \S~\ref{sec:weightreg}. It now simply reads
\begin{equation}
{\tilde f}({x},{v})=\sum_{\ell=0}^{L}\frac{1}{V^{\ell}} 
\sum_i M_i^{\ell}\  G_{\ell}({x}-{x}_{\rm G}^{i,\ell},{v}-{v}_{\rm G}^{i,\ell}),
\label{eq:ffromspladap}
\end{equation}
where $V^{\ell}=V^0 4^{-\ell}=\Delta_{x} \Delta_{v} 4^{-\ell}$ and function
$G_{\ell}$ is the same as function $G$ but with a radius $R_{\ell}=2^{-\ell} R$. To estimate $f_i^L$ from the
ancient set of clouds, one can similarly adapt Eq.~(\ref{eq:fsmoothed}), which
now becomes
\begin{equation}
f({x},{v})=\sum_{\ell=0}^L \frac{W}{V^{\ell}} 
\frac{\sum_i G_{\ell}({x}-{x}_{\rm G}^{i,\ell},{v}-{v}_{\rm G}^{i,\ell}) M_i^{\ell}}
{\sum_i G_{\ell}({x}-{x}_{\rm G}^{i,\ell},{v}-{v}_{\rm G}^{i,\ell})}.
\label{eq:fsmoothedadap}
\end{equation} 
We see in Eq.~(\ref{eq:fsmoothedadap}) that
the interpolation is performed at each level of refinement separately.
In order to have proper normalization, we need to add at the border of each refinement level
a layer of ghost clouds with zero masses but contributing to the 
weight in the denominator, as described in Fig.~\ref{fig:complexityofref}.
%
\subsection{Criteria of refinement}
\label{sec:critraf}
%
Our refinement is by essence set up to follow details of the distribution function where
needed. Consider the distribution function as a surface of equation
$z=f({x},{v})$. The local curvature of this surface  
determines to which extent these details can be reproduced
by our clouds of finite size. If the size $R$ of the
clouds is not small enough compared to the local minimum
curvature radius ${\cal R}_{\rm c}$, 
we expect significant loss of details as well as aliasing effects.
A first natural refinement criterion thus relies on the value
of $R/{\cal R}_{\rm c}$. However, it is expectable, equivalently, that
deconvolution will have trouble to converge when $R/{\cal R}_{\rm c} \ga 1$,
since details cannot be adjusted correctly by the clouds. This motivates
an alternate criterion of refinement, based on local convergence of
the reconstruction. The convergence criterion
has also the advantage of being a quite natural extension of our approach: it guarantees
by definition the quality of the reconstruction, at variance with
the local curvature criterion. This latter indeed relies on the measurement of second
derivatives of $f$, which can be quite noisy. However, as illustrated by \S~\ref{sec:simuraf},
both refinement criteria give very similar results.

While constructing the hierarchy of clouds and sub-clouds, we proceed level by level. 
To create clouds of level $L$, clouds of level $L-1$ are tested as follows:
\begin{enumerate}
\item[{\em (A)}] {\em Local convergence criterion:}
in that case, it is necessary to reconstruct the distribution function simultaneously with
the refinement structure, using  steps (o), (i) and (ii) described in
\S~\ref{sec:refinementprinciple}. Simply, step (i) has to supplemented with the criterion
on local convergence to create the new level of refinement, $L$: 
one sets a list of clouds of level $L-1$ for which the reconstruction 
scheme did not converge, $\delta_i^{L-1} > \delta_{\rm c}^{L-1}$
in Eq.~(\ref{eq:van-citter}), after some fixed
number of iterations, say 10 according to discussion in \S~\ref{sec:deconv}.
In principle, the optimal calculation of  $\delta_{\rm c}^{L-1}$ as a function
of $L$ depends on the noise properties of the reconstruction, but it would
go beyond the scope of the paper to analyze them in detail. We therefore consider
a very simple way of setting the convergence threshold,
\begin{equation}
\delta_{\rm c}^{L}=\delta_{\rm c}^{0}, \quad L \geq 1,
\label{eq:deltac1}
\end{equation}
which ignores possible propagation of errors in the reconstructions from level to level,
at the risk of accumulating small scale artifacts.
For this reason, the value of $\delta_{\rm c}^0$ should in practice
be slightly larger than discussed in \S~\ref{sec:deconv}, to avoid unnecessary
refinement due to fluctuations of the small scale noise.
A good practical value is $\delta_{\rm c}^0 \simeq 0.001$.
\item[{\em (B)}] {\em Local curvature criterion:}
the clouds of level $L-1$ 
for which the local curvature of the phase-space distribution function
is larger than some threshold are refined. The local curvature is estimated at cloud
centers from the old set of clouds (prior to remap), using the Hessian of the 
function $f({x},{v})$ given by Eq.~(\ref{eq:fsmoothedadap}) 
(we do not write it here for simplicity). 
Our refinement criterion is
\begin{equation}
\left(\frac{\Delta_{\rm g}}{2^{L-1}}\right)^2 \max(|\lambda_1|,|\lambda_2|) > F_{\rm curvature},
\label{eq:refcurva}
\end{equation}
where $\lambda_1$ and $\lambda_2$ are the eigenvalues of the Hessian. 
A good practical choice of the refinement parameter 
is $F_{\rm curvature}\simeq 0.1$.
\item[{\em (C)}] {\em Keep stability of the refinement:} at positions of clouds of
level $L-1$ to be refined,
check locally what was the refinement level $L_{\rm old}$ in the old set 
of clouds and impose that $L \leq L_{\rm old}+1$. To do that, we consider the square 
of size $\Delta_{\rm g}/2^{L-1}$ centered
on $({x}_{\rm G}^{i,{L-1}},{v}_{\rm G}^{i,{L-1}})$ and find with
standard quad tree search the set of old clouds intersecting with it. $L_{\rm old}$
corresponds to the maximum refinement level found in this set.
For stability purpose, we should have $L \leq L_{\rm old}+1$.
During runtime, we indeed do not expect the creation of more than one level of refinement per remap.
If that happens, it must be due to some spurious small scale artifact
created at the moment of previous remap, that we do not want to propagate furthermore.
\end{enumerate}
There is a subtlety in our algorithm that we have to mention now. Indeed, 
the criteria explained above are applied to a discrete set of 
sampling points. To make sure that this sampling catches all the features
of the distribution function, one has 
to examine locally conditions (A) and (B) at {\em all} levels up to $L_{\rm old}$, by
creating the corresponding sub-clouds.
Due to the large extension of the clouds and the way $L_{\rm old}$ is computed,
a large number of sub-clouds are in fact unnecessary.
At the end, the hierarchy of clouds has to be ``cleaned up'', 
to keep only the relevant refinement regions, i.e. regions 
verifying criteria (A) or/and (B).
After being cleaned up, the new set of clouds and sub-clouds  is ready 
for successive deconvolutions at various levels, which 
are thus performed twice when convergence criterion is used: once to construct
the hierarchy of clouds and once again after this hierarchy has been
cleared out from unnecessary sub-clouds. 
\section{Explicit 2 dimensional cloud equations}
\label{sec:2deqs}
\begin{equation}
\frac {d \lambda_{11}}{dt}-4\,\phi_{11}
\lambda_{31}-4\,\phi_{21} \lambda_{41} = 0, 
\end{equation}
\begin{equation}
\frac {d \lambda_{22}}{dt} -4\,\phi_{12}
\lambda_{32}-4\,\phi_{22} \lambda_{42} = 0, \\
\end{equation}

\begin{equation}
\frac {d \lambda_{33}}{dt}+2\,\lambda_{13} = 0,
\end{equation}
\begin{equation}
\frac {d \lambda_{44}}{dt}+2\,\lambda_{24} = 0,
\end{equation}
\begin{equation}
\frac {d \lambda_{21}}{dt} -2\,\phi_{12}
\lambda_{31}-2\,\phi_{22} \lambda_{41}-2\,\phi_{11} \lambda_
{32}-2\,\phi_{21} \lambda_{42} = 0,
\end{equation}
\begin{equation}
\frac {d \lambda_{31}}{dt}+\lambda_{11}-2\,
\phi_{11} \lambda_{33}-2\,\phi_{21} \lambda_{43} = 0, 
\end{equation}
\begin{equation}
\frac {d \lambda_{32}}{dt}+\lambda_{12}-2\,
\phi_{12} \lambda_{33}-2\,\phi_{22} \lambda_{43} = 0, 
\end{equation}
\begin{equation}
\frac {d \lambda_{41}}{dt} + \lambda_{21}-2\,
\phi_{11}\lambda_{34}-2\,\phi_{21} \lambda_{44} = 0, 
\end{equation}
\begin{equation}
\frac {d \lambda_{42}}{dt} + \lambda_{22}-2\,
\phi_{12} \lambda_{34}-2\,\phi_{22}\lambda_{44} = 0,
\end{equation}
\begin{equation}
\frac {d \lambda_{43}}{dt}+\lambda_{23}+\lambda_{14} = 0. 
\end{equation}
%
%
%
\end{document}